\documentclass[a4paper,twoside,12pt,titlepage]{mybook}
\usepackage{a4}

\usepackage{amsmath,amsfonts,amssymb,amsthm}
\usepackage{url}
\usepackage{float}
\usepackage{graphicx}
\usepackage{subfigure}
\usepackage[sectionbib]{bibunits}
\usepackage{algorithmic}
\usepackage{setspace}
\usepackage{fancyhdr}
  \renewcommand{\chaptermark}[1]{\markboth{\chaptername \ \thechapter \ \ #1}{}}
  
\usepackage{adfatitlepage}
\usepackage{rotating}
\usepackage[none]{hyphenat}
\usepackage[intoc]{nomen}
\usepackage{pstricks}
\usepackage{array}
\usepackage{substr}
\usepackage[chapter]{algorithm}
\usepackage{url}
\usepackage{verbatim}
\usepackage{cite}
\usepackage{mathrsfs}
\usepackage{multirow}
\usepackage{bm}
\usepackage{lineno}
\usepackage{array}
\usepackage{color}
\usepackage{latexsym}
\usepackage{amsxtra}
\usepackage{acronym}
\usepackage{longtable}
\usepackage{nomencl}
\usepackage{titletoc}
\usepackage{stfloats}
\usepackage{hhline}
\usepackage{tabularx}
\usepackage{makecell}
\usepackage{enumerate}
\usepackage{epstopdf}
\usepackage{epsfig}
\usepackage{hyperref}
\usepackage{pifont}
\usepackage{pdfpages}
\newcommand{\mc}[1]{\mathcal{#1}}

\newcommand{\mbb}[1]{\mathbb{#1}}

\newcommand{\msf}[1]{\mathsf{#1}}

\newcommand{\SNR}{\msf{SNR}}

\newcommand{\defeq}{\triangleq}

\newcommand{\E}{\mathbb{E}}

\newcommand{\Z}{\mathbb{Z}}

\newcommand{\F}{\mathbb{F}}

\newcommand{\iid}{i.\@i.\@d.\ }

\DeclareMathOperator*{\argmax}{arg\,max}
\DeclareMathOperator*{\argmin}{arg\,min}

\newtheorem{remark}{Remark}[chapter]

\newtheorem{defi}{Definition}[chapter]

\newtheorem{theo}{Theorem}[chapter]
\newtheorem{lemm}{Lemma}[chapter]

\newtheorem{prop}{Proposition}[chapter]
\newtheorem{exam}{Example}[chapter]

\makeatletter
\usepackage{etoolbox}
\pretocmd{\tableofcontents}{%
  \if@openright\cleardoublepage\else\clearpage\fi
  \pdfbookmark[0]{\contentsname}{toc}%
}{}{}%
\makeatother

\newcommand*{\QEDA}{\hfill\ensuremath{\blacksquare}}

\makenomenclature

\usepackage{geometry}
\begin{document}
\begin{titlepage}
\begin{center}
\vspace*{1cm}
\Huge \hspace{-15mm}\textbf{Lattice Coding for}\\
\Huge\hspace{-15mm}\textbf{Downlink Multiuser Transmission}\\
\vspace{1.5cm}
\hspace{-15mm}\normalsize\textbf{Min Qiu}
\\
\vspace{2cm}
\normalsize
{\hspace{-15mm}A thesis submitted to the Graduate Research School of\\
\hspace{-15mm}The University of New South Wales\\
\hspace{-15mm}in partial fulfillment of the requirements for the degree of\\
\text{ \ }\\
\hspace{-15mm}\textbf{Doctor of Philosophy}}\\
\vspace{2cm}
\hspace{-15mm}\includegraphics[width=0.4\textwidth]{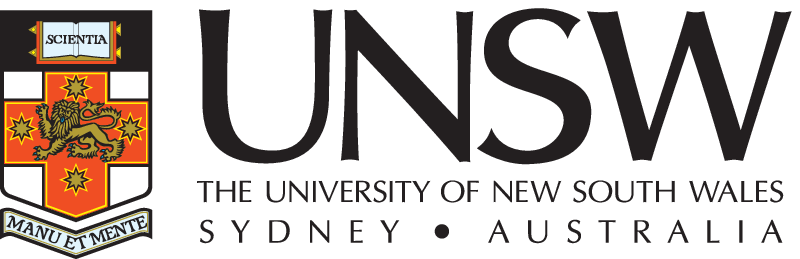}\\
\vspace{2cm}
\textbf{\hspace{-15mm}School of Electrical Engineering and Telecommunications\\
\hspace{-15mm}Faculty of Engineering\\
\hspace{-15mm}The University of New South Wales}\\
\vspace{2cm}
\hspace{-15mm}{February 2019}
\end{center}
\end{titlepage}

\frontmatter
\onehalfspacing
\pagestyle{empty}

\section*{Copyright Statement}

I hereby grant The University of New South Wales or its agents the right
to archive and to make available my thesis or dissertation in whole or part
in the University libraries in all forms of media, now or hereafter known,
subject to the provisions of the Copyright Act 1968. I retain all proprietary
rights, such as patent rights. I also retain the right to use in future works
(such as articles or books) all or part of this thesis or dissertation.

I also authorise University Microfilms to use the abstract of my thesis
in Dissertation Abstract International (this is applicable to doctoral
thesis only).

I have either used no substantial portions of copyright material in my thesis
or I have obtained permission to use copyright material; where permission
has not been granted I have applied/will apply for a partial restriction
of the digital copy of my thesis or dissertation.

\vspace*{1.5cm}
\noindent
\parbox{3in}{Signed \dotfill}

\vspace*{0.5cm}
\noindent
\parbox{3in}{Date  \dotfill}

\vspace{1cm}

\section*{Authenticity Statement}

I certify that the Library deposit digital copy is a direct equivalent of
the final officially approved version of my thesis. No emendation of content
has occurred and if there are any minor variations in formatting, they are
the result of the conversion to digital format.

\vspace*{1.5cm}
\noindent
\parbox{3in}{Signed \dotfill}

\vspace*{0.5cm}
\noindent
\parbox{3in}{Date \dotfill}

\clearpage{\thispagestyle{empty}\cleardoublepage}

\section*{Originality Statement}

I hereby declare that this submission is my own work and to the best
of my knowledge it contains no material previously published or written
by another person, or substantial portions of material which have been
accepted for the award of any other degree or diploma at UNSW or any
other educational institute, except where due acknowledgment is made
in the thesis.  Any contribution made to the research by others, with
whom I have worked at UNSW or elsewhere, is explicitly acknowledged in
the thesis. I also declare that the intellectual content of this thesis
is the product of my own work, except to the extent that assistance from
others in the project's design and conception or in style, presentation
and linguistic expression is acknowledged.

\vspace*{1.5cm}
\noindent
\parbox{3in}{Signed \dotfill}

\vspace*{0.5cm}
\noindent
\parbox{3in}{Date  \dotfill}

\clearpage{\thispagestyle{empty}\cleardoublepage}

\pagenumbering{roman}
\pagestyle{fancy}
\fancyhf{}
\setlength{\headheight}{15pt}
\fancyhead[LE,RO]{\footnotesize \textbf \thepage}
\chapter*{}

\vspace{+5cm}
\begin{center}
    \large \emph{Dedicated to my parents and wife.}
\end{center}

\doublespacing
\chapter*{Abstract}
\addcontentsline{toc}{chapter}{\protect\numberline{}{Abstract}}
In this thesis, we mainly investigate the lattice coding problem of the downlink communication between a base station and multiple users. The base station broadcasts a message containing each user's intended message. The capacity limit of such a system setting is already well-known while the design of practical coding and modulation schemes to approach the theoretical limit has not been fully studied and investigated in the literature. This thesis attempts to address this problem by providing a systematic design on lattice coding and modulation schemes for downlink multiuser communication. The main idea is to exploit the structure property of lattices to harness interference from downlink users. The research work of this thesis can be divided into five parts.

In the first part of our research, we focus on designing a class of lattice codes to approach the capacity of the classical point-to-point communication channel before we address the multiuser systems in the later chapters. A novel encoding structure of our multi-dimensional lattice codes is introduced and this approach is proved to allow our designed codes exhibit symmetry and permutation-invariance properties. By exploring these two properties, the degree distributions and the decoding thresholds of our codes are optimized by using one-dimensional extrinsic information transfer (EXIT) charts, which were mainly used for designing binary linear codes previously.

After the success in point-to-point communications, we move on to multiuser communications based on discrete and finite channel inputs. The second part of our research is to design practical lattice coding schemes for downlink non-orthogonal multiple access (NOMA) without successive interference cancellation (SIC) at the receiver. We first consider the case where the transmitter and receiver have full channel knowledge and propose a framework based on lattice partitions. The individual achievable rate of the proposed framework based on any lattice is derived and its gap to the multiuser capacity is upper bounded by a constant that is only related to the normalized second moment of the underlying lattice.

Next in the third part of our research, we investigate the slow fading scenario where the transmitter does not have full channel state information (CSI). For such a case, the generalization from our previous design with full CSI is non-trivial. Thus, we propose a new scheme for downlink NOMA without SIC by designing coding and modulation schemes based on statistical CSI while the power allocation factors are naturally induced by the design. The individual outage rate is analyzed and its gap to the multiuser outage capacity is proved to be upper bounded by a constant that is universal to the base station power, channel gain, and the number of downlink users.

In the fourth part of our research, we study the problem of downlink communication through block fading channels where the base station does not have CSI. Realizing that our previous two designs in this channel achieve no diversity gain, we propose a class of NOMA schemes by mapping all the users' messages to the same $n$-dimensional algebraic lattices constructed from algebraic number fields. The minimum product distance of the superimposed constellation is analyzed in detailed as it is closely related to the error performance. We show that, even without SIC at the receiver, our scheme can still offer full diversity to each user and provide high coding gain.

Finally, in the last part of our research, we conduct an additional work by designing error-correction codes for ultra-reliable applications such as fibre-optic communication systems and data-storage systems. For this work, we develop a class of product codes with high code rate and low error floor. The unique encoding structure allows the decoder to easily detect and correct more error patterns that contribute to the error floor. Moreover, an efficient post processing technique is proposed to enhance the decoding performance by further lowering the error floor. Theoretical analysis of the error pattern occurrence and the decoding performance is provided.

\doublespacing
\chapter*{Acknowledgments}

First of all, I would like to thank my supervisor, Professor Jinhong Yuan, who has provided me endless support and guidance from the beginning of my Ph.D. study. His deep understanding of the topics, his enthusiasm for the research and his dedication to his students have been a source of inspiration for me. In particular, he always provides me with constructive feedback and insightful suggestions on my work. Most importantly, he taught me how to become an independent researcher. It has truly been fortunate for me to pursue my Ph.D. degree under his supervision.

Second, I would like to thank my co-supervisor, Dr. Lei Yang, for his guidance, many fruitful discussions and technical support on channel coding knowledge and programming. I would never forget that he gave me countless valuable advice when I was struggling in my first year. I would like to thank Dr. Yixuan Xie, who has been supportive to me and provided me a variety of resources to study channel coding techniques. His valuable input on a number of problems which we discussed together, was really helpful in the early stage of my research study. I would also like to thank Dr. Derrick Wing Kwan Ng, for his suggestions on my research and journal writing.

Lots of thanks to Professor Yu-Chih Huang, who had been my host supervisor when I was a visiting Ph.D. student at the National Taipei University. I am very impressed by his immense knowledge on lattices and information theory as well as his endless motivation. I have learned tremendously from his continuous advice and guidance. I am really appreciate his support on both my study and life in Taiwan. We have built long-term research collaboration and we are really enjoy working together.

Many thanks to my colleagues in the wireless communication group of the University of New South Wales. Especially, I want to thank Zhuo Sun, Zhiqiang Wei, Peng Kang, Bryan Liu, Xiaowei Wu, Ruide Li, Yihuan Liao and Shuangyang Li. We have studied together, shared our happiness and frustration together, and helped each other. You really make the Ph.D. journey funny and interesting.

Finally, my deepest appreciation goes to my beloved family for their unconditional love and support. My parents have always been there to help when I had hard time in my study or in my life. I especially would like to thank my beloved wife, Jing Tao, for her unwavering love, encouragement and support. Without her love, all my achievements would be meaningless.

\singlespacing

\chapter*{List of Publications} \label{listofpublications}
\addcontentsline{toc}{chapter}{\protect\numberline{}{List of Publications}}
%

\ifpdf
    \graphicspath{{1_introduction/figures/PNG/}{1_introduction/figures/PDF/}{1_introduction/figures/}}
\else
    \graphicspath{{1_introduction/figures/EPS/}{1_introduction/figures/}}
\fi

\noindent{\large\textbf{Journal Articles:}} \vspace{0.1in}
\begin{enumerate}
\item \textbf{M. Qiu}, L.~Yang, Y. Xie and J.~Yuan, ``On the design of multi-dimensional irregular repeat-accumulate lattice codes,'' {\em IEEE Trans. Commun.}, vol.~66, no.~2, pp. 478--492, Feb. 2018.
\item \textbf{M. Qiu}, Y.-C. Huang, S.-L. Shieh, and J. Yuan, ``A lattice-partition framework of downlink non-orthogonal multiple access without SIC,''
   {\em IEEE Trans. Commun.}, vol.~66, no.~6, pp. 2532--2546, Jun. 2018.
\item \textbf{M. Qiu}, L. Yang, Y. Xie, and J. Yuan, ``Terminated staircase codes for NAND flash memories,''{\em IEEE Trans. Commun.}, vol.~66, no.~12, pp. 5861-5875, Dec. 2018.
\item \textbf{M. Qiu}, Y.-C. Huang, J. Yuan and C.-L. Wang, ``Lattice-partition-based downlink non-orthogonal multiple access without SIC for slow fading channels,'' {\em IEEE Trans. Commun.}, vol.~67, no.~2, pp. 1166-1181, Feb. 2019.
\item \textbf{M. Qiu}, Y.-C. Huang, and J. Yuan, ``Downlink non-orthogonal multiple access without SIC for block fading channels: An algebraic rotation approach,'' {\em IEEE Trans. Wireless Commun.}, accepted, Jun. 2019.

\end{enumerate}
\vspace{0.1in} \noindent{\large\textbf{Conference Articles:}}
\vspace{0.1in}
\begin{enumerate}
\item \textbf{M.~Qiu}, L. Yang, and J. Yuan, ``Irregular repeat-accumulate lattice network codes for two-way relay channels'' in {\em Proc. IEEE Global Commun. Conf. (GLOBECOM)}, Washington, D.C., Dec. 2016, pp. 1-6.
\item \textbf{M Qiu}, L. Yang, Y. Xie, and J. Yuan, ``On the design of multi-dimensional irregular repeat-accumulate lattice codes,'' in { \em Proc. IEEE Symp. Inf. Theory (ISIT)}, Aachen, Jul. 2017, pp. 2598-2602.
\item \textbf{M Qiu}, Y.-C. Huang, S.-L. Shieh, and J. Yuan, “A lattice-partition framework of downlink non-orthogonal multiple access without SIC,'' in { \em Proc. IEEE Global Commun. Conf. (GLOBECOM)}, Singapore, Dec. 2017, pp. 1-6.
\item \textbf{M Qiu}, Y.-C. Huang, J. Yuan and C.-L. Wang, ``Downlink lattice-partition- based non-orthogonal multiple access without SIC for slow fading channels,'' in { \em Proc. IEEE Global Commun. Conf. (GLOBECOM)}, Abu Dhabi, Dec. 2018, pp. 1-6.
    \item \textbf{M Qiu}, Y.-C. Huang, and J. Yuan, ``Downlink NOMA without SIC for fast fading channels: Lattice partitions with algebraic rotations,'' in { \em Proc. IEEE Intern. Commun. Conf. (ICC)}, May 2019, pp. 1-6.
\end{enumerate}

\chapter*{Abbreviations} \label{abbreviations}
\addcontentsline{toc}{chapter}{\protect\numberline{}{Abbreviations}}
\markboth{ABBREVIATIONS}{}

\begin{longtable}[t]{ll}
\textbf{3GPP} \quad\quad&\mbox{3rd Generation Partnership Project} \vspace{0.1in}\\
\textbf{4G} \quad\quad&\mbox{The Fourth-generation} \vspace{0.1in}\\
\textbf{5G} \quad\quad&\mbox{The Fifth-generation} \vspace{0.1in}\\
\textbf{APP} \quad\quad&\mbox{A priori probability} \vspace{0.1in}\\
\textbf{AWGN} \quad\quad&\mbox{Additive white Gaussian noise} \vspace{0.1in}\\
\textbf{BCH} \quad\quad&\mbox{Bose-Chaudhuri-Hocquengham} \vspace{0.1in}\\
\textbf{BDD} \quad\quad&\mbox{Bounded distance decoding} \vspace{0.1in}\\
\textbf{BEC} \quad\quad&\mbox{Binary erasure channel} \vspace{0.1in}\\
\textbf{BER} \quad\quad&\mbox{Bit error rate} \vspace{0.1in}\\
\textbf{BI-AWGN} \quad\quad&\mbox{Binary input additive white Gaussian noise} \vspace{0.1in}\\
\textbf{BS} \quad\quad&\mbox{Base station} \vspace{0.1in}\\
\textbf{BSC} \quad\quad&\mbox{Binary symmetric channel} \vspace{0.1in}\\
\textbf{CDMA} \quad\quad&\mbox{Code-division multiple access} \vspace{0.1in}\\
\textbf{CLC} \quad\quad&\mbox{Convolutional lattice codes} \vspace{0.1in}\\
\textbf{CN} \quad\quad&\mbox{Check node}\vspace{0.1in}\\
\textbf{CRC} \quad\quad&\mbox{Cyclic redundant check} \vspace{0.1in}\\
\textbf{CSI} \quad\quad&\mbox{Channel state information} \vspace{0.1in}\\
\textbf{dB} \quad\quad&\mbox{Decibel} \vspace{0.1in}\\
\textbf{ECC} \quad\quad&\mbox{Error-correction code} \vspace{0.1in}\\
\textbf{eMBB} \quad\quad&\mbox{Enhanced mobile broadband} \vspace{0.1in}\\
\textbf{EXIT} \quad\quad&\mbox{Extrinsic information transfer} \vspace{0.1in}\\
\textbf{GF} \quad\quad&\mbox{Galois field} \vspace{0.1in}\\
\textbf{GLD} \quad\quad&\mbox{General low-density} \vspace{0.1in}\\
\textbf{FD} \quad\quad&\mbox{Full-duplex} \vspace{0.1in}\\
\textbf{FDMA} \quad\quad&\mbox{Frequency-division multiple access} \vspace{0.1in}\\
\textbf{FFT} \quad\quad&\mbox{Fast Fourier Transform} \vspace{0.1in}\\
\textbf{i.i.d.} \quad\quad&\mbox{Independent and identically distributed} \vspace{0.1in}\\
\textbf{ISI} \quad\quad&\mbox{Inter-symbol interference} \vspace{0.1in}\\
\textbf{I/Q}  \quad\quad&\mbox{In-phase/Quadrature} \vspace{0.1in}\\
\textbf{IoT}  \quad\quad&\mbox{Internet-of-things} \vspace{0.1in}\\
\textbf{IRA} \quad\quad&\mbox{Irregular repeat-accumulate} \vspace{0.1in}\\
\textbf{LDLC} \quad\quad&\mbox{Low-density lattice codes} \vspace{0.1in}\\
\textbf{LDPC} \quad\quad&\mbox{Low-density parity-check} \vspace{0.1in}\\
\textbf{LDA} \quad\quad&\mbox{Low-density Construction A} \vspace{0.1in}\\
\textbf{LLR} \quad\quad&\mbox{Log-likelihood ratio} \vspace{0.1in}\\
\textbf{LTE} \quad\quad&\mbox{Long-term evolution} \vspace{0.1in}\\
\textbf{MAP} \quad\quad&\mbox{Maximum a posterior} \vspace{0.1in}\\
\textbf{mMTC} \quad\quad&\mbox{Massive machine type communications} \vspace{0.1in}\\
\textbf{mmWave} \quad\quad&\mbox{Millimeter wave} \vspace{0.1in}\\
\textbf{MIMO} \quad\quad&\mbox{Multiple-input multiple-output} \vspace{0.1in}\\
\textbf{MMSE}  \quad\quad&\mbox{Minimum mean square error} \vspace{0.1in}\\
\textbf{ML}  \quad\quad&\mbox{Maximum-likelihood} \vspace{0.1in}\\
\textbf{MRC} \quad\quad&\mbox{Maximal-ratio combining} \vspace{0.1in}\\
\textbf{MRT} \quad\quad&\mbox{Maximal-ratio transmission} \vspace{0.1in}\\
\textbf{MSE}  \quad\quad&\mbox{Mean squared error} \vspace{0.1in}\\
\textbf{NMSE}  \quad\quad&\mbox{Normalized mean squared error} \vspace{0.1in}\\
\textbf{NOMA} \quad\quad&\mbox{Non-orthogonal multiple access} \vspace{0.1in}\\
\textbf{OFDM} \quad\quad&\mbox{Orthogonal frequency-division multiplexing} \vspace{0.1in}\\
\textbf{OFDMA} \quad\quad&\mbox{Orthogonal frequency-division multiple access} \vspace{0.1in}\\
\textbf{OMA} \quad\quad&\mbox{Orthogonal multiple access} \vspace{0.1in}\\
\textbf{OSTBC} \quad\quad&\mbox{Orthogonal space-time block code} \vspace{0.1in}\\
\textbf{PAM} \quad\quad&\mbox{Pulse amplitude modulation} \vspace{0.1in}\\
\textbf{PDF} \quad\quad&\mbox{Probability density function} \vspace{0.1in}\\
\textbf{PER} \quad\quad&\mbox{Page error rate} \vspace{0.1in}\\
\textbf{QAM} \quad\quad&\mbox{Quadrature amplitude modulation} \vspace{0.1in}\\
\textbf{QoS} \quad\quad&\mbox{Quality-of-service} \vspace{0.1in}\\
\textbf{RA} \quad\quad&\mbox{Repeat-accumulate} \vspace{0.1in}\\
\textbf{RS} \quad\quad&\mbox{Reed-Solomon} \vspace{0.1in}\\
\textbf{SDN} \quad\quad&\mbox{Software defined networks}\vspace{0.1in}\\
\textbf{SER} \quad\quad&\mbox{Symbol error rate} \vspace{0.1in}\\
\textbf{SIC} \quad\quad&\mbox{Successive interference cancellation}\vspace{0.1in}\\
\textbf{SNR}  \quad\quad&\mbox{Signal-to-noise ratio} \vspace{0.1in}\\
\textbf{SINR}  \quad\quad&\mbox{Signal-to-interference-plus-noise ratio} \vspace{0.1in}\\
\textbf{SPA}  \quad\quad&\mbox{Sum-product algorithm} \vspace{0.1in}\\
\textbf{SSD}  \quad\quad&\mbox{solid-state storage device} \vspace{0.1in}\\
\textbf{TDMA} \quad\quad&\mbox{Time-division multiple access} \vspace{0.1in}\\
\textbf{uRLLC} \quad\quad&\mbox{Ultra reliable and low latency communications}\vspace{0.1in}\\
\textbf{VN} \quad\quad&\mbox{Variable node}\vspace{0.1in}\\
\end{longtable}

\chapter*{List of Notations} \label{listofnotations}
\addcontentsline{toc}{chapter}{\protect\numberline{}{List of Notations}}
\markboth{LIST OF NOTATIONS}{}
Scalars, vectors and matrices are written in italic, boldface lower-case and upper-case letters, respectively, e.g., $x$, $\mathbf{x}$ and $\mathbf{X}$. Random variables are written in uppercase Sans Serif font e.g., $\msf{X}$.
\begin{longtable}[t]{ll}
$\mathbf{X}^T$ \quad\quad&\mbox{Transpose of $\mathbf{X}$} \vspace{0.1in}\\
$\mathbf{X}^H$ \quad\quad&\mbox{Conjugate transpose of $\mathbf{X}$ } \vspace{0.1in}\\
$\mathbf{X}^{-1}$ \quad\quad&\mbox{Inverse of $\mathbf{X}$ } \vspace{0.1in}\\
$\mathbf{X}_{i,j}$ \quad\quad&\mbox{The element in the row $i$ and the column $j$ of $\mathbf{X}$ } \vspace{0.1in}\\
$\det\left(\mathbf{X}\right)$ \quad\quad&\mbox{Determinant of $\mathbf{X}$} \vspace{0.1in}\\
$\mbox{tr}\left(\mathbf{X}\right)$ \quad\quad&\mbox{Trace of $\mathbf{X}$} \vspace{0.1in}\\
$|x|$ \quad\quad&\mbox{Absolute value (modulus) of the complex scalar $x$} \vspace{0.1in}\\
$\|\mathbf{x}\|$ \quad\quad&\mbox{The Euclidean norm of a vector $\mathbf{x}$} \vspace{0.1in}\\
$\|\mathbf{X}\|_F$ \quad\quad&\mbox{The Frobenius norm of a matrix $\mathbf{X}$} \vspace{0.1in}\\
$\Lambda$ \quad\quad&\mbox{A lattice} \vspace{0.1in}\\
$\mathbb{R}$ \quad\quad&\mbox{The field of real number} \vspace{0.1in}\\
$\mathbb{C}$ \quad\quad&\mbox{The field of complex number} \vspace{0.1in}\\
$\mathbb{R}^n$ \quad\quad&\mbox{The Euclidean space} \vspace{0.1in}\\
$\mathbb{F}_q$ \quad\quad&\mbox{The finite field of size $q$} \vspace{0.1in}\\
$\mathbb{Z}$ \quad\quad&\mbox{The ring of integers} \vspace{0.1in}\\
$\mathbb{Z}^+$ \quad\quad&\mbox{The ring of positive integers} \vspace{0.1in}\\
$\mathbb{Z}[i]$ \quad\quad&\mbox{The ring of Gaussian integers} \vspace{0.1in}\\
$\mathbb{Z}[\omega]$ \quad\quad&\mbox{The ring of Eisenstein integers} \vspace{0.1in}\\
$\mathbb{H}$ \quad\quad&\mbox{The ring of Hurwitz integers} \vspace{0.1in}\\
$\mathbb{P}\{E\}$ \quad\quad&\mbox{The probability of event $E$ occurs} \vspace{0.1in}\\
$p_{\msf{X}}(x),p(x)$ \quad\quad&\mbox{Probability density function of the random variable $\msf{X}$} \vspace{0.1in}\\
$p_{\msf{X}|\msf{Y}}(x|y),p(x|y)$ \quad\quad&\mbox{Conditional distribution of $\msf{X}$ given $\msf{Y}$} \vspace{0.1in}\\
$p_{\msf{X},\msf{Y}}(x,y),p(x,y)$ \quad\quad&\mbox{Joint distribution of $\msf{X}$ and $\msf{Y}$} \vspace{0.1in}\\
$\left\lceil x\right\rceil^+$ \quad\quad&\mbox{Rounds a real number $x$ to the nearest integer greater than or} \vspace{0.1in}\\
\quad\quad&\mbox{equal to $x$ if $x\geq 0 $ or rounds to 0 for all $x < 0 $} \vspace{0.1in}\\
$\Re(z)$ \quad\quad&\mbox{The real part of a complex number $z$} \vspace{0.1in}\\
$\Im(z)$ \quad\quad&\mbox{The imaginary part of a complex number $z$} \vspace{0.1in}\\
$|\mathcal{S}|$ \quad\quad&\mbox{The cardinality of a set $\mathcal{S}$} \vspace{0.1in}\\
$\mathbf{0}$ \quad\quad&\mbox{The all-zero vector} \vspace{0.1in}\\
$\mathbf{I}_{\mathrm{N}}$ \quad\quad&\mbox{$N$ dimension identity matrix} \vspace{0.1in}\\
$\mathrm{\mathbb{E}}\left[\cdot\right]$ \quad\quad&\mbox{Statistical expectation} \vspace{0.1in}\\
$\text{Vol}(\mathcal{R})$ \quad\quad&\mbox{The volume of a bounded region $\mathcal{R}$ in the Euclidean space} \vspace{0.1in}\\
$\mathcal{B}(r)$ \quad\quad&\mbox{The $n$-dimensional sphere centered at the origin with radius $r$:} \vspace{0.1in}\\
\quad\quad&\mbox{$\mathcal{B}(r) \triangleq \{\mathbf{x} \in \mathbb{R}^n:\|\mathbf{x}\| \leq r \}$} \vspace{0.1in}\\
$\mathcal{N}(\mu,\sigma^2)$ \quad\quad&\mbox{Real Gaussian random variable with mean $\mu$ and variance $\sigma^2$} \vspace{0.1in}\\
$\mathcal{CN}(0,\sigma^2)$ \quad\quad&\mbox{Circularly symmetric complex Gaussian random variable:}\vspace{0.1in} \\
\quad\quad&\mbox{the real and imaginary parts are i.i.d. $\mathcal{N}(0,\sigma^2/2)$} \vspace{0.1in}\\
$\mathcal{CN}(0,\mathbf{K})$ \quad\quad&\mbox{Circularly symmetric Gaussian random vector with}\vspace{0.1in} \\
\quad\quad&\mbox{mean zero and covariance matrix $\mathbf{K}$} \vspace{0.1in}\\
$\ln(\cdot)$ \quad\quad&\mbox{Natural logarithm} \vspace{0.1in}\\
$\log_a(\cdot)$ \quad\quad&\mbox{Logarithm in base $a$} \vspace{0.1in}\\
$\mathrm{diag}\left\{\bm{a}\right\}$ \quad\quad&\mbox{A diagonal matrix with the entries of $\bm{a}$ on its diagonal} \vspace{0.1in}\\
$\lim$ \quad\quad&\mbox{Limit} \vspace{0.1in}\\
$\max\left\{\cdot\right\}$ \quad\quad&\mbox{Maximization} \vspace{0.1in}\\
$\min\left\{\cdot\right\}$ \quad\quad&\mbox{Minimization} \vspace{0.1in}\\
$e^x,\exp(x)$ \quad\quad&\mbox{Natural exponential function} \vspace{0.1in}\\
$\tanh$ \quad\quad&\mbox{Hyperbolic tangent function} \vspace{0.1in}\\
$d_H$ \quad\quad&\mbox{Hamming distance} \vspace{0.1in}\\
$w_H$ \quad\quad&\mbox{Hamming weight} \vspace{0.1in}\\
$d_{E,\min}$ \quad\quad&\mbox{Minimum Euclidean distance} \vspace{0.1in}\\
$d_{P,\min}$ \quad\quad&\mbox{Minimum Product distance} \vspace{0.1in}\\
$\oplus$ \quad\quad&\mbox{Modulo lattice addition} \vspace{0.1in}\\
$\ominus$ \quad\quad&\mbox{Modulo lattice subtraction} \vspace{0.1in}\\
$I(\msf{X};\msf{Y})$ \quad\quad&\mbox{The mutual information between $\msf{X}$ and $\msf{X}$} \vspace{0.1in}\\
$h(\msf{X})$ \quad\quad&\mbox{The entropy of a continous random variable $\msf{X}$} \vspace{0.1in}\\
$H(\msf{X})$ \quad\quad&\mbox{The entropy of a discrete random variable $\msf{X}$} \vspace{0.1in}\\
$\mathcal{S}_1 \setminus \mathcal{S}_2$ \quad\quad&\mbox{Obtain the elements that only belong to set $\mathcal{S}_1$} \vspace{0.1in}\\
$O(N)$ \quad\quad&\mbox{The computational complexity is the order of $N$ operations} \vspace{0.1in}\\
\end{longtable}

\fancyhead[CE]{\footnotesize \leftmark}
\fancyhead[CO]{\footnotesize \rightmark}

\renewcommand{\chaptermark}[1]{%
\markboth{\MakeUppercase{%
\thechapter.%
\ #1}}{}}

\tableofcontents

\listoffigures


\listofalgorithms
\addcontentsline{toc}{chapter}{\protect\numberline{}{List of Algorithms}}



\mainmatter
\doublespacing

\chapter{Introduction}\label{C1:chapter1}
In this chapter, we first introduce the motivation of the research for this thesis before summarizing the principal research problems and the main contributions of the thesis.

\section{Overview of 5G}
With the increasing demands of network access and the explosive growth of smart devices connected to the cellular networks, higher data rate, better quality-of-service (QoS) and more conductivities are required to support these needs. In particular, it is expected that the number of connected devices would reach to about 31.4 billion while the amount of mobile data traffic would rise to 107 exabytes (1 exabytes $= 10^{18}$ bytes) per month in 2023 \cite{EricssonJun18Report}. However, the current fourth generation (4G) cellular network systems have reached their limits and cannot satisfy the future requirements.

The fifth generation (5G) wireless systems are commonly regarded as the enormous breakthrough innovations to the current 4G systems, and thus will revolutionize the way of communication. Most notably, the 5G systems will have three main new features: enhanced mobile broadband (eMBB) offering much higher data rates for data-intensive applications across a wider mobile coverage area, ultra-reliable low latency communications (uRLLC) providing extremely highly reliable communications for strictly latency-sensitive services, and massive machine type communications (mMTC) providing massive connectivity to a massive number of Internet of Things (IoT) devices in a small area \cite{ITUR,TR38.802}. The research and development of 5G technologies have drawn increasingly interests from both academia and industry \cite{DerrickNG17,6824752}. On the path to 5G, a number of techniques such as millimeter wave (mmWave), massive multiple-input multiple-output (MIMO), full-duplex (FD) relaying, software defined networks (SDN) and etc. have been identified as the key technologies by researchers \cite{6736746}. Apart from the aforementioned techniques, new multiple access techniques are also required to support the increasing number of mobile users and to offer better QoS as well as higher spectral efficiency. As such, this thesis aims to give a contribution to new multiple access and coding techniques, addressing the communication problems over multiuser channels.

\section{Motivation}
\subsection{Designing New Channel Coding Schemes}
Most communication theories are built from the basic point-to-point communication where the backbone is channel coding. Channel coding is an essential part to provide reliability to all sorts of communications by protecting the transmitted messages from transmission errors due to noise and interference. The history of channel coding started with Claude Shannon's landmark paper \cite{6773024} in 1948, where the channel capacity was established by means of communication at the highest possible rate with arbitrarily small errors. However, during that time, the channel capacity was thought to be only achieved by using random Gaussian coding, which could arguably be infeasible for practical wireless systems. After six decades of efforts made by many researchers, there are a number of well-known practical coding schemes such as turbo codes \cite{397441,Vucetic:2000:TCP:352869}, low-density parity-check (LDPC) codes \cite{748992,1057683} and polar codes \cite{5075875} that can be easily designed to approach the point-to-point channel capacity. All of these coding schemes have now become parts of the modern communication standards.

Despite the progress being made by those capacity-approaching codes, there are still many important and unsolved problems in coding \cite{7265214}. First, all the codes adopted in current communication standards are binary codes and the most commonly used modulation are quadrature amplitude modulation (QAM) or pulse amplitude modulation (PAM). Although some of the codes have been shown to approach or achieve the capacity of the binary erasure channel (BEC), binary symmetric channel (BSC) and binary additive white Gaussian noise (BI-AWGN) channel, it is difficult for them to approach the unconstrained Shannon limit for which the capacity is not restricted to any signal constellation. Specifically, for high order modulations that are directly coded and mapped into by binary codes, the information loss in demodulation process is unavoidable. The loss can be compensated by using multi-level coding scheme \cite{771140}, however, with high computational complexity and large processing delay. For non-binary codes, there is no loss in the demodulation process since each non-binary coding alphabet is directly mapped into a modulation symbol. Therefore, non-binary codes generally outperform binary codes in the same spectral efficiency. That being said, designing capacity achieving non-binary codes are more challenging compared to binary codes. Second, coding over QAM and PAM modulations exhibit a shaping loss of 1.53 dB \cite{4282117}, corresponding to 0.25 bits/s/Hz/dim loss in the transmission rate. This loss would become more significant when the data rate is higher. Thus, efficient shaping is necessary in order to further improve the spectral efficiency without increasing the channel bandwidth.

It is known that lattices are good for the purposes of both channel coding and shaping \cite{1512416}. Lattice codes, built from lattices, are the Euclidean counterpart of binary linear codes. Lattices are infinite and discrete sets of points. They processes with many nice properties and elegant mathematical structures \cite{conway1999sphere}. The theory behind lattices was born and developed long before they being applied to the realms of communication and signal processing. The idea of employing lattices in channel coding is due to the fact that many nice properties of the lattices can be carried over to solve practical engineering problems. Most notably, it has been proved in \cite{1337105} that there exists a sequence of lattice codes that can achieve the capacity of AWGN channels. This encouraging result illustrates that the ultimate Shannon limit can be achieved with structure codes as opposed to random Gaussian codes. That said, the complexities of the optimal shaping and decoding algorithm therein are formidable and the problem of construct practical capacity-achieving lattice codes are still challenging. Motivated by the success of applying lattices in channel coding, the first part of this thesis is to design practical lattice codes that can approach the unconstrained AWGN channel capacity with moderate decoding complexity.

\subsection{Designing New Multiple Access Schemes}
Different from the point-to-point communication, multiple access techniques are designed to share the channel resources among multiple users. Looking back at the history of mobile network \cite{Goldsmith:2005:WC:993515}, multiple access has changed significantly from frequency-division multiple access (FDMA) in the first generation (1G) network, time-division multiple access (TDMA) in the second generation (2G) network, code-division multiple access (CDMA) in the third generation (3G) network and orthogonal frequency division multiple access (OFDMA) in the current 4G network. These kinds of techniques are all known as orthogonal multiple access (OMA) because the channel resources are divided into orthogonal blocks based on frequency/time/codeword domain and each user is served in one orthogonal resource block exclusively. The benefit of doing so is that the inter-user interference can be avoided. As a result, the multiuser communication problem can be converted into parallel point-to-point communication problems where single-user coding/decoding techniques suffice. However, OMA has low spectral efficiency and cannot reliably operate at the multiuser capacity region in general \cite{tse_book}. Another major drawback of OMA is that the number of served users is strictly limited by the number of orthogonal resources. Thus, it is difficult for conventional OMA to meet the future demands with the explosive growth of mobile users and traffic.

Recently, non-orthogonal multiple access (NOMA) has been proposed \cite{6692652} and is expected to provide higher spectral efficiency, better user fairness, and allows the base station to serve more users \cite{Dai15,DerrickNG17,wei2017fairness}. Unlike OMA, the key idea of NOMA is to allow multiple users to share a given channel resource slot e.g., time/frequency/code and use advanced multiuser detection technique at the receiver to distinguish different users. In addition, it is also possible that NOMA can be integrated with existing multiple access techniques \cite{7510794,7503854,7999275}. For example, a NOMA scheme can be implemented on top of OFDMA where a subcarrier can be allocated to more than one users such that the non-orthogonal transmission occurs in a subcarrier. Although the application of NOMA in cellular networks is relatively new, the principle of NOMA has already been studied in information theory for a long time. Typically, a single-antenna downlink NOMA can be regarded as a case of scalar Gaussian broadcast channel where the transmitter performs superposition coding and the receiver performs successive interference cancellation (SIC) \cite{Cover:2006:EIT:1146355}. In this way, NOMA is capable to operate at the multiuser capacity region. Although the theoretical performance limits of NOMA are well understood, much is still lacking when it comes to practical schemes that are able to approach these limits. In particular, the problem of designing practical downlink NOMA schemes with finite and discrete inputs has received much less attention. Moreover, SIC can introduce a large decoding burden, a long latency and error propagation to the receivers of mobile devices. The problem would be even more pronounced when the number of users participating in the transmission is large. Motivated by the advantages of NOMA and after realizing many successful applications of lattice coding in different communication scenarios \cite{6034734,5605356,Natarajan15}, the main focus of this thesis is dedicated to address the aforementioned limitations to design practical downlink NOMA schemes suitable for different wireless communication scenarios and with performance analysis.

\subsection{Designing New Coding Schemes With Ultra-High Reliability}
For applications that have high reliability and tight delay-constraint requirements, coding with low error floor is often required in order to reduce the number of retransmissions. Typical examples for this scenario can be backhaul communications with optical fibers or data storage systems. For these systems, the error probability requirement is much lower than that for common wireless communication scenario. In general, these systems are required to provide bit error rates (BERs) below $10^{-15}$. Moreover, these systems are also required to support higher data rate and with lower latency. Compared to the channel codes used in long-term evolution (LTE) systems, the rates of the underlying channel codes in theses systems are usually very high, e.g., about 0.9. These requirements
have led to a renewed interest in designing new coding schemes suitable for high data rate, low latency and high reliability applications.

It is well known that LDPC codes and turbo codes are capacity-approaching codes. As such, these codes can be considered for the use in future wireless communications, optical communications and storage as they are able to provide much stronger error-correction capability than the conventional linear block codes, e.g., Hamming codes and  Bose-Chaudhuri-Hocquengham (BCH) codes. However, these codes with irregular degree distributions often exhibit high error floor due to poor minimum codeword distances. On the other hand, LDPC codes with regular degree distributions have better error floor performance but with degraded decoding performance compared to irregular LDPC codes. Recently, spatially coupled (SC) codes such as SC-LDPC codes \cite{5571910,5695130} and SC-turbo codes \cite{8002601,8368318} have been proposed and shown to have remarkable performance in terms of better error floor than their uncoupled counterparts while promising the close-to-capacity performance. Due to the nature of spatial coupling, these codes can be efficiently decoded by using a window decoder, i.e., to decode a portion of coupled codewords by using the component code decoder. In such a way, the latency caused by decoding can be lower than that of decoding a whole codeword block. Despite their capacity-approaching performance, the hardware complexity for implementing these error control systems could limit their applications. First, these codes rely on iterative soft-decision decoding to attain the near-capacity performance. The internal data flow of the iterative decoder, i.e., the rate of routing/storing messages, can exceed the maximum data rate supported by the optical fibre systems. For example, a standard sum-product algorithm can have a 48 Tb/s data flow while an optical transport network can only support 100 Gb/s \cite{Smith12}. Second, soft-decision channel output, i.e., log-likelihood ratios (LLRs), is crucial for the iterative decoder. For solid-state storage devices (SSDs) based on NAND flash memories, the channel
representing NAND flash memory is unique in that only hard-decision channel outputs are available. Soft-decision results
can be indirectly acquired by reading hard-decision outputs
multiple times with different sensing reference voltages. The acquisition of soft channel output is a
costly operation in terms of power consumption and processing latency.

Another family of codes known as product codes have also been considered for storage and optical systems. In particular, product codes with algebraic block codes as component codes are more attractive as they can be encoded and decoded with low-complexity algorithms. Most importantly, these codes with iterative hard-decision decoding can be designed to perform well over the BSC, which is the channel model of many fibre-optic communication systems and storage systems \cite{Smith12,Cho14}. However, the task of designing high rate codes with low error floor is still challenging. Furthermore, it is unclear whether the error floor of general product codes can be analytically and precisely estimated. Motivated by the advantages of the product codes, an additional aim of this thesis is to design a class of product codes along with a new decoder and efficient post processing techniques to offer enhanced error performance and provide a method to compute the error floor of the designed codes \cite{8425763}.

\section{Literature review}
In this section, the related works of this thesis surrounding lattice coding, non-orthogonal multiple access and coding for ultra-reliable applications are discussed and reviewed.

\subsection{Lattice Codes}\label{sec:lattice_code_review}
Extensive research has been conducted on the analytical proving of the capacity-
achieving properties of lattice codes from the information-theoretic perspective. The central line of development in the application of lattices for the AWGN channel originated in the work \cite{29612} and was partially
corrected in \cite{259668}. It was proved in these works that lattice codes can attain the capacity of the AWGN channel under the maximum-likelihood
(ML) decoding, with shaping determined by ``thin'' spherical shells. This peculiar shaping region actually makes the code lose most of its lattice structure and look similar to a random code on a sphere. Moreover, the decision regions of ML decoding are not fundamental regions of the lattices and thus are unbounded. In contrast, lattice decoding amounts to finding the closest lattice point, ignoring the decision boundary of the code. Such an unconstrained search preserves the lattice symmetry in the decoding process and saves complexity. When restricted to lattice decoding, however, it was shown in \cite{641543,651040} that lattice codes can transmit reliably only at rates up to $\frac{1}{2}\log_2(\SNR)$. The loss of ``one'' in this rate formula means significant performance degradation in the low SNR regime. It has been finally proved in \cite{1337105} that the full capacity of the Gaussian channel $\frac{1}{2}\log_2(1+\SNR)$ can be achieved by lattice encoding and decoding. Although the theoretical problem of whether structured codes can achieve capacity was solved, the design of practical lattice codes with close-to-capacity performance is still challenging.

In general, there are two main approaches to construct lattice codes. The first one is to construct lattice codes directly in the Euclidean space. There are two well-known examples: low-density lattice codes (LDLC) \cite{4475389} and convolutional lattice codes (CLC) \cite{5961819}. Another approach is to adapt modern capacity approaching error correction codes to construct lattices, i.e., construct lattices from convolutional codes \cite{6516165,6582523}, LDPC codes \cite{1705007,Tunali15,Boutros16,8122043} and from polar codes \cite{8492454}. Their constructions involve some well-known methods such as Construction $A$ \cite{conway1999sphere} (constructing lattices based on a linear code), Construction $D$ \cite{conway1999sphere} (constructing lattices based on the generator matrices of a series of nested linear codes), and Construction $D'$ \cite{conway1999sphere} (constructing lattices based on the parity check matrices of a series of nested linear codes). These methods allow one to construct lattice codes not only with good error performance inherited from capacity-achieving linear codes, but also having relatively lower construction complexity compared with LDLCs and CLCs. To sum up, most of the aforementioned designs have been shown to approach the Poltyrev limit \cite{312163} (i.e., the channel capacity without either power limit or restrictions on signal constellations) within 1 dB when the codeword length is long enough. In addition, all of these lattices can be decoded with efficient iterative decoding algorithms.

However, for LDLCs, in order to attain the best possible decoding performance, the decoder would have to take the whole probability density functions (PDFs) for processing. This would require a significant amount of memory. As reported in \cite{5961819}, the symbol error rate (SER) of the CLCs is higher than that of LDLCs. Both of these two lattice coding schemes are still difficult to implement in practice due to the use of non-integer lattice constellations. Moreover, the LDPC lattices in \cite{1705007} and the polar lattices in \cite{8492454} involve multilevel coding and multistage decoding due to their construction methods. This poses a much higher delay in encoding and decoding than that of low-density Construction $A$ (LDA) lattices in \cite{8122043}.

Since most of the available designs are based on infinite lattice constellations, their error performances are compared against Poltyrev limit. To put these lattice codes into practice, a power constraint must be satisfied. Moreover, most lattice codes built from LDPC codes have high complexity encoding structures due to the sparseness of their parity-check matrices which in general can lead to high-density generator matrices. Furthermore, most of the Construction $A$, Construction $D$ and Construction $D'$ lattice codes are designed based on one or two-dimensional (real dimension) lattice partitions \cite{6516165,6584536,7124694}. It is understood that this can result in a shaping loss in error performance compared with using higher-dimensional lattice partitions \cite{Shum15}. Constructing codes over multi-dimensional lattices have been considered in \cite{Kositwattanarerk15,Oggier13,Kositwattanarerk13,Khodaiemehr16}. In \cite{Kositwattanarerk15} and \cite{Kositwattanarerk13}, the authors proposed a method for constructing lattices over number fields and have studied their application in wiretap block fading channels. In \cite{Oggier13}, the authors have proposed a lattice construction method to allow Construction A lattices equipped with multiplication, which has potential application in nonlinear distributed computing over a wireless network. In \cite{Khodaiemehr16}, the authors have designed lattices to obtain diversity orders in block fading channels. However, \cite{Kositwattanarerk15,Oggier13,Kositwattanarerk13,Khodaiemehr16} mainly focused on constructing lattices over algebraic number fields with applications to block fading channels while designing lattice codes to approach the unconstrained Shannon limit was not taken into account.

Recently, we have designed irregular repeat-accumulate (IRA) lattice network codes with finite constellations for two-way relay channels (TWRC) in \cite{Qiu16}. The lattice codes are constructed via Construction A on non-binary IRA codes. We have used the extrinsic information transfer (EXIT) charts to optimize the degree distributions of our code ensembles in a bid to minimize the required decoding SNRs. However, this scheme is based on two-dimensional lattice partitions and thus still has a non-negligible performance gap to the unconstrained Shannon limit.

\subsection{Non-Orthogonal Multiple Access}
According to the literature \cite{weisurvey16,Ding17J,8114722,8085125,7676258}, the designs of NOMA are generally categorized into power-domain and code-domain schemes. The main idea of power-domain schemes \cite{6692652,7676258} is that the transmitter superimposes different users' signals sharing the same resource block and the receiver employs SIC to partially or fully cancel out interference. Due to its simplicity and efficiency, the 3rd Generation Partnership Project (3GPP) has proposed a preliminary version of power-domain NOMA terms multiuser superposition transmission (MUST) \cite{TR36.859} for LTE networks. Code-domain NOMA is evolved from the conventional CDMA where low-density sequences are used as signatures of users and efficient message passing algorithms are adopted for joint decoding. Some code-domain NOMA schemes such as low density spreading (LDS) \cite{Beek09}, sparse code multiple access (SCMA) \cite{Nikopour13}, and pattern division multiple access (PDMA) \cite{8352623} have become potential candidates for future uplink multiple access schemes \cite{8316582}. Both power-domain and code-domain NOMA schemes have demonstrated significant gains over conventional OMA schemes and each category has its advantages and disadvantages. For downlink multiuser communications, power-domain NOMA is often considered due to its low decoding complexity as compared to code-domain NOMA.

For power-domain NOMA, extensive research has been conducted to further enhance the performance. This includes designing efficient user pairing \cite{ding17pair}, user scheduling algorithms \cite{di16,Hsu2018VTC}, power allocation optimization for paired users \cite{8345745,Wei17} and system throughput analysis \cite{wei2017performance,WeiCOML,wei2018multiICC}. The benefits of NOMA in various communication scenarios such as MIMO systems \cite{7236924} and physical layer security \cite{7812773} have also been investigated.

Very recently, there have been several designs for power-domain NOMA where discrete inputs over finite constellations are considered. For example, \cite{Choi2016} and \cite{Dong17} investigate the power allocation of two-user NOMA with QAM inputs. In \cite{Fang16}, a downlink multiuser transmission scheme named lattice partition multiple access (LPMA) is proposed, where the concept of Construction $\pi_A$ \cite{7962201} is adopted to partition a two-dimensional lattice into individual constellations. Although it is shown that such scheme can perform well even when the difference between two channel gains is small, the requirement of coding over prime fields makes it less attractive in practice. A simple $K$-user NOMA scheme based on PAM inputs is proposed in \cite{Shieh16} to substantially reduce the burden of decoding at NOMA receivers. In such a scheme, the input distributions are deliberately chosen to be uniformly distributed over some PAM such that the decoder can directly treat interference as noise without severely degrading performance. Theoretical results therein show that this scheme can operate at rate pairs close to the capacity region regardless of the channel parameters, and simulation results further indicate that the actual gaps to the capacity region are much smaller than the theoretical guarantee. However, all the above works assume that the instantaneous channel state information (CSI) is available at the transmitter while the channel gain is constant over a transmission frame. When the transmitter does not have full CSI, the design of NOMA becomes challenging since the optimal user ordering and power allocation all depend on accurate CSI. Although some existing works in the literature \cite{Wei2016NOMA,7361990,7438933,Wei17,7959198,8063934,8327866} have considered NOMA with only statistical CSI at the transmitter, continuous Gaussian inputs are still adopted. To the best of our knowledge, systematic designs of practical NOMA schemes based on discrete and finite input without transmit CSI have not been reported in the literature yet. For block fading channels, the idea of constellation rotation has been adopted in \cite{7880967} to design a two-user downlink NOMA system. More specifically, their design is to optimize the error performance of either one of the two users and only the user whose constellation is optimized can enjoy the diversity gain. Moreover, their approach is based on exhaustive search and thus is of high complexity. In addition, the diversity order they obtained is at most 2 for two users.

\subsection{Channel Coding With Ultra-Reliable Requirements}
In this subsection, we review the previous works on coding for ultra-reliable applications such as flash memories.

Conventional single-level cell (SLC) NAND flash memories only required mild error-correction capabilities for which Hamming codes were sufficient and acceptable in industries \cite{TN-29-63}. Later, stronger error-correction codes (ECCs) such as BCH codes have been widely used for error correction in NAND flash memories \cite{Dolecek17}. With the well-established algebraic coding theory \cite{Lin:2004:ECC:983680}, BCH codes can be explicitly designed to meet the specific requirements including information length requirements, rate requirements and the required number of correctable errors. However, the decoding complexity for BCH codes with length $n$ and error correction capability $t$ is of $O(n + t^2)$. When the code rate is fixed, the complexity grows quadratically with $n$ and $t$ \cite{Cho14}. A special case of BCH codes, i.e., Reed-Solomon (RS) codes have also been applied in flash memories \cite{4671744,6804935}. As RS codes are non-binary and defined over $\text{GF}(q^m)$, they can correct multiple symbols where each symbol contains a number of bits. However, RS codes have a higher computational complexity than binary BCH codes due to the $\text{GF}(q^m)$ operations in encoding and decoding. Note that all the above coding schemes do not have error floor, which is desirable for storage systems.

While BCH codes with existing hard-decision decoding algorithms are still popular in the current design practice \cite{7553563}, some capacity-approaching channel codes such as the LDPC codes \cite{Gallager63low-densityparity-check,6364973,6804932,7416649,7553579,7553518} have been adopted in some flash memory controllers. To attain the best possible performance for LDPC codes, soft information is required for their soft-decision iterative decoding. However, soft reading signals off the NAND flash memory chips would require multiple reads with varying sensing levels. Compared to the hard-decision memory sensing, soft-decision memory sensing introduces longer system latency and more power consumption \cite{5629456}. In addition, the obtained and processed soft information requires more memory space than that for hard information to store. Furthermore, most capacity-approaching LDPC codes exhibit high error floor (bit error rate between $10^{-5}$ and $10^{-6}$) \cite{Richardson03error-floorsof}. Thus, the use of LDPC codes in NAND flash memories poses significant challenges in both code design and hardware implementations.

In order to satisfy the requirements for current NAND flash memory design practice as well as to obtain a higher coding gain than the baseline BCH codes, product codes based on linear block codes are proposed and developed \cite{6118315}. These coding techniques with iterative hard-decision decoding have been shown to provide comparable performance to LDPC codes with hard-input. It has been proved in \cite{7954697} that product codes with BCH component codes can approach the capacity at high-rate regime. Some design examples such as block-wise concatenated BCH codes proposed in \cite{Cho14} and \cite{7192620} have demonstrated strong error-correction capabilities under hard-decision decoding and a low error floor (page error rate (PER) below $10^{-10}$). Other product code schemes for flash memories such as concatenated Raptor codes \cite{Yu14} and Hamming product codes \cite{5645968} also show better performance than their stand-alone counterparts. Another type of product codes called half product codes was investigated in \cite{Emmadi15,8362743} and are shown to have better minimum distance properties than full product codes. Compared to stand-alone BCH codes with the same design code length, these product code schemes with iterative hard-decision decoding have a lower implementation complexity.

Recently, a new class of product codes known as staircase codes was proposed in \cite{Smith12} for high-speed fibre-optic communications. The codes are unterminated and constructed via recursive convolutional coding and block coding while the component codes can be chosen from any conventional ECCs, e.g., Hamming, BCH, RS, etc. The unterminated nature ensures that all the information blocks are protected by both row and column codewords. The staircase code decoder features a sliding-window decoding with hard-decision decoding in each window. Most notably, the simulation result therein shows that the staircase codes with BCH component codes can operate about 0.56 dB away from the BSC limit when BER is at $10^{-15}$. It has been reported in \cite{6787025} that the net coding gain of the staircase codes is competitive with the best known hard-decision decodable codes over a range of overheads. It is also worth mentioning that the error floor can be accurately estimated and analyzed by the proposed union bound technique in \cite{Smith12}. The error floor is mainly due to the error patterns known as stall patterns that cannot be resolved by the decoder with no updated information (similar to the trapping sets in LDPC codes). Extensive research has been carried out to improve the performance of staircase codes. In \cite{hager2017approaching}, an iterative decoding algorithm was developed to reduce the event of miscorrection due to the underlying component code \emph{without cyclic redundant check (CRC)} and thus improve the net coding gain. A post-processing technique based on exhaustive pattern search was proposed in \cite{7905932} to handle stall patterns in order to reduce the error floor. Very recently, an improved staircase code decoder with a low complexity bit-flip algorithm was proposed in \cite{Holzbaur17}. The numerical results therein show that the error floor can be lowered by resolving some of the stall patterns. However, certain stall patterns cannot be solved by their purposed decoding algorithm. Moreover, it is still unclear which stall patterns can be solved definitely by the decoder. Although these unterminated staircase codes demonstrate superior error performance, the unterminated nature is not suitable for general storage applications, such as flash memory devices because an error propagation could cause severe data corruption. Furthermore, direct termination of the staircase codes will leave the last information block only protected by row or column codewords, which results in performance degradation. 

\section{Thesis Outline and Main Contributions}

\subsection{Thesis Organization}
In this subsection, the outline of each chapter in this thesis is given. There are nine chapters in total, including an overview of 5G communication scenarios, the motivation of the research conducted in this research, related works on channel coding and multiuser communications, background information on lattices and wireless communications, details of the conducted research and the conclusion of this thesis.

\textbf{Chapter 1}

This chapter provides an overview of 5G communication scenario and the future requirements. Then, the motivation of this thesis and the relevant works are stated. It also presents the outline and the main contributions of this thesis.

\textbf{Chapter 2}

In this chapter, all the fundamental background knowledge of lattices are presented. Examples with relevant figures are provided to help understand the concept. The materials presented in this chapter will be used throughout the rest of this thesis.

\textbf{Chapter 3}

The basics of modern digital communication systems encompass channel coding, digital modulation, detection are described in this chapter. In addition, different channel models from point-to-point communication to multiuser communication as well as their corresponding channel capacities are also presented.

\textbf{Chapter 4}

In this chapter, we address the problem of communication over classical point-
to-point AWGN channels by designing practical multi-dimensional lattice codes over finite constellations to approach the unconstrained Shannon limit. Full descriptions of how to optimize the decoding threshold and relevant simulation results are presented.

\textbf{Chapter 5}

In Chapter 5, we introduce the proposed lattice-partition framework of downlink NOMA scheme without SIC where the underlying input is based on any $n$-dimensional lattice. Detailed design of coding and constellation for downlink NOMA and the theoretical analysis on the individual achievable rates and their gaps to multiuser capacity are presented in this chapter.

\textbf{Chapter 6}

In Chapter 6, we consider the problem of downlink multiuser communication through NOMA over slow fading channels where the transmitter only has statistical CSI. A novel lattice-partition based downlink NOMA scheme is presented. Detailed explanation on how to design the input distributions for each user based on the average channel condition as well as the theoretical and numerical analysis for the performance of the proposed scheme without SIC are also presented in this chapter.

\textbf{Chapter 7}

In this chapter, we investigate the problem of downlink multiuser communication via NOMA over block fading channels with the transmitter having statistical CSI only. An algebraic rotation approached is adopted to design efficient and practical NOMA scheme to allow each user to obtain higher coding gain and full diversity gain. Theoretical analysis on the performance of the proposed scheme and relevant simulation results are provided in this chapter.

\textbf{Chapter 8}

In this chapter, we present the additional work on designing staircase codes for storage systems. In particular, the proposed code structure and the proposed decoding algorithm to provide enhanced error floor performance are described in detailed. Both theoretical and simulation results are also provided.

\textbf{Chapter 9}

This chapter concludes the thesis by summarizing the main ideas of each chapter and the contributions of all the works conducted during my Ph.D research.

\subsection{Research Contributions}
In what follows, a detailed list of the research contributions in chapters 4-8 are presented.

Chapter 4 presents the design of lattice codes built from Construction A lattices where the underlying linear codes are non-binary irregular repeat-accumulate (IRA) codes. Most importantly, our codes are based on multi-dimensional lattice partitions with \emph{finite constellations}. We propose a novel encoding structure that \emph{adds} randomly generated lattice sequences to the encoder's messages, instead of multiplying lattice sequences to the encoder's messages because most multi-dimensional (more than two dimensions) lattice partitions only form additive quotient groups and lack multiplication operations. We further prove that our approach can ensure that the decoder's messages exhibit permutation-invariance and symmetry properties. With these two properties, the densities of the messages in the iterative decoder can be modeled by Gaussian distributions described by a single parameter. With Gaussian approximation, EXIT charts for our multi-
dimensional IRA lattice codes are developed and used for analyzing the convergence behavior and optimizing the decoding thresholds. Simulation results are provided and show that our codes can approach the unconstrained Shannon limit within 0.46 dB and outperform the previously designed lattice codes with two-dimensional lattice partitions and existing lattice coding schemes for large codeword length.

The results in Chapter 4 have been presented in the following publications:
\begin{itemize}
\item \textbf{M. Qiu}, L.~Yang, Y. Xie and J.~Yuan, ``On the Design of Multi-Dimensional Irregular Repeat-Accumulate Lattice Codes,'' {\em IEEE Trans. Commun.}, vol.~66, no.~2, pp. 478--492, Feb. 2018.

\item \textbf{M.~Qiu}, L. Yang, and J. Yuan, ``Irregular repeat-accumulate lattice network codes for two-way relay channels'' in {\em Proc. IEEE Global Commun. Conf. (GLOBECOM)}, Washington, D.C., Dec. 2016, pp. 1--6.
\item \textbf{M Qiu}, L. Yang, Y. Xie, and J. Yuan, ``On the design of multi-dimensional irregular repeat-accumulate lattice codes,'' in { \em Proc. IEEE Symp. Inf. Theory (ISIT)}, Aachen, Jul. 2017, pp. 2598--2602.
\end{itemize}

In Chapter 5, a novel lattice-partition-based downlink non-orthogonal multiple access framework is presented. This framework is motivated by recognizing the algebraic structure behind the scheme recently proposed by Shieh and Huang in \cite{Shieh16} as a lattice partition in $\mathbb{Z}$; and is in fact a generalization of the scheme in \cite{Shieh16} to any base lattice. The schemes in the proposed framework enjoy many desirable properties such as explicit and systematic design and discrete input distributions. Moreover, the proposed method only requires a limited knowledge of channel parameters. The rates achieved by the proposed scheme with \textit{any base lattice} and with single-user decoding (i.e., without successive interference cancellation) are analyzed and a universal upper bound on the gap to the multiuser capacity is obtained as a function of the normalized second moment of the base lattice. Since the proposed framework has a substantially larger design space than that of \cite{Shieh16} whose base lattice is a one-dimensional lattice, one can easily find instances in larger dimensions that can provide superior performance. Design examples with the base lattices $A_2$, $D_4$, $E_8$, and Construction A lattices, respectively, are provided and both theoretical and simulation results exhibit smaller gaps to the multiuser capacity as dimensions increase.

The results in Chapter 5 have been presented in the following publications:
\begin{itemize}
\item \textbf{M. Qiu}, Y.-C. Huang, S.-L. Shieh, and J. Yuan, ``A Lattice-Partition Framework of Downlink Non-Orthogonal Multiple Access without SIC,''
   {\em IEEE Trans. Commun.}, vol.~66, no.~6, pp. 2532--2546, Jun. 2018.

\item \textbf{M Qiu}, Y.-C. Huang, S.-L. Shieh, and J. Yuan, ``“A lattice-partition framework of downlink non-orthogonal multiple access without SIC,'' in { \em Proc. IEEE Global Commun. Conf. (GLOBECOM)}, Singapore, Dec. 2018, pp. 1--6.
\end{itemize}

In Chapter 6, the problem of downlink non-orthogonal multiple access over {\it slow fading channels} is studied. Full channel state information is assumed at the receivers, while only {\it statistical} CSI is assumed to be available at the transmitter. A novel lattice-partition-based scheme is proposed which, according to statistical CSI, employs discrete inputs from appropriately designed constellations carved from a lattice, rather than continuous Gaussian inputs as used in most existing works. Theoretical analysis shows that for any outage probability smaller than $63.21\%$, which covers almost all the cases of practical interest, the proposed scheme with {\it single-user decoding}, i.e., without successive interference cancellation is able to approach the NOMA outage capacity region within a constant gap, independent of the signal-to-noise ratio and the number of users. Simulation results fortify the effectiveness of the proposed scheme by showing that the approach without SIC can achieve outage rates that are very close to the outage capacity region and the gap becomes even smaller when SIC is employed.

The results in Chapter 6 have been presented in the following publications:
\begin{itemize}
\item \textbf{M. Qiu}, Y.-C. Huang, J. Yuan and C.-L. Wang, ``Lattice-Partition-Based Downlink Non-Orthogonal Multiple Access without SIC for Slow Fading Channels,'' {\em IEEE Trans. Commun.}, vol.~67, no.~2, pp. 1166-1181, Feb. 2019.

\item \textbf{M Qiu}, Y.-C. Huang, J. Yuan and C.-L. Wang, ``Downlink lattice-partition- based non-orthogonal multiple access without SIC for slow fading channels,'' in { \em Proc. IEEE Global Commun. Conf. (GLOBECOM)}, Abu Dhabi, Dec. 2018, pp. 1--6.
\end{itemize}

In Chapter 7, we investigate the problem of downlink NOMA over block fading channels. For the single antenna case, we propose a class of NOMA schemes where all the users' signals are mapped into $n$-dimensional constellations corresponding to the same algebraic lattices from a number field, allowing every user attains full diversity gain with single-user decoding, i.e., no successive interference cancellation. The minimum product distances of the proposed scheme with arbitrary power allocation factor are analyzed and their upper bounds are derived. Within the proposed class of schemes, we also identify a special family of NOMA schemes based on lattice partitions of the underlying ideal lattices, whose minimum product distances can be easily controlled. Our analysis shows that among the proposed schemes, the lattice-partition-based schemes achieve the largest minimum product distances of the superimposed constellations, which are closely related to the symbol error rates for receivers with single-user decoding. Simulation results are presented to verify our analysis and to show the effectiveness of the proposed schemes as compared to benchmark NOMA schemes. Extensions of our design to the multi-antenna case are also considered where similar analysis and results are presented.

The results in Chapter 7 have been presented in the following publications:
\begin{itemize}
\item \textbf{M. Qiu}, Y.-C. Huang, and J. Yuan, ``Downlink Non-Orthogonal Multiple Access without SIC for Block Fading Channels,'' {\em IEEE Trans. Wireless Commun.}, accepted, Jun. 2019.

\item \textbf{M Qiu}, Y.-C. Huang, and J. Yuan, ``Downlink NOMA without SIC for fast fading channels: Lattice partitions with algebraic rotations,'' in { \em Proc. IEEE Intern.  Commun. Conf. (ICC)}, May 2019, pp. 1-6.
\end{itemize}

In Chapter 8, we propose novel terminated staircase codes for ultra-reliable applications such as NAND flash memories. Specifically, we design a rate 0.89 staircase code whose component code is a BCH code, for flash memories with page size of 16K bytes. Different from most conventional unterminated staircase codes, we propose a novel coding structure by performing CRC encoding and decoding on each component codeword including information bits and parity bits. The CRC bits are protected by both row and column codewords. Furthermore, a novel iterative bit flipping algorithm is developed to solve stall patterns and lower the error floor. Based on our design, we perform an improved analysis on the error floor. We prove and show that our proposed decoding algorithm can solve more stall patterns which leads to a lower error floor compared to conventional staircase codes. Numerical results show that our terminated staircase codes outperform the stand-alone BCH codes and the conventional staircase codes.

The results in Chapter 8 have been presented in the following publication:
\begin{itemize}
\item \textbf{M. Qiu}, L. Yang, Y. Xie, and J. Yuan, ``Terminated Staircase Codes for NAND Flash Memories,''{\em IEEE Trans. Commun.}, vol.~66, no.~12, pp. 5861-5875, Dec. 2018.
\end{itemize}

\chapter{Background on Lattices
}\label{C2:chapter2}

\section{Introduction}
In this part, we first introduce the basic concept of lattices and lattice codes. Then, we introduce algebraic number theory that is useful and essential for constructing lattices. We summarize the most important definitions and results without proofs. More details about lattices, lattice codes \cite{conway1999sphere,huang13phd,dipietro:tel-01135575,Zamir15} and algebraic number theory can be found in \cite{Oggier:2004:ANT:1166377.1166378,Oggier:33651,costa2018lattices}, respectively, and the references therein. Here, we also fix most of the notation that will be employed later on. For our purposes, all the concepts below are introduced based on real-dimensional lattices.

\section{Lattices}\label{sec:2_1}
An $n$-dimensional lattice $\Lambda$ is a discrete set of points $\boldsymbol{\lambda}$ in $\mathbb{R}^n$. It is a discrete subgroup that is closed under addition and reflection. That is, for any pair of lattice points, $\lambda_1,\lambda_2 \in \Lambda$, we have $\lambda_1+\lambda_2 \in \Lambda$.

\begin{figure}[ht]
\centering
\includegraphics[width=4.5in]{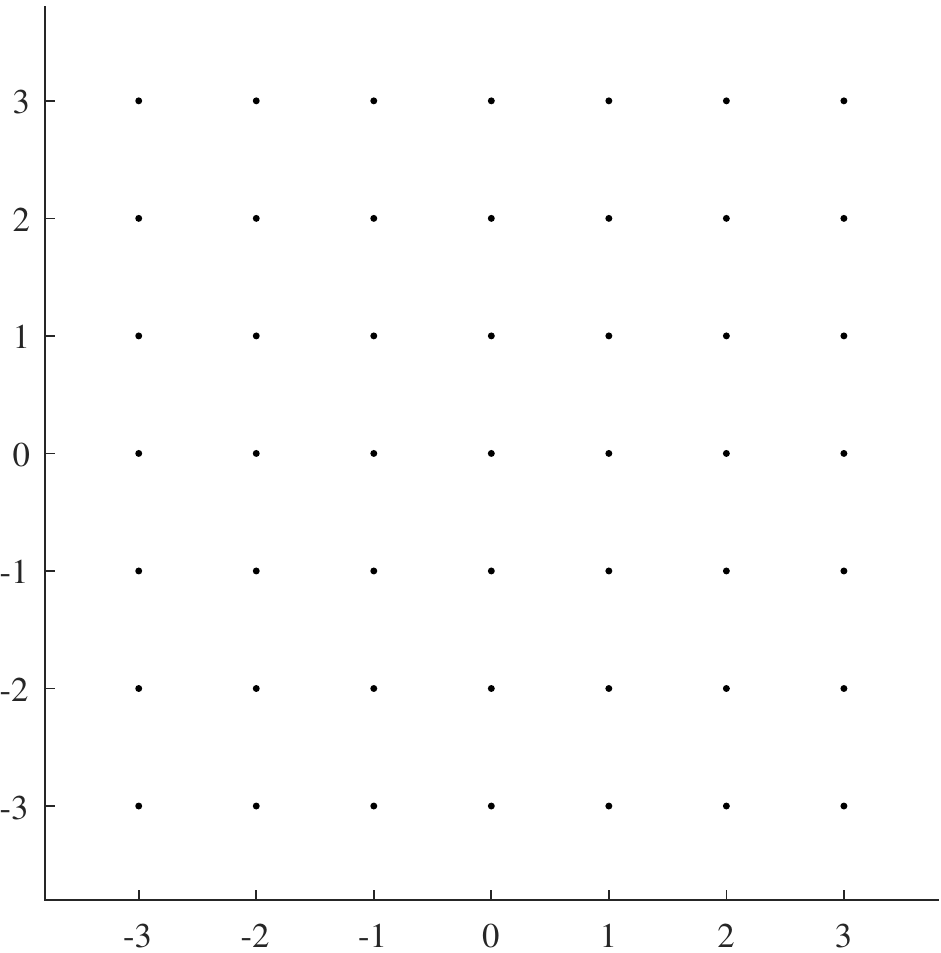}
\caption{Two-dimensional square lattice $\mathbb{Z}^2$.}
\label{fig:z2_lattice}
\end{figure}

\begin{figure}[ht]
\centering
\includegraphics[width=4.5in]{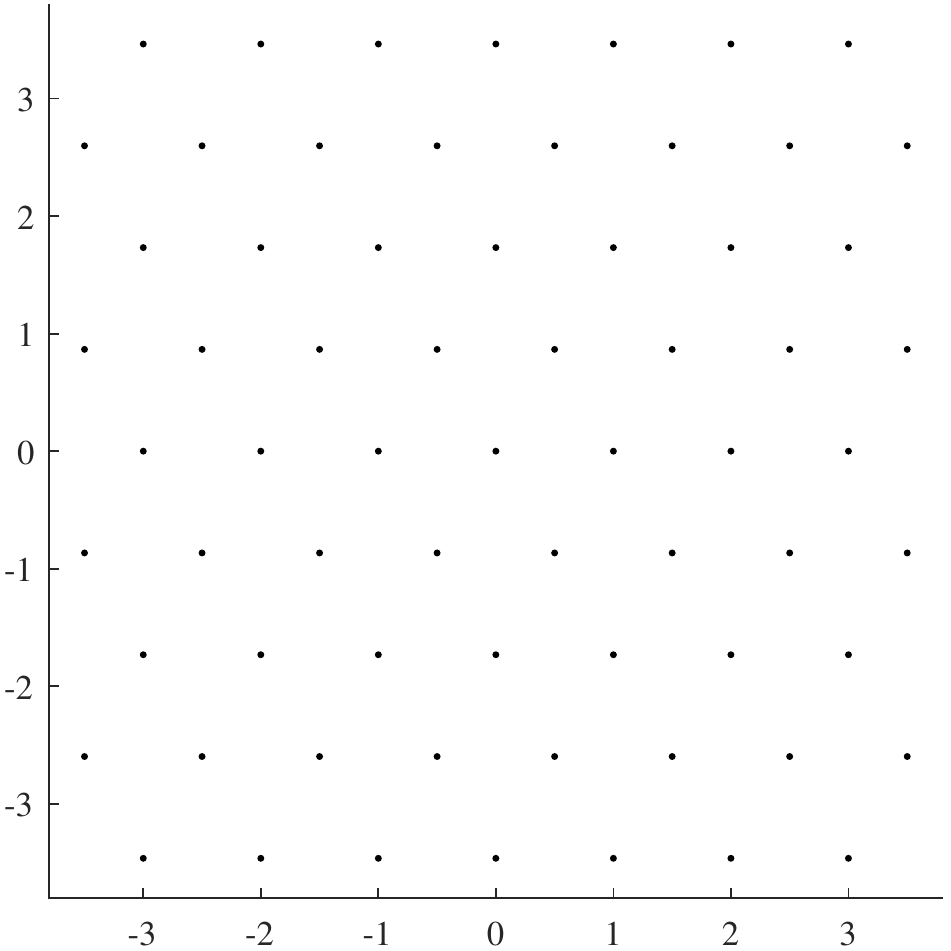}
\caption{Two-dimensional hexagonal lattice $A_2$.}
\label{fig:a2_lattice}
\end{figure}

In Fig. \ref{fig:z2_lattice} and Fig. \ref{fig:a2_lattice}, we provide two examples of two-dimensional lattices, namely square lattice $\mathbb{Z}^2$ and hexagonal lattice $A_2$, respectively. In algebraic number theory, these two lattices are referred to as the ring of Gaussian integer $\mathbb{Z}[i] \triangleq \mathbb{Z}[\sqrt{-1}] = \{a+b\sqrt{-1}: a,b \in \mathbb{Z}\}$ and the ring of Eisenstein integer $\mathbb{Z}[\omega] \triangleq \mathbb{Z}[\frac{-1+\sqrt{-3}}{2}] = \{a+b\left(\frac{-1+\sqrt{-3}}{2}\right): a,b \in \mathbb{Z}\}$, respectively. Now, we present the fundamental concepts of lattices.

\subsection{Lattice Basics}
\begin{defi}(Lattice):
An $n$-dimensional lattice $\Lambda$ is the set of all the linear combinations of $n$ linearly independent vectors $\mathbf{g}_1,\ldots,\mathbf{g}_n \in \mathbb{R}^n$ such that:
\begin{equation}
\Lambda= \left\{ \boldsymbol{\lambda} \in \mathbb{R}^n: \boldsymbol{\lambda}= \sum_{i=1}^n b_i\mathbf{g}_i, \exists (b_1,\ldots,b_n) \in \mathbb{Z}^n \right\}.
\end{equation}
\end{defi}
By the above definition, a lattice always contain the all-zero point $\mathbf{0}$. Moreover, we have restricted our definition to \emph{full-rank} lattices (that is, $n$-dimensional lattices in an $n$-dimensional Euclidean space) because we do not need to treat lower-rank lattices for the purposes of our work.

\begin{defi}(Generator Matrix):
A generator matrix of a lattice $\Lambda$ is a matrix whose rows generate $\Lambda$
\begin{equation}
\mathbf{G}_{\Lambda} = \begin{pmatrix}
\mathbf{g}_1 \\
\vdots \\
\mathbf{g}_n \\
\end{pmatrix} and \; \Lambda= \left\{ \boldsymbol{\lambda} \in \mathbb{R}^n: \boldsymbol{\lambda}=\mathbf{b}\mathbf{G}_{\Lambda}, \exists \mathbf{b} \in \mathbb{Z}^n \right\}.
\end{equation}
\end{defi}

\begin{defi}(Cartesian Product):
The Cartesian product of two lattices $\Lambda_1$ and $\Lambda_2$ of dimensions $n_1$ and $n_2$ is an $n = n_1 + n_2$ dimensional lattice $\Lambda$:
\begin{equation}
\Lambda = \Lambda_1 \times \Lambda_2 = \{(\boldsymbol{\lambda}_1,\boldsymbol{\lambda}_2): \boldsymbol{\lambda}_1 \in \Lambda_1,\boldsymbol{\lambda}_2 \in \Lambda_2\}.
\end{equation}
The generator matrix of this product lattice is a block-diagonal matrix
\begin{equation}
\mathbf{G}_{\Lambda} = \begin{pmatrix}
\mathbf{G}_{\Lambda_1} & 0 \\
0 & \mathbf{G}_{\Lambda_2} \\
\end{pmatrix},
\end{equation}
with the component generator matrices on its diagonal, hence its determinant is the
product of the component determinants
\begin{equation}
\det(\Lambda) = \det(\Lambda_1 \times \Lambda_2) = \det(\Lambda_1)\cdot\det(\Lambda_2).
\end{equation}
\end{defi}

\begin{defi}(Lattice Quantizer):
A lattice quantizer (or the nearest neighbor quantizer) $Q_{\Lambda}(\mathbf{x})$ with respect to $\Lambda$ maps a point $\mathbf{x} \in \mathbb{R}^n$ to its closest lattice point of $\Lambda$ as
\begin{align}\label{eq:1a}
Q_{\Lambda}(\mathbf{x}) &= \boldsymbol{\lambda} \in \Lambda, \|\mathbf{x}-\boldsymbol{\lambda} \| \leq \|\mathbf{x}-\boldsymbol{\lambda}' \|, \boldsymbol{\lambda}'\neq \boldsymbol{\lambda} \in \Lambda, \nonumber \\
&= \argmin_{\boldsymbol{\lambda} \in \Lambda}\| \mathbf{x} -  \boldsymbol{\lambda} \|.  
\end{align}
\end{defi}

\begin{defi}(Fundamental Voronoi Region/Cell):
Given a lattice $\Lambda$, the fundamental Voronoi region/cell of this lattice is defined as
\begin{equation}
\mathcal{V}_{\mathbf{0}}(\Lambda)=\{\mathbf{x}\in\mathbb{R}^n:Q_{\Lambda}(\mathbf{x})=\mathbf{0}\}.
\end{equation}
In other words, $\mathcal{V}_{\mathbf{0}}(\Lambda)$ is the set of all the real vectors $\mathbf{x}$ that are closer (or as close) to the all-zero lattice point than to any other lattice point.
\end{defi}

\begin{defi}(Voronoi Region/Cell):
Given a lattice $\Lambda$, the Voronoi region/cell of this lattice is defined as
\begin{equation}
\mathcal{V}_{\boldsymbol{\lambda}}(\Lambda)=\{\mathbf{x}\in\mathbb{R}^n:Q_{\Lambda}(\mathbf{x})=\boldsymbol{\lambda}\}.
\end{equation}
Similar to the above definition, $\mathcal{V}_{\boldsymbol{\lambda}}(\Lambda)$ is the set of all the real vectors $\mathbf{x}$ that are closer (or as close) to the lattice point $\boldsymbol{\lambda}$ than to any other lattice point.
\end{defi}
Clearly, the Voronoi cell have the following three properties:
\begin{itemize}
\item Each Voronoi cell $\mathcal{V}_{\boldsymbol{\lambda}}(\Lambda)$ is a shift of the fundamental Voronoi cell by $\boldsymbol{\lambda} \in \Lambda$, i.e., $\mathcal{V}_{\boldsymbol{\lambda}}(\Lambda)= \boldsymbol{\lambda}+\mathcal{V}_{\mathbf{0}}(\Lambda)$.
\item The cells do not intersect with each other, i.e., $\mathcal{V}_{\boldsymbol{\lambda}}(\Lambda) \cap \mathcal{V}_{\boldsymbol{\lambda}'}(\Lambda) = \emptyset$ for all $\boldsymbol{\lambda} \neq \boldsymbol{\lambda}'$.
\item The union of the cells covers the whole Euclidean space, i.e., $\bigcup\limits_{\boldsymbol{\lambda} \in \Lambda}\mathcal{V}_{\boldsymbol{\lambda}}(\Lambda) = \mathbb{R}^n$.
\end{itemize}

\begin{defi}(Modulo Operation):
The modulo-lattice operation with respect to $\Lambda$ is defined as
\begin{equation}\label{eq:1c}
\left[\mathbf{x}\right]_{\Lambda} \triangleq\mathbf{x} \;\text{mod}\; \Lambda = \mathbf{x} - Q_{\Lambda}(\mathbf{x}).
\end{equation}
\end{defi}

\begin{figure}[ht]
\centering
\includegraphics[width=4.5in]{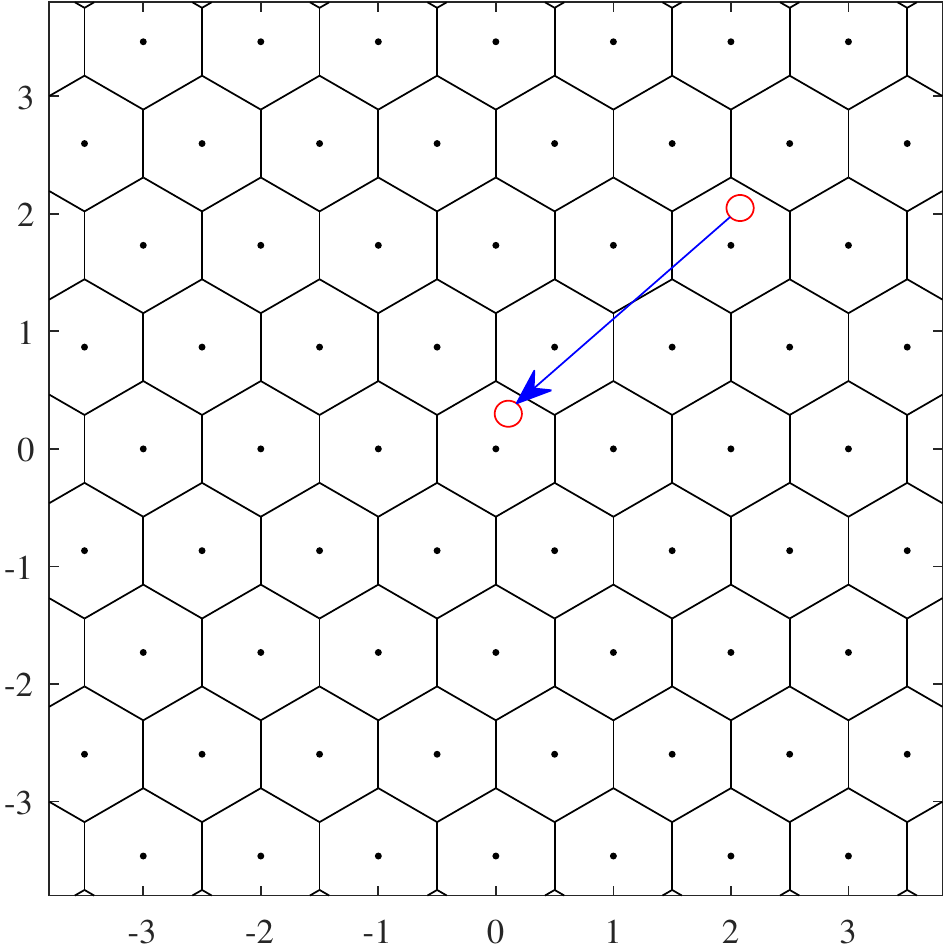}
\caption{Example showing the modulo lattice operations.}
\label{fig:modA2}
\end{figure}

One example that explains the above operations can be found in Fig. \ref{fig:modA2} where the $A_2$ lattice is considered and the circle in the upper right corner represents a vector $\mathbf{x} \in \mathbb{R}^2$. The nearest neighbor quantizer associated with $A_2$ will quantize $\mathbf{x}$ to the lattice point inside the same hexagon with $\mathbf{x}$. The hexagon circumventing the origin is the fundamental Voronoi region of $A_2$. Moreover, the modulo operation will map $\mathbf{x}$ to the corresponding position inside the fundamental Voronoi region as the circle shown in the middle of this figure.

\begin{defi}(Volume):
The volume of a lattice $\Lambda$ with generator matrix $\mathbf{G}_{\Lambda}$ is defined by
\begin{equation}
\text{Vol}(\Lambda)  = |\det(\mathbf{G}_{\Lambda})|.
\end{equation}
Note that the volume of $\Lambda$ is sometimes written as $\text{Vol}(\mathcal{V}_{\mathbf{0}}(\Lambda))$, namely the volume of the fundamental Voronoi region of $\Lambda$.
\end{defi}

\begin{defi}(Minimum Euclidean Distance):
The minimum Euclidean distance of a lattice $\Lambda$ is
\begin{equation}
d_{E,\min}(\Lambda) = \min\limits_{\boldsymbol{\lambda} \in \Lambda}\|\boldsymbol{\lambda} \|.
\end{equation}
\end{defi}
This quantity is closely related to the performance of a lattice for the transmission of information over a real channel with Gaussian noise, which is similar to the minimum distance of a linear code.

\begin{defi}(Kissing Number):
The Kissing number of a lattice $\Lambda$ is the number of lattice points whose norm is the minimum Euclidean distance of $\Lambda$
\begin{equation}
\gamma(\Lambda) = |\{\boldsymbol{\lambda}\in \Lambda: \|\boldsymbol{\lambda} = d_{E,\min}(\Lambda) \| \}|.
\end{equation}
\end{defi}

One may also be interested in counting the number of lattice points that have
any fixed norm and not only the minimum one. These numbers are collected by the
so-called theta series.

\begin{defi}(Theta Series):
Let $\tau = \exp(\pi\sqrt{-1}z)$ for some $z \in \mathbb{C}$ with $\Im(z)\geq 0$; let
$N_m$ be the number of points of a certain lattice $\Lambda$ whose squared Euclidean
norm is $m$. Then the theta series of $\Lambda$ is defined by
\begin{equation}
\Theta_{\Lambda}(z) = \sum_{\boldsymbol{\lambda}\in \Lambda}\tau^{\boldsymbol{\lambda}\boldsymbol{\lambda}^T} = \sum_{m=0}^{\infty}N_m\tau^m.
\end{equation}
\end{defi}
It can be observed that $\gamma(\Lambda) = N_{d_{\min}(\Lambda)}$ according to the above two definitions.

\begin{defi}(Equivalence):
A lattice $\Lambda_1$ is \emph{equivalent} to another lattice $\Lambda_2$ if $\Lambda_2 = \alpha\mathbf{R}  \Lambda_1$, where $\alpha$ is a positive scalar and $\mathbf{R}$ is an orthogonal matrix such that $\mathbf{R}\cdot \mathbf{R}^T = \mathbf{I}_n$ and $\mathbf{I}_n$ is an identity matrix with size $n$.
\end{defi}

\begin{defi}(Sublattice):
A lattice $\Lambda'$ is a sublattice of (nested in) another lattice $\Lambda$ if $\Lambda' \subseteq \Lambda$. 
\end{defi}

\begin{defi}(Lattice Partition):
A lattice partition is formed by
\begin{equation}\label{eq:1e}
\Lambda / \Lambda^{\prime} = \{ \boldsymbol{\lambda} + \Lambda^{\prime}, \boldsymbol{\lambda} \in \Lambda \},
\end{equation}
where $\Lambda$ is the fine lattice and $\Lambda^{\prime}$ is the coarse lattice such that $\Lambda^{\prime}$ is nested in $\Lambda$: $\Lambda^{\prime} \subseteq \Lambda$.
\end{defi}
Note that the lattice partition above forms a quotient group.

\begin{defi}(Coset):
Given the lattice partition $\Lambda/\Lambda'$ and for each $\boldsymbol{\lambda} \in \Lambda$, the set $\boldsymbol{\lambda} + \Lambda^{\prime}$ is a coset of $\Lambda^{\prime}$ in $\Lambda$.
\end{defi}

\begin{figure}[ht]
\centering
\includegraphics[width=4.5in]{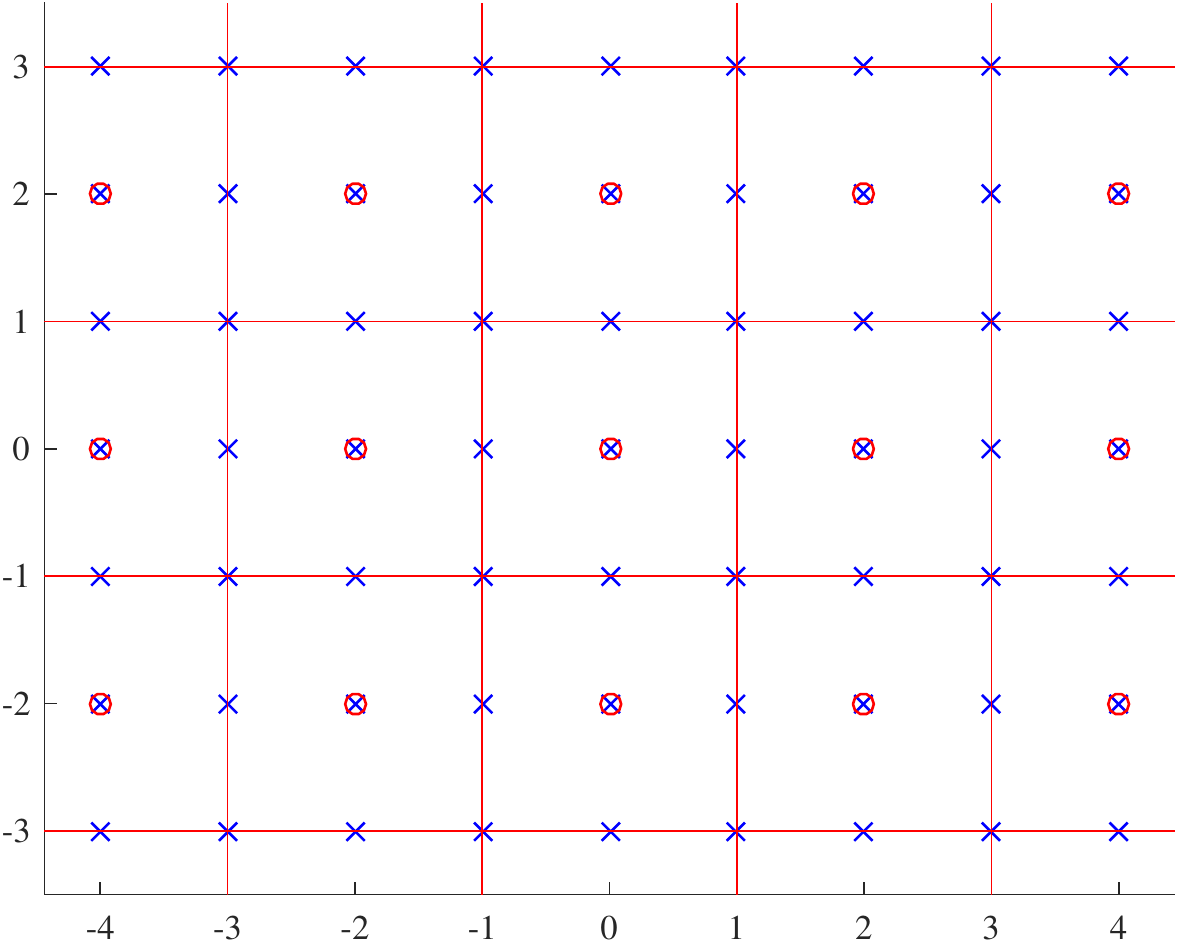}
\caption{Lattice cosets $\mathbb{Z}^2/2\mathbb{Z}^2$.}
\label{fig:z2coset}
\end{figure}

An example of the cosets from lattice partition $\mathbb{Z}^2/2\mathbb{Z}^2$ is shown in Fig. \ref{fig:z2coset}. The fine lattice points are represented by the crosses and the coarse lattice points are represented by the circles. There are four cosets in the Voronoi region of the coarse lattice $2\mathbb{Z}^2$.

\begin{defi}(Coset leader):
Given the lattice partition $\Lambda/\Lambda'$, the point $\boldsymbol{\lambda}\;\text{mod}\; \Lambda^{\prime}$ is called the coset leader of coset $\boldsymbol{\lambda} + \Lambda^{\prime}$.
\end{defi}

\begin{figure}[ht]
\centering
\includegraphics[width=4.5in]{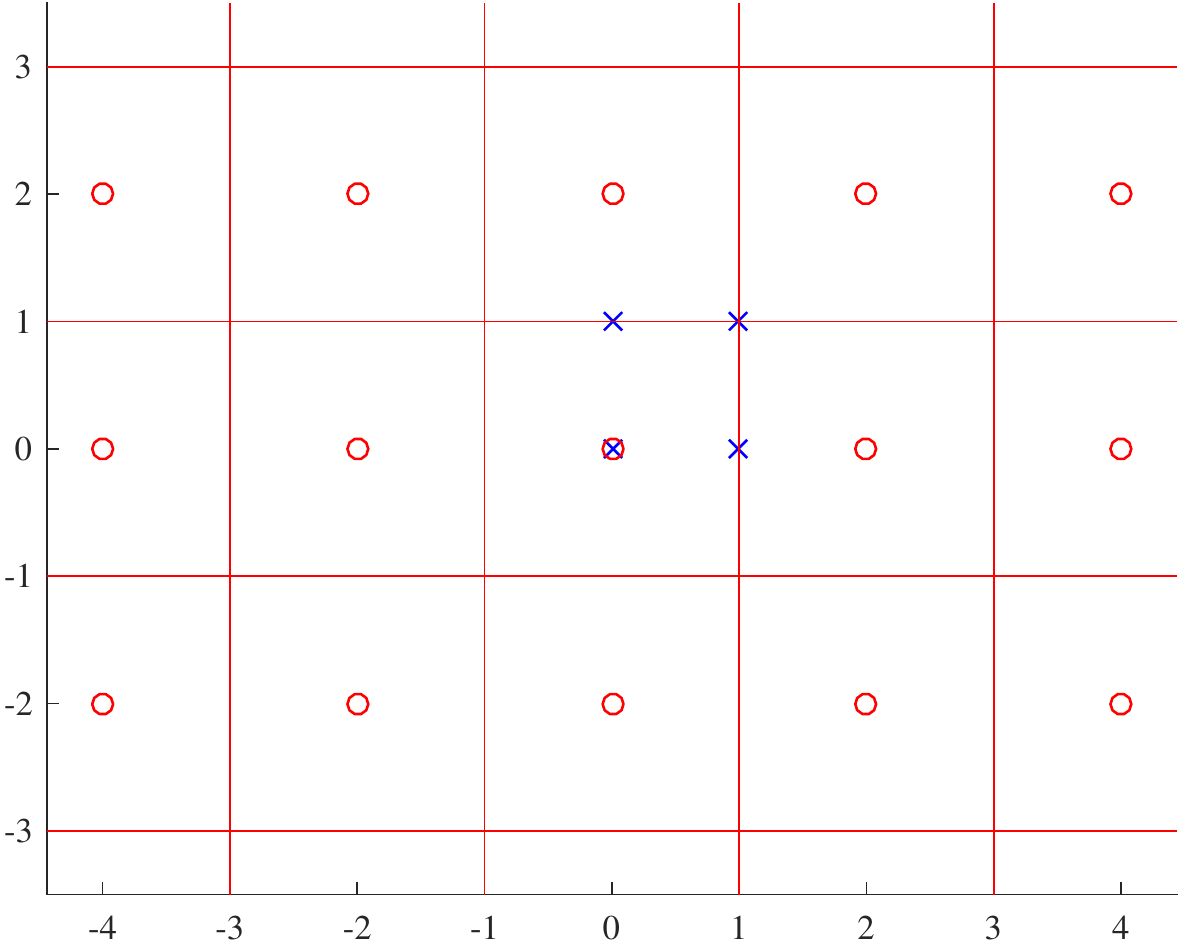}
\caption{Lattice coset leaders $\mathbb{Z}^2/2\mathbb{Z}^2$.}
\label{fig:z2coset_leader}
\end{figure}

An example of the coset leaders of lattice partition $\mathbb{Z}^2/2\mathbb{Z}^2$ is shown in Fig. \ref{fig:z2coset_leader}. The coset leaders are the fine lattice points that lie inside the fundamental Voronoi  region of the coarse lattice. For this case, the coset leaders form a shift version of 4QAM.

\begin{defi}(Nesting Ratio):
Given the $n$-dimensional lattice partition $\Lambda / \Lambda^\prime$, the nested ratio is precisely calculated as
\begin{equation}\label{eq:1f}
|\Lambda / \Lambda^\prime|^{\frac{1}{n}} = \left(\frac{\text{Vol}(\Lambda^\prime)}{\text{Vol}(\Lambda)}\right)^{\frac{1}{n}}.
\end{equation}
\end{defi}

For a pair of nested lattice $\Lambda' \subseteq \Lambda$, we denote the modulo-lattice addition with respect to $\Lambda'$ by ``$\oplus$'' where
\begin{equation}\label{eq:1d}
\boldsymbol{\lambda}_1 \oplus \boldsymbol{\lambda}_2 = (\boldsymbol{\lambda}_1+\boldsymbol{\lambda}_2)\;\text{mod}\; \Lambda^\prime, \; \; \boldsymbol{\lambda}_1,\boldsymbol{\lambda}_2 \in \Lambda.
\end{equation}
Similarly, we denote the modulo-lattice subtraction by ``$\ominus$'' where
\begin{equation}\label{eq:1dss}
\boldsymbol{\lambda}_1 \ominus \boldsymbol{\lambda}_2 = (\boldsymbol{\lambda}_1-\boldsymbol{\lambda}_2)\;\text{mod}\; \Lambda^\prime, \; \; \boldsymbol{\lambda}_1,\boldsymbol{\lambda}_2 \in \Lambda.
\end{equation}

\begin{defi}(Nested Lattice Code):
Given an $n$-dimensional fine lattice $\Lambda$ and an $n$-dimensional coarse
lattice $\Lambda'$, where $\Lambda' \subseteq \Lambda$, an $n$-dimensional nested lattice code (Voronoi code), which we refer to as $\mathcal{L}$, is the set of all coset leaders in $\Lambda$ that lie in the fundamental Voronoi region of the coarse lattice $\Lambda'$
\begin{equation}
\mathcal{L} = \Lambda\cap \mathcal{V}_{\mathbf{0}}(\Lambda')= \{\boldsymbol{\lambda}: Q_{\Lambda'}(\boldsymbol{\lambda}) = 0,\boldsymbol{\lambda} \in \Lambda\}.
\end{equation}
Due to this geometry property, the fundamental Voronoi region $\mathcal{V}_{\mathbf{0}}(\Lambda^\prime)$ is also called the shaping region. Shaping is essential in designing practical lattice codes because a finite section of the lattice points must be selected to satisfy a transmission power constraint for a communication system.
The code rate of this nested lattice code in bits/s/Hz/real dimension is given by
\begin{equation}\label{eq:1g}
R(\Lambda / \Lambda^{\prime}) = \frac{1}{n}\log_2(|\Lambda / \Lambda^\prime|).
\end{equation}
\end{defi}

The fine lattice and the coarse lattice need to be carefully chosen in order to construct reliable lattice coding schemes. In what follows, we provide some definitions on the figures of merit of lattices in terms of packing, covering, quantization, and channel coding.

\subsection{Figures of Merit}\label{sec:fom}

\begin{defi}(Packing Radius):
For a given lattice $\Lambda$, a radius $r>0$ is said to be a packing radius if the set $\Lambda + \mathcal{B}(r)$ is a packing in Euclidean space for all distinct lattice points $\boldsymbol{\lambda} \neq \boldsymbol{\lambda}' \in \Lambda$, we have
\begin{equation}
(\boldsymbol{\lambda}+\mathcal{B}(r))\cap (\boldsymbol{\lambda}'+\mathcal{B}(r)) = \emptyset.
\end{equation}
That is, the spheres do not intersect. The packing radius $r_{\text{pack}}(\Lambda)$ of the lattice is
defined by the largest balls the lattice can pack
\begin{equation}
r_{\text{pack}}(\Lambda) = \sup\{r: \Lambda + \mathcal{B}(r) \; \text{is a packing} \}.
\end{equation}
\end{defi}

\begin{defi}(Effective Radius):
The effective radius of a lattice $\Lambda$, which we denote by $r_{\text{eff}}(\Lambda)$, is defined as the radius such that the corresponding sphere has the same volume as that of the lattice
\begin{equation}
\text{Vol}(\mathcal{B}(r_{\text{eff}}(\Lambda))) = \text{Vol}(\Lambda).
\end{equation}
\end{defi}

\begin{defi}(Covering Radius):
For a given lattice $\Lambda$, a radius $r>0$ is said to be a covering radius if the set $\Lambda + \mathcal{B}(r)$ is a covering of Euclidean space such that
\begin{equation}
\mathbb{R}^n \subseteq \boldsymbol{\lambda}+\mathcal{B}(r).
\end{equation}
That is, each point in space is covered by at least one sphere. The covering radius $r_{\text{cov}}(\Lambda)$ of the lattice is defined as
\begin{equation}
r_{\text{cov}}(\Lambda) = \min\{r: \Lambda + \mathcal{B}(r) \; \text{is a covering} \}.
\end{equation}
\end{defi}

\begin{figure}[ht]
\centering
\includegraphics[width=4.0in]{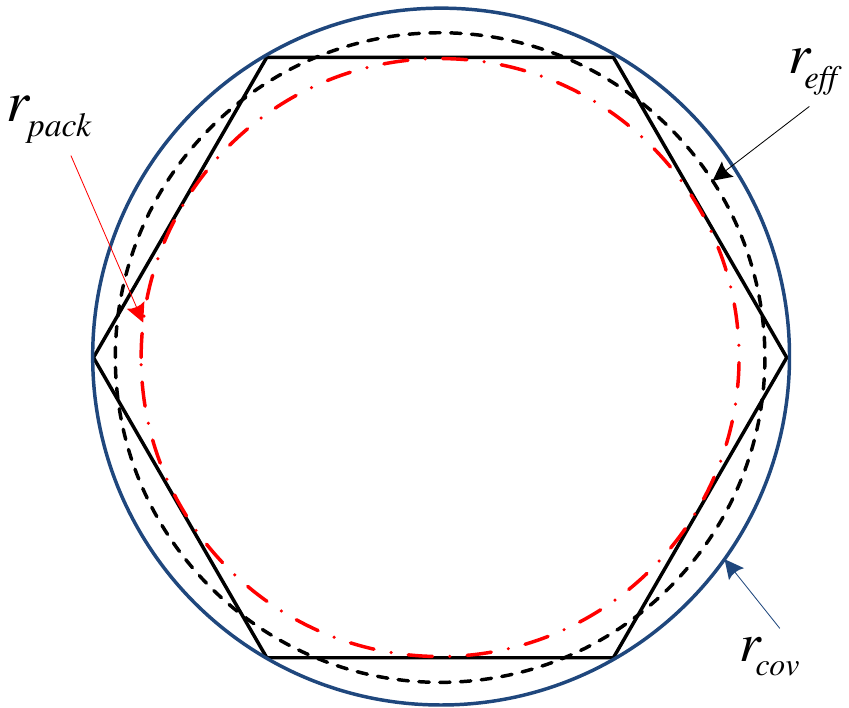}
\caption{Covering radius, effective radius and packing radius of a lattice. The solid hexagon is the Voronoi region of the $A_2$ lattice.}
\label{fig:cep_A2}
\end{figure}

We depict the packing radius, effective radius and covering radius of the $A_2$ lattice in Fig. \ref{fig:cep_A2}. In
this figure, it is obvious that $r_{\text{cov}} \geq r_{\text{eff}} \geq r_{\text{pack}}$.

\begin{defi}(Packing Efficiency):
The packing efficiency of a lattice $\Lambda$ is defined as
\begin{equation}
\rho_{\text{pack}}(\Lambda) = \frac{r_{\text{pack}}(\Lambda)}{r_{\text{eff}}(\Lambda)}.
\end{equation}
The packing efficiency always satisfies
\begin{equation}
0 < \rho_{\text{pack}}(\Lambda)\leq 1.
\end{equation}
\end{defi}

\begin{defi}(Goodness for Packing):
A sequence of lattices is good for packing if it satisfies
\begin{equation}
\rho_{\text{pack}}(\Lambda) = \frac{1}{2}.
\end{equation}
\end{defi}
This is the best known lower bound given by the Minkowski-Hlawka theorem \cite{Roger64}.

\begin{defi}(Covering Efficiency):
The covering efficiency of a lattice $\Lambda$ is defined as
\begin{equation}
\rho_{\text{cov}}(\Lambda) = \frac{r_{\text{cov}}(\Lambda)}{r_{\text{eff}}(\Lambda)}.
\end{equation}
The covering efficiency is by definition not less than 1. However, it goes above 1 for all $n > 1$.
\end{defi}

\begin{defi}(Goodness for Covering):
A sequence of lattices is good for covering if it satisfies
\begin{equation}
\rho_{\text{cov}}(\Lambda) = 1.
\end{equation}
\end{defi}

\begin{defi}(Second moment):
The second moment the lattice $\Lambda$ is defined as the average energy per dimension
of a uniform distribution over the fundamental Voronoi region of $\Lambda$
\begin{equation}\label{eq:sm}
    \sigma^2(\Lambda) = \frac{1}{n \text{Vol}(\Lambda)}\int_{\mc{V}_{\mathbf{0}}(\Lambda)}\|\mathbf{x}\|^2 d \mathbf{x}.
\end{equation}
\end{defi}

\begin{defi}
The normalized second moment (NSM) of the lattice $\Lambda$ is given by
\begin{equation}\label{eq:nsm}
    \psi(\Lambda) = \frac{\sigma^2(\Lambda)}{\text{Vol}(\Lambda)^{\frac{2}{n}}},
\end{equation}
which is invariant to scaling or rotation of $\Lambda$.
\end{defi}

\begin{defi}(Goodness for Quantization):
A sequence of lattices is good for quantization if it satisfies
\begin{equation}
\lim\limits_{n \rightarrow \infty} \psi(\Lambda) = \frac{1}{2 \pi e}.
\end{equation}
\end{defi}
These lattices are also commonly referred to as Rogers good, since it was first
shown by Rogers that such lattices exist \cite{Roger57}.

\begin{defi}(Shaping Gain):
The shaping gain $\gamma_s(\Lambda)$ is defined as the energy gain by achieving the reduction of the average energy of a lattice constellation compared with the constellation points that form an $n$-dimensional cube. It can be calculated as
\begin{equation}\label{eq:sg}
\gamma_s(\Lambda) = \frac{1/12}{\psi(\Lambda)},
\end{equation}
where $\frac{1}{12}$ is the NSM of an $n$-dimensional cubic lattice. A lattice with a smaller normalised second moment is always desirable as its shaping gain is higher. When the dimension approaches infinite, there exists a sequence of lattices that can achieves the optimal shaping gain:
\begin{equation}
\lim_{n \rightarrow \infty}\gamma_s(\Lambda) = \frac{\pi e}{6}.
\end{equation}
\end{defi}

\begin{defi}(Goodness for Coding):
Consider an $n$-dimensional lattice $\Lambda$. A lattice point $\boldsymbol{\lambda} \in \Lambda$ is transmitted through an AWGN channel:
\begin{equation}
\mathbf{y} = \boldsymbol{\lambda}+\mathbf{z},
\end{equation}
where $\mathbf{y}$ is the received signal vector and $\mathbf{z}$ is an $n$-dimensional
independent and identically distributed (i.i.d) Gaussian noise vector with each element $\sim \mathcal{N}(0, \sigma_z^2)$. We define the effective radius of the noise vector by
\begin{equation}
r_z = \sqrt{n\sigma_z^2}.
\end{equation}
The lattice decoder attempts to decodes $\mathbf{y}$ to the nearest lattice point $\boldsymbol{\lambda}$. An error would occur only if $\mathbf{y}$ is outside the Voronoi region of
$\boldsymbol{\lambda}$. Due to the lattice symmetry, this is equivalent to $\mathbf{z}$ leaving the fundamental Voronoi region $\mathcal{V}_{\mathbf{0}}(\Lambda)$. As such, the error probability can be written as
\begin{equation}
P_e(\Lambda,r_z) = \mathbb{P}\{\mathbf{z} \notin \mathcal{V}_{\mathbf{0}}(\Lambda)\}.
\end{equation}
A sequence of lattices is good for coding if for any $r_z < r_{\text{eff}}(\Lambda)$, the error probability satisfies
\begin{equation}
\lim\limits_{n\rightarrow \infty} P_e(\Lambda,r_z) = 0.
\end{equation}
\end{defi}
These lattices are also commonly referred to as Poltyrev good \cite{312163}. The existence of such lattices are shown by Loeliger in \cite{641543}.

\begin{defi}(Dithering):
In quantization theory, as well as in some non-linear processing systems, the term
``dithering'' corresponds to intentional randomization aimed at improving the perceptual
effect of the quantization. In the context of lattice quantization or shaping, dither is an effective means of
guaranteeing desired distortion or power levels, independent of the input statistics. Specifically, let $\msf{D}$ be a random dither statistically independent of a lattice codeword $\mathbf{t} \in \Lambda$, known
to both the transmitter and the receiver, uniformly distributed
over $\mathcal{V}_{\mathbf{0}}(\Lambda^{\prime})$. The dithered codeword
\begin{equation}
\msf{X} = [\mathbf{t} - \msf{D}]\mod \Lambda^{\prime},
\end{equation}
is also uniformly distributed over $\mathcal{V}_{\mathbf{0}}(\Lambda^{\prime})$ and is statistically independent of $\mathbf{t}$. As a result,
\begin{equation}
\frac{1}{n}\E [\|\msf{X} \|] = \sigma^2(\Lambda^{\prime}).
\end{equation}
\end{defi}

Introducing the random dither variable $\msf{D}$ is just a tactic to prove many theorems related to the capacity achieving property of lattice codes \cite{Forney03}. This is analogous to the tactic used in \cite{Elias55} to prove that binary linear block codes can achieve the
capacity of a binary input-symmetric channel, namely the introduction of a random translate $\mathcal{C} + \msf{D}$ of a binary
linear block code $\mathcal{C}$ of length $N$, where $\msf{D}$ is a random binary $N$-tuple which is uniform over $\mathbb{F}_2^N$. Very recently, it is shown that lattice codes can achieve the capacity of Gaussian channels without dithering when $\SNR>1$ \cite{8122043}. On the other hand, dither is still necessary in the low SNR regime \cite{Yona10}.

\section{Lattices from Codes}
Lattices can be seen as the generalization of linear codes over a finite field (Hamming
space) to the Euclidean space. In this perspective, we will present some classical ways
of constructing lattices from linear codes. These strategies are employed
in the literature for the achievement of both theoretical and practical results. Among those construction methods, Construction $A$, Construction $D$ and Construction $D'$ are the most three common approaches.

\subsection{Construction $A$}
Let $\mathcal{C}$ be a linear code over $\mathbb{F}_q$ of length $n$, dimension $k$, and rate $R_{\mathcal{C}} = \frac{k}{n}$. The code is generated via
\begin{equation}
\mathcal{C} = \{\mathbf{x} \in \mathbb{Z}_q^n: \mathbf{x} = [\mathbf{b}\mathbf{G}_{\mathcal{C}}]\hspace{-3mm}\mod q, \mathbf{b} \in  \mathbb{Z}_q^k\},
\end{equation}
where $\mathbf{G}_{\mathcal{C}} \in \mathbb{F}_q^{k \times n}$ is the generator matrix of code $\mathcal{C}$.

The lattice obtained by Construction $A$ is
\begin{equation}
\Lambda = \{\boldsymbol{\lambda} \in \mathbb{Z}^n: \boldsymbol{\lambda} \hspace{-3mm} \mod q \in \mathcal{C}\}.
\end{equation}
Moreover, let $\Phi: \mathbb{F}_q \rightarrow \mathbb{Z}$ be the natural embedding of $\mathbb{F}_q$ into $\mathbb{Z}$. Specifically,
\begin{equation}
\Phi(\mathbb{F}_q) = \left\{-\frac{q-1}{2},\ldots, \frac{q-1}{2}\right\}.
\end{equation}
Another way of describing Construction $A$ is
\begin{equation}
\Lambda = \Phi(\mathcal{C})+q\mathbb{Z}^n = \{\boldsymbol{\lambda} \in \mathbb{Z}^n: \boldsymbol{\lambda} = \Phi(\mathbf{c})+q\mathbf{z}, \exists \mathbf{c} \in \mathcal{C}, \mathbf{z} \in \mathbb{Z}^n\}.
\end{equation}
For this lattice, it can be easily seen that
\begin{equation}\label{eq:relation_ch2}
q\mathbb{Z}\subseteq \Lambda \subseteq \mathbb{Z}.
\end{equation}
The above relationship shown in \eqref{eq:relation_ch2} also that
\begin{equation}
d_{E,\min}(\Lambda) = \min\left\{q, \min_{\mathbf{c} \in \mathcal{C}}\|\Phi(\mathbf{c})\| \right\}.
\end{equation}
The volume of the lattice is given by
\begin{equation}
\text{Vol}(\Lambda) = q^{n-k} = q^{n(1-R_{\mathcal{C}})}.
\end{equation}
It is also possible to build the Construction $A$ lattices in $\mathbb{Z}[i]$ and $\mathbb{Z}[\omega]$ domain by using the embedding of $\mathbb{F}_q \rightarrow \mathbb{Z}[i]$ and $\mathbb{F}_q \rightarrow \mathbb{Z}[\omega]$, respectively.

\subsection{Construction $D$}
Construction $D$ involves chains of nested binary linear codes and is employed
to build lattices with a low-complexity iterative decoding algorithm. 

Consider the chain of nested binary linear codes
\begin{equation}\label{eq:cons_D_chain}
\mathcal{C}_L \subseteq \mathcal{C}_{L-1} \subseteq  \ldots \mathcal{C}_1 \subseteq \mathcal{C}_0,
\end{equation}
where $\mathcal{C}_l$ is a length $n$ dimension $k_l$ code for $l \in \{1,\ldots,L\}$ and $\mathcal{C}_0$ is a length $n$ dimension $n$ code. We denote by $\mathbf{c}_1, \mathbf{c}_2,\ldots, \mathbf{c}_{k_l}$ the $k_l$ vectors of $\mathbb{F}_2$ that generate the $l$-th code. The construction $D$ lattice $\Lambda \subseteq \mathbb{R}^n$ with $L+1$ levels is obtained from
\begin{equation}
\Lambda = \left\{\boldsymbol{\lambda} \in \mathbb{Z}^n: \boldsymbol{\lambda} = \mathbf{z}+\sum_{l=1}^L\sum_{j=1}^{k_l} \beta_j^{(l)}2^{L-l}\mathbf{c}_j,\exists \mathbf{z} \in 2^L\mathbb{Z}^n, \beta_j^{(l)} \in \{0,1\}\right\}.
\end{equation}
Another way of describing Construction $D$ lattices is
\begin{equation}
\Lambda = \mathcal{C}_L+2\mathcal{C}_{L-1}+ \ldots+2^{L-2}\mathcal{C}_2+2^{L-1}\mathcal{C}_1+2^L\mathbb{Z}^n.
\end{equation}
Note that Construction $A$ over $\mathbb{Z}$ with $q = 2$ is a particular case of Construction
$D$ with $L = 1$.

\subsection{Construction $D'$}
Construction $D'$ is dual to Construction $D$. The construction is described by using the parity-check matrices of the component linear codes.

Consider the chain of nested binary linear codes as in \eqref{eq:cons_D_chain}. Every one of the codes is generated by
$r_l = n -k_l$ parity-check equations for $l = 1,2,\ldots,L$. Let $\mathbf{h}_1, \mathbf{h}_2 , \ldots \mathbf{h}_{r_L} \in \mathbb{F}^n_2$
be the equations that
generate the smallest code $\mathbb{C}_L$ and suppose that $\mathcal{C}_l$ is generated by $\mathbf{h}_1, \mathbf{h}_2 , \ldots \mathbf{h}_{r_l}$.
This guarantees that the inclusions in \eqref{eq:cons_D_chain} are respected.

The construction $D'$ lattice $\Lambda \subseteq \mathbb{R}^n$ with $L+1$ levels is obtained from
\begin{equation}
\Lambda = \left\{\boldsymbol{\lambda} \in \mathbb{Z}^n: [\mathbf{h}_j\boldsymbol{\lambda}^T] \mod 2^{l+1}= 0, \forall l = 0,1,\ldots,L, \text{and} \; r_{L-l-1}+1\leq j\leq r_{L-l}\right\}.
\end{equation}

\section{Algebraic Number Theory}\label{sec:2_2}
In this section, we introduce the basic concepts of algebraic number theory. We will present only the relevant definitions and results which lead to algebraic lattice constructions.

Let $\mathbb{Z}$ be the set of rational integers $\{\ldots,-2,-1,-,1,2,\ldots \}$ and let $\mathbb{Q}$ be the set of rational number $\mathbb{Q} = \{\frac{a}{b}|a,b \in \mathbb{Z}, b\neq0 \}$.

\subsection{Elementary Concepts}
In this subsection, we introduce some elementary concepts of algebraic
number theory. We will present only the relevant definitions and results
which lead to algebraic lattice constructions.

\begin{defi}(Group):
Let $\mathcal{G}$ be a set equipped with an internal operation (here we use addition which is enough for the purpose of our work) that combines any two elements $a$ and $b$ to form another element, denoted $a+b$. The set $(\mathcal{G},+)$ is a group if
\begin{itemize}
\item For all $a,b \in \mathcal{G}$, the result of the operation $a+b \in \mathcal{G}$.
\item The operation is associative, i.e., $a+ (b + c) = (a + b)+ c$ for all $a, b, c \in \mathcal{G}$.
\item There exists a neutral element 0, such that $0 + a = a + 0$ for all $a \in \mathcal{G}$.
\item For all $a \in \mathcal{G}$, there exists an inverse $-a \in \mathcal{G}$ such that $a - a =
-a + a = 0$.
\end{itemize}
The group $\mathcal{G}$ is said to be Abelian if $a + b = b + a$ for all $a, b \in \mathcal{G}$, i.e., the internal operation is commutative.
\end{defi}

\begin{defi}(Subgroup):
Let $(\mathcal{G},+)$ be a group and $\mathcal{H}$ be a non-empty subset
of G. We say that $\mathcal{H}$ is a subgroup of $\mathcal{G}$ if $(\mathcal{H},+)$ is a group, where $+$ is
the internal operation inherited from $\mathcal{G}$.
\end{defi}

\begin{defi}(Ring):
Let $\mathcal{A}$ be a set equipped with two internal operations $+$ and $\cdot$. The set $(\mathcal{A},+,\cdot)$ is a ring if
\begin{itemize}
\item $(\mathcal{A},+)$ is an Abelian group.
\item The operation $\cdot$ is associative, i.e., $a \cdot (b \cdot c) = (a \cdot b) \cdot c$ for all
$a, b, c \in \mathcal{A}$ and has a neutral element 1 such that $1 \cdot a = a \cdot 1$
for all $a \in \mathcal{A}$.
\item The operation $\cdot$ is distributive over $+$, i.e., $a\cdot(b+c) = a\cdot b+a\cdot c$
and $(a + b) \cdot c = a \cdot c + b \cdot c$ for all $a, b, c \in \mathcal{A}$.
\item There exists a neutral element 0, such that $0 + a = a + 0$ for all $a \in \mathcal{G}$.
\end{itemize}
The ring $\mathcal{A}$ is commutative if $a \cdot b = b \cdot a$ for all $a, b \in \mathcal{A}$.
\end{defi}

\begin{defi}(Field):
Let $\mathcal{A}$ be a commutative ring. The set $(\mathcal{A},+,\cdot)$ is a field if for all $a \in \mathcal{G}$, there exists a multiplicative inverse $a^{-1} \in \mathcal{G}$ such that $a \cdot a^{-1} =
 1$.
\end{defi}

\begin{defi}(Algebraic Number):
Let $\alpha$ be an element of a field $\mathbb{K}$ containing $\mathbb{Q}$, we say that $\alpha$ is
an algebraic number if it is a root of a monic polynomial (whose leading coefficient is 1)
with coefficients in $\mathbb{Q}$.
\end{defi}

\begin{defi}(Algebraic Integer):
We say that $\alpha \in \mathbb{K}$ is an algebraic integer if it is a root of a monic polynomial (whose leading coefficient is 1)
with coefficients in $\mathbb{Z}$.
\end{defi}

\begin{defi}(Field Extension):
Let $\mathbb{K}$ and $\mathbb{L}$ be two fields. If $\mathbb{K} \subseteq \mathbb{L}$, we say that $\mathbb{L}$
is a field extension of $\mathbb{K}$. We denote it $\mathbb{L}/\mathbb{K}$.
\end{defi}

\begin{defi}(Degree):
Let $\mathbb{L}/\mathbb{K}$ be a field extension. The dimension of $\mathbb{L}$ as
vector space over $\mathbb{K}$ is called the degree of $\mathbb{L}$ over $\mathbb{K}$ and is denoted by
$[\mathbb{L} : \mathbb{K}]$. If $[\mathbb{L} : \mathbb{K}]$ is finite, we say that $\mathbb{L}$ is a finite extension of $\mathbb{K}$.
\end{defi}

\begin{defi}(Number Field):
A number field $\mathbb{K} = \mathbb{Q}(\theta)$ is a field extension of $\mathbb{Q}$ of finite degree, where $\theta$ is an algebraic number and also a primitive element, such that the $\mathbb{Q}$-vector space $\mathbb{K}$ is generated by the powers of $\theta$. If this number field has degree $n$, then $\{1,\theta,\ldots,\theta^{n-1}\}$ is a basis for $\mathbb{K}$.
\end{defi}

\subsection{Embedding}
In this subsection, we will see how a number field $\mathbb{K}$ can be represented, we say embedded, into $\mathbb{C}$.

\begin{defi}(Ring of Integers):
Let $\mathbb{K}$ be a number field of degree $n$. The ring of integers of $\mathbb{K}$, denoted by $\mathcal{O}_\mathbb{K}$, is the set of all algebraic integers in number field $\mathbb{K}$ and has rank $n$ (that is, there exists a basis of $n$ elements over $\mathbb{Z}$).
\end{defi}

\begin{defi}(Integral Basis):
Let $\{\omega_1,\ldots,\omega_n \}$ be a basis of $\mathcal{O}_\mathbb{K}$. If for any element of $\mathcal{O}_\mathbb{K}$ can be uniquely expressed as a linear combination of the basis element, i.e., $\sum_{i=1}^n\alpha_i\omega_i$ with $\alpha_i \in \mathbb{Z}$ for $i = 1,\ldots,n$, we say that $\{\omega_1,\ldots,\omega_n \}$ is an integral basis of $\mathbb{K}$.
\end{defi}

\begin{defi}(Ring Homomorphism):
Let $\mathbb{K}/\mathbb{Q}$ and $\mathbb{L}/\mathbb{Q}$ be two field extensions of $\mathbb{Q}$. We
call $\varphi : \mathbb{K} \rightarrow \mathbb{L}$ a $\mathbb{Q}$–homomorphism if $\varphi$ is a ring homomorphism that satisfies $\varphi(\alpha) = \alpha$ for all $\alpha \in \mathbb{Q}$, i.e., that fixes $\mathbb{Q}$. Recall that if $\mathcal{A}$ and
$\mathcal{B}$ are rings, a ring homomorphism is a map $\psi : \mathcal{A} \rightarrow \mathcal{B}$ that satisfies the following for all $a, b \in \mathcal{A}$.
\begin{itemize}
\item $\psi(a+b) = \psi(a)+\psi(b)$.
\item $\psi(a \cdot b) = \psi(a)\cdot \psi(b)$.
\item $\psi(1) = 1$.
\end{itemize}
\end{defi}

\begin{defi}(Embedding):
For the number field $\mathbb{K} = \mathbb{Q}(\theta)$ with digree $n$, there are $n$ distinct $\mathbb{Q}$-homomorphisms $\sigma_j: \mathbb{K} \rightarrow \mathbb{C}$ which is also called the embedding of $\mathbb{K}$ into $\mathbb{C}$. The embedding is defined by $\sigma_j(\theta) = \theta_j$, where $\theta_j$ are the distinct zeros in $\mathbb{C}$
of the minimum polynomial of $\theta$ over $\mathbb{Q}$.
\end{defi}
For any $\varsigma = a_0 +a_1 \theta + \ldots+ a_{n-1} \theta^{n-1} \in \mathbb{K}$, the embedding of $\varsigma$ into $\mathbb{C}$ is given by
\begin{align}
\sigma_j(\varsigma) = \sigma_j\left(\sum_{i=0}^{n-1} a_i \theta^i \right)
= \sum_{i=0}^{n-1} \sigma_j(a_i) \sigma_j(\theta)^i, j\in \{1,\ldots,n \} .
\end{align}

\begin{defi}(Discriminant):
Let $\{\omega_1,\ldots,\omega_n \}$ be an integral basis of $\mathbb{K}$. We define the discriminant of $\mathbb{K}$ as
\begin{align}
d_\mathbb{K} = \det
\begin{pmatrix}
\sigma_1(\omega_1) & \sigma_2(\omega_1)  & \cdots & \sigma_n(\omega_1)   \\
    \sigma_1(\omega_2) & \sigma_2(\omega_2) & \cdots & \sigma_n(\omega_2)   \\
    \vdots & \vdots & \ddots & \vdots  \\
    \sigma_1(\omega_n) & \sigma_2(\omega_n) & \cdots & \sigma_n(\omega_n)
\end{pmatrix}^2.
\end{align}
\end{defi}

\begin{defi}(Signature):
The signature of $\mathbb{K}$ is denoted by $(r_1,r_2)$ if among those $n = r_1+2r_2$ $\mathbb{Q}$-homomorphisms, there are $r_1$ real $\mathbb{Q}$-homomorphisms, i.e., $\sigma_1,\ldots,\sigma_{r_1}$, and $r_2$ pairs of complex $\mathbb{Q}$-homomorphisms, i.e., $\sigma_{r_1},\ldots,\sigma_n$, where $\sigma_{r_1+r_2+i}$ is the conjugate of $\sigma_{r_1+i}$ for $i \in \{1,\ldots,r_2\}$.
\end{defi}

\begin{defi}(Totally Real Number Field):
A number field is said to be totally real if it has signature $(r_1,r_2) = (n,0)$, i.e., $r_2 = 0$.
\end{defi}

\begin{defi}(Totally Complex Number Field):
A number field is said to be totally complex if it has signature $(r_1,r_2) = (0,\frac{n}{2})$, i.e., $r_1 = 0$.
\end{defi}

\begin{defi}(Algebraic Norm):
The algebraic norm of $\varsigma$ given above is given by $N(\varsigma) = \prod_{i=1}^n \sigma_i(\varsigma)$.
\end{defi}

\begin{defi}\label{def:CE}(Canonical Embedding):
The canonical embedding $\Psi:\mathbb{K} \rightarrow \mathbb{R}^{r_1} \times \mathbb{C}^{r_2}$ is a ring homomorphism defined by
\begin{align}
\Psi(\varsigma) = [\sigma_1(\varsigma),\ldots,\sigma_{r_1}(\varsigma),\sigma_{r_1+1}(\varsigma),\ldots,\sigma_{r_1+r_2}(\varsigma)] \in \mathbb{R}^{r_1} \times \mathbb{C}^{r_2}.
\end{align}
If we identify $\mathbb{R}^{r_1} \times \mathbb{C}^{r_2}$ with $\mathbb{R}^n$, the canonical embedding can be
rewritten as $\Psi:\mathbb{K} \rightarrow \mathbb{R}^n$
\begin{align}\label{eq:embed_R}
\Psi(\varsigma) &= [\sigma_1(\varsigma),\ldots,\sigma_{r_1}(\varsigma),\Re(\sigma_{r_1+1}(\varsigma)),\Im(\sigma_{r_1+1}(\varsigma)), \nonumber \\
&\ldots,\Re(\sigma_{r_1+r_2}(\varsigma)),\Im(\sigma_{r_1+r_2}(\varsigma))] \in \mathbb{R}^n.
\end{align}
\end{defi}
The canonical embedding gives a geometrical representation of a number
field, the one that will serve our purpose.

\subsection{Algebraic Lattices}
We are now ready to introduce algebraic lattices. The definition of
canonical embedding (Definition \ref{def:CE}) establishes a one-to-one correspondence
between the elements of an algebraic number field of degree
$n$ and the vectors of the $n$-dimensional Euclidean space. The final step
for constructing an algebraic lattice is given by the following result.

\begin{defi}(Algebraic Lattice):
Let $\{\omega_1,\ldots,\omega_n \}$ be an integral basis of $\mathbb{K}$. An algebraic lattice $\Lambda  = \Psi(\mathcal{O}_\mathbb{K})$ is a lattice in $\mathbb{R}^{r_1} \times \mathbb{C}^{r_2} \cong \mathbb{R}^n$ with a generator matrix
\begin{align}
&\mathbf{G}_{\Lambda} = \begin{pmatrix}
\Psi(\omega_1) \\
\vdots \\
\Psi(\omega_n) \\
\end{pmatrix}  \nonumber \\
& \overset{\eqref{eq:embed_R}}=  \begin{pmatrix}
\sigma_1(\omega_1),\ldots,\sigma_{r_1}(\omega_1),\Re(\sigma_{r_1+1}(\omega_1)),\Im(\sigma_{r_1+1}(\omega_1)),\ldots,\Im(\sigma_{r_1+r_2}(\omega_1)) \\
\vdots \\
\sigma_1(\omega_n),\ldots,\sigma_{r_1}(\omega_n),\Re(\sigma_{r_1+1}(\omega_n)),\Im(\sigma_{r_1+1}(\omega_n)),\ldots,\Im(\sigma_{r_1+r_2}(\omega_n))  \\
\end{pmatrix}
\end{align}
\end{defi}
The volume of the lattice is given by
\begin{equation}
\text{Vol}(\Lambda) = 2^{-2r_2}|d_{\mathbb{K}}|,
\end{equation}
where $d_{\mathbb{K}}$ is the discriminant of $\mathbb{K}$.
Before going further, let us take some time to emphasize the correspondence
between a lattice point $\boldsymbol{\lambda} \in \Lambda \subset \mathbb{R}^n$ and an algebraic integer in
$\mathcal{O}_\mathbb{K}$. A lattice point is of the form
\begin{align}
\boldsymbol{\lambda}  &= [\lambda_1,\ldots,\lambda_{r_1},\lambda_{r_1+1},\ldots,\lambda_{r_1+r_2}] , \nonumber \\
& = [\sigma_1(\varsigma),\ldots,\sigma_{r_1}(\varsigma),\Re(\sigma_{r_1+1}(\varsigma)),\Im(\sigma_{r_1+1}(\varsigma)),\ldots,\Re(\sigma_{r_1+r_2}(\varsigma)),\Im(\sigma_{r_1+r_2}(\varsigma))] \nonumber \\
& = \Psi(\varsigma),
\end{align}
for some $\varsigma = \sum_{i=1}^n \alpha_i \omega_i \in \mathcal{O}_{\mathbb{K}}$ with $\alpha_i \in \mathbb{Z}$ for $i = 1,\ldots,n$.

\begin{defi}(Ideal):
Let $\mathbb{K}$ be a number field of degree $n$ and
$\mathcal{O}_{\mathbb{K}}$ its ring of integers. An ideal $\mathcal{I} \subseteq \mathcal{O}_{\mathbb{K}}$ is that for every $\alpha \in \mathcal{O}_{\mathbb{K}}$ and $b \in \mathcal{I}$ we have $ab \in \mathcal{I}$,
briefly $a\mathcal{I} \subset \mathcal{I}$ and $b\mathcal{O}_{\mathbb{K}} \subset \mathcal{I}$.
\end{defi}

\begin{defi}(Principal ideal):
An ideal $\mathcal{I}$ is called principal if $\mathcal{I}= \alpha\mathcal{O}_{\mathbb{K}}$ for some algebraic integer $\alpha$, in this case we also
denote $\mathcal{I} = (\alpha)$.
\end{defi}
For a principal ideal $\mathcal{I} = (\alpha)\mathcal{O}_{\mathbb{K}}$ of $\mathcal{O}_{\mathbb{K}}$, its norm is computed as $N(\mathcal{I}) = |N(\alpha)|$. Otherwise, if it is not principal, the norm is the cardinality $|\mathcal{O}_{\mathbb{K}}/\alpha|$.

\begin{defi}(Ideal Lattice):
For a totally real number field $\mbb{K}$ of degree $n$ and an ideal $\mathcal{I} \subseteq \mathcal{O}_\mathbb{K}$ with an integral basis $\{\beta_1,\ldots, \beta_n\}$, the corresponding ideal lattice is given by $\Lambda = \Psi(\mathcal{I})$ which has the generator matrix
\begin{equation}\label{eq:G_ideal}
\mathbf{G}_{\Lambda} = \begin{pmatrix}
\Psi(\beta_1) \\
\Psi(\beta_2) \\
\vdots \\
\Psi(\beta_n) \\
\end{pmatrix} \cdot \begin{pmatrix}
\sqrt{\sigma_1(\varsigma)} & 0  & \cdots & 0   \\
    0 & \sqrt{\sigma_2(\varsigma)} & \cdots & 0   \\
    \vdots & \vdots & \ddots & \vdots  \\
    0 & 0 & \cdots & \sqrt{\sigma_n(\varsigma)}
\end{pmatrix}.
\end{equation}
\end{defi}
We can think of the diagonal matrix in \eqref{eq:G_ideal} as
a pre-fading, used to stretch an algebraic lattice into another, such as
the $\mathbb{Z}^n$ lattice.

\begin{defi}(Diversity):
A scheme is said to achieve a diversity order of $n$ if the average error probability satisfies
\begin{equation}
\lim_{\SNR \rightarrow \infty} \frac{\log(P_e(\SNR))}{\log(\SNR)} = -n.
\end{equation}
The diversity of an $n$-dimensional lattice $\Lambda$ is defined by
\begin{equation}
div(\Lambda) = \min_{\mathbf{0}\neq \boldsymbol{\lambda} \in \Lambda} |\{i | \lambda_i \neq 0, i = 1,\ldots,n \}|.
\end{equation}
\end{defi}

\begin{defi}(Minimum Product Distance):
Let $\Lambda$ be a lattice in $\mathbb{R}^n$. If $\Lambda$ has diversity $l\leq n$, we define its minimum product distance by
\begin{equation}
d_{p,\min}(\Lambda) = \min_{\boldsymbol{\lambda} \neq \boldsymbol{\lambda}'\in \Lambda} \prod |\lambda_i - \lambda'_i|,
\end{equation}
or equivalently, since we may consider the distance of $\boldsymbol{\lambda} = [\lambda_1,\ldots,\lambda_n]$ from the origin, by
\begin{equation}
d_{p,\min}(\Lambda) = \min_{\mathbf{0} \neq \boldsymbol{\lambda}\in \Lambda} \prod |\lambda_i|,
\end{equation}
where both products are taken over the $l$ non-zero components of the vectors.
\end{defi}

It is shown in \cite{485720,681321} that codes carved from ideal lattices of totally real number fields attain the full diversity. Moreover, the minimum product distance of the codes thus can be easily guaranteed by the norm of the ideal $\mathcal{I}$.

\section{Summary}
In this chapter, we present the background materials on lattices. The main points presented in this chapter are summarized as follows.
\begin{itemize}
\item We quickly overview basic lattice definitions and some properties, with the main
intention of giving some simple examples and getting used with some objects
which will be used in the later chapters.
\item We introduce the figures of merit of lattices to show how ``good'' a lattice can be.
\item We present several classical methods of constructing lattices from linear codes.
\item We provide some basic knowledge and definitions on algebraic number theory related to lattice construction.
\item We also show the construction for several families of lattices that based on algebraic number field.
\end{itemize}

\chapter{Wireless Communications and Channel Coding}\label{C3:chapter3}

%
\section{Introduction}
In this chapter, we first introduce some basics of wireless communications, including different types of channel models, channel characteristics and the capacity. We then introduce the fundamental background of coding theories and techniques from point-to-point channels to multi-antenna channels. The materials in this chapter serve as the technical guidelines to provide the necessary background to understand the works in the later chapters. The contents are summarized from \cite{Cover:2006:EIT:1146355,tse_book,Richardson:2008:MCT:1795974,johnson_2009,Lin09} and no new results are presented.

\section{Binary Input Memoryless Channels}
We start with the point-to-point communication channel model with binary channel inputs.

\begin{defi}(Discrete Channel):
A discrete channel is one that transmits a symbol $x$ from a discrete set $\mathcal{X} = \{X_1,\ldots,X_l\} $, known as the source alphabet, and returns a symbol $y$ from
another (possibly different) discrete alphabet, $\mathcal{Y} = \{Y_1,\ldots,Y_m \}$.
\end{defi}

A communication channel can be modeled as a random process. For a given
symbol $x[t]$ transmitted at time $t$, such that $x[t]$ is one of the symbols from the set
$\mathcal{X}$, i.e. $x[t] = X_j \in \mathcal{X}= \{X_1,\ldots,X_l\}$, the channel transition probability $p(y|x) =
p(y = Y_j |x = X_j )$ gives the probability that the returned symbol $y[t]$ at time $t$ is
the symbol $Y[t]$ from the set $\mathcal{Y}$ , i.e. $y[t] = Y_j \in \mathcal{Y}= \{Y_1,\ldots,Y_m \}$.

\begin{defi}(Memoryless Channel):
A channel is said to be memoryless if the channel output at any time instant depends only on the input
at that time instant, not on previously transmitted symbols. More precisely, for
a sequence of transmitted symbols $\mathbf{x} = [x[1], \ldots, x[N] ]$ and received symbols
$\mathbf{y} = [y[1], \ldots, y[N] ]$, a memoryless channel is therefore completely described by its input and output
alphabets and the conditional probability distribution $p(x[t]|y[t])$ for each input–
output symbol pair.
\begin{equation}
p(\mathbf{y}|\mathbf{x}) = \prod_{t=1}^N p(x[t]|y[t]).
\end{equation}
\end{defi}
The discrete memoryless channels are considered in this thesis.

\subsection{Binary Erasure Channel}
In the BEC, the channel input at time $t$ is binary, i.e., $x[t] \in \{0,1\}$. The corresponding channel output $y[t]$
takes on values in the alphabet $\{0, 1, ?\}$, where $?$ indicates an erasure. Each transmitted
bit is either erased with probability $\epsilon$, or received correctly: $y[t] = \{x[t], ?\}$ and $\mathbb{P}\{y[t] = ?\} = \epsilon$. Erasure occurs for each $t$ independently.

It is easy to see that the capacity of the BEC is \cite[Chapter 3.1]{Richardson:2008:MCT:1795974}
\begin{equation}
C_{\text{BEC}}(\epsilon) = 1-\epsilon,
\end{equation}
in bits per channel use.

The BEC can be used to model data networks, where packets either
arrive correctly or are lost due to buffer overflows or excessive delays \cite[Chapter 3]{Richardson:2008:MCT:1795974}.

\subsection{Binary Symmetric Channel}
In the BSC, the channel input at time $t$ is binary, i.e., $x[t] \in \{0,1\}$. The corresponding channel output $y[t]$
is also binary, i.e., $y[t] \in \{0, 1\}$. Each transmitted
bit is either flipped with probability $\epsilon$, or received correctly. The parameter $\epsilon$ is called the crossover probability of the channel. Moreover, the output-symmetric property leads to
\begin{align}
&p(y[t] = 0|x[t] = 1) = p(y[t] = 1|x[t] = 0) = \epsilon, \\
&p(y[t] = 0|x[t] = 0) = p(y[t] = 1|x[t] = 1) = 1 - \epsilon.
\end{align}

The capacity of the BSC is \cite[Chapter 1.2.3]{johnson_2009}
\begin{equation}
C_{\text{BSC}}(\epsilon) = 1+(\epsilon \log_2 \epsilon + (1 - \epsilon) \log_2(1 - \epsilon)),
\end{equation}
in bits per channel use.

The BSC channel can be used to model the communication channel in optical fibre \cite{Smith12} as well as general storage systems where the channel outputs are in the form
of hard-decision results \cite[Section II]{Cho14}. This model will appear in the work of designing product codes for storage systems in Chapter 8.

\subsection{Binary Additive White Gaussian Noise Channel}
In the BI-AWGN channel, the channel input at time $t$ is $x[t] \in \{+1,-1\}$. The corresponding channel output $y[t]$
is real-valued. More precisely, the input-output relationship is described by
\begin{equation}
y[t] = x[t]+z[t],
\end{equation}
where $z[t]\sim \mathcal{N}(0, \sigma^2)$ is a Gaussian random variable with zero mean and variance $\sigma^2$. Each transmitted bit is corrupted by AWGN. The probability density function for $z$ is
\begin{equation}
p_{\msf{Z}}(z) = \frac{1}{\sqrt{2\pi\sigma^2}}e^{-\frac{z^2}{2\sigma^2}}.
\end{equation}

The capacity of the BI-AWGN chanel is \cite[Example 4.38]{Richardson:2008:MCT:1795974}
\begin{equation}
C_{\text{BI-AWGN}}(\sigma) = \int_{-1}^{+1}\frac{\sigma}{\sqrt{2\pi}(1-y^2)}e^{-\frac{(1-\sigma^2\tanh^{-1}(y))^2}{2\sigma^2}}\log_2(1+y)dy,
\end{equation}
in bits per channel use.

\subsection{Unconstrained Additive White Gaussian Noise Channel}
In contrast to the BI-AWGN channel, we define the unconstrained AWGN channel as follows.
In the unconstrained AWGN channel, the input is not restricted to any signal constellation. Given the power constraint of the input signal $P$, the capacity is \cite[Summary 5.1]{tse_book}
\begin{equation}\label{ch3_AWGN}
C_{\text{AWGN}}(\sigma) = \frac{1}{2}\log_2\left(1+\frac{P}{\sigma^2}\right),
\end{equation}
in bits per channel use.

This model is the channel model considered in Chapter 4.


\section{Fading Channels}
We now look some general channel models beyond the above basic models. In a typical wireless communication scenario, the transmitted signal will be affected by both AWGN and fading attenuations. The general term fading is used to describe fluctuations in the envelope of a transmitted radio signal. Based on the variations of the channel strength over time and over frequency, it is generally divided into two types \cite[Chapter 2]{tse_book}:
\begin{itemize}
\item Large-scale fading, due to path loss of signal as a function of distance
and shadowing by large objects such as buildings and hills. This occurs as
the mobile moves through a distance of the order of the cell size, and is
typically frequency independent.
\item Small-scale fading, due to the constructive and destructive interference of the
multiple signal paths between the transmitter and receiver. This occurs at the
spatial scale of the order of the carrier wavelength, and is frequency dependent.
\end{itemize}
Large-scale fading is more relevant to issues such as
cell-site planning. Small-scale multipath fading is more relevant to the design
of reliable and efficient communication systems, which is considered in this thesis.

Now we introduce some important characteristics of the wireless channel.

\begin{defi}(Coherence Bandwidth):
Coherence bandwidth is a statistical measurement of the range of frequencies such that the approximate maximum bandwidth or frequency interval over which two frequencies of a signal are likely to experience comparable or correlated fading \cite[Chapter 3.3.2]{Goldsmith:2005:WC:993515}.
\end{defi}

\begin{defi}(Coherence Time):
The coherence time of a channel is defined as the time interval after which the channel
impulse response decorrelates \cite[Chapter 3.3.3]{Goldsmith:2005:WC:993515}.
\end{defi}

Based on the coherence bandwidth, the wireless channel model can be further divided into frequency-selective channels and flat fading channels.

\subsection{Frequency-Selective Fading Channel}
If the bandwidth of the transmitted signal is larger than the channel coherence bandwidth, then the channel amplitude
values of the received signal at frequencies separated by more than the coherence bandwidth are roughly independent. Thus, the channel
amplitude varies widely across the signal bandwidth. In this case, the channel is called frequency-selective \cite[Chapter 3.3.2]{Goldsmith:2005:WC:993515}. When
this occurs, the received signal includes multiple versions of the transmitted waveform which are
attenuated (faded) and delayed in time, and hence the received signal is distorted. As a result
of that, the channel induces inter symbol interference (ISI). The input-output relationship of the frequency-selective channel is given by
\begin{equation}
y[t] = \sum_{l = 0}^{L-1}h_lx[t-l]+z[t],
\end{equation}
where in this case the channel has $L$ taps.

\subsection{Flat Fading Channel}
If the wireless channel has a constant channel gain and the bandwidth of the transmitted signal is less than channel coherence bandwidth, the channel is usually referred to as flat fading \cite[Chapter 2.3.2]{tse_book}. In other words, fading across the entire
signal bandwidth is highly correlated, i.e. the fading is roughly equal across the entire signal bandwidth. The input-output relationship of the channel is given by
\begin{equation}
y[t] = h[t]x[t]+z[t],
\end{equation}
where $h[t]$ is the fading coefficient at time $t$. Here, we do
not specify the dependence between the fading coefficients $h[t]$ at different
times $t$.

Based on the channel coherence time, the wireless channel model can be further divided into slow fading channels, block fading channels and fast fading channels \cite[Chapter 2.3.1]{tse_book}.

\subsection{Slow Fading Channel}
Slow fading arises when the coherence time of the channel is much larger than the delay requirement of the application. In this regime, the amplitude and phase change imposed by the channel can be considered constant over the transmission duration of the packet \cite[Chapter 5.4.1]{tse_book}. This channel is also sometimes referred to as quasi-static fading. The input-output relationship is given by
\begin{equation}
y[t] = hx[t]+z[t],
\end{equation}
where the channel coefficient $h$ remains constant for $t = 1,\ldots,T$, i.e., the whole data packet length.

Slow fading is considered when we design new multiple access scheme in Chapter 5.



\subsection{Fast Fading Channel}
Fast fading occurs when the coherence time of the channel is much smaller than the delay requirement of the application. In this case, the amplitude and phase change imposed by the channel varies considerably over the period of use \cite[Chapter 5.4.5]{tse_book}. The channel model at time $t$ is given by
\begin{equation}\label{eq:ff_model}
y[t] = h[t]x[t]+z[t].
\end{equation}

\subsection{Block Fading Channel}
Block fading occurs when the channel coherence time is smaller than the data packet length. However, the channel is constant for a number of symbol interval. The channel model is given by
\begin{equation}\label{eq:bf_model}
y[t] = h[t]x[t]+z[t],
\end{equation}
where $h[t] =h_l$ remains constant over the $l$-th coherence period of $T$ symbols and is i.i.d. across different coherence periods \cite[Chapter 5.4.5]{tse_book}.

This model is considered in our research work in Chapter 7.

%

\section{Non-Orthogonal Multiple Access}
In this section, we introduce multiuser communications. In particular, we will focus on the downlink communication scenario where a single transmitter (the base-station) attempts to communication information simultaneously to multiple users. The fundamental concept of NOMA is to realize the downlink multiple access technologies from the power domain. The key enabling technologies for current power-domain NOMA is based on two principles, namely, superposition coding and SIC \cite[Chapter 15.6.3]{Cover:2006:EIT:1146355}.

\subsection{Superposition Coding}
At the transmitter side, the transmit signal is the (linear) superposition of the signals of all the users. It was first proposed in \cite{1054727} and was proved to be optimal compared to time-sharing. It is one of the fundamental building blocks of coding schemes to achieve the capacity on the scalar Gaussian broadcast channel. In fact, it has been shown that superposition coding is capable of achieving the capacities of general degraded broadcast channel \cite{Cover:2006:EIT:1146355}.

\begin{defi}(Degraded Broadcast Channel):
Consider a broadcast channel with one input alphabet $\mathcal{X}$ and
two output alphabets, $\mathcal{Y}_1$ and $\mathcal{Y}_2$. This broadcast channel is said to be physically degraded if the channel transition probability satisfies \cite[Chapter 15.6.2]{Cover:2006:EIT:1146355}
\begin{equation}
p(y_1, y_2|x) = p(y_1|x)p(y_2|y_1).
\end{equation}
\end{defi}

Compared to orthogonalization schemes, superposition coding can provide a very
reasonable rate to the strong user, while achieving close to the single-user bound for the weak user. Intuitively, the strong
user, being at a high SNR, is degree-of-freedom limited and superposition coding allows it to use the full degrees of freedom of the channel while being
allocated only a small amount of transmit power, thus causing small amount of interference to the weak user. In contrast, an orthogonal scheme has to
allocate a significant fraction of the degrees of freedom to the weak user to
achieve near single-user performance, and this causes a large degradation in
the performance of the strong user.

\subsection{Successive Interference Cancellation}
SIC plays an important role in achieving the capacities of downlink NOMA. For an SIC receiver, it first decodes other users' signals one by one based on a decoding order before decoding its own signal. Upon finishing decoding one user's signal, the receiver subtracts it from the received signal. As a result, the interference can be successfully removed and the achievable data rate is improved. In general, users with better channel conditions can perform SIC to mitigate the inter-user interference. Due to its advantages, SIC is also employed in practical systems such as CDMA \cite{298053} and vertical-bell laboratories layered space-time (V-BLAST) \cite{738086}.

However, there are several potential practical issues in using SIC in a wireless system \cite[Discussion 6.1]{tse_book}.
\begin{itemize}
\item Complexity scaling with the number of users: In the downlink, the use of SIC at the mobile means that it now has to decode
information intended for some of the other users. Then the complexity at each mobile
scales with the number of users in the cell.

\item Error propagation: Capacity analysis assumes error-free decoding. However, with actual codes, decoding errors do occur. Once an error occurs for
a user, this error can propagate to the decoders for all the users later in the SIC decoding order. This will affect the decoding error probabilities of the network.

\item Imperfect channel estimation: SIC order heavily relies on accurate channel estimation. Imperfect channel estimation will affect the SIC ordering and may lead to SIC failure. In such a case, error propagation will also occur.

\item Analog-to-digital quantization error: When the received powers of
the users are very disparate, the analog-to-digital (A/D) converter needs
to have a very large dynamic range, and at the same time, enough
resolution to quantize accurately the contribution from the weak signal.
\end{itemize}

\subsection{Multiuser Capacity Region}
Consider a $K$-user scalar Gaussian broadcast channel where both the transmitter and receivers have full CSI. The baseband channel model is
\begin{equation}
y_k[t] = h_kx[t]+z_k[t], k = 1,\ldots,K, t = 1,\ldots,T,
\end{equation}
where $y_k[t]$ is the $k$-th user's received signal at time $t$; $h_k$ is the channel coefficient of the channel between the base station and the $k$-th user; $x[t]$ is the superposition coded message broadcasted by the base station and satisfies the total power constraint at the base station, i.e., $\frac{1}{T}\sum_{t=1}^T|x[t]|^2 \leq P$; and $z_k[t] \sim \mathcal{N}(0,\sigma^2)$ is the i.i.d. AWGN experienced at user $k$. Without loss of generality, we assume that the channel gain follows
\begin{equation}
|h_1|^2 \geq \ldots \geq |h_K|^2.
\end{equation}
The multiuser capacity region is the closure of the rate tuple \cite[Chapter 14.5]{Goldsmith:2005:WC:993515}
\begin{equation}\label{ch3_bc_cap}
R_k = \frac{1}{2}\log_2\left( 1+\frac{P_k|h_k|^2}{\sum_{i=1}^{k-1}P_i|h_k|^2+\sigma^2} \right), k = 1,\ldots,K,
\end{equation}
in bits/s/Hz/real dimension for all possible splits $\sum_{k=1}^K P_k = P$ of the total power at the base station, where $P_k \in [0,P]$ is the power allocation for user $k$.

In contrast to OMA such as TDMA, the rate region is given by
\begin{equation}
R_k' = \frac{\alpha_k}{2}\log_2\left( 1+\frac{P|h_k|^2}{\sigma^2} \right), k = 1,\ldots,K,
\end{equation}
where $\alpha_k \in [0,1]$ is the time-sharing parameter for user $k$ and $\sum_{k=1}^K\alpha_k$ = 1. The TDMA region is strictly inside the capacity region of the broadcast channel.

\section{Linear Deterministic Model}
Here, we introduce the linear deterministic model \cite{Avestimehr11} which is used for modeling our downlink broadcast channel. The deterministic model allows us to characterize the capacity region of the broadcast channel approximately by considering an appropriate finite-field model of the broadcast channel. Coding schemes can then be designed according to the insight obtained from this relatively simple model. As we will see later that the deterministic model is employed for solving complex downlink communication problems in Chapter 5 and Chapter 6.

\subsection{Modeling Signal Strength}
Consider the real scalar Gaussian model for a point to point link
\begin{equation}\label{eq:dmod_1}
y = hx+z,
\end{equation}
where $x,y,h \in \mathbb{R}$ correspond to the channel input, output and the channel gain, respectively; $z \sim \mathcal{N}(0,1)$ is the noise; and the transmitter has the power constraint $\mathbb{E}[|x|^2]\leq 1$. Note that here both of the transmit power and noise power are normalized to 1. The channel gain $h$ is related to SNR, i.e., $|h| = \sqrt{\SNR}$.

The model in \eqref{eq:dmod_1} can be written as
\begin{align}
y &= 2^{\frac{1}{2}\log_2\SNR}\sum_{i=1}^{+\infty}x(i)2^{-i}+\sum_{i=-\infty}^{+\infty}z(i)2^{-i} \label{eq:dmod_2} \\
& \overset{(a)}= 2^{\frac{1}{2}\log_2\SNR}\sum_{i=1}^{+\infty}x(i)2^{-i}+\sum_{i=1}^{+\infty}z(i)2^{-i} \label{eq:dmod_3} \\
& \overset{(b)} \approx 2^n\sum_{i=1}^{+\infty}x(i)2^{-i}+\sum_{i=1}^{+\infty}(x(i+n)+z(i))2^{-i} \label{eq:dmod_4},
\end{align}
where $(a)$ follows by assuming the noise has a peak power equal to 1 and $(b)$ follows that
\begin{equation}\label{eq:dmod_5}
n = \lceil \frac{1}{2}\log_2\SNR\rceil^+.
\end{equation}
If the 1 bit of the carry-over from the second summation to the first summation in \eqref{eq:dmod_4} is ignored, the point-to-point Gaussian channel can be approximated as a pipe that only passes the bits above the noise level and truncates the bits below the noise level. Therefore, think of the transmitted signal $x$
as a sequence of bits at different signal levels, with the highest
signal level in being the most significant bit and the lowest
level being the least significant bit. As such, the receiver can see the $n$ most significant bits of $x$ without any
noise and the rest are not seen at all.

The capacity of this deterministic channel is thus described by \eqref{eq:dmod_5}. This capacity is within $\frac{1}{2}$-bit approximation of the capacity of the AWGN channel in \eqref{ch3_AWGN}. In the case of complex Gaussian channel, $n = \lceil \log_2\SNR\rceil^+$ and the approximation is within 1-bit of the true AWGN capacity \cite{Avestimehr11}.

\subsection{Modeling Broadcast}
Consider the real scalar Gaussian broadcast channel with two receivers. The received SNR at receiver $k$ is denoted by $\SNR_k$ for $k = 1,2$ and we assume $\SNR_1 \geq \SNR_2$ without loss of generality. The Gaussian broadcast channel is deterministically modeled as follows:
\begin{itemize}
\item Receiver 2 (weak user) receives only the most significant $n_2$ bits of $x$.
\item Receiver 1 (strong user) receives only the most significant $n_1$ bits of $x$, and $n_1 > n_2$.
\end{itemize}
The $n_2$ bits in the deterministic model can be decoded by both users while the remaining $n_1-n_2$ bits can only be decoded by the strong user. The capacity of this model is then given by \cite{Avestimehr11}
\begin{equation}
n_k = \lceil \frac{1}{2}\log_2\SNR_k\rceil^+, k = 1,2.
\end{equation}
The gap between the capacity region of the deterministic model and that of the Gaussian broadcast channel model \eqref{ch3_bc_cap} is within 1 bit for each user. However, this is only the
worst-case gap and in the typical case where channel difference is large, the gap is much smaller than 1 bit.


\section{Channel Coding}
Shannon's 1948 work shows that it is possible to transmit digital data with arbitrarily high reliability,
over noise-corrupted channels, by encoding the digital message with an error
correction code prior to transmission and subsequently decoding it at the receiver \cite{6773024}. The transmitted symbols may be corrupted in some way
by the channel, and it is the function of the error correction decoder to use the
added redundancy to determine the transmitted message despite the
imperfect reception. In this section, we introduce some codes design for the AWGN channel.

First, we give some useful definitions \cite[Chapter 1.4]{Richardson:2008:MCT:1795974} in the following.

\begin{defi}(Code):
A code $\mathcal{C}$ of length $n$ and cardinality $M$ over a field $\mathbb{F}_q$ is a
collection of $M$ elements from $\mathbb{F}_q^n$, i.e.,
\begin{equation}
\mathcal{C} = \{x[1], . . . , x[M]\}, x[m] \in \mathbb{F}_q^n, 1 \leq m \leq M.
\end{equation}
The code is linear if for any $x[m]\neq x[k]$, we have $x[m]+x[k]\mod q \in \mathcal{C}$. The code rate is given by
\begin{equation}
R(\mathcal{C}) = \frac{\log_2(M)}{n}.
\end{equation}
It is measured in bits per transmitted symbol.
\end{defi}

For this thesis, we only consider linear codes.

\begin{defi}(Hamming Weight)
The Hamming weight of a codeword $\mathbf{u}$, which we denote by $w_H(\mathbf{u})$, is equal to the number of
non-zero symbols in $\mathbf{u}$, i.e., the cardinality of the support set.
\end{defi}

\begin{defi}(Hamming Distance)
Given two codewords $\mathbf{u}$ and $\mathbf{v}$, the Hamming distance
of a pair $(\mathbf{u},\mathbf{v})$, which we denote by $d_H(\mathbf{u},\mathbf{v})$, is the number of positions in which $\mathbf{u}$
differs from $\mathbf{v}$. We have
\begin{equation}
d_H(\mathbf{u},\mathbf{v}) = d_H(\mathbf{u}-\mathbf{v}, 0) = w_H(\mathbf{u}-\mathbf{v}).
\end{equation}
Further, $d(\mathbf{u},\mathbf{v}) = d(\mathbf{v},\mathbf{u})$
and $d(\mathbf{u},\mathbf{v}) \geq 0$, with equality if and only if $\mathbf{u} = \mathbf{v}$.
\end{defi}

In what follows, we give some specific codes related to the work in this thesis.

\subsection{BCH Codes}
BCH codes are a class of cyclic codes such that a cyclic shift of a codeword is still a valid codeword \cite[Chapter 3.3]{Lin09}. BCH codes are specified in terms of the roots of their generator polynomials in finite fields.

Given two positive integers $m$ and $d$ such that $d \leq 2^m - 1$, a primitive narrow-sense BCH code over the Galois field $\mathbb{F}_2$ with code length $n = 2^m -1$ and minimum distance at least $d$ is constructed by the following method.

Let $\alpha$ be a primitive element of $\mathbb{F}_{2^m}$. For any positive integer $i$, let $\phi_i(x)$ be the minimal polynomial of $\alpha^i$. The generator polynomial of the BCH code is defined as the least common multiple (LCM)
\begin{equation}
\mathbf{g}(x) = \text{LCM}\{\phi_1(x),\ldots,\phi_{d-1}(x)\}.
\end{equation}
The error correction capability is $t = \lfloor \frac{d-1}{2}\rfloor$ \cite[Chapter 6.2]{Lin:2004:ECC:983680}.

For any information sequence $\mathbf{m} = [m_0,\ldots,m_{k-1}]$, the polynomial representation of $\mathbf{m}$ is
\begin{equation}
\mathbf{m}(x) = m_0 + m_1x + \ldots + m_{k-1}x^{k-1}.
\end{equation}
The polynomial representation of codeword is then generated by
\begin{align}
\mathbf{c}(x) &= \mathbf{m}(x)\mathbf{g}(x) 
 = c_0 + c_1x + \ldots + c_{n-1}x^{n-1},
\end{align}
where $\mathbf{c} = [c_0,\ldots,c_{n-1}]$ is the codeword.

BCH codes can be efficiently decoded by Berlekamp–Massey algorithm \cite{Berlekamp:2015:ACT:2834146}, which realizes the bounded distance decoding (in Section \ref{sec:BDD_Decoding}).

\subsection{Low Density Parity-Check Codes}
LDPC codes are a class of linear block codes with near-capacity performance \cite{Richardson:2008:MCT:1795974,Lin09,johnson_2009}. As their name suggests, LDPC codes are block codes with parity-check matrices that contain only a very small number of non-zero entries. This sparseness is essential for an iterative decoding complexity that increases only linearly with the code length \cite[Chapter 5]{Lin09}. The parity-check matrix of an LDPC code can be represented by a Tanner graph \cite{1056404}.

\begin{defi}(Tanner Graph):
The Tanner graph consists of two sets of nodes: nodes for the
codeword bits (called variable nodes (VNs)), and nodes for the parity
-check equations
(called check nodes (CNs)). An edge joins a variable node to a check node if that bit is
included in the corresponding parity-check equation and so the number of edges
in the Tanner graph is equal to the number of 1s in the parity-check matrix.
\end{defi}
An $m\times n$ parity-check matrix $\mathbf{H}$ can be represented by a Tanner graph with $n$ VNs and $m$ CNs.

An LDPC code parity-check matrix is called $(w_H(c), w_H(r))$-regular if each code
bit is contained in a fixed number of $w_H(c)$ of parity checks and each parity-check
equation contains a fixed number $w_H(r)$ of code bits. In other words, the parity-check matrix has $w_H(c)$ column weights for each column and $w_H(r)$ row weights for each row. The code rate $R$ for a regular LDPC code is bounded as \cite[Chapter 5.1.1]{Lin09}
\begin{equation}
R \geq 1-\frac{w_H(c)}{w_H(r)},
\end{equation}
with equality when the parity-check matrix is full rank.

For irregular LDPC codes, the parameters $w_H(c)$ and $w_H(r)$ vary with the columns and rows. It is more useful to specify the degree distribution of the VN and the CN, denoted by $\alpha(x)$ and $\beta(x)$, respectively. The polynomials have the form \cite[Chapter 5.1.2]{Lin09}
\begin{align}
\alpha(x) &= \sum_{i=1}^I \alpha_ix^{i-1} \nonumber \\
\beta(x) &= \sum_{j=1}^J \beta_jx^{j-1},
\end{align}
where $\alpha_i$ denotes the fraction of all edges connected to degree-$i$ VNs; $\beta_j$ denotes the fraction of all edges connected to degree-$j$ CNs; $I$ is the maximum VN degree; and $J$ is the maximum CN degree.
The code rate for an irregular LDPC code is bounded as \cite[Chapter 5.1.2]{Lin09}
\begin{equation}
R \geq 1- \frac{\int_0^1\alpha(x)dx}{\int_0^1\beta(x)dx}.
\end{equation}
The irregular LDPC codes have better decoding threshold than their regular counterparts. The decoding is generally performed by sum-product decoding described in Section \ref{sec:SPA}.

\subsection{Repeat-Accumulate Codes}
Repeat-accumulate (RA) codes are a specific class of serially concatenated
codes in which the outer code is a rate-$1/q$ repetition code (repeating $q$ times) and the inner code
is a convolutional code with generator $1/(1 + D)$. A $1/(1 + D)$ convolutional
code simply outputs the sum of the current input bit and the previous
output bit over $\mathbb{F}_2$, i.e. it provides a running sum of all past inputs and so is often called an
accumulator. These two component codes give repeat-accumulate codes their
name. RA codes are a simple class of turbo-like codes \cite{divsalar1998coding} as they are built from fixed convolutional codes interconnected with random interleavers. Here, the interleaver
is placed between the inner and outer codes to improve the minimum Hamming distance and provide an interleaver gain for the turbo-like codes \cite{Vucetic:2000:TCP:352869}.

The irregular repeat-accumulate (IRA) codes generalize the RA codes in
that the repetition rate may differ for each of the information bits and that the repeated bits are combined by a combiner and then are sent through the accumulator. IRA codes provide two important advantages over
RA codes. First, they allow flexibility in the choice of the repetition rate for each
information bit so that high-rate codes may be designed. Second, their irregularity
allows operation closer to the capacity limit \cite{Jin00}.

Similar to LDPC codes, the family of RA codes can be represented by a Tanner graph. The code rate can also be determined from the degree distributions of VN and CN. The advantage of RA codes is that they have a much lower encoding complexity than LDPC codes while achieving comparable performance to LDPC codes. RA codes can also be decoded by using either sum-product algorithm or BCJR algorithm \cite[Chapter 6.2]{johnson_2009}.

\subsection{Extended Codes and Subcodes}
Code extension is commonly used to provide a better minimum Hamming distance at the cost of lowering the code rate. Take an $(n,k,t,d_{\min})$ BCH code as an example, where $n$ is the codeword length, $k$ is the length of the bits to be encoded, $t$ is the error correction capability and $d_{\min}$ denotes the minimum Hamming distance of the code. A singly-extended BCH code is obtained through an
additional parity bit $p_1$, formed by adding all coded bits over $\mathbb{F}_2$. In this case, the original codeword $\mathbf{c} = [c_1,\ldots,c_n]$ becomes $\mathbf{c}' = [c_1,\ldots,c_n,p_1]$. The singly-extended BCH code has the new parameters $(n+1,k,t,d_{\min}+1)$. On the other hand, a doubly-extended
BCH code has two additional parity bits, denoted by $p_1$ and
$p_2$, such that
\begin{align}
[c_1+c_3+\ldots+c_{n-1}+p_1] \mod2 = 0, \\
[c_2+c_4+\ldots+c_{n}+p_2] \mod2 = 0,
\end{align}
i.e., the parity bits perform checks separately on odd and even
bit positions. The doubly extension yields an $(n+2,k,t,d_{\min}+1)$ BCH code.

As an alternative to extending the code, one may
employ a subcode of the original BCH code. For example,
the singly-extended BCH code behaves similarly to the even weight
subcode of the BCH code, which is obtained by
multiplying its generator polynomial by $(1+ x)$. As such, the resultant code has the parameters $(n,k-1,t,d_{\min}+1)$. The doubly-extended
BCH code behaves similarly to the BCH subcode
where odd and even coded bits separately sum to zero. This
subcode is obtained by multiplying the generator polynomial
by $(1+x)^2$ and becomes $(n,k-2,t,d_{\min}+1)$. Note that this subcode is not cyclic \cite[Remark 1]{hager2017approaching}. Compared to the code extension, subcodes have more rate loss because
\begin{equation}
 R_{\text{sub}} = \frac{k-a}{n} < R_{\text{ext}} = \frac{k}{n+a},
\end{equation}
for any $a>0$, where $a$ represents the change in the number of bits.

\subsection{Product Codes}
Let $\mathcal{C}_1$ be a binary $(n_1,k_1,d_{\min,1})$ linear block code, and $\mathcal{C}_2$ be a binary $(n_2,k_2,d_{\min,2})$ linear
block code, where $n_i$, $k_i$, and $d_{\min,i}$ represent code $\mathcal{C}_i$'s
length, dimension, and minimum distance, respectively for $i \in \{1,2\}$. A code with $n_1n_2$ symbols can be constructed by making a rectangular
array of $n_1$ columns and $n_2$ rows in which every row is a codeword in $\mathcal{C}_1$ and every
column is a codeword in $\mathcal{C}_2$. One code array or code block consists of $k_1k_2$ information symbols and $n_1n_2 - k_1k_2$ parity-check symbols.
Since the rows (or columns) are codewords in $\mathcal{C}_1$ (or $\mathcal{C}_2$), the sum of two corresponding
rows (or columns) in two code arrays is a codeword in $\mathcal{C}_1$ (or in $\mathcal{C}_2$). The resultant product code $\mathcal{P}(\mathcal{C}_1,\mathcal{C}_2)$
form a two-dimensional $(n_1n_2,k_1k_2,d_{\min,1}d_{\min,2})$ linear block code \cite[Chapter 3.5.1]{Lin09}.

Let $\mathbf{H}_i \in \mathbb{F}_2^{(n_i-k_i)\times n_i}$
be the parity-check matrix of the binary $(n_i, k_i, d_{\min,i})$
linear code $\mathcal{C}_i$ for $i=1,2$. The product code $\mathcal{P}(\mathcal{C}_1,\mathcal{C}_2)$ based on $\mathcal{C}_1$ and $\mathcal{C}_2$ is defined as
\begin{equation}\label{eq:prod_code}
\mathcal{P}(\mathcal{C}_1,\mathcal{C}_2) = \left\{\mathbf{X} \in \mathbb{F}_2^{n_1\times n_2}|\; [\mathbf{X}^T\mathbf{H}_1^T]\hspace{-4mm} \mod \hspace{-1mm} 2 = \mathbf{0}^{n_2 \times (n_1-k_1)}, [\mathbf{X}\mathbf{H}_2^T ] \hspace{-4mm}\mod \hspace{-1mm} 2 = \mathbf{0}^{n_1 \times (n_2-k_2)} \right\},
\end{equation}
where each column and row of the codeword $\mathbf{X}$ is a valid codeword of $\mathcal{C}_1$ and $\mathcal{C}_2$, respectively.


\section{Decoding}
In this section, we introduce some common decoders for decoding linear block codes. We only present the main idea of the decoder while the detailed decoding algorithm is omitted.

\subsection{Bounded Distance Decoding (BDD)}\label{sec:BDD_Decoding}
Consider the transmission of a $t$-error correcting code codeword $\mathbf{c} \in \mathcal{C}$
over a binary channel. The error vector
introduced by the channel is denoted by $\mathbf{z}$. Applying BDD to the received word $\mathbf{r} = \mathbf{c} + \mathbf{z}$ results in
\begin{align}\label{eq:BDD}
\text{BDD}(\mathbf{r}) = \left\{\begin{array}{ll}
\mathbf{c}, 
&\text{if} \; d_H(\mathbf{r},\mathbf{c}) = w_H(\mathbf{z}) \leq t\\
\mathbf{c}' \in \mathcal{C} &\text{if} \; w_H(\mathbf{z}) > t \; \text{and} \; d_H(\mathbf{r},\mathbf{c}') \leq t\\
\text{Error}
&\text{otherwise} \\
\end{array} \right.
\end{align}
Note that the second case in \eqref{eq:BDD} corresponds to an undetected error
or miscorrection. If the channel is BEC, the second case will not happen. The BDD decoder is also known as the hard-decision decoder for decoding conventional linear block codes such as Hamming codes and BCH codes.

\subsection{Maximum-Likelihood (ML) Decoding}
The ML decoder always choose the codeword that is most likely to have
produced the received vector $\mathbf{y}$. Specifically, given a received vector $\mathbf{y}$ and a codebook $\mathcal{C}$, the ML decoder will choose
the codeword $\mathbf{c}$ that maximizes the probability $p(\mathbf{y}|\mathbf{c})$. The ML decoder returns
the decoded codeword $\hat{\mathbf{c}}^{\text{ML}}$ according to the rule \cite[Chapter 1.5]{Richardson:2008:MCT:1795974}
\begin{equation}\label{eq:ML_rule}
\hat{\mathbf{c}}^{\text{ML}} = \argmax\limits_{\mathbf{c} \in \mathcal{C}}p(\mathbf{y}|\mathbf{c}).
\end{equation}
In the AWGN channel, the ML decoding is equivalent to finding the codeword that has the minimum Euclidean distance to the received codeword
\begin{equation}
\hat{\mathbf{c}}^{\text{ML}} = \argmin\limits_{\mathbf{c} \in \mathcal{C}}\|\mathbf{c} - \varphi^{-1}(\mathbf{y}) \|,
\end{equation}
where $\varphi^{-1}(.)$ denotes the demodulation function. The Viterbi decoding algorithm \cite{1054010} was later shown to be an ML decoding algorithm \cite{1450960}. 

\subsection{Maximum A Posteriori (MAP) Decoding}
Given a received vector $\mathbf{y}$ and a codebook $\mathcal{C}$, a MAP decoder or block-MAP decoder chooses the codeword $\mathbf{c}$
that maximizes the a posteriori probability $p(\mathbf{c}|\mathbf{y})$ for $\mathbf{c}$. The MAP decoding rule is given by \cite[Chapter 1.5]{Richardson:2008:MCT:1795974}
\begin{align}
\hat{\mathbf{c}}^{\text{MAP}} &= \argmax\limits_{\mathbf{c} \in \mathcal{C}}p(\mathbf{c}|\mathbf{y}) \nonumber \\
& = \argmax\limits_{\mathbf{c} \in \mathcal{C}}\frac{p(\mathbf{y}|\mathbf{c})p(\mathbf{c})}{p(\mathbf{y})} \;\; (\text{by Bayes's rule})\nonumber \\
& \overset{(a)}= \argmax\limits_{\mathbf{c} \in \mathcal{C}}p(\mathbf{y}|\mathbf{c})p(\mathbf{c}),
\end{align}
where $p(\mathbf{c})$ is the a priori probability of choosing codeword $\mathbf{c}$ and $(a)$ follows that $p(\mathbf{y})$ can be treated as a normalizing constant. If each codeword is equally likely to have been sent, then the MAP decoding rule is equivalent to the ML decoding rule
\begin{align}
\hat{\mathbf{c}}^{\text{MAP}} &= \argmax\limits_{\mathbf{c} \in \mathcal{C}}p(\mathbf{y}|\mathbf{c})p(\mathbf{c}) \nonumber \\
& = \argmax\limits_{\mathbf{c} \in \mathcal{C}}p(\mathbf{y}|\mathbf{c})\frac{1}{|\mathcal{C}|} \nonumber \\
& = \argmax\limits_{\mathbf{c} \in \mathcal{C}}p(\mathbf{y}|\mathbf{c})  \nonumber \\
&\overset{\eqref{eq:ML_rule}}= \hat{\mathbf{c}}^{\text{ML}}
\end{align}
The MAP decoding can also be done on a symbol by symbol
basis. The symbol-MAP decoder will choose the most probable symbol, in our
case bit, for each transmitted symbol (even if the set of chosen bits does not make
up a valid codeword). A symbol-MAP decoder chooses the symbol $\hat{c}_i^{\text{MAP}}$ according to the rule:
\begin{align}
\hat{c}_i^{\text{MAP}} = \argmax\limits_{c_i \in \{0,1\}}p(c_i|\mathbf{y}).
\end{align}
An efficient algorithm for performing symbol-MAP decoding is the BCJR algorithm \cite{1055186}.

\subsection{Sum-Product Decoding}\label{sec:SPA}
The sum-product algorithm (SPA) \cite{910572} is also sometimes called the belief-propagation (BP) algorithm \cite{PEARL1986241}.
It is an iterative soft-input and soft-output decoder which accepts the probability for each received bit as
input and compute the probability of each received bit being one or zero after each decoding iterations by exchanging extrinsic information.


For the code that can be represented by a Tanner graph with VN and CN on each side, it can be thought of as a collection of VN decoders concatenated through an interleaver to a collection of CN decoders.
The VN and CN decoders work cooperatively and iteratively to estimate the log-likelihood ratio (LLR) for each code bit. The LLR of a binary value $x$ is defined as
\begin{align}
L(x) \triangleq \ln \left(\frac{p(x=0)}{p(x=1)}\right).
\end{align}
The VNs process their
inputs and pass extrinsic information up to their neighboring CNs; the
CNs then process their inputs and pass extrinsic information down to
their neighboring VNs; and the procedure repeats, starting with the
VNs. After a preset maximum number of iterations of
this VN/CN decoding round, or after some stopping criterion has been met (e.g., the parity-check equations are satisfied), the
decoder computes (estimates) the LLRs from which decisions on the codeword bits are made. When the cycles in the code graph are large or the graph is cycle free, the estimates will be very accurate and
the decoder will have near-optimal (MAP) performance \cite[Chapter 5.4]{Lin09}.

Let $\mu_{x \rightarrow f}(x)$ denote the message sent from node $x$ of VN to node $f$ of CN in the operation of the SPA. Let $\mu_{f \rightarrow x}(x)$ denote the message sent from node $f$ of CN to node $x$ of VN. Also, let $\mathcal{S}(v)$ denote the set of neighbors of a given node $v$ in a Tanner graph. The message computations performed by the SPA can be expressed as follows \cite{910572}:

VN to CN update:
\begin{equation}
\mu_{x \rightarrow f}(x) = \prod_{h \in \mathcal{S}(x) \setminus \{f \}} \mu_{h \rightarrow x}(x),
\end{equation}
where $\mathcal{S}(x) \setminus \{f \}$ is the set of neighboring nodes to $x$, which excludes node $f$.

CN to VN update:
\begin{equation}
\mu_{f \rightarrow x}(x) = \sum_{\sim\{x\}}\left(F(X)\prod_{y\in\mathcal{S}(f) \setminus \{x\}}\mu_{y \rightarrow f}(y) \right),
\end{equation}
where $\mathcal{S}(f) \setminus \{x \}$ is the set of neighboring nodes of $f$ without node $x$; $X = \mathcal{S}(f)$ is the set of arguments of the probability mass function $F$; and $\sum_{\sim\{x\}}$ means the summation is taken over all nodes without node $x$.

In contrast with the error-rate curves for classical codes – e.g., BCH
codes with a BDD decoder,
the error-rate curves for iteratively decoded codes generally have a region in which
the slope decreases as the channel SNR increases (or, for a BEC, as the input erasure probability decreases). The sharp transition region on the curve is generally referred to as the waterfall region of the error-rate curve \cite[Chapter 1.9]{Richardson:2008:MCT:1795974}.

\section{Performance Analysis of Channel Coding}
For a given code and decoder, one would like to know for which channel
noise levels the decoder will be able to correct the errors and
for which it will not. In most cases, the ensemble \cite[Definition 1.15]{Richardson:2008:MCT:1795974} of all possible codes with certain parameters (for
example, a certain degree distribution) will be evaluated rather than a particular choice of code
having those parameters.

\begin{defi}(Ensemble):
Consider the code is over the field $\mathbb{F}_q$. We denote by $\mathcal{C}(n,M)$
the ensemble of codes of length $n$ and cardinality $M$. There ere
are $nM$ degrees of freedom in choosing a code, one degree of freedom for each
component of each codeword. The ensemble consists of all $q^{nM}$ possible codes
of length $n$ and cardinality $M$.
\end{defi}

In what follows, we describe two techniques to design and analyze the performance of a modern code ensemble, such as LDPC code ensembles and turbo-like code ensembles \cite{Vucetic:2000:TCP:352869,{Richardson:2008:MCT:1795974}}.

\subsection{Density Evolution}
When very long codes are considered, the extrinsic LLRs passed between
the component decoders can be assumed to be independent and identically
distributed. Under this assumption, the expected iterative decoding performance
of a particular ensemble can be determined by tracking the evolution of these
PDFs through the iterative decoding process, a technique
called density evolution (DE) \cite[Chapter 3.9]{Richardson:2008:MCT:1795974}.

DE can be used to find the maximum level of channel
noise which is likely to be corrected by a particular code ensemble. A recursive function is used to track the expected residual graph evolution throughout the iterative decoding process. The decoding threshold is the point over which the error probability cannot drop to zero even after an infinite number of iterations. The derivation of the DE based on the following properties \cite[Chapter 7.2]{johnson_2009}.
\begin{itemize}
\item Symmetry: The output of the channel is symmetric, e.g. for binary input, if $p(y|x = 1) =
f (y)$ then $p(y|x = -1) = f (-y)$. As such, the LLRs output by the iterative decoder
are also symmetric.
\item All-zeros codeword: Using the symmetric condition above, the iterative decoding
performance can be shown to be independent of the codeword transmitted.
This result allows the performance of a code-decoder pair to be modelled by
sending only the all-zeros codeword.
\item Cycle-free graphs: As the codeword length goes to infinity, the ensemble
average performance of the iterative decoder approaches that of decoding on
a cycle-free graph.
\item Concentration: With high probability, a randomly chosen code from an
ensemble will have an iterative decoding performance close to the
average performance of that ensemble.
\end{itemize}
To find
the optimal degree distributions in the sense of the minimum threshold for a fixed code rate,
a global optimization algorithm that searches the
space of degree polynomials, is required on top of the DE algorithm.

\subsection{Extrinsic Information Transfer Chart}
As an alternative to DE, the EXIT
chart technique is a graphical tool for estimating the decoding threshold of
a code ensemble \cite{957394}. The technique not only simplifies the DE process by representing the extrinsic information transferred between component codes
by a single parameter, but also provides some intuition regarding the dynamics and convergence
properties of an iteratively decoded code.

The idea behind EXIT charts begins with the fact that the VN decoders and CN decoders work cooperatively and iteratively
to make bit decisions, with the metric of interest generally improving with each
half-iteration. A transfer curve plotting
the input metric versus the output metric can be obtained both for the VN decoders
and for the CN deciders. Further, since the output metric for one processor is the input metric
for its companion decoder, one can plot both transfer curves on the same axes,
but with the abscissa and ordinate reversed for one decoder. Such a chart aids
in the prediction of the decoding threshold of the ensemble of codes characterized
by given VN and CN degree distributions: the decoding threshold is the
$\SNR$ or $\epsilon$ at which the transfer curve of the VN decoders just touches the curve of the CN decoders, precluding
convergence of the two decoders. Similar to DE, decoding-threshold
prediction via EXIT charts assumes a graph with no cycles, an infinite codeword
length, and an infinite number of decoding iterations.

We will employ the EXIT chart technique in our code design in Chapters 4-5.

\section{Summary}

In this chapter, we present some basic background materials on wireless communications and channel coding which are closely related to and required by the research work in the thesis. The main points presented in this chapter are summarized as follows.
\begin{itemize}
\item We start with the introduction of point-to-point binary memoryless channel, including BEC, BSC, BI-AWGN and unconstrained AWGN channels. These channels are basic models for many coding designs.
\item We provide some basic knowledge of different kinds of fading channels and their characteristics. 
\item We give background knowledge on NOMA, a promising way of downlink multiuser transmission. Two essential ingredients of NOMA as well as its performance limit, namely multiuser capacity region are also presented.
\item We briefly describe different types of channel coding schemes, ranging from conventional algebraic codes and the modern capacity-approaching iterative decoded codes.
\item For the above coding schemes, we also introduce the corresponding decoders by presenting the main ideas behind them.
\item We briefly introduce two techniques DE and EXIT chart that are used for analyzing the average decoding performance of a code ensemble.
\end{itemize}

\chapter{Design of Multi-Dimensional Irregular Repeat-Accumulate Lattice Codes}\label{C4:chapter4}

\section{Introduction}

Lattice are effective arrangements of equally spaced points in Euclidean space. They have attracted considerable attentions in the coding community because their appealing algebraic structures can be efficiently exploited for encoding and decoding. In this chapter, we present our detailed design for a class of multi-dimensional lattice codes to attain the near-capacity performance for power constrained point-to-point channels, before we focus on the coding scheme design for multiuser downlink channels in the next few chapters.

\subsection{Problem Statement}
In light of the previous work mentioned in Section \ref{sec:lattice_code_review}, we aim to design new multi-dimensional lattice codes to further approach the unconstrained AWGN channels. That being said, directly extending our previous design in \cite{Qiu16} where the codes are based on two-dimensional lattice partitions to multi-dimensional lattice partitions is very challenging. There are two fundamental reasons why this is the case. First, in the previous setting, we employed a two-dimensional lattice partition to form a quotient ring which is isomorphic to a finite field. However, most multi-dimensional lattice partitions form additive quotient groups where addition is the only group operation. If we use multi-dimensional lattice partitions in our previous design, the multiplication between two lattice points cannot be performed on additive groups. Second, simply removing the multiplication in the encoding structure will prevent us from analysing and optimizing the multi-dimensional IRA lattice codes effectively. In the previous design, the encoder's messages are multiplied by some randomly generated sequences so that the permutation-invariant property \cite{Bennatan06} can be obtained. Under this property, the analysis and optimization of our lattice codes can be significantly simplified. It is possible to remove all the operations of multiplying random sequences to allow the use of multi-dimensional lattice partitions. However, the permutation-invariance property will not hold in this case. As a result, the densities of the messages in the iterative decoder can only be represented by a multivariate Gaussian distribution. This will lead to an extremely high complexity for our design and analysis.

\subsection{Main Contributions}
For this work, we aim to design multi-dimensional IRA lattice codes with \emph{finite constellations} to further approach the unconstrained Shannon limit. This is different from most lattice codes which are based on infinite constellations in the literature. Even though these codes have been shown to approach the Poltyrev limit within 1 dB, it is still unclear whether these codes with power constraint can approach the unconstrained Shannon limit within 1 dB. In order to practically approach the unconstrained Shannon limit, we must optimize the degree distribution of our codes based on constellations, detection methods and decoding algorithms. Furthermore, we continue to use Construction A as it has been proved to be a simple and powerful tool for constructing capacity-achieving lattice codes according to the literature. The main contributions of our work are summarized as below:
\begin{itemize}
\item We designed a class of lattice codes with finite constellations based on multi-
dimensional lattice partitions. More specifically, we proposed a novel encoding structure that adds random lattice sequences to the encoder's messages (output of the interleaver, combiner and accumulator). In addition, we introduced a constraint on the random lattice sequences in our encoder and proved that the constraint can lead to linearity of our codes. Since no multiplication is required in our encoder, our design can be directly applied to any lattices of any dimensions.

\item We investigated the optimal degree distributions of our lattice codes, aiming at approaching the unconstrained Shannon limit. We proved and showed that our encoding structure can produce permutation-invariant and symmetric effects in the densities of the decoder's messages (soft information propagated in the iterative docoder). These two properties enable to use a Gaussian distribution characterised by a single parameter to model the soft information propagated inside the iterative decoder. Under this condition, we used a two-dimensional EXIT chart to analyse the convergence behaviour of the iterative decoder. With EXIT charts, we designed a set of lattice codes for different target code rates with the minimum decoding threshold.

\item Numerical results are provided and show that our designed and optimised lattice codes can approach the unconstrained Shannon limit within 0.46 dB. We demonstrate that our lattice codes not only outperforms previously designed lattice codes in \cite{Qiu16} with two-dimensional lattice partitions, but also have less coding loss compared with the existing lattice coding schemes in \cite{Boutros14,Boutros16,Khodaiemehr17,8122043,4475389} for large codeword length, i.e., a codeword has more than 10,000 symbols.

\end{itemize}

\section{Multi-Dimensional IRA Lattice Codes}
In this section, we present the proposed multi-dimensional IRA lattice codes. We consider the channel to be a complex AWGN channel where the input is non-binary, which means asymmetric-output in general. For this channel, different transmitted symbols have different error resistance to the non-binary AWGN noise. Thus the decoding errors for different symbols are different.

\subsection{IRA Lattices Construction}
We begin with the construction of our lattice codes. The lattice codes are constructed via Construction A \cite{conway1999sphere}. The error performance of Construction A lattices heavily depends on the underlying error correction codes. Thus, we choose IRA codes as they have been shown to have capacity-approaching performance in AWGN channels and has lower encoding complexity than that of general LDPC codes \cite{Jin00,Chiu10,7008249,7124694,Yang15,Qiu17}.

In this work, we use the conventional Construction A method to a more generic case which is not merely limited to two-dimensional lattices. Denote a non-binary IRA codes over $\text{GF}(p^M)$ by $\mathcal{C}$, where $p$ is a prime number and $M$ is a positive integer. The IRA encoder takes length $K$ input messages and produces length $N$ codewords. Here, $K \leq N$ and all the encoding operations are over $\text{GF}(p^M)$. We denote the Construction A lattice by $\Lambda_{\mathcal{C}}$. It is generated via:
\begin{equation}\label{eq:31}
\Lambda_{\mathcal{C}} = \{ \boldsymbol{\lambda} = \phi(\mathcal{C})+\xi\mathcal{R}^N \},
\end{equation}
where $\xi \in \mathcal{R}$ and $\mathcal{R}$ is a lattice; $\phi(.)$ is a homomorphism mapping function that maps each codeword component to the elements in the lattice partition:
\begin{equation}\label{eq:31a}
\phi:\mathbb{F}_p^M\rightarrow \mathcal{R}/\xi\mathcal{R}.
\end{equation}
Note that $N$ in (\ref{eq:31}) should be a multiple of $M$ in (\ref{eq:31a}).

It is also noteworthy that in conventional Construction A, $\mathcal{R}$ can be any principal ideal domains (PID) such as rational integers $\mathbb{Z}$ and Gaussian integers $\mathbb{Z}[i]$. In that case, the lattice partition forms a quotient ring that is isomorphic to a finite field. In most cases where $\mathcal{R}$ is a multi-dimensional lattice, the lattice partition forms a quotient group \cite{Oggier13}.

In (\ref{eq:31a}), the $\mathcal{R}$-lattice is partitioned into $p^M$ numbers of cosets where each coset has a coset leader. For designing \emph{finite} constellations, only coset leaders are used in transmission to satisfy the power constraint requirement. Therefore, using (\ref{eq:1g}), the information rate $R$ for this Construction A lattice is
\begin{equation}\label{eq:aa3}
R = \frac{K}{N}\cdot\frac{1}{n}\log_2(p^M),
\end{equation}
where $n$ is the dimension of the $\mathcal{R}$-lattice.

We now present a specific design example of using the $D_4$ lattice via Construction A. According to \cite{conway1999sphere}, the $D_4$ lattice is a four-dimensional lattice which has the highest sphere packing density in the four-dimensional space. It is defined as:
\begin{equation}\label{eq:32}
D_4 = \bigg\{(x_1,x_2,x_3,x_4) \in \mathbb{Z}^4: \sum_{i=1}^4 x_i \in 2 \mathbb{Z}\bigg\}.
\end{equation}
It has the generator matrix in the integer lattice form:
\begin{equation}\label{eq:33}
G_{D_4}= \begin{bmatrix}
-1 & -1 & 0 & 0 \\
1 & -1 & 0 & 0 \\
0 & 1 & -1 & 0  \\
0 & 0 & 1 & -1 \\
\end{bmatrix}.
\end{equation}

As explained in Section \ref{sec:fom}, we use the NSM as the goodness to measure the shaping performance of the lattices. By (\ref{eq:nsm}), we calculate the NSM for $D_4$ is about $0.0766$. Then using (\ref{eq:sg}) we can see that $D_4$ can provide a shaping gain about $0.3657$ dB over the four dimensional cubic lattice.

According to \cite{Natarajan15}. the $D_4$ lattice can be identified as \emph{Hurwitz quaternion integers}:
\begin{equation}\label{eq:34}
\mathbb{H} = \bigg\{a+bi+cj+dk |a,b,c,d \in \mathbb{Z} \;\text{or}\; a,b,c,d \in \mathbb{Z} + \frac{1}{2} \bigg\},
\end{equation}
where $\{1,i,j,k\}$ is the basis of the number system for representing Hurwitz integers.
Addition in $\mathbb{H}$ is component wise whereas multiplication is non-commutative and defined based on the following relations:
\begin{equation}\label{eq:35}
i^2 = j^2 = k^2 = ijk = -1.
\end{equation}

Given $A = a+bi+cj+dk$, the norm of $A$ is:
\begin{equation}\label{eq:36}
N(A) = a^2+b^2+c^2+d^2 \in \mathbb{Z}.
\end{equation}

Consider the following example. In (\ref{eq:31a}), if we let $\xi = 1+2i$, then the homomorphism mapping function becomes:
\begin{equation}\label{eq:37}
\phi:\mathbb{F}_5^2 \rightarrow \mathbb{H}/(1+2i)\mathbb{H}.
\end{equation}
Note that this lattice partition can be further expressed as:
\begin{align}
\mathbb{H}/(1+2i)\mathbb{H} &= \boldsymbol{\lambda}/(1+2i)D_4, \;\; (\boldsymbol{\lambda} \in D_4)\nonumber \\
& \overset{(a)}= \boldsymbol{\lambda} - Q_{(1+2i)D_4}\left( \boldsymbol{\lambda} \right) \nonumber \\
& \overset{(b)}= \boldsymbol{\lambda} - (1+2i)Q_{D_4}\left( \frac{\boldsymbol{\lambda}}{(1+2i)}\right),
\end{align}
where (a) follows Eq. (\ref{eq:1c}) and (b) follows \cite[Eq. (2.43)]{Zamir15}. The multiplication and division here should follow quaternion arithmetic \cite{Smith03}. For the quantizer $Q_{D_4}$, we follow the approach in \cite{Conway82} to develop the quantization algorithm of finding the closest $D_4$ lattice point to an arbitrary point in $\mathbb{R}^4$. The quantization algorithm has a lower computational complexity compared with ML decoding. It is very useful in the scenario where we perform the $D_4$ lattice partitions. The cardinality of this partition can be calculated by using (\ref{eq:36}) as $N(1+2i)^2 = 25$. In this way, the $D_4$ lattice is partitioned into 25 cosets. Even though $\mathbb{H}$ is a PID \cite{Huang17}, we only have the group homomorphism as the multiplication for $\mathbb{H}$ is non-commutative.

Now we compare the mutual information of the uniform input distribution over the coset leaders of the $D_4$ lattice partition with that of a two-dimensional lattice to see the performance gain introduced by the multi-dimensional lattices. In this work, the two-dimensional square lattice $\mathbb{Z}^2$ is set to be a benchmark for performance comparison. Note that a finite portion of the $\mathbb{Z}^2$ lattice is known as a quadrature amplitude modulation (QAM). The $\mathbb{Z}^2$ lattice can be identified as Gaussian integers $\mathbb{Z}[i] = \{a+bi:a,b \in  \mathbb{Z} \}$. For fair comparison, we partition both lattices in a way such that the information rates for both lattice partitions are the same.

\begin{figure}[ht!]
	\centering
\includegraphics[width=3.4in,clip,keepaspectratio]{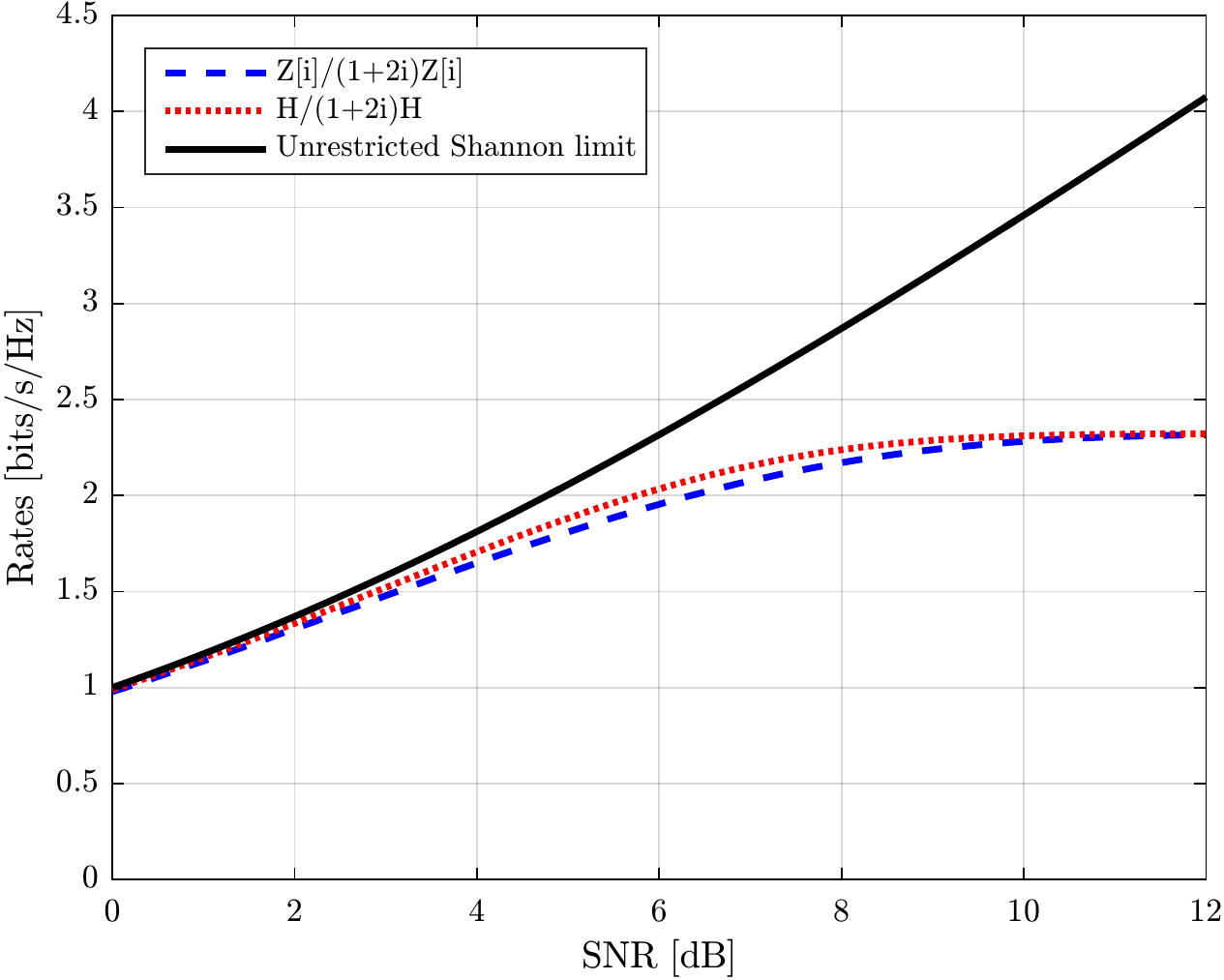}
\caption{Uniform input capacities of $\mathbb{H}/(1+2i)\mathbb{H}$ and $\mathbb{Z}[i]/(1+2i)\mathbb{Z}[i]$.}
\label{fig:capacity_CH4}
\end{figure}

We consider the examples of lattice partitions $\mathbb{H}/(1+2i)\mathbb{H}$ and $\mathbb{Z}[i]/(1+2i)\mathbb{Z}[i]$, where both partitions yield the same information rate. This is because using (\ref{eq:1g}) we can obtain the information rates for $D_4$ and $\mathbb{Z}^2$ as $\frac{1}{2}\log_2(25)$ and $\log_2(5)$, respectively. Here the $D_4$ lattice can be deemed as a two-dimensional complex lattice while the $\mathbb{Z}^2$ lattice is a one-dimensional complex lattice. Therefore the dimensions $n$ in (\ref{eq:1g}) for both lattices are 2 and 1, respectively. In other words, the $\mathbb{Z}[i]$ lattice requires one time slot to transmit its lattice point where the $D_4$ lattice requires two time slots to transmit a $D_4$ lattice point.

Given SNR values, the unconstrained Shannon limit for the AWGN channel is plotted in Fig. \ref{fig:capacity_CH4} along with the capacities of the $D_4$ lattice and the $\mathbb{Z}^2$ lattice. As observed from Fig. \ref{fig:capacity_CH4}, the curve for the $D_4$ lattice always lies above that for the $\mathbb{Z}^2$ lattice. Therefore, under the same information rate, we can construct $D_4$ lattice partition based IRA lattice codes that require lower decoding SNR than any IRA lattice codes based on the $\mathbb{Z}^2$ lattice partitions. This is due to the advantage of shaping gain.


\subsection{IRA Lattice Encoder}\label{IRAE}
\begin{figure}[ht!]
	\centering
\includegraphics[width=3.8in,clip,keepaspectratio]{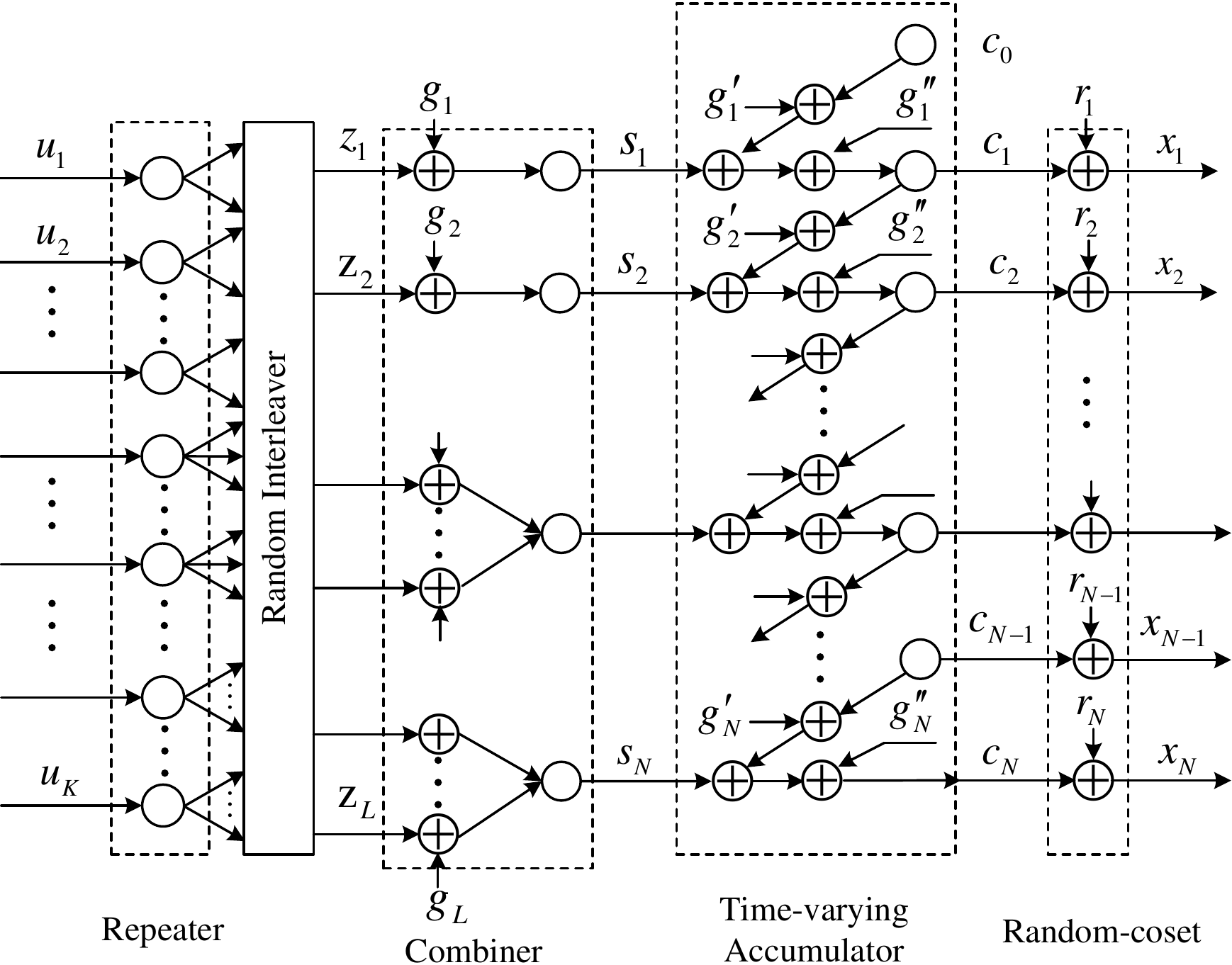}
\caption{Block diagram of the IRA lattice encoder.}
\label{fig:encoder_CH4}
\end{figure}

Here we show our proposed encoder design. The block diagram of the IRA lattice encoder is depicted in Fig. \ref{fig:encoder_CH4}.
First of all, the input to the encoder is a length $K$ message $\mathbf{u}=[u_1,u_2,\ldots,u_K]^T$, where each element $u_k$ for $k = 1,2,\ldots, K$ is taken from the set of coset leaders $\Psi = \{\psi_0, \psi_1, \ldots, \psi_{p^M-1}\}$. This message $\mathbf{u}$ is then fed into a repeater and repeated according to a discrete distribution of $f_1,f_2,\ldots,f_I$, where $f_i \geq 0$ for $i = 1, 2, \ldots ,I$ and $\sum_if_i=1 $. The number $f_i$ represents the fraction of message symbols are repeated by $i$ times. The maximum repeating times is $I$ times, where $I \geq 2$, thus $f_1 = 0$. After repeating, the total number of symbols becomes $L = K\sum_iif_i$.

Next, the repeated symbols are passed into a random interleaver. We denote the interleaved sequence by $\mathbf{z} = [z_1,z_2,\ldots,z_L]^T$. A randomly generated sequence with the same length $\mathbf{g} = [g_1,g_2,\ldots,g_L]^T$ is added to the interleaved sequence $\mathbf{z}$ via $\mathbf{z} \oplus \mathbf{g}$ in an element-wise manner, where ``$\oplus$'' is the modulo-lattice addition defined in (\ref{eq:1d}). Note that each element of $\mathbf{g}$ is randomly and uniformly chosen from the set of coset leaders $\Psi$ such that a linear code constraint is met, which will be introduced later.

The resultant symbols are combined according to a discrete distribution of $b_1,b_2,\ldots,b_J$, where $b_j \geq 0$ for $j = 1,2,\ldots,J$ and $\sum_jb_j=1 $. Here the number $b_j$ represents the fraction of message symbols that are obtained from combining $j$ symbols from the output of the interleaver and the corresponding $j$ addition factors in $\mathbf{g}$. After combining, the message sequence becomes a length $N$ sequence denoted by $\mathbf{s} = [s_1,s_2,\ldots,s_N]^T$, where $N = L\sum_j j b_j$. For $n = 1,...,N$, each symbol $s_n$ is calculated as:
\begin{equation}\label{eq:38}
s_n = (z_{a_n} \oplus g_{a_n}) \oplus \ldots \oplus (z_{a_n+j_n-1} \oplus g_{a_n+j_n-1}),
\end{equation}
where $z_{a_n}$ and $z_{a_n+j_n-1}$ represent the first and last interleaved symbols input to the $n$-th combiner, respectively; $g_{a_n}$ and $g_{a_n+j_n-1}$ are the addition factors with respect to $z_{a_n}$ and $z_{a_n+j_n-1}$; $j_n \in \{1,2,\ldots,J \}$ represents the number of symbols to be combined at the $n$-th combiner; $a_n$ is the index of the first interleaved symbol input to the $n$-th combiner. Note that the combiner is to combine the interleaved messages in order to satisfy the code rate requirement.

The combined message sequence $\mathbf{s}$ is passed into a time-varying accumulator which features a time-varying transfer function determined by two randomly generated lattice sequences $\mathbf{g^\prime} = [g^\prime_1,g^\prime_2,\ldots,g^\prime_N]^T$ and $\mathbf{g^{\prime\prime}} = [g^{\prime\prime}_1,g^{\prime\prime}_2,\ldots,g^{\prime\prime}_N]^T$. All the elements in both sequences are uniformly distributed over the set of coset leaders $\Psi$ such that a linear code constraint is met, which will be introduced later. The output message of the time-varying accumulator is denoted by $\mathbf{c} = [c_1,c_2,\ldots,c_N]^T$. The $n$-th symbol $c_n$, where $n = 1,2,\ldots,N$, is generated by
\begin{equation}\label{eq:39}
c_n = (s_n \oplus (c_{n-1} \oplus g^\prime_n)) \oplus g^{\prime\prime}_n,
\end{equation}
where the initial condition is given as $c_0 = 0$. Here $c_0$ is a dummy parity that is fixed to $0$ and will not be transmitted. It is also noteworthy that the random vectors $\mathbf{g}$, $\mathbf{g}^\prime$ and $\mathbf{g}^{\prime\prime}$ in the encoding structure introduce and realize the permutation-invariance property on \emph{all edges} of a Tanner graph as shown in Fig. \ref{fig:tg} and will be discussed in Section \ref{msi}.

Finally, the output of the accumulator $\mathbf{c}$ adds a random-coset vector $\mathbf{r}$ with length $N$ and become the coded lattice sequence $\mathbf{x}$:
\begin{equation}\label{eq:40}
\mathbf{x} = \mathbf{c} \oplus \mathbf{r}.
\end{equation}
Elements of $\mathbf{r}$ are uniformly distributed over the set of coset leaders $\Psi$. Before transmission, the average energy of codeword symbols is normalised to 1.

Note that although the four lattice sequences $\mathbf{g}$, $\mathbf{g^{\prime}}$, $\mathbf{g^{\prime\prime}}$ and $\mathbf{r}$ are random, they are assumed to be known at both transmitters and receivers prior to transmission. Furthermore, the underlying linear codes for our Construction A lattices can be either systematic or nonsystematic non-binary IRA codes.

\subsection{The Linearity of IRA Lattice Codes}
It can be noticed that our proposed lattice encoding structure is different from previous designs. More specifically, instead of using the modulo-lattice multiplication between encoder messages and random lattice sequences in \cite{Qiu16}, we use a different approach by introducing the ``$\oplus$'' operation in the encoding process. However, this difference introduced non-linearity to our codes if $\mathbf{g}$, $\mathbf{g^{\prime}}$ and $\mathbf{g^{\prime\prime}}$ are totally independent, which is not appealing for low complexity decoding. To address this issue, we introduce a constraint on these random sequences to ensure the codes are linear.
\begin{prop}\label{linear}
The multi-dimensional IRA lattice codes are linear if the $n$-th output element from the encoder satisfies the following conditions:
\begin{equation}\label{eq:LC}
g_{a_n} \oplus \ldots \oplus g_{a_n+j_n-1} \oplus g_n^{\prime} \oplus g_n^{\prime\prime}  = 0.
\end{equation}
\end{prop}
\emph{\quad Proof: } Please refer to Appendix \ref{appendix:1}.\QEDA

Note that this equation has $j_n+2$ elements. We randomly choose any $j_n+1$ elements out of these $j_n+2$ elements to be random and uniformly distributed over the set of coset leaders $\Psi$. The last element is then determined by Eq. (\ref{eq:LC}). One can also notice that the linearity condition excludes the random-coset vector $\mathbf{r}$. This is because the random-coset vector is independent of the encoder's messages and is always removed before decoding. If the random-coset vector is included in the condition, the output-symmetric effect in the non-binary AWGN channel will vanish.

\subsection{Tanner Graph}\label{chep:tn}
\begin{figure}[ht!]
	\centering
\includegraphics[width=2.2in,clip,keepaspectratio]{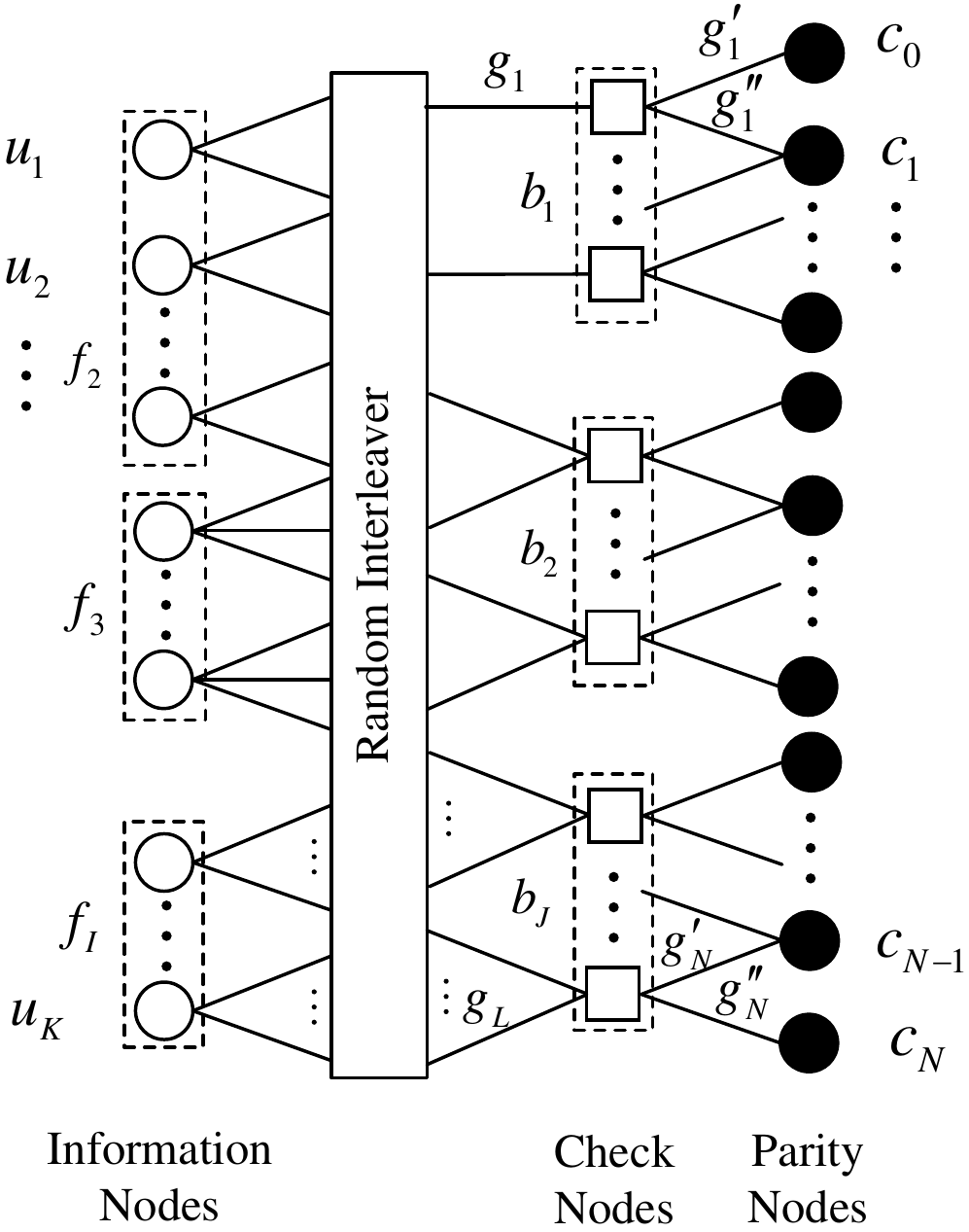}
\caption{Tanner graph of the IRA lattice codes.}
\label{fig:tg}
\end{figure}

Similar to conventional binary IRA codes in \cite{Jin00}, our multi-dimensional IRA lattice codes can be represented by a Tanner graph as shown in Fig. \ref{fig:tg}.

The Tanner graph is a bipartite graph with variable nodes and check nodes. In the figure, variable nodes are represented by circles while check nodes are represented by squares. There are $N+K$ variable nodes on the Tanner graph. The $K$ variable nodes that placed on the left, are called \emph{information nodes}. They represent the $K$ repeaters in the encoder. The degree distribution of information nodes with degree $i$ is denoted by $f_i$ in the figure. This means that the fraction of information nodes are connected to $i$ check nodes. Note that the random interleaver here introduces randomness in the edges between information nodes and check nodes. This randomness can prevent short cycles in the Tanner graph which leads to a better decoding performance \cite{Johnson05}. On the right of the Tanner graph, there are $N$ variable nodes which are called \emph{parity nodes}, representing the output $\mathbf{c}$ from the time-vary accumulator. In the middle of the Tanner graph, there are $N$ check nodes, representing $N$ combiners. The degree distribution of check nodes with degree $j+2$ is denoted by $b_j$ which represents the fraction of check nodes connected to $j$ information nodes and 2 parity nodes. Note that the random-coset vector $\mathbf{r}$ is removed before performing decoding, thus it is not shown in the Tanner graph.

Now consider the $n$-th check node with degree $j+2$, according to (\ref{eq:38}), (\ref{eq:39}) and the Tanner graph in Fig. \ref{fig:tg}, the parity-check equation at the $n$-th check node is
\begin{align}\label{eq:41}
&(z_{a_n} \oplus g_{a_n}) \oplus \cdots \oplus (z_{a_n+j_n-1} \oplus g_{a_n+j_n-1})\oplus \nonumber \\
& (c_{n-1} \oplus g^\prime_n) \oplus ( c_n^{-1}  \oplus g^{\prime\prime}_n) = 0,
\end{align}
where $c_n^{-1} \oplus c_n = 0$. Note that in the Tanner graph, $c_0$ is a dummy bit and will not be transmitted.

We decompose the elements on the left hand side of Equation (\ref{eq:41}) into two vectors:
\begin{equation}\label{eq:42}
\mathbf{t}_n = [z_{a_n}, \ldots ,z_{a_n+j_n-1}, c_{n-1}, c_n^{-1}].
\end{equation}
\begin{equation}\label{eq:43}
\mathbf{h}_n = [g_{a_n}, \ldots ,g_{a_n+j_n-1}, g^\prime_n, g^{\prime\prime}_n] .
\end{equation}
The first vector $\mathbf{t}_n$ represents the symbols coming from the variable nodes connected to the $n$-th check node. More specifically, $z_{a_n}, \ldots ,z_{a_n+j_n-1}$ are from information nodes while $c_{n-1}$ and $c_n^{-1}$ are from parity nodes. The second vector $\mathbf{h}_n$ represents the addition factors on the corresponding edges of the $n$-th check nodes as shown in Fig. \ref{fig:tg}.

\subsection{IRA Lattice Decoder}
As shown in Section \ref{chep:tn}, the multi-dimensional IRA lattice codes have a Tanner graph representation. Therefore, we can employ a modified belief prorogation (BP) decoding algorithm to decode our lattice codes.

The decoder attempts to recover the source message $\mathbf{u}$ from the noisy observation of the AWGN channel output $\mathbf{y} = \mathbf{x}+\mathbf{n}_z $, where $\mathbf{n}_z \thicksim \mathcal{CN}(0,\sigma_{ch}^2) $ denotes the complex AWGN noise. Before decoding, we first need to calculate the symbol-wise a posterior probability (APP) of each coset leader and for each lattice codeword component $x_n$, which is written as:
\begin{equation}\label{eq:app1}
P(x_n|y_n) = \frac{p(y_n|x_n)p(x_n)}{p(y_n)}, \,\, \text{for}\, \, n = 1,2,\ldots,N.
\end{equation}
For the sake of simplicity, We let
\begin{equation}\label{eq:app3}
P_{\psi_k}[n] = P(x_n = \psi_k|y_n),
\end{equation}
where $k = 0,1,\ldots,p^M-1$ and $\psi_k$ is the $k$-th coset leader. Since the transmitted codeword symbol is $x_n = c_n+r_n$, where $r_n$ is uniformly distributed over $\Psi$, thus the distribution for $x_n$ is also uniform over $\Psi$. Therefore, Eq. (\ref{eq:app3}) can be written as
\begin{equation}\label{eq:app5}
P_{\psi_k}[n] = \frac{P(y_n|x_n = \psi_k)}{\sum_{k=0}^{p^M-1} P(y_n|x_n = \psi_k)},
\end{equation}
where
\begin{equation}\label{eq:app4}
P(y_n|x_n = \psi_k) = \frac{1}{\sqrt{2\pi\sigma_{ch}^2}}\exp\left(-\frac{\|y_n-\sqrt{\text{SNR}}\psi_k\|^2}{2\sigma_{ch}^2}\right),
\end{equation}
In this way, we have $\sum_{k=0}^{p^M-1}P_{\psi_k}[n]=1$.

In (\ref{eq:app4}), $\psi_k$ and $y_n$ both are vectors with length equal to the dimension of the lattice. In our design example, $\psi_k$ is a $D_4$ lattice point with four dimensions. We perform the symbol-wise maximum-likelihood detection. Considering that practical systems can only transmit and receive one two-dimensional signal at each time slot, the detection is a joint detection for two two-dimensional signals.

We denote the APP vector by $\mathbf{P}[n]$ where
\begin{equation}\label{eq:app6}
\mathbf{P}[n] = [P_{\psi_0}[n],P_{\psi_1}[n],\ldots,P_{\psi_{p^M-1}}[n]]^T.
\end{equation}

Then the above APP vectors are fed into a coset remover to obtain the APP vectors with respect to $\mathbf{c}$ in (\ref{eq:40}) as the message before adding the random-coset vector $\mathbf{r}$. We denote the APP vector after removing coset by $\mathbf{P}^\prime[n]$:
\begin{align}\label{eq:app2}
\mathbf{P}^\prime[n] & = \mathbf{P}(c_n|y_n) \nonumber \\
&= [P_{\psi_0 \ominus r_n}[n],P_{\psi_1  \ominus r_n}[n],\ldots,P_{\psi_{p^M-1}  \ominus r_n}[n]]^T.
\end{align}
where $\ominus$ is defined in (\ref{eq:1dss}). The resultant APP vector $\mathbf{P}^\prime[n]$ is then passed into a BP decoder.

The decoder updates the information between check nodes and variable nodes in an iterative manner. We denote the message from the $m$-th variable node to the $n$-th check node by $\bm{r}(m,n)$. The message passed from the $n$-th check node to the $m$-th variable node is denoted by $\bm{l}(n,m)$. Both vectors are probability vectors with dimension $p^M$. Use the Tanner graph in Fig. \ref{fig:tg}, we let $\mathcal{A}(m)$ and $\mathcal{B}(n)$ represent the set of check nodes connected to the $m$-th variable node and the set of variable nodes adjacent to the $n$-th check node, respectively. Without the loss of generality, let the index of information nodes be from $1$ to $K$ and the index of parity nodes be from $(K+1)$ to $(K+N)$ of the variable nodes. The decoding steps can be summarized in the following.

\emph{1) Initialization step:}
According to the Tanner graph in Fig. \ref{fig:tg}, the channel output must go through the parity nodes first. Thus for all edges $(m,n)$ between the parity nodes and the check nodes in the Tanner graph, the initial message $\bm{r}(m,n)$ is the channel APP in (\ref{eq:app2}):
\begin{align}\label{eq:de1}
\bm{r}(m,n) = \mathbf{P}^\prime[m-K], \,\,\,\,\text{for}\,\, &m = K+1,\ldots,K+N \nonumber \\
&n = 1,2,\ldots,N.
\end{align}

For all edges $(m,n)$ between the information nodes and the check nodes in the Tanner graph, we let
\begin{align}\label{eq:de1a}
r_k(m,n) = \frac{1}{p^M}, \,\,\,\, \text{for}\,\, &k =0,1,\ldots,p^M-1 \nonumber \\
&m = 1,2,\ldots,K.
\end{align}


\emph{2) Update the check nodes to variable nodes messages:}
For all edges $(m,n)$ that connected to the $n$-th check node, generate the probability vector $\bm{l}(n,m)$ with its $k$-th element given by
\begin{equation}\label{eq:de2}
l_k(n,m) = \sum_{\substack{t_1,\ldots,t_{j_n-1} \in \Psi \\ \bigoplus_{i=1}^{j_n-1}(t_i \oplus h_i) \oplus \psi_k \oplus (h_{j_n})=0}}\prod_{i=1}^{j_n-1}r_{t_i}^{(i)},
\end{equation}
where $\bigoplus$ is the summation performed by $\oplus$; $j_n$ is the degree of the $n$-th check node; $\bm{r}^{(1)},\ldots,\bm{r}^{(j_n-1)}$ are the incoming messages from all the connected variable nodes except the $m$-th variable node, i.e., $\{\bm{r}(m^\prime,n):m^\prime \in \mathcal{B}(n) \setminus \{m\} \}$; $t_1,\ldots,t_{j_n-1}$ are the lattice symbols from the associated variable nodes; $h_{j_1},h_{j_2},\ldots,h_{j_n-1}$ are the addition factors on the corresponding edges and $h_{j_n}$ denotes the addition factor for the edge $(m,n)$. Note that the calculations of the check node messages are different from that in conventional IRA decoding as the parity-check equations and the associated arithmetic are different.

\emph{3) Update the variable nodes to check nodes messages: }
For all edges $(m,n)$ between the variable nodes and the check nodes in the Tanner graph, generate the probability vector $\bm{r}(n,m)$ with the $k$-th element given by
\begin{equation}\label{eq:de3}
r_k(m,n) = \frac{\gamma^{(n)}_k\prod_{i=1}^{j_m-1}l_k^{(i)}}{\sum_{k^\prime=0}^{p^M-1} \gamma^{(n)}_{k^\prime}\prod_{i=1}^{j_m-1}l_{k^\prime}^{(i)}},
\end{equation}
where $j_m$ denotes the degree of the $m$-th variable node; $\bm{l}^{(1)},\ldots,\bm{l}^{(j_m-1)}$ denote the incoming messages from all the connected check nodes except the $n$-th check node, i.e., $\{\bm{l}(n^\prime,m):n^\prime \in \mathcal{A}(m) \setminus \{n\} \}$; $\gamma^{(n)}_k = r_k(m,n)$ in (\ref{eq:de1}) for $m = K+1,\ldots,K+N$ when the messages are from parity nodes to the $n$-th check node and $\gamma^{(n)}_k = r_k(m,n)$ in (\ref{eq:de1a}) for $m = 1,\ldots,K$ when the messages are from information nodes to the $n$-th check node.

\emph{4) Stopping condition: }
For each iteration, make the hard decision on the $m$-th variable node by calculating
\begin{equation}\label{eq:de4}
\hat{\delta}_n = \argmax_k\frac{\gamma^{(n)}_k\prod_{i=1}^{j_m}l_k^{(i)}}{\sum_{k^\prime=0}^{p^M-1} \gamma^{(n)}_{k^\prime}\prod_{i=1}^{j_m}l_{k^\prime}^{(i)}},
\end{equation}
for $n = 1,2,\ldots,K+N$. It contains information from all the connected edges. If the hard decision results  $\hat{\delta}_1,\hat{\delta}_2,\ldots,\hat{\delta}_{K+N}$ satisfy the parity-check equations in (\ref{eq:41}) or a predetermined maximum number of iterations is reached, then stop; otherwise go to Step 2).

The calculation in (\ref{eq:de2}) has a very high computational complexity if the cardinality of the lattice partition $p^M$ is very large. We follow \cite{Richardson01} to employ DFT and IDFT in our lattice decoding process to reduce the complexity.

First we need to introduce some important notations which will be used in the rest of this chapter. Define a probability vector as $\boldsymbol{\rho} = [\rho_{\psi_0}, \rho_{\psi_1},\ldots,\rho_{\psi_{p^M-1}}]$ representing the probability of a lattice point being $\psi_0, \psi_1, \ldots, \psi_{p^M-1}$. In addition, the probability vector must satisfy $\rho_{\psi_k} \geq 0$ and $\sum_{k=0}^{p^M-1}\rho_{\psi_k} = 1$. Given a probability vector $\boldsymbol{\rho}$ and $\chi \in \Psi$, we define the $\oplus \chi$ operation as the following
\begin{equation}\label{eq:de5}
 \boldsymbol{\rho}^{\oplus\chi}=[\rho_{\psi_0 \oplus \chi}, \rho_{\psi_1\oplus \chi},\ldots,\rho_{\psi_{p^M-1}\oplus \chi}].
\end{equation}

Now consider the expression in (\ref{eq:de2}), an equivalent expression can be written as
\begin{equation}\label{eq:de6}
\bm{l} = \bigg[\bigotimes_{i=1}^{j_n-1}\left(\bm{r}^{(i)}\right)^{\ominus h_i} \bigg]^{\ominus h_{j_n}},
\end{equation}
where $\bm{l}$ is the vector that contains elements $l_k$, $k = 0,1,\cdots, p^M-1$ in (\ref{eq:de2}) and the ``$\bigotimes$'' operator performs the modulo-lattice convolution between two vectors. It produces a vector whose $k$-th component is:
\begin{equation}\label{eq:de7}
[\bm{r}^{(1)} \otimes \bm{r}^{(2)}]_k = \sum_{\chi \in \Psi} r_{\chi}^{(1)}\cdot r_{\psi_k \ominus \chi}^{(2)}, \,\, \text{for}\, k =0,1,\ldots,p^M-1.
\end{equation}
This convolution can be evaluated by using $M$-dimensional DFT and IDFT \cite{Dudgeon84}. In this way, (\ref{eq:de6}) can be evaluated as
\begin{equation}\label{eq:de8}
\bm{l} = \Bigg[\text{IDFT} \bigg[\prod_{i=1}^{j_n-1}\text{DFT}\left( \left( \bm{r}^{(i)}\right)^{\ominus h_i}\right)  \bigg]\Bigg]^{\oplus h_{j_n}},
\end{equation}
where the multiplication of the DFT vectors is performed in a component-wise manner. A further reduction in complexity of implementation can be obtained by using fast Fourier transform and inverse fast Fourier transform algorithms.

\subsection{Complexity of IRA lattice codes}
In this subsection, the complexity of our multi-dimensional IRA lattice codes will be investigated and compared to that of the IRA lattice codes with two-dimensional lattice partitions. Note that both lattice codes are built from Construction A. The underlying linear code for our design is over $\mathbb{F}_p^2$ while the linear codes for the design with two-dimensional lattices is over $\mathbb{F}_p$ \cite{Qiu16}.

For encoding, the computational complexity is the same as that of our previous design. As it can be seen from Fig. \ref{fig:encoder_CH4}, the computational complexity of repeating, interleaving, combining and accumulating process does not change with the cardinality of the coset leaders. However, the storage is of $O(p^2)$ for storing the lookup table for the modulo lattice operation while for \cite{Qiu16} is $O(p)$.

Next, we focus on the complexity of symbol-wise detection. For an ML detector, the detection is based on the entire constellation. Thus, for a two-dimensional constellation with size $p$, the computational complexity is in the order of $O(p)$. In our design, we have a four-dimensional constellation with size $p^2$, the computational complexity is $O(2p^2)$. The ``2'' here is due to the joint detection for two two-dimensional symbols. The computational complexity of the nonbinary BP decoding is in the order of $O(p\log_2 p)$ when FFT is employed for check node calculations \cite{Ganepola08}. For our decoder to decode lattice codes with four-dimensional lattice partitions, the complexity is $O(p^2\log_2 p^2)$. Compared with our previous coding scheme with two-dimensional lattice partitions, the complexity of the code design in this work is $2p$ times higher. Note that here we do not include the discussion of complexity contribution of check nodes and variable nodes since they will be optimized and varied in our designed and are not fixed. Furthermore, the memory usage is associated with the non-zero elements of the parity-check matrix. Thus, the required memory can only be determined specifically case by case.

For Construction A lattices, it has been shown in \cite{1337105} that the finite field size of the underlying linear code has to be large enough to achieve the capacity. Therefore, we have traded the complexity to attain better performance by introducing multi-dimensional lattice partitioned in our design.

\section{Design and Analysis of Multi-dimensional IRA Lattice Codes}
In this work, the analysis of our multi-dimensional IRA lattice codes focus on the average behaviour of randomly selected codes from an ensemble of codes. First, let $\alpha_i$ be the fraction of interleaver's edges that connected to the information nodes with degree $i$ and let $\beta_j$ be the fraction of interleaver's edges that are connected to the check nodes with degree $j+2$. Recall in Section \ref{IRAE} that $i = 2,3,\ldots,I$ and $j = 1,2,\ldots,J$. The additional ``2'' here means every check node has two deterministic connections from the connected parity nodes as shown in Fig. \ref{fig:tg}. Following \cite{Jin00}, the edge degree distributions of our multi-dimensional IRA lattice codes can be written as
\begin{equation}\label{eq:lam}
\alpha(x) = \sum_{i=2}^I\alpha_ix^{i-1}.
\end{equation}
\begin{equation}\label{eq:rho}
\beta(x) = \sum_{j=1}^J\beta_jx^{j-1}.
\end{equation}
Given $\alpha$, $\beta$, the type of lattice $\mathcal{R}$ and the scaling factor $\xi$ in (\ref{eq:31a}), we define an $(\alpha, \beta, \xi,\mathcal{R})$ ensemble as the
set of our multi-dimensional IRA lattice codes obtained via Construction A.

\subsection{Modeling the Decoder's Message Distributions}\label{msi}
In our multi-dimensional IRA lattice codes, the soft information propagated in the iterative decoder can be modeled by a multi-dimensional LLR vector. Even though APP is used in our iterative decoder, it is common to use LLR in EXIT chart analysis. Note that APP and LLR are different but equivalent representations of the decoder's soft information. In order to track the convergence behaviour of the iterative decoding, multi-dimensional EXIT charts may be required. However, developing these EXIT chart functions can be very difficult. To deal with this challenge, the new encoding structure is proposed. We will prove that using this structure, the densities of the messages in BP decoder can attain permutation-invariance and symmetry properties. With these two properties, the densities of the decoder's messages can be represented as a single parameter. In this way, our method only needs to track one-dimensional variables rather than the true densities of the multi-dimensional LLR vectors. In addition, the symmetry property enables to use all-zero lattice codeword assumption in the \emph{EXIT chart analysis}. As such, the expression of mutual information in the EXIT chart analysis can be simplified.

We first introduce some useful definitions and notations in the following.

\subsection{Preliminaries}
Following the definition in \cite{Li03}, we define the LLR values for a given probability vector $\boldsymbol{\rho}$ as
\begin{equation}\label{eq:LLR}
\omega_{\psi_k} = \ln \left(\frac{\rho_{\psi_0}}{\rho_{\psi_k}} \right), \,\text{for}\, k = 0,1,\ldots,p^M-1.
\end{equation}
It is intuitive that $\omega_{\psi_0} = 0$.

The $p^M$-dimensional LLR vector is then defined as $\boldsymbol{\omega} = [\omega_{\psi_0}, \omega_{\psi_1},\ldots,\omega_{\psi_{p^M-1}}]^T$. Note that unlike most LLR definitions, we include the element $\omega_{\psi_0}$ in the LLR vectors as it is associated with our analysis of permutation-invariance which will be introduced shortly. When we apply the $\oplus \chi$ operation defined in (\ref{eq:de5}) on the LLR value $\omega_{\psi_k}$, we have:
\begin{equation}\label{eq:LLR1}
\omega_{\psi_k}^{\oplus \chi} = \ln \left(\frac{\rho_{\psi_0 \oplus \chi}}{\rho_{\psi_k \oplus \chi}} \right) = \omega_{\psi_k \oplus \chi} - \omega_{\psi_0 \oplus \chi}.
\end{equation}

Following from above, a $p^M$-dimensional \emph{probability-vector random variable} is defined as
$\mathbf{P} = [P_{\psi_0},P_{\psi_1},\ldots,P_{\psi_{p^M-1}}]^T$ that only takes valid probability values. The associated $p^M$-dimensional \emph{LLR-vector random variable} is defined as $\mathbf{W} = [W_{\psi_0},W_{\psi_1},\ldots,W_{\psi_{p^M-1}}]^T$.

Now we introduce the definitions of the symmetry and permutation-invariance properties and explain how we can achieve these properties.

\subsection{ Symmetry}
Recall in Section \ref{IRAE}, we add a random-coset vector $\mathbf{r}$ at the end of the encoder. The random-coset elements are randomly chosen and uniformly distributed over the set of coset leaders $\Psi$. Thus we have the following theorem.

\begin{theo}\label{OS}
Adding a random-coset vector $\mathbf{r}$ to the encoder output $\mathbf{c}$, where $\mathbf{r}$ is uniformly distributed over $\Psi$, can produce the output-symmetric effect in non-binary input AWGN channels.
\end{theo}

\emph{\quad Proof: } Please refer to Appendix \ref{appendix:4}.\QEDA

Similar to the non-binary LDPC codes in \cite{Bennatan06}, the LLR random vectors are symmetric under the output-symmetric effect. The symmetry property of an LLR random vector is defined as follows.

\begin{defi}
Given an LLR random vector $\mathbf{W}$ and an $r \in \Psi$, $\mathbf{W}$ is symmetric if and only if $\mathbf{W}$ satisfies
\begin{equation}\label{eq:SS}
\text{Pr}[\mathbf{W} = \boldsymbol{\omega} ] = e^{\omega_{\psi_k}}\text{Pr}[\mathbf{W} = \boldsymbol{\omega}^{\oplus r}]
\end{equation}
for all LLR vectors $\boldsymbol{\omega}$ and all $r \in \Psi$.
\end{defi}

With this property, the probability of decoding error is equal for any transmitted codeword \cite{Bennatan06}. In other words, the symmetry property removes the dependence of the decoder's LLRs on transmitted codewords \cite{Richardson01}. Therefore, we can use all-zero lattice codewords in our EXIT chart analysis.

\subsection{ Permutation-Invariance}\label{PI}
We start with the definition of permutation-invariance \cite[Section 2.6]{Severini05} on a probability-vector random variable. Then we will show that our approach can achieve this property under our proposed structure.

\begin{defi}
A probability-vector random variable $\mathbf{X} = [X_0,X_1,X_2 \ldots]$ is permutation-invariant if for any permutation $\varpi$ of the indices such that the random vector $\varpi(\mathbf{X}) = [X_{\varpi(0)},X_{\varpi(1)},X_{\varpi(2)},\ldots]$ is distributed identically with $\mathbf{X}$.
\end{defi}

Under this property, all the random variables in $\mathbf{X}$ are identically distributed (but may not be independent). Therefore, changing the order of the elements in $\mathbf{X}$ will not change the distribution of $\mathbf{X}$ .

Recall in Section \ref{IRAE}, our codes have three randomly generated sequences added to the encoder's messages. This leads to a symbol level permutation (the permutation from a coset leader to another coset leader) on the messages. The densities of these messages can be shown to have the permutation-invariance property. Now, we have the following theorem:

\begin{theo}\label{PERI}
Given a $p^M$-dimensional probability-vector random variable $\mathbf{P}$ and a $\chi \in \Psi$, the random vector $\mathbf{P}^{\oplus \chi} = [P_{\psi_0 \oplus \chi}, P_{\psi_1\oplus \chi},\ldots,P_{\psi_{p^M-1}\oplus \chi}]$ is identically distributed with $\mathbf{P}$. Therefore $\mathbf{P}$ is permutation-invariant.
\end{theo}

\emph{\quad Proof: } Please refer to Appendix \ref{appendix:2}.\QEDA

This theorem can be carried over straightforwardly to LLR representation. Thus we have the following lemma:

\begin{lemm}\label{PERIW}
Let $\mathbf{W}= [W_{\psi_0},W_{\psi_1},\ldots,W_{\psi_{p^M-1}}]^T$ be an LLR-vector random variable such that $W_{\psi_k} = \ln \left(\frac{P_{\psi_0}}{P_{\psi_k}} \right), \,\text{for}\, k = 0,1,\ldots,p^M-1$. If $\mathbf{P}$ is permutation-invariant, then $\mathbf{W}$ is also permutation-invariant.
\end{lemm}

\emph{\quad Proof: } Please refer to Appendix \ref{appendix:3}.\QEDA

Therefore, under the BP decoding, the messages passed within the Tanner graph of our codes satisfy all the symmetry and permutation-invariance properties.

\subsection{Gaussian Approximation}
With the symmetry and permutation-invariance properties, the $p^M$-dimensional LLR can be modeled using a multivariate Gaussian distribution \cite{Bennatan06}:
\begin{equation}\label{eq:ga}
f_{\mathbf{W}}(\boldsymbol{\omega})=\frac{1}{(2\pi)^{\frac{p^M}{2}}|\boldsymbol{\Sigma}|^{\frac{1}{2}}}\exp\left(-\frac{1}{2}(\boldsymbol{\omega}-\mathbf{m})^T\boldsymbol{\Sigma}^{-1}(\boldsymbol{\omega}-\mathbf{m})\right),
\end{equation}
with mean vector $\mathbf{m}$ and covariance matrix $\boldsymbol{\Sigma}$ given by
\begin{equation}\label{eq:ga1}
\mathbf{m}=
\begin{bmatrix}
    \frac{\sigma^2}{2}       \\
     \frac{\sigma^2}{2}        \\
    \vdots \\
     \frac{\sigma^2}{2}
\end{bmatrix}
\,\text{and}\,\,\boldsymbol{\Sigma}=
\begin{bmatrix}
    \sigma^2 & \frac{\sigma^2}{2} & \cdots & \frac{\sigma^2}{2}   \\
    \frac{\sigma^2}{2} & \sigma^2 & \cdots & \frac{\sigma^2}{2}   \\
    \vdots & \vdots & \ddots & \vdots  \\
    \frac{\sigma^2}{2} & \cdots & \cdots & \sigma^2
\end{bmatrix}.
\end{equation}
More specifically, $m_i = \frac{\sigma^2}{2}$ for $i = 1,2,\ldots,p^M$, and $\boldsymbol{\Sigma}_{i,j} = \sigma^2$ if $i=j$ and $\frac{\sigma^2}{2}$ otherwise. As a result, the density of the $p^M$-dimensional LLR is completely described by a single parameter $\sigma$. It is worth mentioning that our definition of LLR random vector is $p^M$-dimensional rather than $p^M-1$ in the literature. This is because the $\oplus \chi$ operation will change the position of $W_{\psi_0}$. Thus we need to use a $p^M$-variate Gaussian distribution to model the $p^M$-dimensional LLR.

\subsection{Convergence Analysis}
EXIT charts track the mutual information between the transmit lattice symbol $u$ and the LLR random vector $\mathbf{W}$. With the all-zero lattice codeword assumption, the mutual information can be evaluated according to \cite{Bennatan06}
\begin{equation}\label{eq:exit1}
I(u;\mathbf{W}) = 1-E\Bigg[\log_{p^M}\left(\sum_{i=0}^{p^M-1}e^{-w_i} \right) \Bigg|u = 0 \Bigg],
\end{equation}
where $\mathbf{W}$ is modeled by (\ref{eq:ga}) and (\ref{eq:ga1}). Thus, the mutual information is a function of the single parameter $\sigma$. For simplicity, we let $J(\sigma) = I(u;\mathbf{W})$ as every value of $\sigma$ corresponds to a value of $I(u;\mathbf{W})$. Since the mapping is bijective, we can also define the inverse function $J(.)^{-1}$ to obtain $\sigma$ when given $I(u;\mathbf{W})$.

In the EXIT chart analysis, variable nodes are treated as a component decoder while the combiners and the time-varying accumulator together is treated as another decoder. As such, we compute the variable-node decoder (VND) curve and the check-node decoder (CND) curve. The argument of each curve is denoted as $I_A$ and the value of the curve is denoted as $I_E$, representing a priori input and the extrinsic output of each component decoder. The details of obtaining the transfer functions will be explained next.

\subsection{EXIT Function for VND}
For a variable node with $i_m$ degrees, the output mutual information of the VND for this type of variable nodes is given by \cite{Brink03}:
\begin{equation}\label{eq:VND}
I_{E,VND}(I_A,i_m)\approx J\left( \sqrt{(i_m-1)}J^{-1}(I_A)  \right).
\end{equation}
For a given VN degree distribution $(i,\alpha_i)$, the EXIT function for the VND of the entire IRA code is:
\begin{equation}\label{eq:VND1}
I_{E,VND}(I_A) =  \sum_{i=2}^I \alpha_iI_{E,VND}(I_A;i).
\end{equation}

\subsection{EXIT Function for CND}
For a check node with degree $j_n$, we use a numerical method to obtain the approximated EXIT functions as there is no closed-form expression in the literature.

For a given $I_A$, we obtain the corresponding parameter using $\sigma = J^{-1}(I_A)$. Then the input a priori LLR vectors are generated according to (\ref{eq:ga}) and (\ref{eq:ga1}). For a given SNR, generate the all-zero lattice codeword, three random sequences $\mathbf{g}$, $\mathbf{g^\prime}$, $\mathbf{g^{\prime\prime}}$, a random-coset vector $\mathbf{r}$ and an AWGN channel noise sequence with variance of $\sigma_{ch}^2$. We calculate the channel APPs by following (\ref{eq:app1}) to (\ref{eq:app2}) and then substitute the results into (\ref{eq:LLR}) to obtain the channel input LLR $\mathbf{W}_{ch}$. Given $\mathbf{g}$, $\mathbf{g^\prime}$, $\mathbf{g^{\prime\prime}}$, $\mathbf{r}$, $j_n$ and $\mathbf{W}_{ch}$, we perform BP decoding with one iteration to produce the output LLR. The $I_{E,CND}(I_A)$ associated with the check node degree $j_n$ is obtained by substituting the output LLR into (\ref{eq:exit1}).

For a given CN degree distribution $(j,\beta_j)$, the EXIT function for the CND of the entire IRA code can be obtained by:
\begin{equation}\label{eq:CND}
I_{E,CND}(I_A,\sigma_{ch}) =  \sum_{j=1}^J \beta_jI_{E,CND}(I_A;j,\sigma_{ch}).
\end{equation}

\subsection{Design Examples}
Based on our EXIT functions, we now employ the EXIT chart curve fitting technique \cite{Brink03} to find the optimal CN and VN degree distributions such that the area between the CN curve and the VN curve is minimized. First, we carefully select an appropriate CN degree distribution. Then, we fit the EXIT curve of VND to CND by using linear programming to optimize the degree distribution for VN. Next, we update the CN degree distribution based on the optimized VN degree distribution. The optimization for the degree distribution of CN and VN are carried out in an iterative manner. Note that we have set the minimum gap between the VND curve and the CND curve to be greater than zero but not too large, e.g., 0.0001. In this way, the produced VND curve do not intersect with the CND curve and both curves create a narrow tunnel. The number of optimization iteration is set to 10 as more iterations does not improve the optimization results further.

\begin{figure}[ht!]
	\centering
\includegraphics[width=3.2in,clip,keepaspectratio]{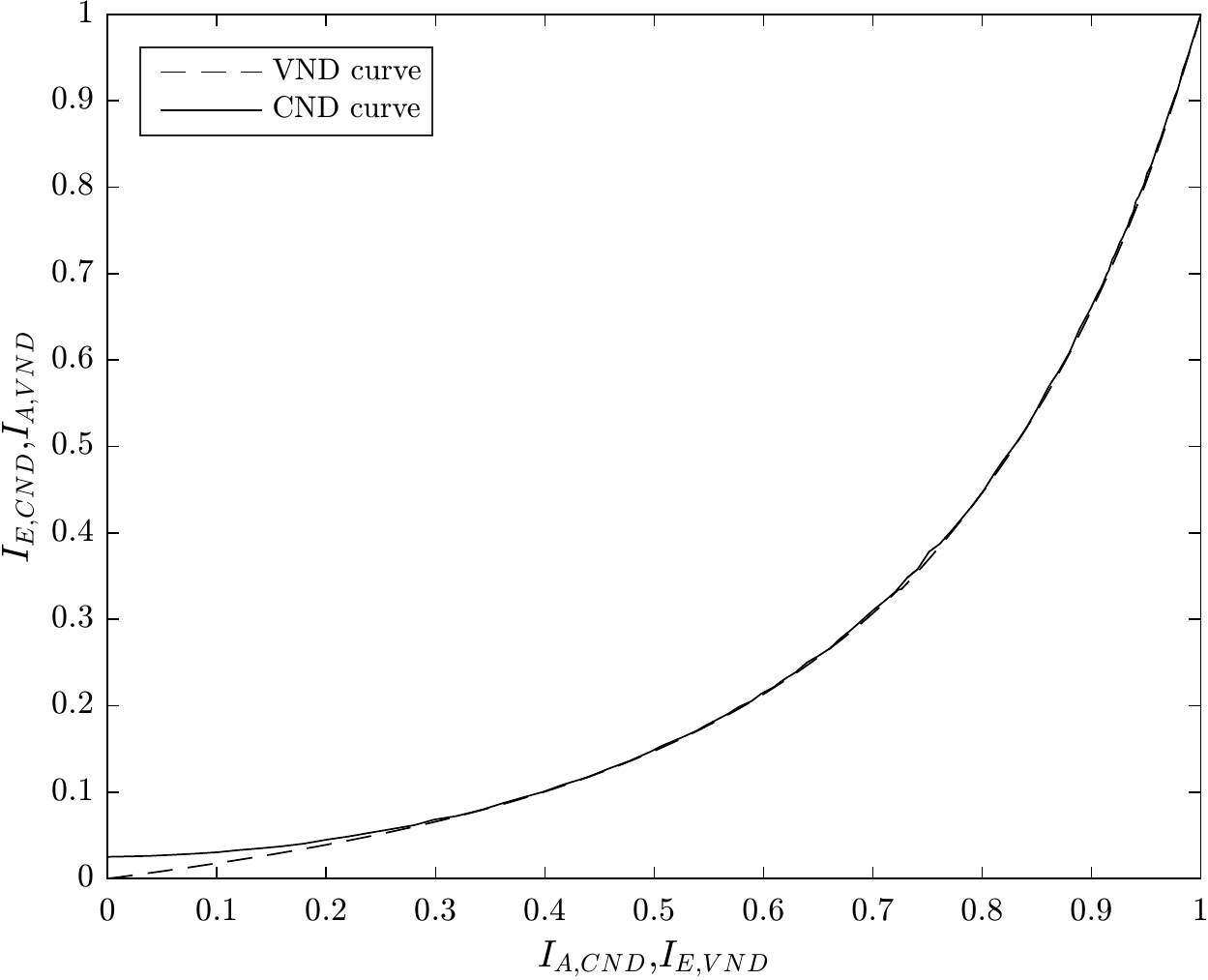}
\caption{EXIT Chart of optimized degree distributions for the rate $\frac{2}{3}$ multi-dimensional IRA lattice code.}
\label{fig:EXIT_CH4}
\end{figure}

An example of an EXIT chart for our multi-dimensional IRA lattice codes over $\mathbb{H}/(1+2i)\mathbb{H}$ with code rate of $\frac{2}{3}$ is illustrated in Fig. \ref{fig:EXIT_CH4}.
In our design, the portion of degree 1 CN must not be too small in order to ensure the decoder works in the first few iterations because our codes are nonsystematic \cite{Brink03}. From Fig. \ref{fig:EXIT_CH4}, we can see that the VND curve literally touches the CND curve for the range $[0,1]$, which guarantees successful convergence and accurate decoding threshold.

We have adopted the proposed approach in designing the $\left(\alpha, \beta, 1+2i,\mathbb{H} \right)$-lattice ensemble with three code rates $\frac{3}{4}$, $\frac{2}{3}$ and $\frac{1}{2}$. The degree distributions and the decoding thresholds are shown in Table \ref{Table:CH4_2}.

\begin{table*}
\caption{Optimal degree distributions and decoding thresholds of $\left(\alpha, \beta, 1+2i,\mathbb{H} \right)$-lattice ensemble with various code rates}
\label{Table:CH4_2}
\centering
\begin{tabular}{|c|c|c|c|}
\hline
 Rates & Thresholds & Degree Distributions: $(i,\alpha_i)$ for VNs, $(j,\beta_j)$ for CNs \\ \hline
  $\frac{3}{4}$ & 4.47 dB & \begin{tabular}{c}
  $\alpha$: (2,0.288274), (3,0.265333), (7,0.188119),\\ (13,0.123885), (15,0.134389)\\
  $\beta$: (1,0.055556), (3,0.944444)\\
  \end{tabular}  \\ \hline
 $\frac{2}{3}$ & 3.31 dB & \begin{tabular}{c}
  $\alpha$: (2,0.240605), (3,0.231215), (7,0.081754), \\(8,0.190942), (19,0.175951), (20,0.079534)\\
  $\beta$: (1,0.053861), (3,0.946139)\\
\end{tabular}  \\ \hline
 $\frac{1}{2}$ & 1.26 dB & \begin{tabular}{c}
  $\alpha$: (2,0.163689), (3,0.170788), (8,0.120858), \\(9,0.148837), (19,0.038618), (20,0.088323), (34,0.268886)\\
  $\beta$: (1,0.054328), (3,0.945672)\\
\end{tabular}    \\ \hline
\end{tabular}
\end{table*}
As shown in the table, the optimized CN distributions are degree 1 and degree 3 because this pair of CN distributions have the lowest optimization complexity and the minimum decoding threshold for the three code rates. We have also designed our codes with other pairs of CN distributions, but their performance is not much better than the code with only degree 1 and degree 3 CNs.

\begin{table*}[]
  \centering
 \caption{Comparisons of coding schemes}\label{table3}
\begin{tabular}{|c|c|c|c|}
\hline
 Coding schemes  & n [symbols]   & Coding loss [dB] & Gap [dB] \\  \hline
 GLD lattices \cite{Boutros14}  & 1,000  & 1.3  &  N/A\\  \hline
 \multirow{2}{*}{LDA lattices \cite{8122043}}  & 1,000  & 1.36 &  N/A \\
 \hhline{~---}
   & 10,000  & 0.7 &  N/A \\ \hline
  \multirow{3}{*}{LDA lattices  \cite{Boutros16}}  & 10,008  & 0.55 &  1.05 \\
  \hhline{~---}
    & 100,008  & 0.36 &  0.9 \\
    \hhline{~---}
   & 1,000,008  & 0.3 &  0.8 \\ \hline
 \multirow{3}{*}{ LDLCs \cite{4475389} }       & 1,000  & 1.5  & N/A \\
   \hhline{~---}
          & 10,000  & 0.8  & N/A \\
  \hhline{~---}
          & 100,000  & 0.6  & N/A \\ \hline
\multirow{2}{*}{ QC-LDPC lattices \cite{Khodaiemehr17}}  & 1,190  &  2 & N/A \\
  \hhline{~---}
    & 30,000  &  1.5 & N/A \\ \hline
\multirow{3}{*}{    IRA lattices } & 1,000  & 1.5 &  1.7 \\
  \hhline{~---}
      & 10,000  & 0.6 &  0.8 \\
  \hhline{~---}
      & 100,000  & 0.3 &  0.46 \\ \hline
\end{tabular}
\end{table*}

\section{Simulation Results}
\begin{figure}[ht!]
	\centering
\includegraphics[width=3.53in,clip,keepaspectratio]{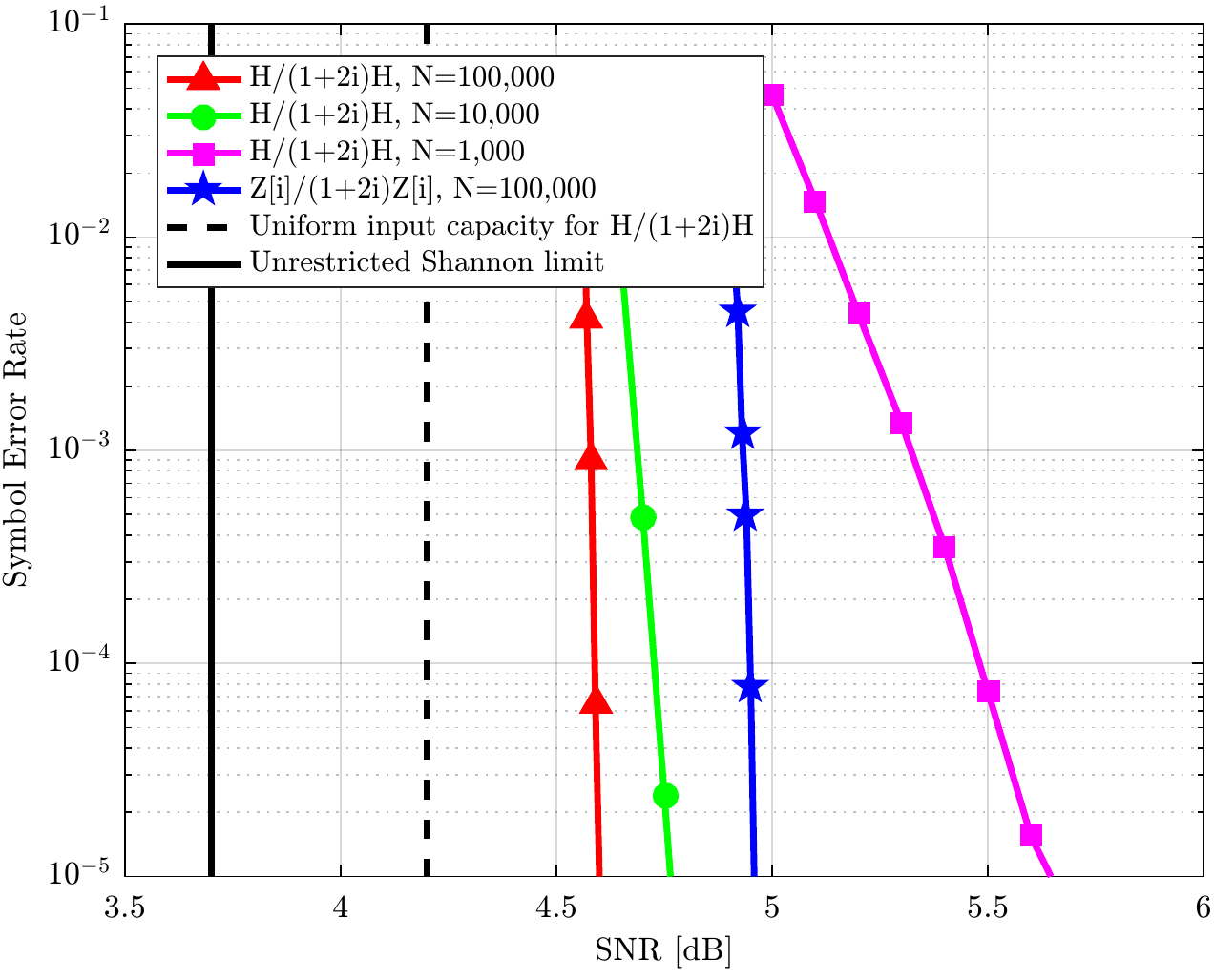}
\caption{Symbol error rate performance of rate $\frac{3}{4}$ codes.}
\label{fig:SER_C4_4}
\end{figure}

\begin{figure}[ht!]
	\centering
\includegraphics[width=3.53in,clip,keepaspectratio]{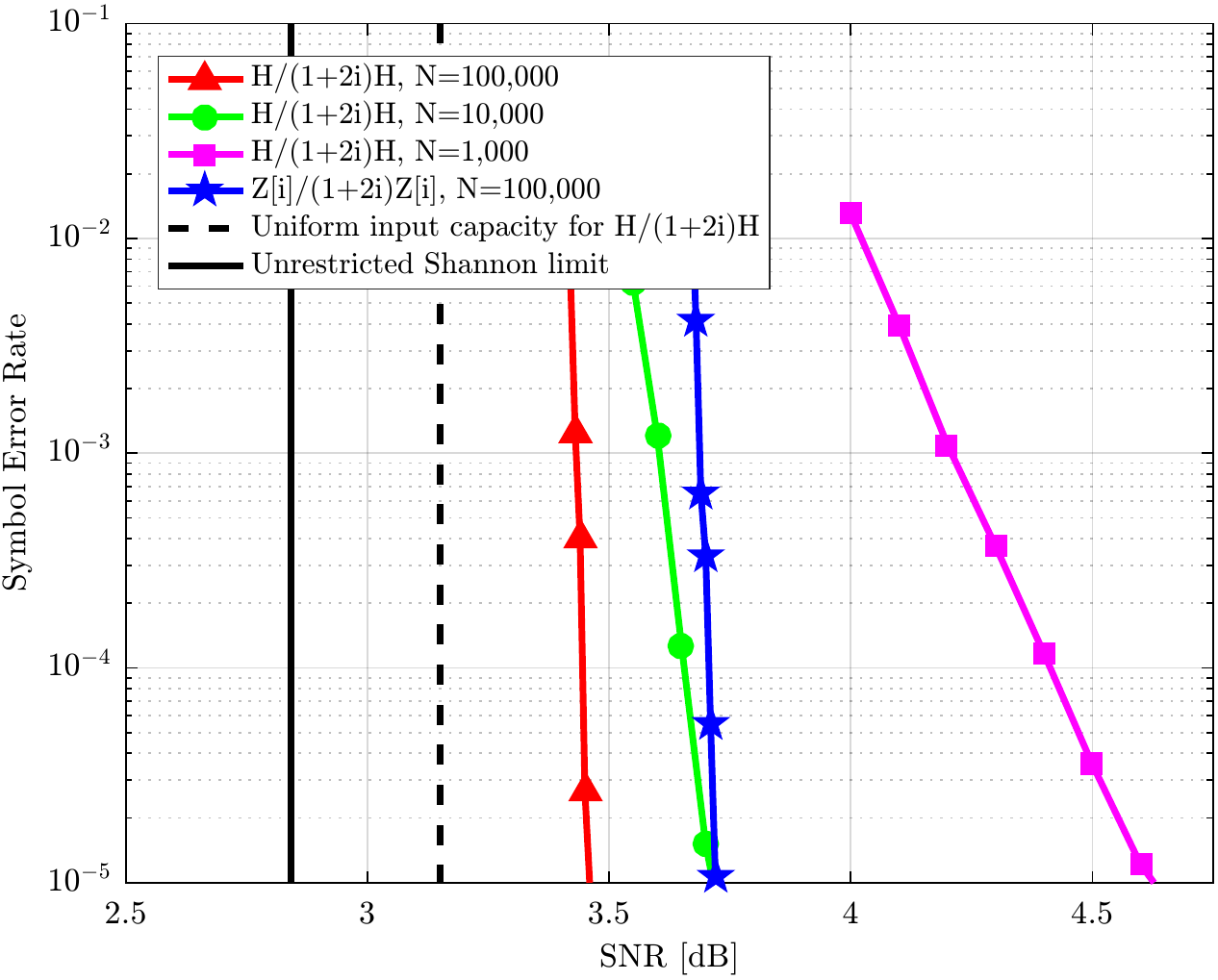}
\caption{Symbol error rate performance of rate $\frac{2}{3}$ codes.}
\label{fig:SER_C5_4}
\end{figure}

\begin{figure}[ht!]
	\centering
\includegraphics[width=3.53in,clip,keepaspectratio]{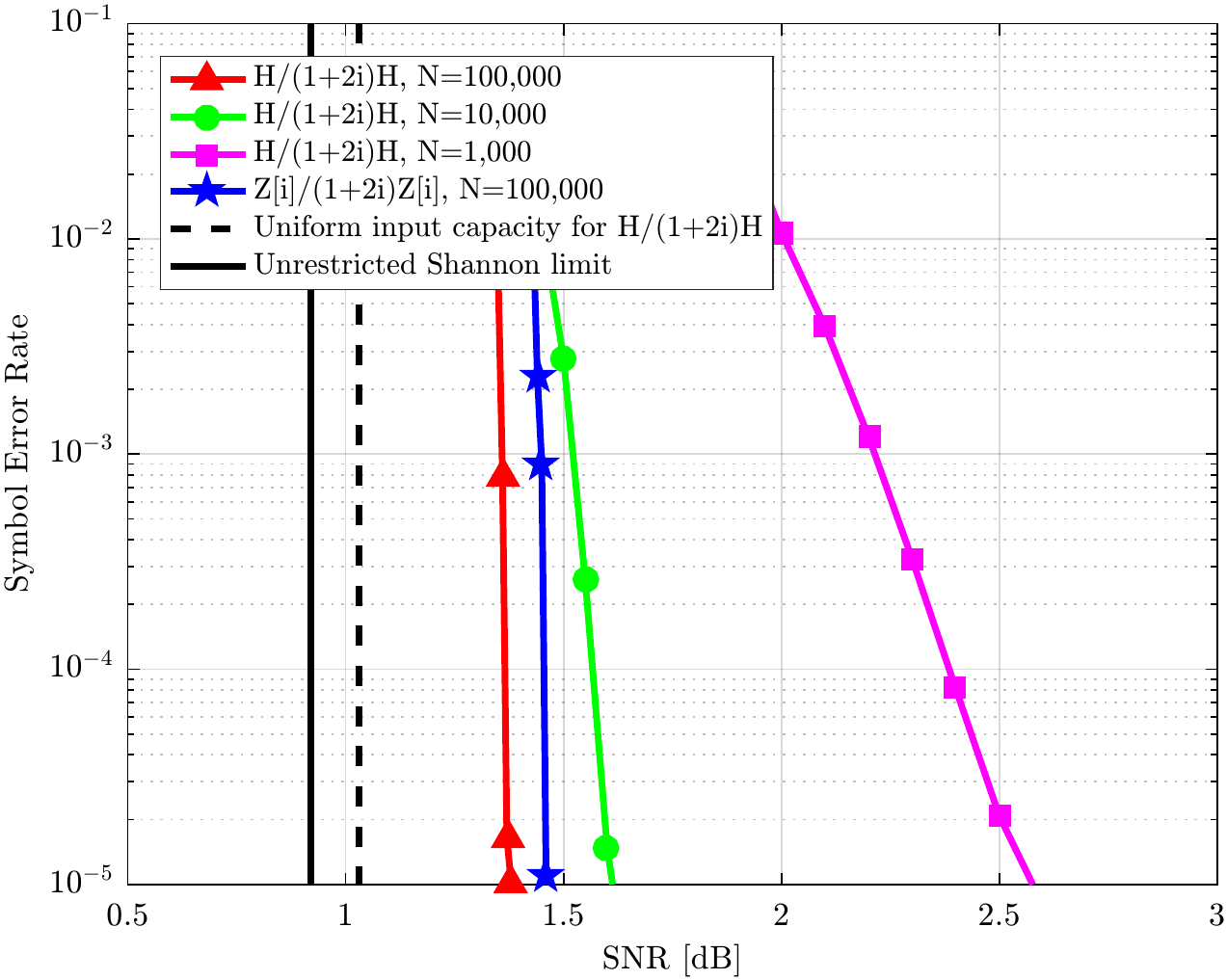}
\caption{Symbol error rate performance of rate $\frac{1}{2}$ codes.}
\label{fig:SER_C6_4}
\end{figure}

In this subsection, we present our simulation results for our multi-dimensional IRA lattice codes over $\mathbb{H}/(1+2i)\mathbb{H}$. In order to evaluate the average behavior of our codes, we randomly generated a codeword from the $(\alpha, \beta, 1+2i,\mathbb{H})$ ensemble and randomly select the values for $\mathbf{g}$, $\mathbf{g^\prime}$, $\mathbf{g^{\prime\prime}}$ and $\mathbf{r}$ in every channel realization. Since our coding scheme is based on \emph{finite constellations} with power constraint, the performance for three designed code rates $\frac{3}{4}$, $\frac{2}{3}$ and $\frac{1}{2}$ is measured in terms of symbol error rate (SER) versus SNR, which are depicted in Fig. \ref{fig:SER_C4_4}, Fig. \ref{fig:SER_C5_4} and Fig. \ref{fig:SER_C6_4}, respectively. Based on these designed code rates, the corresponding information rates are calculated by using (\ref{eq:aa3}) as $R_1 = 1.741$ bits/s/Hz, $R_2 =1.548$ bits/s/Hz and $R_3 =1.161$ bits/s/Hz, respectively. The corresponding unconstrained Shannon limit and uniform input capacity for each information rate are plotted in each figure. Additionally, we also show the SER performance for the previously designed IRA lattice codes over $\mathbb{Z}[i] /(1+2i) \mathbb{Z}[i]$ in all the figures for comparison because both partitions result in the same information rate. In our simulations, we set the codeword length to be 1,000, 10,000 and 100,000 symbols whereas the corresponding step sizes for SNR are 0.1 dB, 0.05 dB and 0.01 dB, respectively. The maximum number of decoding iterations was set to be 200.


In Fig. \ref{fig:SER_C4_4}, the unconstrained Shannon limit for $R_1$ is 3.70 dB. In this case, we observe that the gap to the unconstrained Shannon limit at the SER of $10^{-5}$ is 0.90 dB for our rate $\frac{3}{4}$ $D_4$-partition-based lattice code and 1.28 dB for the code in \cite{Qiu16}. Thus, our newly designed four-dimensional IRA lattice code is 0.38 dB better than the lattice code with two-dimensional lattice partitions. The unconstrained Shannon limit for $R_2$ is 2.84 dB. As shown in Fig. \ref{fig:SER_C5_4}, the gap between our lattice code and the unconstrained Shannon limit is 0.62 dB. For the code in \cite{Qiu16}, the gap is 0.88 dB. Therefore, the proposed lattice code is 0.26 dB better. Fig. \ref{fig:SER_C6_4} shows that the gap to the unconstrained Shannon limit is further reduced to 0.46 dB for our rate $\frac{1}{2}$ four-dimensional IRA lattice code. Our code is 0.1 dB better than the rate $\frac{1}{2}$ two-dimensional lattice code in \cite{Qiu16}. To this end, our proposed codes have lower decoding thresholds than that of the codes in \cite{Qiu16} but with higher encoding and decoding complexities.

Now we compare our designed lattice codes with the lattice coding schemes from \cite{Boutros14,Boutros16,Khodaiemehr17,8122043,4475389} for the same codeword length. Since these schemes are based on infinite constellations, their performances are measured in terms of gap to the Poltyrev limit which can be considered as coding loss \cite[Section VI-B]{Boutros16}. To obtain the coding loss in our lattice coding scheme, we measure the gap to uniform input capacity. The comparisons are listed in Table \ref{table3}, showing the simulation results which are reported for each scheme in the appropriate reference, including codeword length, coding loss and the gap to unconstrained Shannon limit when SER is at $10^{-5}$.

From Figs. \ref{fig:SER_C4_4}-\ref{fig:SER_C6_4}, one can observe that our code with rate $\frac{1}{2}$ have the smallest \emph{coding loss}. To be more specific, the coding loss for our lattice codes with $N=100,000$, $N=10,000$ and $N=1,000$ when SER is at $10^{-5}$ is about 0.3 dB, 0.6 dB and 1.5 dB. From Table \ref{table3}, it can be seen that our coding scheme outperforms all of these schemes for large codeword length, i.e., $N \geq 10,000$. When the codeword length is 1,000, our code is about 0.2 dB worse compared with LDA lattices \cite{8122043} and GLD lattices \cite{Boutros14} because of the probability of short cycles are higher when the codeword length is small. Since our goal is to design capacity-approaching lattice codes, thus we mainly focus on the codes with large codeword length, i.e., $N \geq 10,000$. Note that the direct comparison of encoding and decoding complexities for lattice codes with infinite constellations and our codes with finite constellations may not be fair and thus is omitted.

It is also worth noting that the waterfall regions of our multi-dimensional lattice codes are within 0.14 dB to the predicted decoding thresholds as shown in Table \ref{Table:CH4_2} for various code rates. Therefore, it is evident that the proposed EXIT chart analysis for our multi-dimensional lattice codes is effective.

\section{Summary}
In this chapter, we designed new multi-dimensional IRA lattice codes with finite constellations. Most compellingly, we proposed a novel encoding structure and proved that our codes can attain the permutation-invariance and symmetry properties in the densities of the decoder's messages. Under these properties, we used two-dimensional EXIT charts to analyze the convergence behavior of our codes and to minimize the decoding threshold. Our design can employ any higher-dimensional lattice partitions. Numerical results show that our designed and optimized lattice codes can achieve within 0.46 dB of the unconstrained Shannon limit and outperform existing lattice coding schemes for large codeword length.

\chapter{A Lattice-Partition Framework of Downlink NOMA without SIC}\label{C5:chapter5}

\section{Introduction}
In the previous chapter, we discuss how to design multi-dimensional lattice codes for power constrained point-to-point channels. In this chapter, we start dealing with the design and analysis of lattice coding schemes for downlink multiuser communication systems.

With the increasing demands of network access and the continuous growth of smart devices connected to the cellular networks, the current fourth generation systems have reached their limit and cannot meet the future requirements such as higher data rates, massive connectivity, and/or higher spectral efficiency. To this end, it is imperative to develop new multiple access techniques. Recently, non-orthogonal multiple access has drawn considerable attention due to its capability of providing high system throughput and massive connectivity while maintaining user fairness. As such, NOMA has been recognized as a promising technique for the next generation wireless communications \cite{Saito13,DerrickNG17,Dai15,Ding17,Ding17J,Lien17}.

\subsection{Main Contributions}
In this work, we continue the quest of designing \textit{downlink} NOMA schemes that can be efficiently decoded with low decoding complexity and latency. In particular, we follow the footsteps of \cite{Shieh16} to devise coding schemes that can be decoded with single-user decoding i.e., without SIC, for downlink NOMA. Our scheme exploits the structural property of the lattices to harness inter-user interference while taking advantage of higher shaping gains from multi-dimensional lattices. The main contributions of our work are summarized as follows.

\begin{itemize}
\item
We generalize the scheme in \cite{Shieh16} to a general lattice partition framework for downlink NOMA without SIC. This is done by revisiting the two-step approach adopted in \cite{Shieh16}. In the first step, the corresponding linear deterministic model \cite{Avestimehr11} is investigated and an optimal input distribution is derived. This optimal distribution is then translated into a uniform distribution over a properly chosen PAM constellations for the original model. Our generalization is based on the observation that a PAM constellation can be regarded as isomorphic to a lattice partition of the one-dimensional lattice $\mbb{Z}$. With this algebraic structure, we propose a general lattice partition framework which allows us to use a lattice partition of \textit{any lattice in any dimension} as constellation. This substantially enlarges the design space and subsumes the scheme in \cite{Shieh16} as a special case.

\item
The achievable rate of the proposed scheme is then analyzed and its gap to multiuser capacity region is upper bounded by a function of the normalized second moment (which will be defined later) of the base lattice. This upper bound is universal in the sense that it is independent of the channel parameter and the number of users participating in the transmission. We would like to emphasize that similar to the scheme in \cite{Shieh16}, the proposed framework only requires a limited knowledge of channel parameters (which will be clearly defined in Chapter~\ref{sysm_5}) rather than full channel state information. In addition, we extend our design to $K$-user downlink NOMA as well as provide its capacity gap analysis and prove that the upper bound of the capacity gap does not scale with $K$.

\item
Based on the derived bound, we compute the gap to the capacity of the proposed scheme with the base lattice chosen from some well-known lattices such as $\mbb{Z}_2$, $A_2$, $D_4$, $E_8$, and Construction A lattices. While generating these design examples, for handling the crucial issue of breaking ties, an efficient method is also discussed. The results show that as the dimension increases, one can find good lattices such that the gap shrinks. We then provide simulation results to show that the actual gap to the capacity region can be much smaller than the upper bound, which confirms that the proposed framework is capable of operating very close to the multiuser capacity region with only single user decoding at each user.

\end{itemize}

%
%

\section{System Model}\label{sysm_5}
\begin{figure}[ht!]
	\centering
\includegraphics[width=3.43in,clip,keepaspectratio]{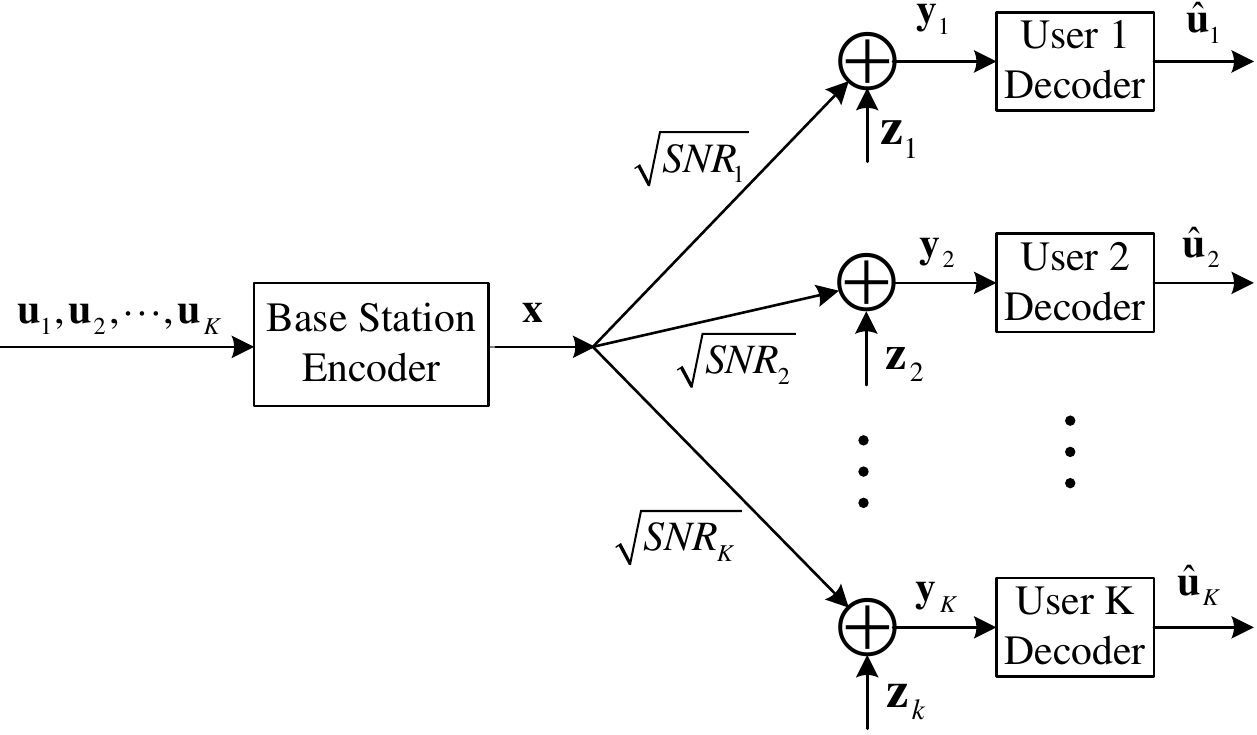}
\caption{The system model of the downlink NOMA.}
\label{fig:system_mod_5}
\end{figure}

In this work, we consider a downlink NOMA system that operates in a narrow band of frequencies and all the users in the system are assumed to be experiencing flat fading. For wideband systems, this model can be obtained by employing orthogonal frequency division multiplexing that efficiently transforms a frequency-selective fading channel into multiple flat fading ones. According to \cite{1054727}, the problem of downlink NOMA can be modeled as the Gaussian broadcast channel where a base station would like to broadcast messages $\mathbf{u}_1,\ldots,\mathbf{u}_K$ to users $1,\cdots,K$, respectively, as shown in Fig. \ref{fig:system_mod_5}.

In our system setting, the base station and all users are equipped with a single antenna\footnote{For extension to multiple antenna case, one may use a zero-forcing beamforming scheme to null out interference and convert the problem into multiple single antenna problems as in \cite{Geraci16}.} and work in a half-duplex mode. In addition, we assume that each receiver knows the phase of its channel and can compensate the phase, i.e., we consider coherent detection in \cite[Chapter 3]{tse_book}. In this way, the complex channel model can be transformed into two real channels with real channel gains and real noise (i.e., our channel model with $n=2$). Hence, our channel model encompasses the one with complex inputs and outputs as a special case. As shown in Fig. \ref{fig:system_mod_5}, the base station first encodes $\mathbf{u}_1,\cdots,\mathbf{u}_K$ into a codeword $\mathbf{x} \in \mbb{R}^n$ with power constraint $\mathbb{E}[\|\mathbf{x}\|^2]\leq n$, i.e., we use $n$ real channel jointly. The received signals arrive at user $k$ is given by
\begin{equation}\label{eq:sm1_5}
\mathbf{y}_k = \sqrt{\SNR_k}\mathbf{x}+\mathbf{z}_k,  \quad k \in \{ 1,2,\cdots K \},
\end{equation}
where $\mathbf{z}_k \thicksim \mathcal{N}(0,I)$ is the additive white Gaussian noise (AWGN) experienced at the $k$-th user and $\SNR_k$ is the $k$-th user's signal-to-noise ratio (SNR), representing the real channel gain. This models a general communication problem over a flat fading channel where each receiver knows its own channel state information. In our setting, the \emph{full} channel state information is not available at the transmitter; instead, only a \emph{quantized} version of $\SNR_k$, i.e., $\left\lceil \frac{1}{2}\log_2(\SNR_k)\right\rceil^+$ for $k = \{1,\ldots,K\}$ are available at the transmitter. Upon receiving, user $k$ attempts to decode $\mathbf{u}_k$ from $\mathbf{y}_k$. The achievable rate and the capacity region are defined in the usual information-theoretic manner (see \cite{Cover:2006:EIT:1146355} for example).

Although the goal of this work is to develop transmission schemes for the $K$-user case, in what follows, we first discuss and analyze the proposed scheme for the two-user case for the sake of simplicity. Despite being a special case, the code design and the analysis for the two-user case captures all the essences of the proposed framework and provide significant insights about the main concepts of the proposed framework, which allows us to substantially simplify the discussion for the $K$-user case presented in Section \ref{sec:K_user}.

\section{Downlink NOMA based on Multi-dimensional lattices without SIC}\label{sec:proposed_CH5}
In this section, we first review the deterministic model in \cite{Avestimehr11} for a two-user downlink NOMA and the scheme in \cite{Shieh16}. We then propose our general framework for downlink NOMA without SIC. Some design examples and analysis of achievable rates are then presented.

\subsection{The Deterministic Approach to Downlink NOMA}\label{Assumption_5}
\begin{figure}[ht!]
	\centering
\includegraphics[width=2in,clip,keepaspectratio]{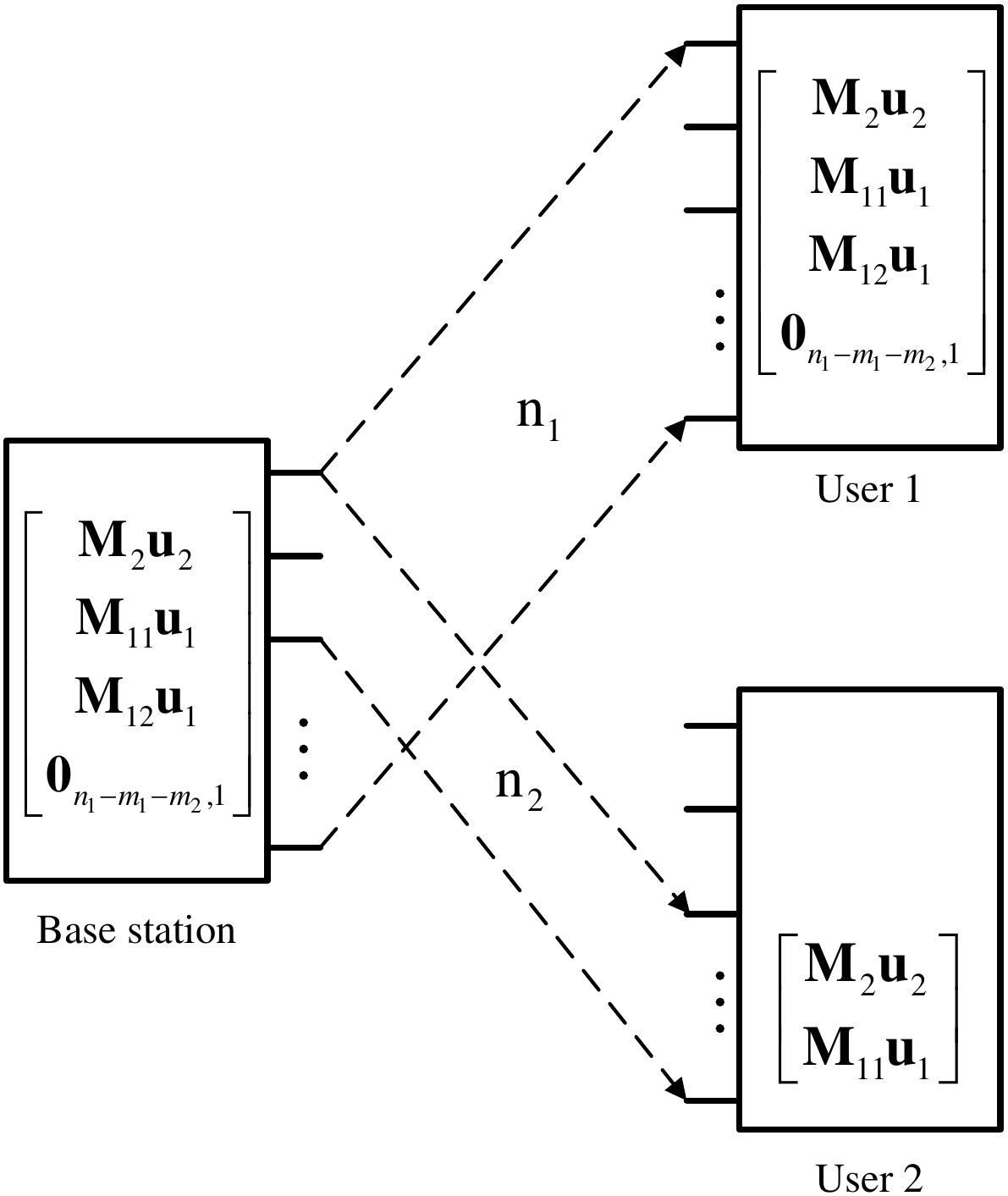}
\caption{The deterministic model for two-user case.}
\label{fig:d_model_5}
\end{figure}

The deterministic model is used for modeling downlink NOMA for two-user case. The analysis for the linear deterministic model will provide guidance and significant insights into to the original downlink NOMA model. By applying the deterministic model, the original Gaussian broadcast channel can be modeled as a pipe with two links. As shown in Fig. \ref{fig:d_model_5}, the essential idea is that the pipe only passes the bits above the noise level while truncating the bits below the noise level. We define $n_i \defeq \left\lceil \frac{1}{2}\log_2(\SNR_i)\right\rceil,$ for $i = \{1,2\}$ and we know $n_1 > n_2$. Note that we have assumed $\SNR_i \geq 1$ so that the ``$+$'' sign in \cite[Eq. (10)]{Avestimehr11} can be dropped. The base station broadcasts $n_1$ bits to both users. The maximum number of bits that user 1 can successfully receive and decode is $n_1$ while user 2 can only receive and decode up to $n_2$ bits as there are $n_1-n_2$ bits which are below the noise level and get shifted out at user 2.

We denote the number of transmitted bits intended for user $i$ by $m_i$ for $i \in \{1,2\}$, which must be a non-negative integer. As no SIC is employed in our scheme, even though user 1 can receive $n_1$ bits information, it will treat the other $m_2$ bits as interference. Therefore, the deterministic rate pair $(m_1,m_2)$ must satisfy the following:
\begin{align}
m_1+m_2&\leq n_1.\label{eq:d2}\\
m_2 &\leq n_2.\label{eq:d3}
\end{align}

A capacity-achieving scheme in \cite{Shieh16} for the deterministic model is given as follows. Following the notation defined in Chapter ``List of Notations'' of this thesis, we denote the message vectors $\mathbf{u}_1\in\mbb{F}_2^{m_1}$ and $\mathbf{u}_2\in\mbb{F}_2^{m_2}$. The message vector $\mathbf{u}_i$ where $i\in\{1,2\}$, is passed into encoding matrix $\mathbf{G}_i$ of size $n_1\times m_i$ and becomes the encoded message via $\mathbf{v}_i=\mathbf{G}_i\mathbf{u}_i$. We define
\begin{equation}\label{eq:r11_5}
r_{11}\defeq\min\{m_1,n_2-m_2\},
\end{equation}
and
\begin{equation}\label{eq:r12_5}
r_{12}\defeq \max\{m_1+m_2-n_2,0\},
\end{equation}
where $r_{11} + r_{12} = m_1 $ and $r_{11} + m_2 = n_2 $. Considering the rate pair $(m_1,m_2)$ satisfying (\ref{eq:d2}) and (\ref{eq:d3}), we choose
\begin{equation}\label{eq:m1}
\mathbf{G}_1= \begin{bmatrix}
\mathbf{0}_{m_2,m_1} \\
\mathbf{M}_{11} \\
\mathbf{M}_{12} \\
\mathbf{0}_{n_1-m_1-m_2,m_1} \\
\end{bmatrix},
\end{equation}
and
\begin{equation}\label{eq:m2}
\mathbf{G}_2= \begin{bmatrix}
\mathbf{M}_2 \\
\mathbf{0}_{n_1-m_2,m_2} \\
\end{bmatrix},
\end{equation}
where $\mathbf{M}_{11}$, $\mathbf{M}_{12}$, and $\mathbf{M}_2$ are full rank binary matrices with size $r_{11} \times m_1$, $r_{12} \times m_1$, and $m_2 \times m_2$, respectively. In this way, the transmitted codeword becomes a length $n_1$ column vector
\begin{equation}\label{eq:m3}
\mathbf{x}=\mathbf{v}_1+\mathbf{v}_2 = \mathbf{G}_1\mathbf{u}_1+\mathbf{G}_2\mathbf{u}_2=\begin{bmatrix}
\mathbf{M}_2\mathbf{u}_2 \\
\mathbf{M}_{11}\mathbf{u}_1 \\
\mathbf{M}_{12}\mathbf{u}_1 \\
\mathbf{0}_{n_1-m_1 - m_2,1} \\
\end{bmatrix},
\end{equation}
where all the operations above are over $\mathbb{F}_2$. One can follow the derivation in \cite{Shieh16} to show that this scheme is capacity achieving for the downlink NOMA channel. From the second equality of \eqref{eq:m3}, it can be noticed that the signals of two users are assigned to different rows of $\mathbf{x}$, representing different power levels in the original downlink NOMA model. Under this assumption, $\mathbf{M}_{12}\mathbf{u}_1$ is considered as under the noise level and thus it is not received by user 2. In this case, there are $r_{12}$ bits get truncated.

In \cite{Shieh16}, Shieh and Huang translate the above scheme from the deterministic model into the coding scheme for the Gaussian model. The scheme therein translates each $\mathbf{G}_i$, $i\in\{1,2\}$ into a pulse amplitude modulation (PAM) scheme and scales each user's signal by a power allocation factor. They show that this simple scheme can approach the capacity region of the downlink NOMA channel within a constant gap even without SIC.

\subsection{Proposed Lattice Framework for Downlink NOMA without SIC}\label{LOMA_5}
We first note that the scheme in \cite{Shieh16} corresponds to the superposition of properly scaled PAM constellations with size $2^{r_{12}}, 2^{r_{11}}$, and $2^{m_2}$. Since an $M$-ary PAM constellation is isomorphic to the coset decomposition $\mbb{Z}/M\mbb{Z}$, it is not difficult to show that the scheme in \cite{Shieh16} is in fact isomorphic to the one-dimensional lattice partition chain $\mbb{Z}/2^{r_{12}}\mbb{Z}/2^{m_1}\mbb{Z}/2^{m_1+m_2}\mbb{Z}$. However, the hypercube shaping induced by this partition chain provides no shaping gain and it is well known that lattices with better shaping gain exist in higher dimensions. We therefore consider jointly modulating signals onto constellations with $n$ real dimensions.

For any pair $(m_1,m_2)$ satisfying \eqref{eq:d2} and $\eqref{eq:d3}$, our scheme makes use of any lattice partition chain $\Lambda/2^{m_1}\Lambda/2^{m_1+m_2}\Lambda\defeq\Lambda_s$. The restriction of having partition orders being powers of 2 is merely for practical purpose and it can be lifted. In our proposed scheme, we choose a complete set of coset leaders of $\Lambda/2^{m_1}\Lambda$ to be the constellation for user 1. This gives us the constellation $\mc{C}_1$ which is isomorphic to $\Lambda/2^{m_1}\Lambda$ and has cardinality $2^{nm_1}$. Similarly, we choose a complete set of coset leaders of $\Lambda/2^{m_2}\Lambda$ to form the constellation $\mc{C}_2$ of user 2. This constellation is isomorphic to $\Lambda/2^{m_2}\Lambda$ and has cardinality $2^{nm_2}$.

Following (\ref{eq:m3}), the proposed scheme first encodes a length $M_i$ signal $\mathbf{u}_i, i \in \{1,2\}$ into a codeword $\mathcal{E}_i(\mathbf{u}_i)$ via the encoding function $\mathcal{E}_i$ and then bijectively maps every $nm_i$ bits from the codeword onto constellation $\mathcal{C}_i$ to obtain $\mathbf{v}_i \in \mathcal{C}_i$. The transmitted signal is then given by
\begin{align}\label{eqn:x_def_5}
    \mathbf{x} &= \beta\left(\left[\mathbf{v}_1+2^{m_1}\mathbf{v}_2-\mathbf{d}\right]_{\Lambda_s}\right) \nonumber \\
    &\in \beta\left(\left[\mc{C}_1+2^{m_1}\mc{C}_2-\mathbf{d}\right]_{\Lambda_s}\right) = \beta\mc{C},
\end{align}
where $\mathbf{d}\in\mc{V}(\Lambda_s)$ is a deterministic dither\footnote{In the literature of lattice codes, the dither could be random or deterministic vectors known at both the transmitter and receiver. For the proposed framework, deterministic dithers suffice. Typically, in practice, the dither is a deterministic vector such that the overall constellation is zero mean and has the minimum transmit power.}, $\mc{C}\defeq\left[\mc{C}_1+2^{m_1}\mc{C}_2-\mathbf{d}\right]_{\Lambda_s}$ is the combined constellation, and $\beta$ is to ensure $\mbb{E}[\|\mc{C}\|^2]\leq n$. We note that
\begin{equation}
    \left[\mc{C} + \mathbf{d}\right]_{\Lambda_s} = \left[\mc{C}_1+2^{m_1}\mc{C}_2\right]_{\Lambda_s},
\end{equation}
corresponds to a complete set of coset leaders of $[\Lambda/2^{m_1}\Lambda+2^{m_1}(\Lambda/2^{m_2}\Lambda)]_{\Lambda_s}  = \Lambda/2^{m_1+m_2}\Lambda$. Thus, the combined constellation has the cardinality $2^{n(m_1+m_2)}$ and preserves the structure of lattice $\Lambda$.

We emphasis here that following the definition in (\ref{eq:r11_5}) and (\ref{eq:r12_5}), $\mc{C}_1$ can be further decomposed into $\mc{C}_{11}$ and $\mc{C}_{12}$ as opposed to $r_{11}$ and $r_{12}$, which will come in handy when analyzing the achievable rates. More specifically, let us consider the lattice partition chain $\Lambda/2^{r_{12}}\Lambda/2^{m_1}\Lambda$ where $\mc{C}_{11}$ and $\mc{C}_{12}$ are isomorphic to $\Lambda/2^{r_{11}}\Lambda$ and $\Lambda/2^{r_{12}}\Lambda$, respectively. $\mc{C}_1$ can then be represented as
\begin{equation}\label{eq:C1de_5}
    \mc{C}_1 = \left[\mc{C}_{12} + 2^{r_{12}}\mc{C}_{11}\right]_{2^{m_1}\Lambda}.
\end{equation}

\begin{remark}
    From \eqref{eqn:x_def_5}, one can see that the proposed scheme naturally induces a power allocation scheme from the lattice partition. Unlike power allocation schemes adopted by conventional NOMA schemes, the power allocation induced by our proposed scheme makes sure that the overall constellation preserves a lattice structure. In this way, our scheme can exploit the lattice structure to harness inter-user interference.
\end{remark}

\begin{remark}\label{remark2}
It is worth noting that our proposed scheme in fact belongs to a larger framework in which one picks a complete set of coset leaders of $\Lambda/2^{m_1+m_2}\Lambda$ as the overall constellation. The above choice, $[\mc{C}_1+2^{m_1}\mc{C}_2]_{\Lambda_s}$, ensures that the overall constellation has the smallest power within this family; therefore, after normalization for fitting the power constraint, it will have the largest minimum distance for user 1's signal. Another reasonable choice is to simply let the overall constellation be $\mc{C}_1+2^{m_1}\mc{C}_2$, which resembles the conventional superposition coding.
\begin{equation}\label{eqn:x_def2}
    \mathbf{x}'' = \beta''\left(\mathbf{v}_1+2^{m_1}\mathbf{v}_2-\mathbf{d}\right)\in \beta''\left(\mc{C}_1+2^{m_1}\mc{C}_2-\mathbf{d}\right).
\end{equation}
This constellation will have a larger power than $[\mc{C}_1+2^{m_1}\mc{C}_2]_{\Lambda_s}$ and hence will result in a smaller minimum distance for user 1 after normalization. However, since we do not perform modulo $\Lambda_s$, the distance between cluster centers (each corresponds to an element in $2^{m_1}\mc{C}_2$) will be larger than that in $[\mc{C}_1+2^{m_1}\mc{C}_2]_{\Lambda_s}$. Therefore, this choice will result in a better performance for user 2 by sacrificing the performance of user 1. In what follows, we analyze the performance of $[\mc{C}_1+2^{m_1}\mc{C}_2]_{\Lambda_s}$ solely for simplicity and leave the exploration of other choices of coset leaders as future work.
\end{remark}

\subsection{An Extension to K-User Case}\label{sec:K_user}
Similar to \cite{Shieh16}, we can generalize the proposed framework to the $K$-user case. We first investigate the corresponding deterministic model in the following.

We denote $n_1, n_2,\cdots,n_K$ to be the channel capacity from the base station to user $1,2,\cdots,K$, respectively. In this model, we assume $\SNR_1 > \SNR_2 > \cdots > \SNR_K$. We define $n_i \defeq \left\lceil \frac{1}{2}\log_2(\SNR_i)\right\rceil$ and denote the number of transmitted bits intended for user $i$ by $m_i$ for $i \in \{1,2,\cdots, K\}$. The deterministic rate tuple $(m_1,m_2,\cdots,m_K)$ must satisfy the following constraints:
\begin{align}
m_1+m_2+\cdots+ m_K&\leq n_1,\label{eq:dK2_5} \\
m_2 +\cdots+ m_K &\leq n_2,\label{eq:dK3_5} \\
&\vdots \nonumber \\
m_K &\leq n_K.\label{eq:dK4_5}
\end{align}

We start with the analysis for the first (strongest) user. From (\ref{eq:dK2_5}) and (\ref{eq:dK3_5}), we combine users $2,3,\cdots,K$ into a super-user demanding $m_2^\prime = \sum_{i=2}^K m_i$ bits and with channel capacity $n_2$. Now the problem is reduced to a two-user case. Therefore, to analyze the achievable rate for user 1, one can directly follow our approach as described in Section \ref{Assumption_5}.

Now we analyze the achievable rate for user $k>1$. At user $k$'s channel, we have the capacity constraint as follows:
\begin{equation}\label{eq:dKk_Ch5}
m_k +\cdots+ m_K \leq n_k.
\end{equation}
For this case, we treat users $1,2,\cdots,k-1$ as a super-user demanding $m_1^\prime = \sum_{i=1}^{k-1}m_i$ bits and users $k,k+1,\cdots,K$ as another super-user demanding $m_k^\prime = \sum_{i=k}^Km_i$ bits. The problem can again be deemed as a two-user case. We thus choose the rate pairs $(m_1^\prime, m_k^\prime)$ that satisfies:
\begin{align}
m_1^\prime+m_k^\prime &\leq n_1,\label{eq:dKk15}\\
m_k^\prime &\leq n_k.\label{eq:dKk25}
\end{align}
In this way, the same approach from Section \ref{Assumption_5} can be used for analyzing this case. We then define
\begin{align}\label{eq:dKk35}
r_{11}^\prime &\defeq\min\{m_1^\prime,n_k-m_k^\prime\} \nonumber \\
&= \min\left\{\sum_{i=1}^{k-1}m_i,n_k-\sum_{j=k}^Km_j\right\},
\end{align}
and
\begin{align}\label{eq:dKk45}
r_{1k}^\prime &\defeq \max\{m_1^\prime+m_k^\prime-n_k,0\} \nonumber \\
&= \max\left\{\sum_{i=1}^K m_i-n_k,0\right\},
\end{align}
where $r_{11}^\prime + r_{1k}^\prime = m_1^\prime$ and $r_{11}^\prime + m_k^\prime = n_k$. According to the deterministic model, for user $k$, there are $r_{1k}^\prime$ number of bits are considered as under noise level and thus get truncated. Similarly, we can follow the steps in (\ref{eq:m1}) and (\ref{eq:m2}) to obtain the capacity-achieving scheme $\mathbf{G}_1^\prime$ and $\mathbf{G}_k^\prime$ for generating capacity-achieving input distributions.

Similar to our two-user case in Section \ref{LOMA_5}, we translate the scheme from the deterministic model into the scheme for the Gaussian model. For any rate tuple $(m_1,m_2,\cdots,m_K)$ within the capacity region of the $K$-user linear deterministic model, we construct a lattice partition chain $\Lambda/2^{m_1}\Lambda/2^{m_1+m_2}\Lambda/\cdots/2^{\sum_{i=1}^K m_i}\Lambda \triangleq \Lambda_{s}$ where $\Lambda$ is an $n$-dimensional lattice. For each $k$, the individual constellation $\mc{C}_k$ is isomorphic to $\Lambda/2^{m_k}\Lambda$ and has cardinality $2^{nm_k}$. The transmitter first encodes the message $\mathbf{u}_k \in \F_2^{M_k}$ into a codeword $\mathcal{E}_k(\mathbf{u}_k)$ via the encoding function $\mathcal{E}_k$ and then bijectively maps every $nm_k$ bits from the codeword onto constellation $\mc{C}_k$ to obtain $\mathbf{v}_k \in \mc{C}_k$ and then transmit
\begin{align}
\mathbf{x} &= \beta\left(\left[\mathbf{v}_1+\sum_{k=2}^{K}2^{\sum_{i = 1}^{k-1}m_i}\mathbf{v}_k-\mathbf{d} \right]_{\Lambda_{s}}\right) \nonumber \\
&\in \beta\left(\left[\mc{C}_1+\sum_{k=2}^{K}2^{\sum_{i = 1}^{k-1}m_i}\mc{C}_k-\mathbf{d} \right]_{\Lambda_{s}}\right),
\end{align}
where $\mathbf{d} \in \mc{V}(\Lambda_{s})$ is a dither and $\beta$ is a normalize factor to ensure $\E[\|\mc{C} \|^2] \leq n$. Similar to the two-user case, the combined constellation has the cardinality $2^{n\sum_{i=1}^K m_i}$ and preserves the structure of lattice $\Lambda$.

\section{Analysis of Achievable Rates and their Gaps to the Multiuser Capacity Region}
In this section, we analyze the achievable rates of the proposed scheme under single-user decoding (without SIC) and their gaps to the capacity region. We first present the main result of this work as follows.

\begin{prop}\label{main_result5}
In the $K$-user downlink NOMA, regardless of the channel SNR, the gap between the individual rate achieved by the proposed scheme and the multiuser capacity region in bits per real dimension is upper bounded by
 \begin{align}
\Delta_1  &< 1 + \frac{1}{2}\log_2 2\pi e \left(5\psi(\Lambda)\right),\label{eq:main_1_5} \\
    \Delta_k &< 1 + \frac{1}{2}\log_2 2\pi e \left(6\psi(\Lambda)\right),\; \text{$k\in\{2,\ldots,K\}$.}\label{eq:main_K}
\end{align}
Moreover, this gap only depends on the NSM of the base lattice $\Lambda$ and does not scale with $K$.
\end{prop}
In what follows, we provide the proof for the $K=2$ case. This simplest case will allow us to explain all the important ingredients of our analysis. The proof for the general $K$-user case will be deferred to Appendix~\ref{Kuser_5}. After proving the results, we then provide some analysis for the proposed scheme constructed over some well-known lattices and compare the required complexity with some existing NOMA schemes.

\subsection{Analysis of the Two-User Case}\label{GapRate}
We first consider $K=2$. Let $\msf{V}_1$ and $\msf{V}_2$ be random variables uniformly distributed over $\mc{C}_1$ and $\mc{C}_2$, respectively, and  $\msf{X}=\beta(\left[\msf{V}_1+2^{m_1}\msf{V}_2-\mathbf{d}\right]_{\Lambda_s})$ corresponding to the channel input random variable. Following the relationship given in (\ref{eq:sm1_5}), we define $\msf{Y}_1$ and $\msf{Y}_2$ to be the random variables corresponding to the received signal at users 1 and 2, respectively.

To analyze the achievable rate of the strong user without SIC, we define $\Lambda_1\defeq \sqrt{\SNR_1}\beta\Lambda$ and bound the mutual information as follows,
\begin{align}\label{eqn:rate_1_GBC_5}
    &I(\msf{V}_1;\msf{Y}_1)= h(\msf{Y}_1) - h(\msf{Y}_1|\msf{V}_1) \nonumber \\
    &= [h(\msf{Y}_1) - h(\msf{Y}_1|\msf{X})]- \left[h\left(\msf{Y}_1|\msf{V}_1\right)- h(\msf{Y}_1|\msf{X})\right] \nonumber \\
    & \overset{(a)}{=} [h(\msf{Y}_1) - h(\msf{Y}_1|\msf{X})]
    - \left[h\left(\msf{Y}_1|\msf{V}_1\right)- h(\msf{Y}_1|\msf{V}_1,\msf{V}_2)\right] \nonumber \\
    &= I(\msf{X};\msf{Y}_1) - I(\msf{V}_2;\msf{Y}_1|\msf{V}_1) \nonumber \\
    &\geq I(\msf{X};\msf{Y}_1) - H(\msf{V}_2|\msf{V}_1) \nonumber \\
    &\overset{(b)}{=} I(\msf{X};\msf{Y}_1) - H(\msf{V}_2) \nonumber \\
    &\overset{(c)}{\geq} nm_1 - \frac{n}{2}\log_2 2\pi e \left(\text{Vol}(\Lambda_1)^{-\frac{2}{n}} + \psi(\Lambda_1)\right),
\end{align}
where $(a)$ is because the mapping between $\msf{X}$ and $(\msf{V}_1,\msf{V}_2)$ is bijective, $(b)$ follows from the independence of $\msf{V}_1$ and $\msf{V}_2$, and $(c)$ follows from the lower bound of the mutual information between a discrete random input $\msf{X}$ and its noisy observation $\msf{Y}$, which is established in Appendix~\ref{appendix:lemma}.

Since the gap between the achievable rate pairs in \eqref{eq:d2} and \eqref{eq:d3} and the multiuser capacity region is at most 1 bit per real dimension \cite{Avestimehr11}, we have the gap of user 1's achievable rate to the multiuser capacity is at most
\begin{equation}\label{eqn:gap_1}
    \Delta_1 = 1+\frac{1}{2}\log_2 2\pi e \left( \text{Vol}(\Lambda_1)^{-\frac{2}{n}} + \psi(\Lambda)\right),
\end{equation}
bits per real dimension, where we have used the fact that $\psi(\Lambda)$ is invariant to scaling.

To bound $\text{Vol}(\Lambda_1)$, we first need to find the analytical expression of the overall scaling factor $\sqrt{\SNR_1}\beta$. Lemma~\ref{lma:power} in Appendix~\ref{appendix:lemma} establishes an upper bound on the power required by lattice constellation, which shows that there exists a fixed dither $\mathbf{d}$ such that the resulting $\mc{C}$ has
\begin{equation}\label{eq:lemma_f}
    \frac{1}{n} \mbb{E}[\|\mc{C}\|]^2 \leq \left(|\mc{C}| \text{Vol}(\Lambda)\right)^{\frac{2}{n}}\psi(\Lambda),
\end{equation}
where $|\mc{C}|$ outputs the cardinality of $\mc{C}$ following from (\ref{eq:1f}).
We then establish a lower bound for the scaling factor for user 1 as
\begin{equation}
\sqrt{\SNR_1} \beta > \sqrt{\frac{1}{4\text{Vol}(\Lambda)^{\frac{2}{n}} \psi(\Lambda)}},
\end{equation}
in \eqref{eq:sc1} in Appendix \ref{pf_user1_gap}. By plugging this bound into \eqref{eqn:gap_1}, we conclude that the gap between user 1's achievable rate and the multiuser capacity in bits per real dimension is lower bounded by
\begin{align}
\Delta_1  < 1 + \frac{1}{2}\log_2 2\pi e \left(5\psi(\Lambda)\right),
\end{align}
which completes the proof of \eqref{eq:main_1_5} in Proposition \ref{main_result5}. The detail is shown in \eqref{eq:gap_12} in Appendix \ref{pf_user1_gap}.

For the weak user, we can write $\msf{V}_1=\left[\msf{V}_{12}+ 2^{r_{12}}\msf{V}_{11}\right]_{2^{m_1}\Lambda}$ by following \eqref{eq:C1de_5}. The choice of parameters $r_{11}$ and $r_{12}$ suggested in the deterministic model ensures that $\msf{V}_{12}$ is under the noise level. This leads to
\begin{align}\label{eqn:rate_2_GBC_ch5}
    &I(\msf{V}_2;\msf{Y}_2) = h(\msf{Y}_2) - h(\msf{Y}_2|\msf{V}_2) \nonumber \\
    &= \left[h(\msf{Y}_2) - h\left(\msf{Y}_2|\left[2^{r_{12}}\msf{V}_{11}+2^{m_1}\msf{V}_2\right]_{\Lambda_s}\right)\right] \nonumber \\
    &\;\;\;\;- \left[h(\msf{Y}_2|\msf{V}_2)-h\left(\msf{Y}_2|\left[2^{r_{12}}\msf{V}_{11}+2^{m_1}\msf{V}_2\right]_{\Lambda_s}\right)\right] \nonumber \\
    &\overset{(a)}{=} \left[h(\msf{Y}_2) - h\left(\msf{Y}_2|\left[2^{r_{12}}\msf{V}_{11}+2^{m_1}\msf{V}_2\right]_{\Lambda_s}\right)\right]\nonumber \\
    &\;\;\;\;- \left[h(\msf{Y}_2|\msf{V}_2)-h\left(\msf{Y}_2|\msf{V}_{11},\msf{V}_2\right)\right] \nonumber \\
    &= I(\left[2^{r_{12}}\msf{V}_{11}+2^{m_1}\msf{V}_2\right]_{\Lambda_s};\msf{Y}_2) - I(\msf{V}_{11};\msf{Y}_2|\msf{V}_2) \nonumber \\
   & \geq I(\left[2^{r_{12}}\msf{V}_{11}+2^{m_1}\msf{V}_2\right]_{\Lambda_s};\msf{Y}_2) - H(\msf{V}_{11}|\msf{V}_2) \nonumber \\
   & \overset{(b)}{=} I(\left[2^{r_{12}}\msf{V}_{11}+2^{m_1}\msf{V}_2\right]_{\Lambda_s};\msf{Y}_2) - H(\msf{V}_{11}),
\end{align}
where $(a)$ is due to the bijective mapping between $\left[2^{r_{12}}\msf{V}_{11}+2^{m_1}\msf{V}_2\right]_{\Lambda_s}$ and $(\msf{V}_{11},\msf{V}_2)$, and $(b)$ again follows from the independence of $\msf{V}_{11}$ and $\msf{V}_2$.

To further bound $I(\left[2^{r_{12}}\msf{V}_{11}+2^{m_1}\msf{V}_2\right]_{\Lambda_s};\msf{Y}_2)$, we note that the effective noise is $\sqrt{\SNR_2}\beta\msf{V}_{12}+\msf{Z}_2$. We thus scale $\msf{Y}_2$ by
\begin{equation}
    \gamma=\sqrt{\frac{n}{\SNR_2\beta^2\mbb{E}[\|\msf{V}_{12}\|^2]+n}},
\end{equation}
to make the effective noise $\msf{Z}'_2 = \gamma (\sqrt{\SNR_2}\beta\msf{V}_{12}+\msf{Z}_2)$ with $\mbb{E}[\|\msf{Z}'_2\|^2]=n$.
The equivalent communication channel then becomes $\msf{Y}'_2 = \msf{X}'_2 + \msf{Z}'_2$ where
\begin{equation}
    \msf{X}'_2 = \gamma\sqrt{\SNR_2} \beta \left[2^{r_{12}}\msf{V}_{11}+2^{m_1}\msf{V}_2\right]_{\Lambda_s},
\end{equation}
and $\msf{Y}'_2 = \gamma\msf{Y}_2$. One can then again apply the lower bound of the mutual information between a discrete random input and its noisy version shown in Appendix~\ref{appendix:lemma} to obtain
\begin{equation}
    I(\msf{V}_2;\msf{Y}_2) \geq n m_2 - \frac{n}{2}\log_2 2\pi e \left(\text{Vol}(\Lambda_2)^{-\frac{2}{n}} + \psi(\Lambda_2)\right),
\end{equation}
where $\Lambda_2\defeq \gamma\sqrt{\SNR_2}\beta 2^{r_{12}}\Lambda$.

We again establish a lower bound for the scaling factor for user 2 as
\begin{equation}
\gamma\sqrt{\SNR_2}\beta 2^{r_{12}} > \sqrt{\frac{1}{5\text{Vol}(\Lambda)^{\frac{2}{n}}\psi(\Lambda)}},
\end{equation}
in \eqref{eq:sc2_5} from Appendix \ref{pf_user2_gap} and plug this into \eqref{eqn:gap_2} to obtain
\begin{align}
\Delta_2 < 1 + \frac{1}{2}\log_2 2\pi e \left(6\psi(\Lambda)\right),
\end{align}
which completes the proof of \eqref{eq:main_K} for $K = 2$ in Proposition \ref{main_result5}. The detail is in \eqref{eq:gap_22} in Appendix \ref{pf_user2_gap}.

\begin{remark}\label{333_5}
From the derived results, one can see that the gaps to the capacity region can be upper-bounded by a function proportional to the logarithm of the NSM of the base lattice. This indicates that better results can be obtained by using lattices with smaller NSM. This is not surprising at all since smaller NSM means that the shape of the fundamental Voronoi region is closer to an $n$-dimensional ball, which results in better shaping. Moreover, it is well-known that the NSM reduces as the dimension increases \cite{Zamir15}; hence, it is beneficial to construct codes with large dimension. Additionally, we have used $n_i - \frac{1}{2}\log_2(\SNR_i) < 1$ for dropping the dependency of $\SNR_i$. The gap can be shrunk if one tailors the bounds specifically for the actual $\SNR$. For example, when $\SNR$ is a power of 2, by following similar steps above, one obtains  $\Delta_1 = 1 + \frac{1}{2}\log_2 2\pi e \left(2\psi(\Lambda)\right)$ and $\Delta_2 =  1+\frac{1}{2}\log_2 2\pi e \left(3\psi(\Lambda)\right)$ bits. Moreover,
the 1 bit we added in both $\Delta_1$ and $\Delta_2$ is for bounding the difference between the Gaussian capacity region and the capacity region of the linear deterministic model for any ($\SNR_1,\SNR_2$) universally. Note that the capacity difference between the two models is a function of the $\SNR$ parameters and cannot be larger than 1 bit \cite{Avestimehr11}. Therefore, we consider the worst-case scenario to attain an upper bound for any $\SNR$ parameters universally. In most of the cases, this capacity difference is much smaller than 1 bit and thus the gap can be further shrunk. Our simulation in Chapter~\ref{SIM_5} will confirm this observation that the actual gaps are usually much smaller than $\Delta_1$ and $\Delta_2$ derived here. This observation also applies to the $K$-user case presented in Appendix~\ref{Kuser_5}.
\end{remark}

\subsection{Analysis of the Capacity Gap for Certain Lattices}\label{sec:Z2_E8}
In this subsection, we first compute the gaps to the capacity region for the proposed scheme constructed over some well-known lattices including $\mathbb{Z}^2$, $A_2$, $D_4$ and $E_8$. $\mbb{Z}^2$ is a two-dimensional lattice corresponding to the Cartesian product of two $\mbb{Z}$. Plugging the corresponding NSM into \eqref{eq:gap_12} and \eqref{eq:gap_22} results in the gap upper bounds as shown in Table~\ref{table1}. We note that since $\mbb{Z}^2$ and $\mbb{Z}$ have the same algebraic structure, the gap in bit per real channel for $\mbb{Z}^2$ is going to be identical to that for $\mbb{Z}$, which has been analyzed in \cite{Shieh16}\footnote{When comparing the results obtained here and that in \cite{Shieh16}, one observes that the gaps for user 1 are indeed identical. However, there is a slight difference in the gaps for user 2. In fact, the analysis for user 2 in \cite{Shieh16} contains a typo and thus the gap should again be identical to $\Delta_2$ here.}. We then compute the gap for the hexagonal lattice $A_2$ (the densest packing lattice in $\mbb{R}^2$), the checkerboard lattice $D_4$ (the densest packing lattice in $\mbb{R}^4$), and the Gosset lattice $E_8$ (the densest packing lattice in $\mbb{R}^8$). Their NSM and the gap upper bounds are shown in Table~\ref{table1}.
\begin{table}[tbp]
\centering
 \caption{Gap upper bounds of $\mbb{Z}^2$, $A_2$,$D_4$, $E_8$ and $\Lambda_{\vartheta}$.}\label{table1}
\begin{tabular}{|c|c|c|c|}
\hline
 $\Lambda$  &   $\psi(\Lambda) $ &$\Delta_1$ (bits)  & $\Delta_2$ (bits) \\  \hline
 $\mathbb{Z}^2$   &   $\frac{1}{12}$  & 2.4156 & 2.5471 \\  \hline
 $A_2$          &  $\frac{5}{36\sqrt{3}}$  & 2.3878 &2.5193 \\ \hline
  $D_4$         &  $\frac{13}{120\sqrt{2}}$  & 2.3548 &2.4864 \\ \hline
   $E_8$        &  $\frac{929}{12960}$   & 2.3069 &2.4385 \\ \hline
   $\Lambda_{\vartheta}$        &  $\frac{1}{2\pi e}$   & 2.1620 &2.2925 \\ \hline
\end{tabular}
\end{table}

We then choose the base lattice from the family of Construction A lattices and provide an analysis on the gap of the achievable rate to the multiuser capacity region. Construction A is known for its ability to produce optimal lattices in many senses including packing, covering, channel coding, and shaping \cite{1512416}.
\begin{defi}(Construction A \cite{conway1999sphere}):
Let $\vartheta$ denote a non-binary $(n,k)$ linear code over $\F_p$, where $p$ is a prime number. The Construction A lattice $\Lambda_{\vartheta}$ is then generated via:
\begin{equation}\label{eq:constrA}
\Lambda_{\vartheta} \triangleq \{\boldsymbol{\lambda}_{\vartheta} = \phi(\vartheta) + p\Z^n\},
\end{equation}
where $\phi(.)$ is the natural mapping that maps each codeword component to an element in $\mbb{Z}_p$.
\end{defi}

In our analysis, we focus on using lattices from a random Construction A lattice ensemble specified by $(n,k,p)$. This ensemble is obtained via Construction A by lifting a random $p$-ary linear code to the Euclidean space. The random linear code is generated via a generator matrix $\mathbf{G}_\vartheta \in \Z_p^{k \times n}$ where the entries are i.i.d. and uniformly distributed over $\F_p$. It is shown in \cite{1512416} that within this ensemble, there exists a sequence of lattices that can attain the smallest NSM. In what follows, we use such lattices as base lattices for our proposed scheme and compute the capacity gap.

Given an $n$-dimensional sphere $\mc{B}$ with radius $r$, the volume of the sphere is given by $\text{Vol}(\mc{B}) = V_nr^n$, where $V_n$ is the volume of an $n$-dimensional sphere with unit radius. We define the effective radius $r_{\text{eff}}(\Lambda_{\vartheta})$ as the radius of an $n$-dimensional sphere $\mathcal{S}$ which has the same volume as $\Lambda_{\vartheta}$ such that
\begin{equation}
r^2_{\text{eff}}(\Lambda_{\vartheta}) \triangleq \frac{\text{Vol}(\Lambda_{\vartheta})^{\frac{2}{n}}}{V_n^{\frac{2}{n}}}.
\end{equation}
We know that the sphere $\mathcal{S}$ has the smallest second moment of all $n$-dimensional lattices with volume $\text{Vol}(\Lambda_{\vartheta})$. Thus, the second moment of the Construction A lattice can be lower bounded by:
\begin{equation}
\sigma^2(\Lambda_{\vartheta}) \geq \sigma^2(\mathcal{S}) = \frac{r^2_{\text{eff}}}{n+2}.
\end{equation}
The NSM of the Construction A lattice is then lower bounded by:
\begin{equation}
\psi(\Lambda_{\vartheta}) \geq \frac{1}{V_n^{\frac{2}{n}}(n+2)}.
\end{equation}
We know that (see for example \cite{Ordentlich16} and \cite{8122043}) when the dimension $n$ is large enough, there exists a sequence of Construction A lattices whose Voronoi region is arbitrarily close to an $n$-dimensional sphere where almost all the points of the constellation lie with close to the surface of the sphere. In this case, the second moment of the Construction A lattices can be arbitrarily close to the second moment of the sphere $\mathcal{S}$. Therefore, the NSM of our Construction A lattice becomes
\begin{align}
\psi(\Lambda_{\vartheta}) &= \lim_{n \rightarrow \infty }\frac{1}{V_n^{\frac{2}{n}}(n+2)}  \nonumber \\
&\overset{(d)}= \lim_{n \rightarrow \infty } \frac{n}{2 \pi e (n+2)} = \frac{1}{2 \pi e},
\end{align}
where (d) follows from
\begin{equation}
\lim_{n \rightarrow \infty } V_n^{\frac{2}{n}} = \frac{2 \pi e}{n}.
\end{equation}
Now using this sequence of lattices as our base lattices $\Lambda$ and substituting our results into (\ref{eq:gap_12}) and (\ref{eq:gap_22}) result in $\Delta_1 = 2.1620$ and $\Delta_2 = 2.2925$, respectively. Similar to Remark \ref{333_5}, when $\SNR_1$ and $\SNR_2$ are powers of 2, the capacity gaps $\Delta_1$ and $\Delta_2$ can be further reduced.

\subsection{Complexity Comparison}
Now we compare the complexity on transmitters and receivers of our proposed scheme to that of the conventional NOMA with SIC. Specifically, we focus on the complexity at the transmitter and receiver. However, most of the work on NOMA is based on the achievable rate by assuming Gaussian inputs, which is by no means practical \cite{arXiv:1706.08805}. We thus consider the downlink multiuser superposition transmission (MUST) in \cite{TR36.859} as a conventional NOMA that adopts the current LTE standard modulation, e.g., QAM modulation. Compared to MUST, our scheme has a lower complexity at the receiver as SIC is removed. Most importantly, when the number of users increases, the complexity at the receiver grows with the number of users for conventional NOMA while our scheme still maintains single user decoding complexity. Compared to a recently proposed power domain NOMA scheme in \cite{Shieh16}, both the schemes do not involve SIC but our scheme has a higher decoding complexity due to the fact that larger dimensional constellations are considered. Note that the proposed framework subsumes the scheme in \cite{Shieh16} as a special case with $n=1$.

At the transmitter, the complexity for our scheme is higher than that of \cite{TR36.859} and \cite{Shieh16}. We would like to emphasize that there is no shaping gain in the schemes of \cite{TR36.859} and \cite{Shieh16} while our scheme is constructed over higher-dimensional lattices with a higher shaping gain. There is always a price to pay in having a shaping gain and our scheme is with no exception. In other words, as compared to \cite{TR36.859} and \cite{Shieh16}, our scheme trades the complexity for better performance by introducing multi-dimensional lattice partitioned in our design.

\section{Design Examples and Simulation Results}\label{SIM_5}
In this section, we provide numerical and simulation results for the proposed schemes constructed over the lattices discussed in Chapter~\ref{sec:Z2_E8}. While constructing constellations from lattice partition chains, the first crucial issue one may encounter is to handle ties when a coset of $\Lambda_s$ in $\Lambda$ has more than one minimum-norm element. In this case, the mapping between source information and lattice constellations is not bijective, resulting in ambiguity in decoding. In what follows, we first introduce an algorithm for handling ties and then present simulation results.

\subsection{Handling the Ties of Cosets}
Consider a pair of nested lattice $2^m\Lambda \subseteq \Lambda$. Applying \eqref{eq:1c}, \eqref{eq:1e}, and \cite[Eq. (2.43)]{Zamir15}, we have
 \begin{align}
 \boldsymbol{\lambda} / 2^m \Lambda &=\boldsymbol{\lambda} - Q_{2^m \Lambda}(\boldsymbol{\lambda}) \nonumber \\
 &= \boldsymbol{\lambda} - 2^m Q_{\Lambda}\left(\frac{\boldsymbol{\lambda}}{2^m}\right), \;\;\;\; \boldsymbol{\lambda} \in \Lambda.
 \end{align}
Thus, the problem boils down to quantizing the lattice points to the fine lattice. We follow the approach in \cite{Conway82} to develop a modified algorithm for quantizing $\mathbb{Z}^2$, $A_2$, $D_4$ and $E_8$. As partition algorithms design is not the focus of this work, we thus omit the detail algorithms here but only point out the main difference for ties handling.

A tie occurs when an arbitrary point $\mathbf{x} = (x_1,x_2,\cdots,x_n) \in \mathbb{R}^n$ is close to more than one coarse lattice points. Let us consider $\mathbb{Z}^n$ for example. To find the closest lattice point in $\mathbb{Z}^n$ to $\mathbf{x}$, we have
\begin{equation}
Q_{\mathbb{Z}^n}(\mathbf{x}) = (Q_{\mathbb{Z}}(x_1),Q_{\mathbb{Z}}(x_2),\cdots,Q_{\mathbb{Z}}(x_n)).
\end{equation}
The quantization for each dimension is independent. Now we have a rule to handle the tie as follows. For a random integer $N$ and $i = 1,2,\cdots,n$,
\begin{equation}
Q_{\mathbb{Z}}(x_i) = N, \;\;\;\; N-0.5 < x_i \leq N+0.5.
\end{equation}
We can directly apply this rule for quantizing $\mathbb{Z}^2$. Note that following \cite{Conway82}, quantization for $A_2$, $D_4$ and $E_8$ can be done through finding the closest point to $\mathbb{Z}^n$, thus the ties are already handled in that step.

We adopt the developed lattice quantizers in our lattice partitions and perform simulation which is shown in Section \ref{chap:2} and Section \ref{chap:3}.

\subsection{Achievable Rate Simulation: Two-User Case}\label{chap:2}
\begin{figure}[ht!]
	\centering
\includegraphics[width=3.2in,clip,keepaspectratio]{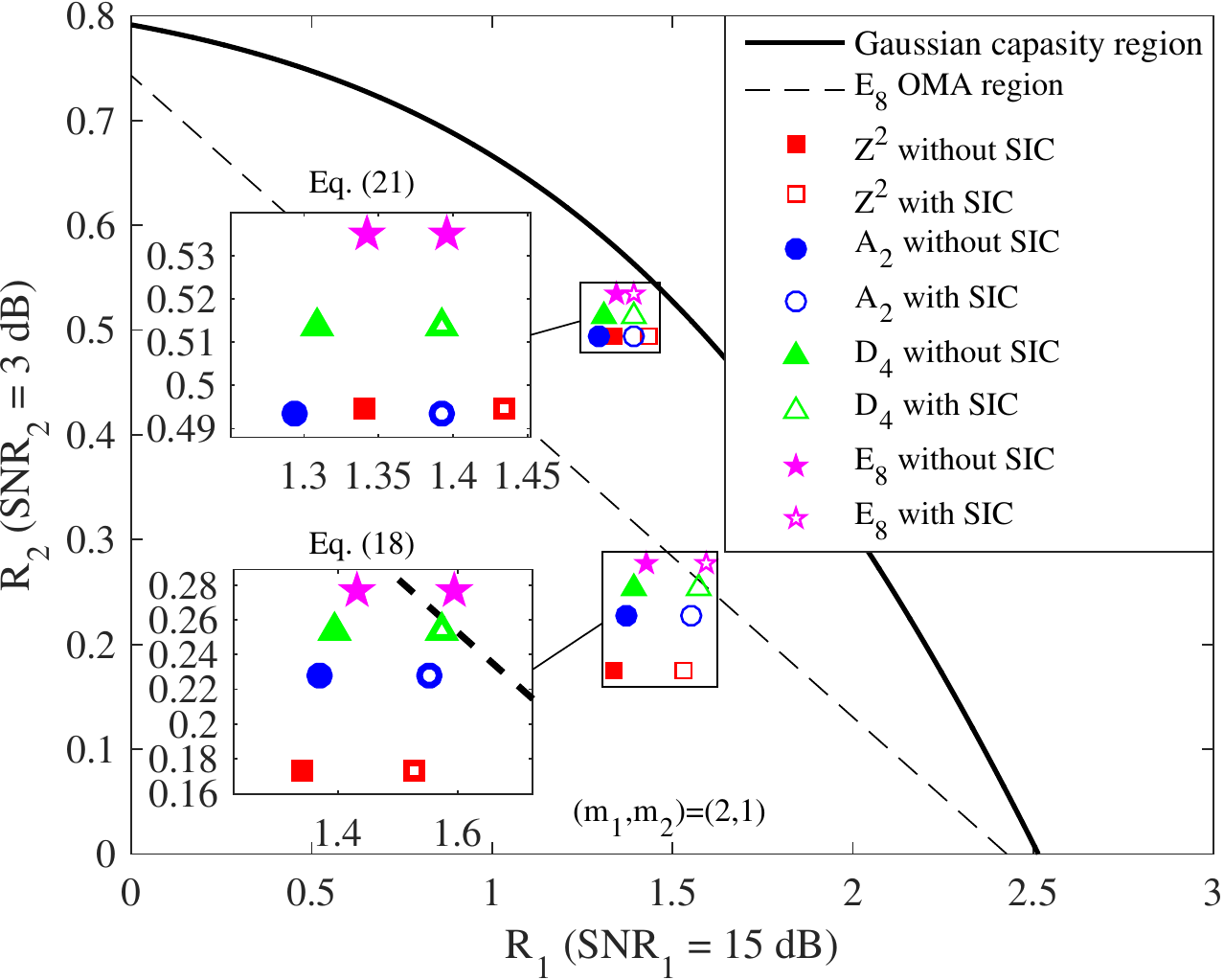}
\caption{The achievable rate pairs of downlink NOMA based on $\mathbb{Z}^2$, $A_2$ ,$D_4$ and $E_8$ with $\SNR_1 = 15$ dB and $\SNR_2 = 3$ dB. }
\label{fig:4}
\end{figure}

\begin{figure}[ht!]
	\centering
\includegraphics[width=3.2in,clip,keepaspectratio]{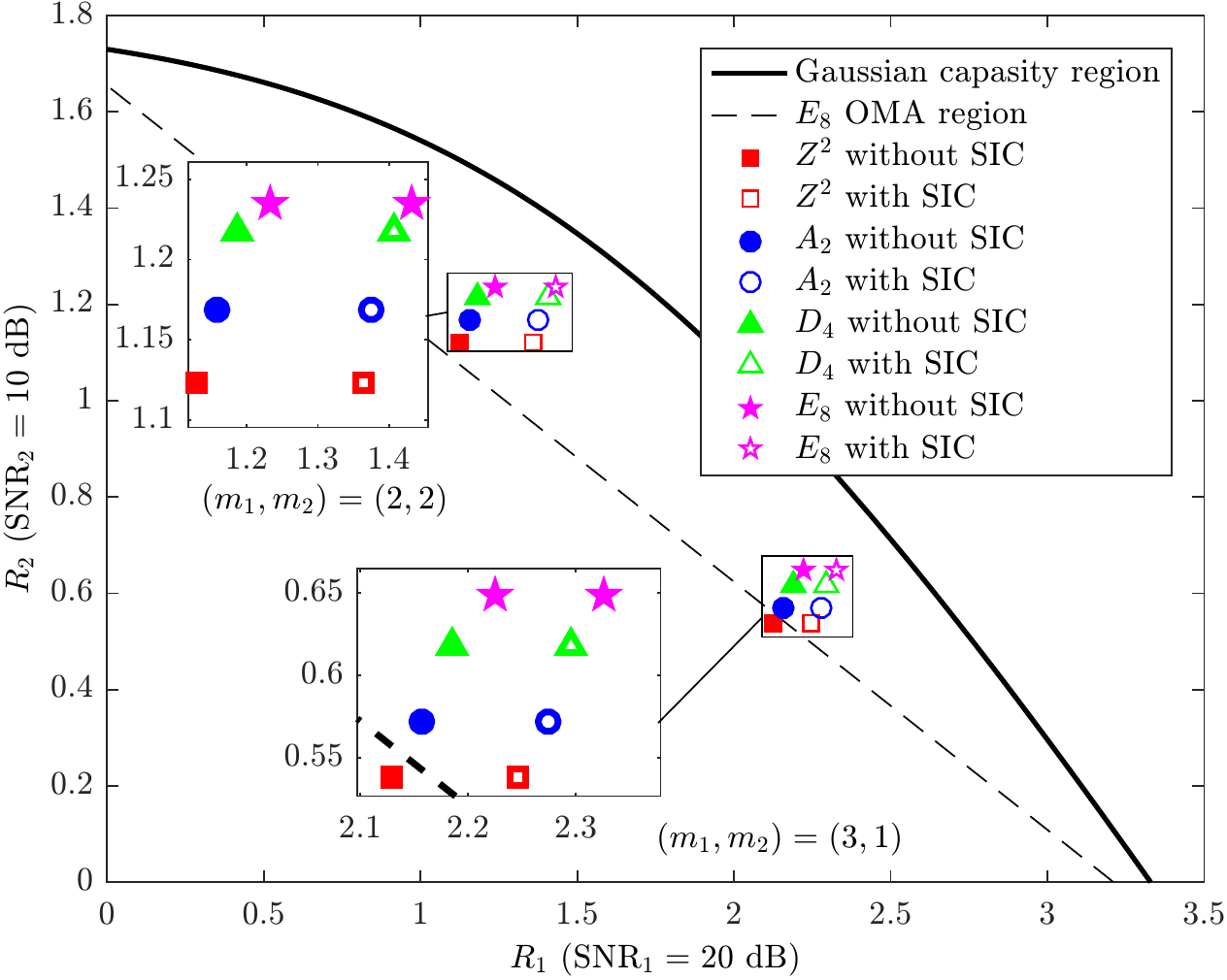}
\caption{The achievable rate pairs of downlink NOMA based on $\mathbb{Z}^2$, $A_2$, $D_4$ and $E_8$ with $\SNR_1 = 20$ dB and $\SNR_2 = 10$ dB. }
\label{fig:5}
\end{figure}

\begin{figure}[ht!]
	\centering
\includegraphics[width=3.2in,clip,keepaspectratio]{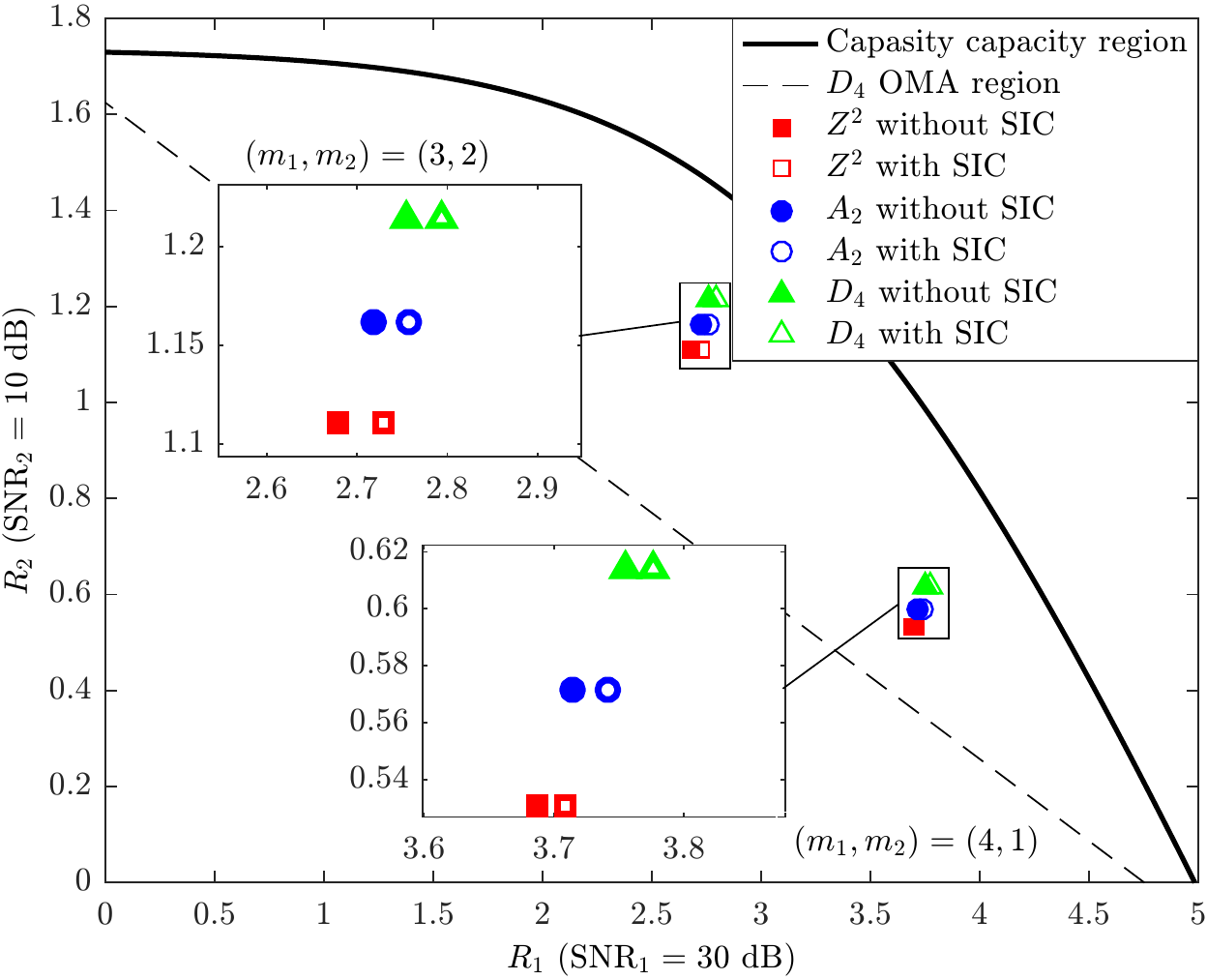}
\caption{The achievable rate pairs of downlink NOMA based on $\mathbb{Z}^2$, $A_2$ and $D_4$ with $\SNR_1 = 30$ dB and $\SNR_2 = 10$ dB. }
\label{fig:6}
\end{figure}

We first present some simulation results for our downlink NOMA scheme of two-user case in Figs.~\ref{fig:4}-\ref{fig:6}. Different from our theoretical analysis, when constructing the transmitted signal in \eqref{eqn:x_def_5}, we use a fixed dither to reduce the overall power consumption of the underlying constellation. This is a common approach in practical simulation. We evaluate rate pairs $(R_1, R_2)$ achieved by our scheme with SIC, and that without SIC for two users by performing Monte Carlo simulation with averaging over at least $10^6$ channel realizations. The corresponding choices of $(m_1,m_2)$ are shown in the figures. Note that the achievable rate pairs for either $m_1 = 0$ or $m_2 = 0$ are not included as these are considered as single user cases. We also plot the capacity region which is obtained by Gaussian input distributions and the OMA time-sharing region obtained by time-sharing between two practical schemes with constellations carved from $E_8$ lattices in all the figures. The actual gaps between the achievable rates obtained from the simulations and the capacity regions in Figs.~\ref{fig:4}-\ref{fig:6}, are given in Tables \ref{tbl:parameter_BC_15_3}-\ref{tbl:parameter_BC_30_10}, respectively, where $(\tilde{\Delta}_1, \tilde{\Delta}_2)$ denotes the pair of the actual gaps to the multiuser capacity region.

From Tables \ref{tbl:parameter_BC_15_3}-\ref{tbl:parameter_BC_30_10}, it can be seen that the simulated gaps to multiuser capacity are smaller than the theoretical results in Table I. This is because the theoretical gaps are upper bounds derived to drop the dependency of the actual $\SNR$ parameters. The actual gaps are obtained by Monte-Carlo simulations which involve actual $\SNR$ parameters. As a result, the actual gap varies with the $\SNR$ parameters. Moreover, as discussed in Remark \ref{333_5}, we have added 1 bit in our derived theoretical gaps to bound the capacity differences between the linear deterministic model and the downlink NOMA model. Therefore, the simulated gaps are usually much smaller than the theoretical bounds.


\begin{table}
\begin{center}
  \caption{Parameters for $\mathbb{Z}^2$, $A_2$, $D_4$ and $E_8$ in Fig. \ref{fig:4}}
  \label{tbl:parameter_BC_15_3}
\begin{tabular}{|c|c|c|c|}
  \hline
  $(m_1, m_2)$ & $\Lambda$ & $(\tilde{\Delta}_1, \tilde{\Delta}_2)$ & SIC \\
  \hline
  & \multirow{2}{*}{$\mathbb{Z}^2$} & (0.903, 0.405) & NO   \\
       \hhline{~~--}
        &  & (0.714, 0.343) & YES \\
        \hhline{~---}
        & \multirow{2}{*}{$A_2$} & (0.783, 0.343) & NO \\
         \hhline{~~--}
   (2, 1)     &  & (0.598, 0.279) & YES \\
     \hhline{~---}
        & \multirow{2}{*}{$D_4$} & (0.711, 0.307) & NO \\
         \hhline{~~--}
        &  & (0.532, 0.246) & YES  \\
          \hhline{~---}
        & \multirow{2}{*}{$E_8$} & (0.634, 0.274) & NO \\
         \hhline{~~--}
        &  & (0.471, 0.216) & YES  \\
      \hline
\end{tabular}
\end{center}
\end{table}
\begin{table}
\begin{center}
  \caption{Parameters for $\mathbb{Z}^2$, $A_2$, $D_4$ and $E_8$ in Fig. \ref{fig:5}}
  \label{tbl:parameter_BC_20_10}
\begin{tabular}{|c|c|c|c|c|c|c|c|}
  \hline
  $(m_1, m_2)$ & $\Lambda$ & $(\tilde{\Delta}_1, \tilde{\Delta}_2)$ & SIC \\
  \hline
   & \multirow{2}{*}{$\mathbb{Z}^2$} & (0.588, 0.446) & NO   \\
    \hhline{~~--}
        &  & (0.470, 0.363) & YES \\
        \hhline{~---}
        & \multirow{2}{*}{$A_2$} & (0.518, 0.392) & NO \\
         \hhline{~~--}
    (3, 1)    &  & (0.401, 0.311) & YES \\
    \hhline{~---}
        & \multirow{2}{*}{$D_4$} & (0.432, 0.328) & NO \\
         \hhline{~~--}
        &  & (0.322, 0.250) & YES  \\
        \hhline{~---}
        & \multirow{2}{*}{$E_8$} & (0.354, 0.269) & NO \\
         \hhline{~~--}
        &  & (0.252, 0.195) & YES  \\ \hline
        \hhline{~---}
      & \multirow{2}{*}{$\mathbb{Z}^2$} & (0.785, 0.376) & NO   \\
       \hhline{~~--}
        &  & (0.551, 0.285) & YES \\
        \hhline{~---}
        & \multirow{2}{*}{$A_2$} & (0.678, 0.319) & NO \\
         \hhline{~~--}
   (2, 2)     &  & (0.459, 0.233) & YES \\
   \hhline{~---}
        & \multirow{2}{*}{$D_4$} & (0.570, 0.261) & NO \\
         \hhline{~~--}
        &  & (0.350, 0.174) & YES  \\
        \hhline{~---}
        & \multirow{2}{*}{$E_8$} & (0.488, 0.226) & NO \\
         \hhline{~~--}
        &  & (0.291, 0.144) & YES  \\
    \hline
\end{tabular}
\end{center}
\end{table}
\begin{table}
\begin{center}
  \caption{Parameters for $\mathbb{Z}^2$, $A_2$ and $D_4$ in Fig. \ref{fig:6}}
  \label{tbl:parameter_BC_30_10}
\begin{tabular}{|c|c|c|c|}
  \hline
  $(m_1, m_2)$ & $\Lambda$ & $(\tilde{\Delta}_1, \tilde{\Delta}_2)$ & SIC \\
  \hline

  &    \multirow{2}{*}{$\mathbb{Z}^2$} & (0.683, 0.495) & NO   \\
      \hhline{~~--}
        &  & (0.609, 0.483) & YES \\
        \hhline{~---}
     (4, 1)   & \multirow{2}{*}{$A_2$} & (0.605, 0.437) & NO \\
     \hhline{~~--}
        &  & (0.579, 0.421) & YES \\
        \hhline{~---}
        & \multirow{2}{*}{$D_4$} & (0.511, 0.369) & NO \\
        \hhline{~~--}
        &  & (0.491, 0.357) & YES  \\ \hline
       & \multirow{2}{*}{$\mathbb{Z}^2$} & (0.870, 0.377) & NO   \\
       \hhline{~~--}
        &  & (0.821, 0.362) & YES \\
        \hhline{~---}
     (3, 2)   & \multirow{2}{*}{$A_2$} & (0.740, 0.314) & NO \\
     \hhline{~~--}
        &  & (0.702, 0.302) & YES \\
        \hhline{~---}
        & \multirow{2}{*}{$D_4$} & (0.605, 0.250) & NO \\
        \hhline{~~--}
        &  & (0.565, 0.237) & YES  \\
  \hline
\end{tabular}
\end{center}
\end{table}

In Fig.~\ref{fig:4}, we consider the low SNR regime where $\SNR_1 = 15$ dB and $\SNR_2 = 3$ dB, which correspond to $n_1 = 2$ and $n_2 = 1$, respectively. Here, we can see that the actual gaps $(\tilde{\Delta}_1,\tilde{\Delta}_2)$ between the capacity region and the rates achieved by our proposed are much smaller than the derived upper bounds. For example, the proposed scheme with $\mathbb{Z}^2$, which can be regarded as applying the scheme in \cite{Shieh16} independently twice, can operate within 1 bit to the capacity region. Moreover, our scheme with $A_2$, $D_4$ and $E_8$ achieve rate pairs better than that achieved by $\mbb{Z}^2$. In particular, the proposed scheme constructed over $E_8$ has the highest achievable rates due to the fact that $E_8$ has the highest shaping gain compared with $\mathbb{Z}^2$, $A_2$ and $D_4$. Simulation results for the case of $(\SNR_1, \SNR_2)= (20,10)$ dB and $(\SNR_1, \SNR_2)= (30,10)$ dB are provided and shown in Fig.~\ref{fig:5} and Fig.~\ref{fig:6}, respectively. Similar observations can be made for these settings. Here, we can see that our scheme employing $E_8$ lattice partition can approach the multiuser capacity region within 0.5 bits. Apart from that, another observation in Fig.~\ref{fig:5} is that, the gap in $R_1$ between our scheme with and without SIC becomes larger when $R_2$ is larger. This is due to the fact that user 2 introduces strong interference to user 1, which leads to rate loss in $R_1$. However, when the channel conditions are in huge difference as shown in Fig.~\ref{fig:6}, our scheme without SIC can operate very closed to the schemes with SIC even though user 2 has strong interference, i.e., the case of $(m_1,m_2) = (3,2)$. This result is even more favourable for NOMA as NOMA can attain higher gain when the channel difference is large. Note that the scheme with $E_8$ is not shown in Fig.~\ref{fig:6} mainly due to the larger constellations size that introduces high computational complexity in the simulation.

Note that in Fig.~\ref{fig:4}, we have considered two instances in our proposed framework. These two cases have the same target rate pairs, i.e., $(m_1,m_2) = (2,1)$. The only difference is that the first case has the modulo operation after superposition coding as described in \eqref{eqn:x_def_5} while the second case directly sends the superimposed signal as described in \eqref{eqn:x_def2}. Interestingly, for this case, the rate pairs achieved by these two designs are quite different and deserve some discussions.

First of all, we would like to emphasize that the second design also belongs to our proposed framework. It just corresponds to a different choice of coset leaders as discussed in Remark \ref{remark2}. Secondly, comparing $\mathbf{x}$ in \eqref{eqn:x_def_5} to $\mathbf{x}''$ in \eqref{eqn:x_def2}, the modulo operation in $\mathbf{x}$ ensures that the overall combined constellation has less power and thus the scaling $\beta$ is going to be larger than $\beta''$ for $\mathbf{x}''$, resulting in a larger minimum distance for user 1's signal. This is confirmed by noting that in Fig.~\ref{fig:4}, the rate pairs corresponding to the first design have larger $R_1$ (roughly 0.2 bit larger) than that corresponding to the second design. On the other hand, $\mathbf{x}''$ directly performs superposition without modulo operation provides a larger distance between cluster centers (each cluster corresponds to an element in $2^{m_1}\mc{C}_2$), resulting in a larger minimum distance for user 2's signal. This is confirmed again by noting that in Fig.~\ref{fig:4}, the rate pairs corresponding to the second design have larger $R_2$ (roughly 0.2 bit larger) than that corresponding to the first design. Thirdly, from Fig.~\ref{fig:4}, one observes that the second design is able to provide rate pairs outside the OMA region while the first one cannot. However, in the high SNR regime, the aforementioned rate difference becomes negligible and both designs are able to outperform OMA. Thus, in Fig.~\ref{fig:5} and Fig.~\ref{fig:6}, we only plot the first design for the sake of brevity. Last but not least, we note that the rate loss in $R_2$ for case 1 can be reduced by properly selecting coset leaders. A typical example is the performance achieved by $\mathbb{Z}^2$. Since there is no power saving by performing modulo operation on $\mathbb{Z}^2$ \cite[footnote 9]{Forney03}, the performance differences are only caused by coset leader selections that affects the Euclidean distance between each coset. The selection of coset leaders within our proposed framework is itself an interesting problem that deserves a full investigation and is beyond the scope of this work. In this work, we analyze the performance of the proposed framework with design corresponding to \eqref{eqn:x_def_5} to show that there exists at least one design within our proposed framework that can operate very close to the capacity region and leave the coset leader selection problem as a potential future work.

\subsection{Achievable Rate Simulation: Three-User Case}\label{chap:3}
\begin{figure}[ht!]
	\centering
\includegraphics[width=3.43in,clip,keepaspectratio]{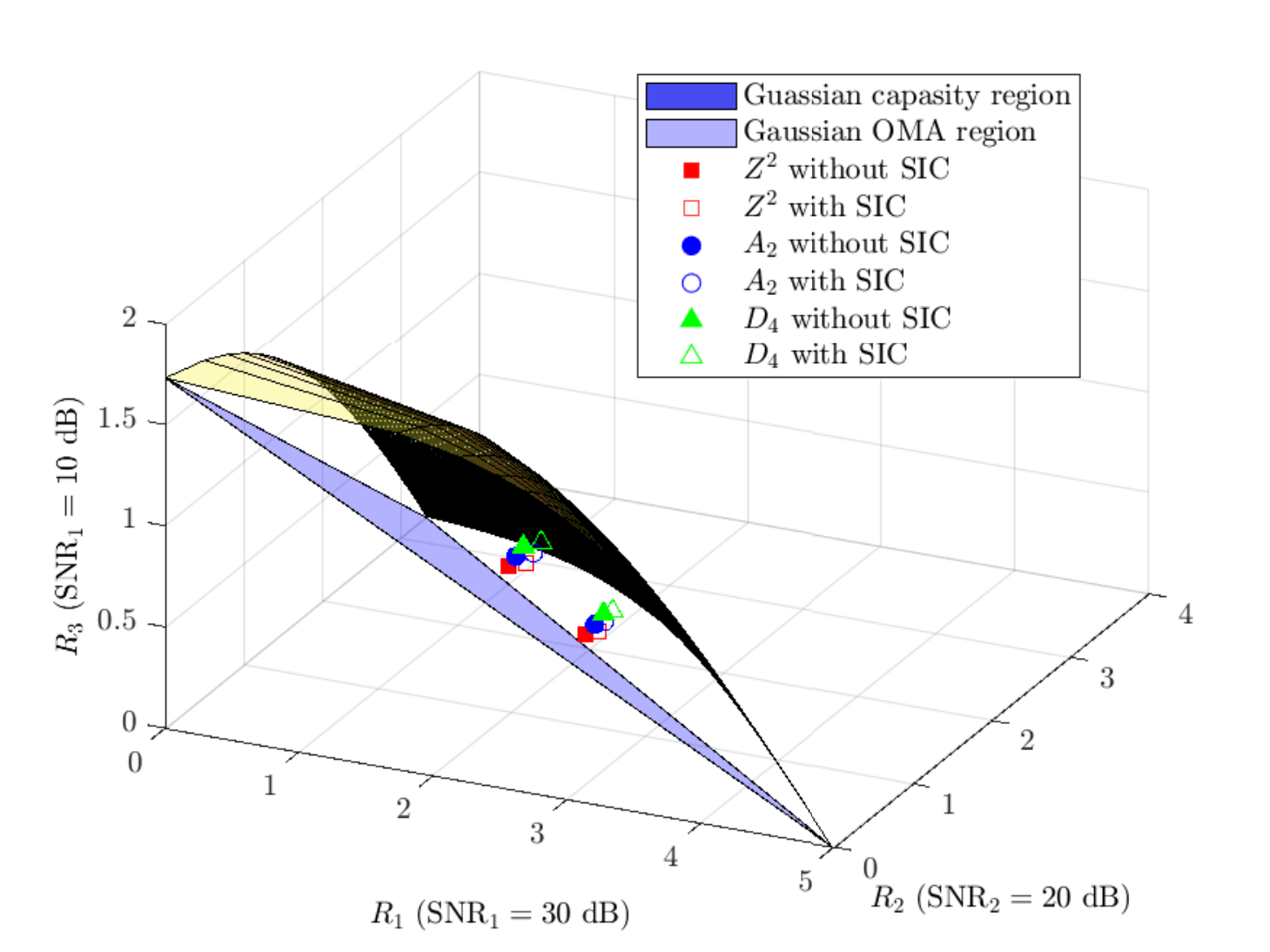}
\caption{The achievable rate tuples of downlink NOMA based on $\mathbb{Z}^2$, $A_2$ and $D_4$ with $(\SNR_1,\SNR_2,\SNR_3) = (30,20,10)$ dB.}
\label{fig:3DNOMA_5}
\end{figure}

Now we present the simulation result for three-user case to demonstrate the performance of our design for $K$-user downlink NOMA. We consider the three-user case with $\SNR_1 = 30$ dB, $\SNR_2 = 20$ dB and $\SNR_3 = 10$ dB, which corresponds to $(n_1,n_2,n_3) = (5,4,2)$. We again use Monte Carlo simulation to evaluate some of the achievable rate tuples $(R_1,R_2,R_3)$ for our scheme with and without SIC based on $\mathbb{Z}^2$, $A_2$ and $D_4$ lattices. When SIC decoder is applied, the strong users (users 1 and 2) first decode and subtracts the signals of the user having smaller SNR than itself, starting from the weakest user. Note that the achievable rate tuples for the proposed scheme with $E_8$ lattices are not simulated due to the huge computational complexity. The results are shown in Fig. \ref{fig:3DNOMA_5} along with the Gaussian capacity region and Gaussian OMA time-sharing region.

Akin to the two-user case, it can be seen that among the considered designs, $D_4$ achieves the highest rate tuples. In addition, all the achievable rate tuples of our proposed schemes lie outside the OMA region, indicating that even without SIC, our scheme outperforms any OMA scheme. Furthermore, the results show that our scheme can approach the multiuser capacity region within a small gap even though the number of users increases.

\subsection{Error Probability Simulation}
Now we build the coded system for our downlink NOMA scheme. In our simulation, each encoding function $\mathcal{E}_i$ mentioned in Section \ref{LOMA_5} consists of two parts, namely a linear code over $GF(2^{nm_i})$ followed by a bijective mapping that maps each $GF(2^{nm_i})$ onto $\mathcal{C}_i$. We note that other popular coded modulation techniques such as bit interleaved coded modulation and multilevel coding can also be adopted. In this work, we adopt coding over $GF(2^{nm_i})$ due to its best performance among these techniques. The codes over $GF(2^{nm_i})$ used in the following simulations are constructed via the code design technique proposed in \cite{8066336}.

To give an illustration on the performance of our scheme with coded systems, we select one rate pair that is achievable by our scheme and design the codes based on it. More specifically, we choose the case in Fig. \ref{fig:4} and select the target rate pair to be the achievable rate pair of $D_4$ lattices, that is $(R_1, R_2)=(1.3954,0.2542)$ bits per real dimension. This results in the code rates $0.6977$ and $0.2542$ for user 1 and user 2, respectively. Then we use extrinsic information transfer (EXIT) charts to design length 10,000 non-binary irregular repeat-accumulate (IRA) codes over the lattice partitions of $\mathbb{Z}^2/2\mathbb{Z}^2$, $\mathbb{Z}^2/4\mathbb{Z}^2$, $A_2/2A_2$, $A_2/4A_2$, $D_4/2D_4$ and $D_4/4D_4$, respectively. The design details can be found in \cite{8066336} and thus are omitted here. Note that the codes are optimized particularly for point-to-point AWGN channel and have been shown to have the near-capacity performance in \cite{8066336}. However, it is entirely possible that the codes are not optimal for downlink NOMA where interference is present. The code design problem for downlink NOMA where the structure of the interference is taken into account is an interesting problem in its own right and is clearly beyond the scope of this work.

\begin{figure}[ht!]
	\centering
\includegraphics[width=3.2in,clip,keepaspectratio]{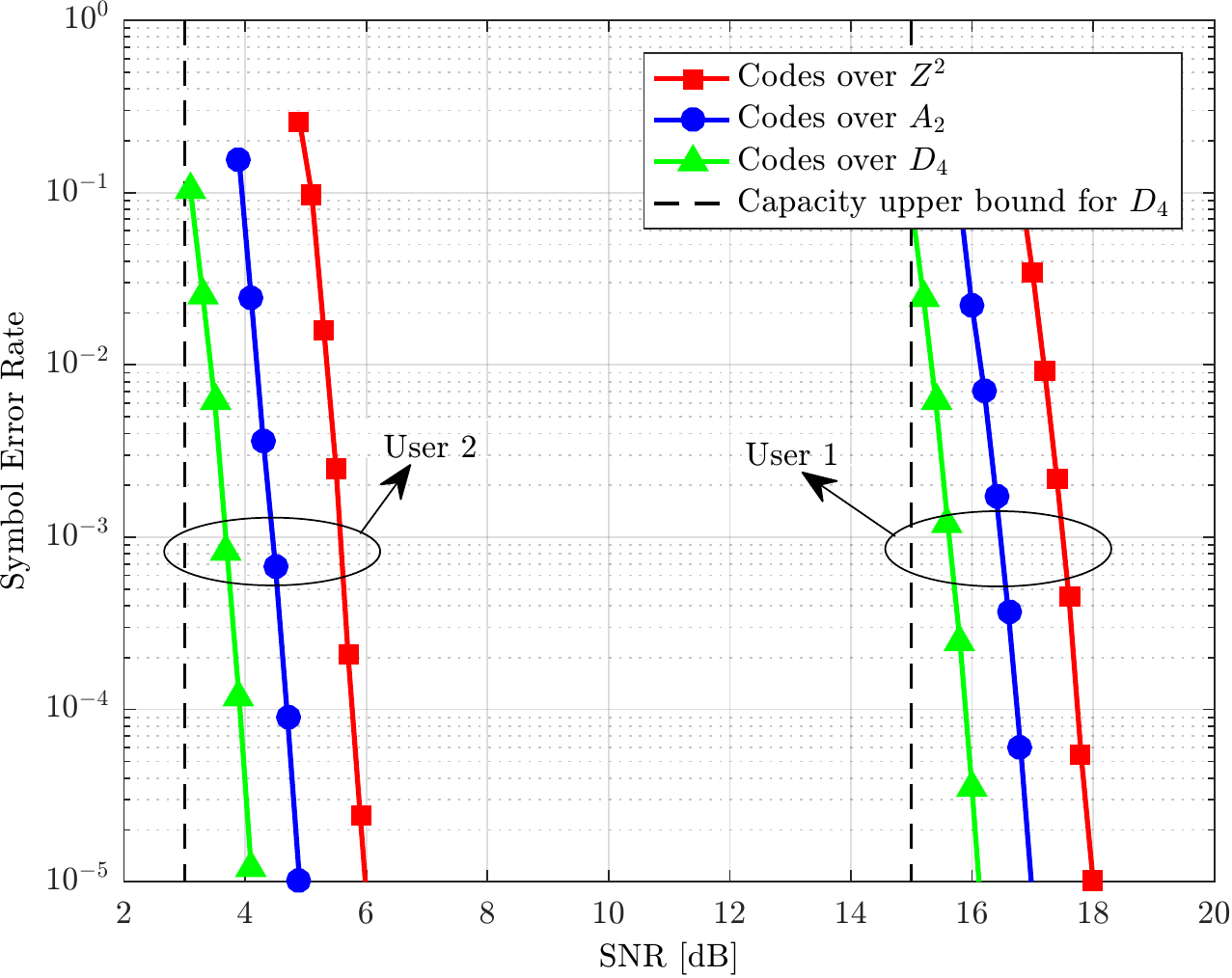}
\caption{Symbol error rate performance of coded system for $(R_1,R_2)=(1.3954,0.2542)$ bits/real dim.}
\label{fig:SER}
\end{figure}

To perform the simulation, the source message is encoded into the non-binary codes via the encoding function $\mathcal{E}_i$ described in Section \ref{LOMA_5} and then bijectively mapped the codeword element onto the constellation. The corresponding SER versus SNR results are presented in Fig. \ref{fig:SER}. The Shannon limit for using the proposed constellation carved from $D_4$ lattice is also plotted by dash line in the figure. No SIC is performed.

From Fig. \ref{fig:SER}, we can see that it would require about 3 dB more power for codes over $\mathbb{Z}^2$ to achieve the same rate as for $D_4$ when the SER is at $10^{-5}$. Codes over $A_2$ are about 1 dB better than codes over $\mathbb{Z}^2$. Most notably, codes over $D_4$ require 1 dB less power to attain the achievable rate of $D_4$. The SER results agree with the achievable rate results, showing that better performance can be attained by using constellations carved from lattices with smaller NSM. The performance loss shown in the figure is mainly due to finite codeword length and the suboptimal code design in the sense that the codes are in fact optimized for point-to-point AWGN channels rather than downlink NOMA.

\section{Summary}\label{sec:conclude}
Guided by the corresponding linear deterministic model, we have developed a general lattice framework of downlink NOMA without SIC. We have analyzed the rates achieved by the proposed framework with any lattice as base lattice and provided an upper bound on the gap between the rates and the capacity region as a function of the NSM of the base lattice. For some well-known lattices such as $A_2$, $D_4$, $E_8$, and Construction A lattices, the gap upper bounds have been evaluated and have been shown to shrink as the dimension increases. Simulation results have shown that the actual gaps can be much smaller than the derived theoretic bounds, which have further demonstrated the capability of our proposed scheme to achieve the near-capacity performance in downlink NOMA.

\chapter{Lattice-Partition-Based Downlink NOMA without SIC for Slow Fading
Channels}\label{C6:chapter6}

\section{Introduction}
For power-domain NOMA, it has been known for quite a while \cite{Cover:2006:EIT:1146355} that the capacity region can be achieved by superimposing codewords with Gaussian distribution at the transmitter and adopting SIC at the receivers. Hence, most of the works in the literature assume the above capacity-achieving scheme and focus largely on determining user pairing and power allocation among paired users and among resource blocks for maximizing system throughput, see for example \cite{7542601,7937794,7959169,7972963,7971899,7974731,8352621}. However, Gaussian signalling is deemed impractical (if not impossible) and SIC imposes a burden on receivers as each receiver has to (partially) decode others' codewords before decoding its own codeword, which increases decoding complexity and latency.

In chapter 5, we have introduced several works on NOMA with practical discrete inputs \cite{Choi2016,Dong17,Fang16,Shieh16,8254176,8291591}. However, having instantaneous CSI (or AWGN channel) as assumed in these works is unlikely to be realistic for scenarios where the feedback links may be costly and limited \cite{tse_book}. Statistical CSI is a more appropriate assumption for such cases and can be deemed as the worst case scenario for these applications \cite[Section I]{5208539}. Although some existing works in the literature \cite{7361990,7438933,Wei17,7959198,8063934,8327866} have considered NOMA with only statistical CSI at the transmitter, continuous Gaussian inputs and SIC are still adopted. To the best of our knowledge, the designs and analyses for NOMA without SIC based on discrete inputs with statistical CSI have not been reported in the literature yet. Hence, further investigation on this case is called for.

\subsection{Main Contributions}
In this work, we consider the problem of designing practical schemes for NOMA systems under slow fading channel where only statistical CSI is available at the transmitter and full CSI is available at the receivers. We focus on bridging the gap between theory and practice. In particular, we develop a lattice-partition-based downlink NOMA scheme which adopts discrete inputs according to statistical CSI and can be efficiently decoded with single-user decoding, i.e., without SIC. The main contributions of our work are summarized as follows.
\begin{itemize}
\item
We propose a novel scheme for the $K$-user downlink NOMA system over slow fading channels. Similar to \cite{Shieh16,8291591}, we first look into the corresponding linear deterministic model \cite{Avestimehr11} and then translates the results back to the NOMA model. As a result, the proposed scheme is a systematic design which determines the modulation, coding rate, and power allocation for each user, according to the statistical CSI and the target outage probability. Our scheme adopts discrete constellations carved carefully from lattices, and hence admits discrete input distributions. By leveraging the structure of interference induced by the proposed signalling, the proposed scheme can be efficiently decoded with single-user decoding, i.e., no SIC is required. We would like to emphasize that this is a non-trivial generalization of \cite{Shieh16,8291591} since in the present work, the transmitter only has statistical CSI while the schemes in \cite{Shieh16,8291591} hinge entirely on (quantized) instantaneous CSI. To overcome this, we design the constellations according to {\it average channel condition} and derive suitable code rates as a function of outage probabilities.

\item
We rigorously show that for any outage probability below $63.21\%$ and any rate tuple lying inside the NOMA outage capacity region, there is an instance of our proposed scheme that can achieve that rate tuple to within a constant gap. Specifically, the derived upper bound of the gap is universal for every signal-to-noise ratio (SNR) and any number of users, and this upper bound is found to be a function of the base lattice adopted in construction only. To the best of our knowledge, this is for the first time that the NOMA outage capacity region can be closely approached with discrete inputs and single-user decoding.

\item
Monte Carlo simulation is conducted to demonstrate that the achievable outage rate tuples of our scheme is very close to the NOMA outage capacity region, even with single-user decoding at each user. The gap to the NOMA outage capacity region is shown to be even smaller if SIC is adopted at each strong user. Furthermore, we also provide simulations for a practical set-up of our scheme where off-the-shelf LDPC codes are employed on top of the underlying lattice constellations. The results reaffirm that our proposed scheme without SIC significantly outperforms OMA-type schemes.
\end{itemize}

%

\section{System Model}\label{sysm}
In this work, we consider a downlink NOMA system where a base station wishes to broadcast messages to $K$ users, each of which experiences independent slow fading. The base station and all users are equipped with a single antenna and work in a half-duplex mode.
\begin{figure}[ht!]
	\centering
\includegraphics[width=3.42in,clip,keepaspectratio]{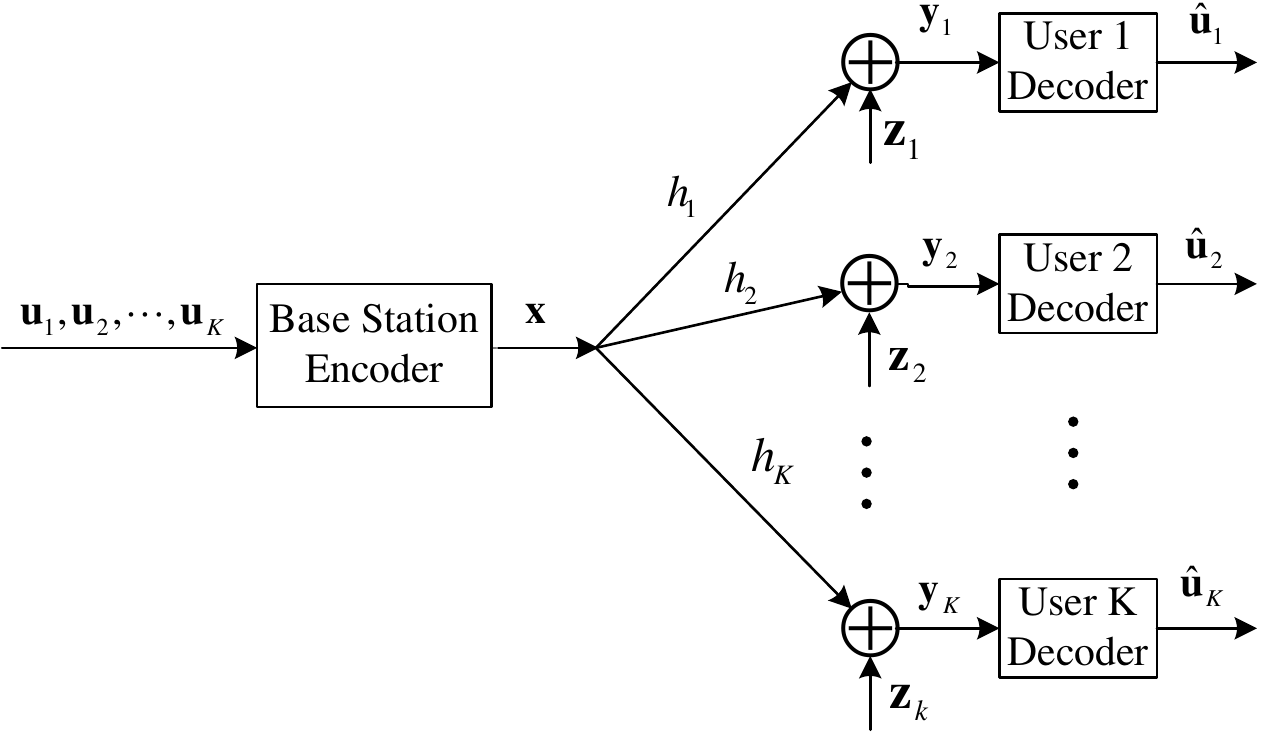}
\caption{The system model of the $K$-user downlink NOMA over slow fading channels.}
\label{fig:1}
\end{figure}
As depicted in Fig. \ref{fig:1}, the base station first jointly encodes the binary messages $\mathbf{u}_1,\ldots,\mathbf{u}_K$ intended for users $1,\ldots,K$ into a codeword where each codeword symbol $\mathbf{x} \in \mbb{R}^n$ is with power constraint $\mathbb{E}[\|\mathbf{x}\|^2]\leq n$, i.e., we use $n$ real channels jointly. We denote by $h_k$ the instantaneous channel coefficient from the base station to user $k$, where $h_k$ is a realization of $\msf{H}_k$ whose channel gain $|\msf{H}_k|^2$ has an inverse cumulative distribution function (ICDF) $F_k(\epsilon)$ for $\epsilon\in[0,1]$. The received signal arrives at user $k$ is then given by
\begin{equation}\label{eq:sm1}
y_k[i] = h_k\sqrt{P}x[i]+z_k[i], \;  k= 1,\ldots, K,\; i =1,\ldots, n
\end{equation}
where $P$ is the total power constraint at the base station and $z_k[.] \thicksim \mathcal{N}(0,1)$ is AWGN experienced at user $k$. We assume that the transmitter only has the knowledge of statistical CSI of $\msf{H}_k$, $k\in\{1,\ldots,K\}$, while each user $k$ knows the realization $h_k$, i.e., instantaneous CSI. Hence, each receiver can compensate the phase of the channel by coherent detection \cite[Ch. 3.1.2]{tse_book}, justifying the real channel model in \eqref{eq:sm1}.

An outage event occurs at user $k$ when this user cannot successfully decode its own message $\mathbf{u}_k$. We denote by $\epsilon_k$ the required outage probability for user $k$ and a rate $R_k$ is said to be achievable with $\epsilon_k$ outage if under this rate, the outage probability for user $k$ is not greater than $\epsilon_k$. The outage capacity region under $(\epsilon_1,\ldots ,\epsilon_K) \in [0,1]^K$ is then defined as the closure of the set of all outage rate tuples $(R_1,\ldots,R_K)$ under the input power constraint. Without loss of generality, we assume that the user ordering follows $F_1(\epsilon_1) \geq  \ldots\geq F_K(\epsilon_K)$ throughout this work. The outage capacity has been characterized in \cite{5208539} and is summarized as follows.
\begin{theo}\cite[Th. 1]{5208539}\label{the:1}
Given the required outage probability vector $(\epsilon_1,\ldots ,\epsilon_K) \in [0,1]^K$, the $k$-th user's outage capacity in bits per real dimension contains every rate tuple $(R_1,\ldots,R_K)$ such that there exist power allocation factors $(\alpha_1,\ldots,\alpha_K)\in [0,1]^K$ satisfying $\sum_{k=1}^K\alpha_k = 1$ results in $R_k<C_k$ where
\begin{equation}\label{eq:def1}
C_k \defeq \frac{1}{2}\log_2\left(1+\frac{F_k(\epsilon_k)\alpha_kP}{F_k(\epsilon_k)\left(\sum_{i=1}^{k-1}\alpha_i\right)P+1} \right),
\end{equation}
for all $k\in\{1,\ldots,K\}$.
\end{theo}
It should be noted that the outage capacity region based on the user ordering by sorting the value of $F_k(\epsilon_k)$ such that $F_1(\epsilon_1) \geq  \ldots\geq F_K(\epsilon_K)$ is larger than any outage capacity region based on an arbitrary user ordering \cite{5208539}.

In this work, we consider that each user's channel follows a Rayleigh distribution for demonstration and therefore $F_k(\epsilon) = \mathbb{E}[|\msf{H}_k|^2]F(\epsilon)$ for $k \in \{1,\ldots,K\}$, where $F(\epsilon) = -\ln(1-\epsilon)$ for $\epsilon \in[0,1]$. Although the analysis is based on Rayleigh fading, our proposed scheme in fact will work for any fading channel whose channel gain has a finite mean.

\section{Proposed Lattice-Partition-Based Downlink NOMA Scheme}\label{sec:proposed}
In this section, we present the proposed lattice-partition-based scheme for downlink NOMA over slow fading channel. In particular, we use the deterministic model \cite{Avestimehr11} as a tool to approximate our downlink NOMA model. We first investigate the corresponding linear deterministic model for the two-user case in Chapter~\ref{Assumption} and then generalize the discussion to the $K$-user case in Chapter~\ref{sec:dmodel_k}. The schemes and observations made for the deterministic model provide significant insights into designing schemes for the original model, which is presented in Chapter~\ref{LOMA}.

Before proceeding, we note that unlike existing work in the literature adopting this deterministic approach, here the noise level at each receiver is determined by its fading realization, which is oblivious to the transmitter. To overcome this, we design the constellations according to statistical CSI and rely on the analysis in Chapter~\ref{sec:rate_analysis} or simulations in Chapter~\ref{SIM} for picking appropriate code rates.

\subsection{Deterministic Model for Two-User Downlink NOMA over Fading Channels}\label{Assumption}
We use the linear deterministic model \cite{Avestimehr11} to approximate the two-user downlink NOMA for a given fading channel realization. The main idea behind the linear deterministic model is to model the broadcast channel links as bit pipes that only pass to each user the bits above its noise level and truncate the bits below the noise level.

First, we define user $k$'s average signal-to-noise power ratio (SNR) including the base station power, average channel gain and noise variance as
\begin{equation}\label{eq:CSI}
\overline{\SNR}_k \triangleq \E[|\msf{H}_k|^2]P.
\end{equation}
We emphasize that as slow fading is considered, user $k$ will only get to experience one channel realization $h_k$ of $\msf{H}_k$. Since the transmitter only has the knowledge of statistical CSI, our scheme will particularly make use of the statistical CSI through $\overline{\SNR}_k$. Let $\bar{n}_k \defeq \left\lceil \frac{1}{2}\log_2(\overline{\SNR}_k)\right\rceil^+$ for $k = \{1,2\}$ and assume that $\overline{\SNR}_1 \geq \overline{\SNR}_2$. Here, $\bar{n}_k$ is the maximum number of bits that user $k$ expects to receive as if the instantaneous SNR is $\overline{\SNR}_k$. The base station broadcasts $\bar{n}_1$ bits to both users. Let non-negative integers $m_1$ and $m_2$ represent the number of transmitted bits intended for user 1 and user 2, respectively, satisfying
\begin{align}
m_1+m_2 &\leq \bar{n}_1,\label{eq:m1n1}\\
m_2 &\leq \bar{n}_2.\label{eq:m2n2}
\end{align}
Since $\overline{\SNR}_1 \geq \overline{\SNR}_2$, the base station treats user 2 as the weak user and thus places user 2's bits above user 1's bits in the deterministic model. Note that the bits in a higher level in the deterministic model means that the corresponding signals get allocated more power in the original downlink NOMA model.

When the instantaneous SNR of user 1 is $\overline{\SNR}_1$, user 1 can receive $\bar{n}_1$ bits and hence the noise level observed by user 1 is below $m_1$. When the instantaneous SNR of user 2 is $\overline{\SNR}_2$, the noise level observed by user 2 is below $m_2$. As $m_1$ is placed below $m_2$ in the deterministic model, user 1's bits can be decomposed into two parts based on the noise level observed by user 2. Specifically, we let $m_1 = r_{11}+r_{12}$ where
\begin{align}
r_{11}&\defeq\min\{m_1,\bar{n}_2-m_2\}, \\
r_{12}&\defeq \max\{m_1+m_2-\bar{n}_2,0\},
\end{align}
are the number of bits that are above and below the noise level of user 2, respectively. Thus, user 2 can receive $m_2+r_{11} = \bar{n}_2$ bits in total.

The above case illustrates the expected performance when the instantaneous channel gain equals to the average channel gain. We now consider the case for a given channel realization $h_k \in \msf{H}_k$. We define $n_k \defeq \left\lceil \frac{1}{2}\log_2(|\hat{h}_k|^2\overline{\SNR}_k)\right\rceil^+$ for $k \in \{1,2\}$, where $|\hat{h}_k|^2$ is the normalized channel gain such that
\begin{equation}\label{eq:h_hat}
\hat{h}_k = \frac{h_k}{\sqrt{\E[|\msf{H}_k|^2]}}.
\end{equation}
Here, $n_k$ is the maximum bits that user $k$ can receive under the channel realization $h_k$. Thus, $\max\{\bar{n}_k - n_k,0 \}$ represents the number of bits that are overtransmitted for user $k$. Based on the noise level observed by user 1, we decompose $ m_1 = r_{11,1} + r_{12,1}$ where
\begin{align}
r_{11,1}&\defeq \max \{ m_1-\max\{\bar{n}_1-n_1,0\},0\},\label{eq:r11}\\
r_{12,1}&\defeq \min\{m_1,\max\{\bar{n}_1-n_1,0\}\},\label{eq:r12}
\end{align}
are user 1's message bits above and under its observed noise level, respectively. The channel fading first starts to affect the least significant bits corresponding to the signals with the lowest signal power, i.e, those corresponding to $r_{12,1}$.

For user 2's channel, we similarly decompose $m_1 = r_{11,2} + r_{12,2}$ with
\begin{align}
r_{11,2}&\defeq  \min \{ m_1,\max\{n_2 - m_2,0\} \},\label{eq:r13}\\
r_{12,2}&\defeq  \max\{m_1+ \min\{m_2 - n_2,0\},0\},\label{eq:r14}
\end{align}
where $r_{11,2}$ and $r_{12,2}$ represent user 1's message bits above and under the noise level, respectively, from user 2's observation. Since the noise level can be above $m_1$, we also decompose $m_2 =r_{21,2} + r_{22,2}$ with
\begin{align}
r_{21,2}&\defeq \min\{m_2,n_2\},\label{eq:r15}\\
r_{22,2}&\defeq  \max\{0,m_2-n_2\},\label{eq:r16}
\end{align}
where $r_{21,2}$ and $r_{22,2}$ represent user 2's message bits above and under the noise level, respectively, observed by user 2.

\begin{figure}[ht!]
	\centering
\includegraphics[width=2.3in,clip,keepaspectratio]{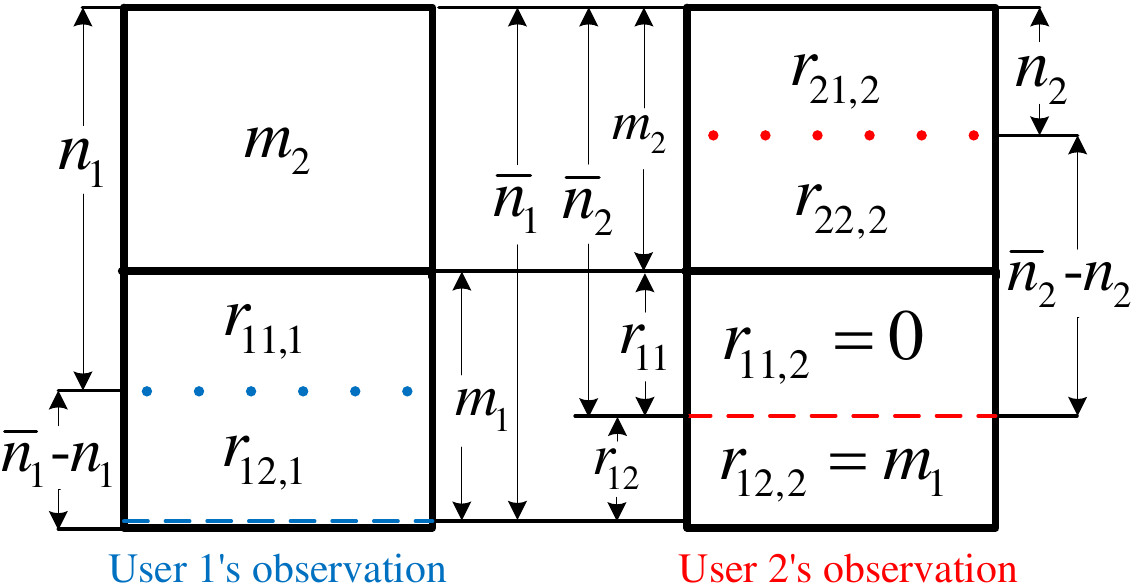}
\caption{An illustration of the defined parameters in \eqref{eq:r11}-\eqref{eq:r16}.}
\label{fig:dm}
\end{figure}

To better understand the relationships between the above variables, an example for the case where $m_2 < n_1<\bar{n}_1$ and $0 < n_2<m_2$ is provided in Fig. \ref{fig:dm}.

In this figure, the dotted lines represent the noise levels based on an instant channel while the dash lines represent the noise levels based on the statistical CSI. For user 1, the total number of bits will be treated as under the noise level, is exactly the number of overtransmitted bits in this instantaneous fading channel. Since $n_1>m_2$, user 1's message bits that are above the noise level is precisely $r_{11,1} = n_1-m_2 = m_1-(\bar{n}_1-n_1)$ as indicated in \eqref{eq:r11}. For user 2, the channel fading starts affecting the least significant bits from $r_{11}$ to $r_{22,2}$. In this case, user 1's bits are all under the noise level and hence $r_{12,2} = m_1$ as it is placed under $m_2$ in the deterministic model. The number of bits that can be received by user 2 is limited by $n_2$, i.e., $r_{21,2} = n_2$ as indicated in \eqref{eq:r15}.

Based on the above definitions, the following relationships can be easily verified.
 \begin{itemize}
\item[i)] $r_{21,2} + r_{11,2} = \min\{n_2,m_1+m_2 \}$;
\item[ii)] $r_{22,2} + r_{12,2} = \max\{m_1+m_2 -n_2,0 \}$;
\item[iii)] $r_{22,2}+r_{11} = \bar{n}_2 - n_2$, when $0\leq n_2 < \bar{n}_2$;
\item[iv)]  $r_{11,2} = 0$, $r_{12,2} = m_1$, $0<r_{22,2}\leq m_2$, $0\leq r_{21,2}< m_2$, when $0\leq n_2 < m_2$;
\item[v)] $r_{22,2} = 0$, $r_{21,2} = m_2$, $0 \leq r_{11,2} \leq m_1$, $0\leq r_{12,2}\leq m_1$, when $m_2\leq n_2$.
\end{itemize}
The fact i) shows the total number of bits that user 2 can successfully receive; ii) shows the total number of bits that will be treated as noise and get truncated at user 2; iii) shows the number of overtransmitted bits for user 2 when the instantaneous SNR is smaller than $\overline{\SNR}_2$; iv) means that only $n_2$ bits are above the noise level when the intended bits for user 2 is larger than $n_2$; and finally v) means that user 2 can receive the whole $m_2$ when its intended rate is smaller than $n_2$.

\subsection{The $K$-User Case}\label{sec:dmodel_k}
Similar to the two-user case, we define the average SNR for user $k$ as in \eqref{eq:CSI} and denote $n_k \defeq \left\lceil \frac{1}{2}\log_2(|\hat{h}_k|^2\overline{\SNR}_k)\right\rceil^+$ to be the maximum number of bits received by user $k$ for $k \in \{1,2,\ldots,K\}$ when the instantaneous SNR equals to the average SNR. Without loss of generality, we assume $\overline{\SNR}_1 \geq \overline{\SNR}_2 \geq \ldots \geq \overline{\SNR}_K$ \emph{throughout Chapter 6}. We then denote the number of transmitted bits intended for user $k$ by $m_k$ for $k \in \{1,2,\ldots, K\}$. The deterministic rate tuple $(m_1,m_2,\ldots,m_K)$ must satisfy the following constraints:
\begin{align}
m_1+m_2+\ldots+ m_K&\leq \bar{n}_1,\label{eq:dK2} \\
m_2 +\ldots+ m_K &\leq \bar{n}_2,\label{eq:dK3} \\
&\;\; \vdots \nonumber \\
m_K &\leq \bar{n}_K.\label{eq:dK4}
\end{align}

We start with the analysis for the first (strongest) user. From (\ref{eq:dK2}) and (\ref{eq:dK3}), we can reduce the problem into a two-user case by combining users $2,3,\ldots,K$ into a super-user, demanding $m_{2\rightarrow K} \triangleq \sum_{i=2}^K m_i$ bits and the maximum number of bits received is $\bar{n}_2$. Therefore, to analyze the received bits for user 1, one can directly follow our approach as described in Section \ref{Assumption}.

Now we analyze the received bits and interference for user $k>1$. At user $k$'s channel, we have the rate constraint as follows:
\begin{equation}\label{eq:dKk}
m_k +\ldots+ m_K \leq \bar{n}_k.
\end{equation}
For this case, we treat users $1,2,\ldots,k-1$ as a super-user, demanding $m_{1 \rightarrow k-1} \triangleq \sum_{i=1}^{k-1}m_i$ bits and users $k,k+1,\ldots,K$ as another super-user demanding $m_{k\rightarrow K} \triangleq \sum_{i=k}^Km_i$ bits. The problem can again be deemed as a two-user case. We thus choose the rate pairs $(m_{1\rightarrow k-1}, m_{k \rightarrow K})$ that satisfies:
\begin{align}
m_{1\rightarrow k-1} + m_{k \rightarrow K} &\leq \bar{n}_1,\label{eq:dKk1} \\
m_{k \rightarrow K} &\leq \bar{n}_k.\label{eq:dKk2}
\end{align}
In this way, the same approach from Section \ref{Assumption} can be used for analyzing this case. As $m_{1\rightarrow k-1}$ is placed below $m_{k \rightarrow K}$ in the deterministic model, it can then be decomposed into $m_{1\rightarrow k-1} = r_{11}^* + r_{12}^*$ based on the noise level observed by user $k$, where
\begin{align}
r_{11}^* &\defeq\min\{m_{1\rightarrow k-1},\bar{n}_k-m_{k \rightarrow K}\} \nonumber \\
&= \min\left\{\sum_{i=1}^{k-1}m_i,\bar{n}_k-\sum_{j=k}^Km_j\right\},\label{eq:dKk3} \\
r_{12}^* &\defeq \max\{m_{1\rightarrow k-1}+m_{k \rightarrow K}-\bar{n}_k,0\} \nonumber \\
&= \max\left\{\sum_{i=1}^K m_i-\bar{n}_k,0\right\}, \label{eq:dKk4}
\end{align}
are the number of bits above and below noise level, respectively. Thus, user $k$ can receive $r_{11}^* + m_{k \rightarrow K} = \bar{n}_k$ bits in total.

To analyze the case for a random channel realization, we similarly define $n_k$ and $\hat{h}_k$ as in \eqref{eq:h_hat} for $k \in \{1,\ldots,K\}$. For user 1 and user $K$, the analysis can be followed from \eqref{eq:r11}-\eqref{eq:r16} by treating the rest of the users as a superuser.

For user $1<k<K$, we treat users $1,2,\ldots,k-1$ as a super-user demanding $m_{1\rightarrow k-1}$ like the above case but treat users $k+1,k+2,\ldots,K$ as another super-user demanding $m_{k+1 \rightarrow K} \triangleq \sum_{i=k+1}^{K}m_i$ bits. Based on the noise level observed by user $k$, we decompose $m_{1\rightarrow k-1} = r^*_{11,k} + r^*_{12,k}$, where
\begin{align}
r^*_{11,k}&\defeq  \min \{ m_{1\rightarrow k-1},\max\{n_k - m_{k+1 \rightarrow K}-m_k,0\} \} \nonumber \\
&= \min \left\{ \sum_{i=1}^{k-1}m_i,\max\{n_k-\sum_{i=k}^{K}m_i,0\} \right\},\label{eq:r13k}\\
r^*_{12,k}&\defeq  \max\{m_{1\rightarrow k-1}+ \min\{m_{k+1 \rightarrow K}+m_k - n_k,0\},0\} \nonumber \\
&= \max \left\{ \sum_{i=1}^{K}m_i-n_k,0\right\} ,\label{eq:r14k}
\end{align}
are the first super-user's message bits above and under noise level, respectively, from user $k$'s observation. Since the noise level can be above $m_{1\rightarrow k-1}$, we also decompose $m_k =r_{k1,k} + r_{k2,k}$, where
\begin{align}
r_{k1,k}&\defeq \min\{m_k,\max\{m_k-(m_k+m_{k+1 \rightarrow K}-n_k),0\}\} \nonumber \\
 &= \min\left\{m_k,\max\left\{n_k-\sum_{i=k+1}^{K}m_i,0\right\}\right\},\label{eq:r15k}\\
r_{k2,k}&\defeq  \max\{0,\min\{m_k,m_k+m_{k+1 \rightarrow K}-n_k\}\} \nonumber \\
 &=\max\left\{0,\min\left\{m_k,\sum_{i=k}^{K}m_i-n_k\right\}\right\} \label{eq:r16k},
\end{align}
are user $k$'s message bits above and under noise level, respectively, observed at user $k$.

Based on the above definitions, the following facts can be easily verified.
 \begin{itemize}
\item[vi)] $r_{k1,k}+ r^*_{11,k} = \min\{n_k,m_{1\rightarrow k-1}+m_k \}$;
\item[vii)] $r_{k2,k} + r^*_{12,k} = \max\{m_{1\rightarrow k-1}+m_k -n_k,0 \}$;
\item[viii)] $r_{k2,k}+r^*_{11} = \bar{n}_k - n_k$, when $m_{k+1 \rightarrow K} \leq n_k < \bar{n}_k$;
\item[ix)]  $r^*_{11,k} = 0$, $r^*_{12,k} = m_{1\rightarrow k-1}$, $0<r_{k2,k}\leq m_k$, $0\leq r_{k1,k}< m_k$, when $m_{k+1 \rightarrow K} \leq n_k < m_{k+1 \rightarrow K}+m_k$;
\item[x)] $r_{k2,k} = 0$, $r_{k1,k} = m_k$, $0 < r^*_{11,k} \leq m_{1\rightarrow k-1}$, $0\leq r^*_{12,k}\leq m_{1\rightarrow k-1}$, when $m_{k+1 \rightarrow K}+m_k<n_k$.
\end{itemize}
The fact vi) shows the total number bits that user $k$ can successfully receive; vii) shows the total number of bits that will be treated as noise and get truncated at user $k$; viii) shows that the number of overtransmitted bits for user $k$ when the instantaneous channel gain is smaller than the average channel gain; ix) means that all $\sum_{i=1}^{k-1}m_i$ bits are below noise level when the number of intended bits for users $k,\ldots,K$ is larger than $n_k$; and finally x) means that user $k$ can receive the whole $m_k$ when the sum of the intended rates of users $k,\ldots,K$ is smaller than $n_k$.

\subsection{Translating Back to the Downlink NOMA Model}\label{LOMA}
We now translate the above scheme for the deterministic model into the coding scheme for the $K$-user downlink NOMA over slow fading channels.

For any rate tuple $(m_1,\ldots,m_K)$ satisfying \eqref{eq:dK2}-\eqref{eq:dK4}, our scheme makes use of the lattice partition chain with any base lattice $\Lambda/2^{m_1}\Lambda/2^{m_1+m_2}\Lambda/\ldots/2^{\sum_{i=1}^K m_i}\Lambda$
$\defeq\Lambda_s$.
The restriction of having partition orders being powers of 2 is merely for practical purpose and it can be lifted. We emphasize here that the selections of $m_1,\ldots,m_K$ are based on the statistical CSI at the transmitter. In the proposed scheme, we use the coset leaders from the above lattice partitions as constellations and code over $N$ uses of the constellations in order to establish reliable communication. Specifically, the encoding and decoding process are summarized as follows.

\subsection{Encoding}\label{sec:enc}
For $k \in \{1,\ldots,K\}$, the binary source messages $\mathbf{u}_k$ of length $M_k$ is encoded into a length $N_k$ binary codeword $\mathcal{E}_k(\mathbf{u}_k)$ via the channel encoding function $\mathcal{E}_k(\cdot)$. The modulated signal $\mathbf{v}_k$ is of length $N$ where each entry is obtained by bijectively mapping every $nm_k$ bits from user $k$'s codeword $\mathcal{E}_k(\mathbf{u}_k)$ onto user $k$'s constellation $\mc{C}_k$ which is a complete set of coset leaders of the lattice partition $\Lambda/2^{m_k}\Lambda$ with cardinality $2^{nm_k}$. Note that we assume that each user has the same packet size as users with the smaller packet size can use zero padding. The overall transmission rate in bits per real dimension is
\begin{equation}\label{eq:code_rate}
R_k = \frac{M_k}{N_k}m_k= \frac{M_k}{Nn}.
\end{equation}
The transmitted signal is then given by
\begin{align}\label{eqn:x_def}
\mathbf{x} &= \beta\left(\left[\mathbf{v}_1+\sum_{k=2}^{K}2^{\sum_{i=1}^{k-1}m_i}\mathbf{v}_k-\mathbf{d} \right]_{\Lambda_{s}}\right) \nonumber \\
&\in \beta\left(\left[\mc{C}_1+\sum_{k=2}^{K}2^{\sum_{i=1}^{k-1}m_i}\mc{C}_k-\mathbf{d} \right]_{\Lambda_{s}}\right)=\beta\mc{C},
\end{align}
where $\mathbf{d}\in\mc{V}(\Lambda_s)$ is a deterministic dither known at both transmitter and the receiver and it is to ensure that the overall constellation $\mc{C}$ is zero-mean and has the minimum transmit power; and $\beta$ is a normalize factor to ensure $\mbb{E}[\|\mathbf{x}\|^2]\leq n$.

\subsection{Decoding}
Consider the decoding procedure at user $k$, $k \in \{1,\ldots,K\}$ with its received signal given in \eqref{eq:sm1}. The decoder attempts to decode $\mathbf{u}_k$ from the received signal $\mathbf{y}_k$ via a decoding function $\mathcal{D}_k(\cdot)$ by treating other users' signals as noise, i.e., a single-user decoder. An outage occurs when the decoded signal $\mathcal{D}_k(\mathbf{y}_k) \neq \mathbf{u}_k$. We note that the received signal can also be decoded via a SIC decoder which first decodes users $K,\ldots,k+1$'s codewords before decoding its own codeword. Both single-user decoding and SIC decoding will be included in simulations for comparison. However, the analysis in what follows focuses solely on single-user decoding as it is one of the main motivation of this work.

An example of the proposed scheme is presented as follows.
\begin{exam}\label{ex}
We provide an illustrative example for our proposed coding scheme. Consider a two-user downlink NOMA where the average SNRs are $(\overline{\SNR}_1,\overline{\SNR}_2) = (30,18)$ dB, which result in $(\bar{n}_1,\bar{n}_2) = (5,3)$. Assume that the intended rates for users 1 and 2 are $(m_1,m_2) = (4,1)$, satisfying \eqref{eq:m1n1} and \eqref{eq:m2n2}. When the base lattice is a two-dimensional $\mathbb{Z}^2$, the modulations $\mc{C}_1$ and $\mc{C}_2$ are $\mathbb{Z}^2/16\mathbb{Z}^2$ and $\mathbb{Z}^2/2\mathbb{Z}^2$, which correspond to 256-QAM and 4-QAM, respectively.
\end{exam}

\begin{remark}
It is noteworthy that the overall constellation $\mc{C}$ corresponds to a complete set of coset leaders of the coarse lattice $2^{\sum_{i=1}^K m_i}\Lambda$ because $[\Lambda/2^{m_1}\Lambda+\ldots + 2^{n\sum_{i=1}^{k-1} m_i}(\Lambda/2^{m_K}\Lambda)]_{\Lambda_s} = \Lambda/2^{\sum_{i=1}^K m_i}\Lambda$. In addition, the proposed scheme naturally induces power allocation factors from the lattice partition chain which is determined based on statistical CSI. Similar to \cite{8291591}, the power allocation induced by our proposed scheme ensures that the combined constellation still preserves the structure of lattice $\Lambda$. In this way, our scheme can exploit the lattice structure to harness inter-user interference.
\end{remark}

\section{Analysis of the Outage Rates and Their Gaps to Multiuser Outage Capacity}\label{sec:rate_analysis}
Guided by the linear deterministic model, we first analyze the individual achievable rate of the proposed scheme without SIC for a fading channel realization. The lower bound on the individual outage rate for a given outage probability, is then obtained based on the individual achievable rate. Finally, the gaps between our outage rates and the multiuser outage capacity are investigated. Throughout the chapter, unless otherwise specified, we use the term ``outage rate" to denote the outage rate achieved by the proposed scheme for brevity.

We now present the main results of this work as follows.
\begin{prop}\label{main_result_rate}
In the $K$ user downlink NOMA over slow fading channel, given the statistical CSI at the transmitter, the outage rate tuple of our scheme without SIC $(R_1,\ldots,R_K)$ is lower bounded by
 \begin{align}
R_1  &>   \min\left\{m_1,m_1+\frac{1}{2}\log_2\left(F(\epsilon_1)\right)\right\}-1-\Psi,\label{eq:main_rate_1} \\
R_k &> \min\left\{m_k,\frac{1}{2}\log_2\left(F(\epsilon_k)\overline{\SNR}_k\right)- \sum_{i=k+1}^{K}m_i\right\}-\Psi, \nonumber \\
& \text{for} \; 1<k<K, \label{eq:main_rate_k} \\
R_K &> \min\left\{m_K,\frac{1}{2}\log_2\left(F(\epsilon_K)\overline{\SNR}_K\right)\right\}-\Psi,\label{eq:main_rate_2}
\end{align}
where $\Psi\defeq \frac{1}{2}\log_2 2\pi e \left(18 \psi(\Lambda)\right)$ and it is determined by the NSM of the base lattice.
\end{prop}
The proof is given in Section \ref{sec:mu}. One can see that the lower bound of the outage rate is a function of the required outage probability. In practice, the acceptable outage probabilities are typically not very large. Hence, we restrict our discussion to the outage probabilities smaller than $63.21\%$, which cover almost all the cases of practical interest. As shown in Appendix \ref{apx:3} and Lemma \ref{lma:pout}, by choosing this number, the multiuser outage capacity is bounded away from $+\infty$ for a given SNR and a required outage probability, which is useful when we analyze the capacity gap in Section \ref{sec:gap}. Under this assumption, we can show that for any rate tuple lying inside the outage capacity region given in Theorem~\ref{the:1}, there is an instance of our proposed scheme that can achieve that rate tuple to within a constant gap. In particular, we have the following proposition whose proof is in Section \ref{sec:gap}.

\begin{prop}\label{main_result}
Given a required outage probability satisfying $\epsilon_k < 0.6321$ for $k = 1,\ldots,K$, for any $(C_1,\ldots,C_K)$ lying on the boundary of the NOMA outage capacity region, one can always find an outage rate tuple $(R_1,\ldots,R_K)$ achieved by our scheme such that $C_k-R_k<\Delta_k$ where
 \begin{align}
\Delta_1  &= 2.5+\Psi,\label{eq:main_1} \\
\Delta_k &= 1.5 - \frac{1}{2}\log_2(F(\epsilon_k))+\Psi ,\; \text{for} \; 1<k<K \label{eq:main_k} \\
\Delta_K &= 1+\Psi.\label{eq:main_2}
\end{align}
\end{prop}

\begin{remark}\label{333}
First, the upper bounds shown above are universal for all $\overline{\SNR}_k$ and does not scale with $K$. Interestingly, one observes that the bounds for users 1 and $K$ are also universal for all outage probabilities. In addition, similar to the results in \cite{Shieh16,8291591}, one can also see that the upper bound is also a function proportional to the logarithm of the NSM of the base lattice, indicating that smaller gaps can be obtained by using lattices with better shaping. For example,
for a 3-user case with $(\epsilon_1,\epsilon_2,\epsilon_3) = (0.05,0.05,0.05)$, $(\Delta_1,\Delta_2,\Delta_3)$ are $(4.8396,5.9821,3.3396)$ and $(4.5850,5.7275,3.0850)$ by using $\mathbb{Z}$ lattice and the optimal lattice whose NSM are $1/12$ and $1/2\pi e$, respectively.
Last but not least, we stress that as evident in the proofs in what follows, many loose bounds based on the worst case scenario are used for making the results universal.
Our simulation results in Section \ref{SIM} will show that the gaps are usually much smaller than the derived upper bound.
\end{remark}

\subsection{Analysis of the Individual Outage Rate}\label{sec:mu}
For user 1, we follow the definition in (\ref{eq:r11}) and (\ref{eq:r12}) to decompose $\mc{C}_1$ into $\mc{C}_{11,1}$ and $\mc{C}_{12,1}$ as opposed to $r_{11,1}$ and $r_{12,1}$. Let us consider the lattice partition chain $\Lambda/2^{r_{12,1}}\Lambda/2^{m_1}\Lambda$ where $\mc{C}_{11,1}$ and $\mc{C}_{12,1}$ are isomorphic to $\Lambda/2^{r_{11,1}}\Lambda$ and $\Lambda/2^{r_{12,1}}\Lambda$, respectively. Thus, $\mc{C}_1$ can then be represented as
\begin{equation}\label{eq:C1de}
    \mc{C}_1 = \left[\mc{C}_{12,1} + 2^{r_{12,1}}\mc{C}_{11,1} \right]_{2^{m_1}\Lambda}.
\end{equation}

Similarly, for user $1<k\leq K$, the super-user's constellation $\mc{C}_{1 \rightarrow k-1}$ and its own constellation $\mc{C}_k$ can be decomposed according to \eqref{eq:r13}-\eqref{eq:r16} and \eqref{eq:r13k}-\eqref{eq:r16k}.
\begin{align}
    \mc{C}_{1 \rightarrow k-1} &= [\mc{C}^*_{12,k} + 2^{r^*_{12,k}}\mc{C}^*_{11,k}]_{2^{m_{1 \rightarrow k-1}}\Lambda}.\label{eq:C1de1} \\
    \mc{C}_k &= \left[\mc{C}_{k2,k} + 2^{r_{k2,k}}\mc{C}_{k1,k}\right]_{2^{m_k}\Lambda}.\label{eq:C2de}
\end{align}

Let $\msf{V}_1, \msf{V}_2, \ldots, \msf{V}_K$ be random variables uniformly over $\mc{C}_1, \mc{C}_2, \ldots, \mc{C}_K$, respectively, and let $\msf{X}=\beta\left(\left[\msf{V}_1+\sum_{k=2}^{K}2^{\sum_{i=1}^{k-1}m_i}\msf{V}_k-\mathbf{d}\right]_{\Lambda_{s}}\right)$ be the input random variable. Following the relationship given in \eqref{eq:sm1}, we define $\msf{Y}_1,\msf{Y}_2, \ldots,\msf{Y}_K$ to be the random variables corresponding to the received signal at users $1, 2,\ldots,K$, respectively.

\emph{1): }
For user 1, we can write $\msf{V}_1=\left[\msf{V}_{12,1}+ 2^{r_{12,1}}\msf{V}_{11,1}\right]_{2^{m_1}\Lambda}$ according to \eqref{eq:C1de}. As user 1 treats users $2,\ldots,K$ as a super-user, we can define $\msf{V}_{2\rightarrow K} \triangleq \sum_{k=2}^{K}2^{\sum_{i=1}^{k-1}m_i}\msf{V}_k$ as the random variable associated with the super-user. To analyze the achievable rate of our scheme without SIC for an instant channel realization $h_1$, we bound the mutual information as follows.
\begin{align}\label{eqn:rate_1_GBC}
    &I(\msf{V}_1;\msf{Y}_1)= h(\msf{Y}_1) - h(\msf{Y}_1|\msf{V}_1) \nonumber \\
    &=h(\msf{Y}_1) - h(\msf{Y}_1|[2^{r_{12,1}}\msf{V}_{11,1} +2^{m_1}\msf{V}_{2\rightarrow K} -\mathbf{d}_2 ]_{\Lambda_s}) \nonumber \\
    &- [h(\msf{Y}_1\big|\msf{V}_1)- h(\msf{Y}_1|[2^{r_{12,1}}\msf{V}_{11,1}+2^{m_1}\msf{V}_{2\rightarrow K} -\mathbf{d}_2 ]_{\Lambda_s})] \nonumber \\
    &\overset{(\ref{eqn:rate_1_GBC}.a)}= h(\msf{Y}_1) - h(\msf{Y}_1\big|[2^{r_{12,1}}\msf{V}_{11,1} +2^{m_1}\msf{V}_{2\rightarrow K} -\mathbf{d}_2]_{\Lambda_s}) \nonumber \\
    &- [h(\msf{Y}_1|\msf{V}_1)- h(\msf{Y}_1|\msf{V}_{11,1},\msf{V}_{2\rightarrow K}) ] \nonumber \\
    &= I([2^{r_{12,1}}\msf{V}_{11,1}+2^{m_1}\msf{V}_{2\rightarrow K} -\mathbf{d}_2 ]_{\Lambda_s};\msf{Y}_1)-h(\msf{Y}_1|\msf{V}_1) \nonumber \\
    &+h(\msf{Y}_1|\msf{V}_1,\msf{V}_{2\rightarrow K})+[h(\msf{Y}_1|\msf{V}_{11,1},\msf{V}_{2\rightarrow K}) 
    -h(\msf{Y}_1|\msf{V}_1,\msf{V}_{2\rightarrow K})] \nonumber \\
    &=I([2^{r_{12,1}}\msf{V}_{11,1}+2^{m_1}\msf{V}_{2\rightarrow K} -\mathbf{d}_2 ]_{\Lambda_s};\msf{Y}_1) \nonumber \\
    &-I(\msf{V}_{2\rightarrow K};\msf{Y}_1|\msf{V}_1)+I(\msf{V}_{12,1};\msf{Y}_1|\msf{V}_{11,1},\msf{V}_{2\rightarrow K}) \nonumber \\
    &\geq I([2^{r_{12,1}}\msf{V}_{11,1}+2^{m_1}\msf{V}_{2\rightarrow K} -\mathbf{d}_2 ]_{\Lambda_s};\msf{Y}_1) 
    - I(\msf{V}_{2\rightarrow K};\msf{Y}_1|\msf{V}_1) \nonumber \\
    &\geq I([2^{r_{12,1}}\msf{V}_{11,1}+2^{m_1}\msf{V}_{2\rightarrow K} -\mathbf{d}_2 ]_{\Lambda_s};\msf{Y}_1) 
    - H(\msf{V}_{2\rightarrow K}|\msf{V}_1) \nonumber \\
    &\overset{(\ref{eqn:rate_1_GBC}.b)} = I([2^{r_{12,1}}\msf{V}_{11,1}+2^{m_1}\msf{V}_{2\rightarrow K} -\mathbf{d}_2 ]_{\Lambda_s};\msf{Y}_1) 
    - H(\msf{V}_{2\rightarrow K}),
\end{align}
where $2^{r_{12,1}}\msf{V}_{11,1}+2^{m_1}\msf{V}_{2\rightarrow K-1}$ is indeed the parts above the noise level for user 1 according to the deterministic model; $\mathbf{d}_2$ is a fixed dither decomposed from $\mathbf{d}$; $(\ref{eqn:rate_1_GBC}.a)$ is due to the bijective mapping between $[2^{r_{12,1}}\msf{V}_{11,1}+2^{m_1}\msf{V}_{2\rightarrow K}  -\mathbf{d}_2 ]_{\Lambda_s}$ and $(\msf{V}_{11,1},\msf{V}_{2\rightarrow K})$, and $(\ref{eqn:rate_1_GBC}.b)$ follows the independence between each $\msf{V}_k$ for $k \in \{1,\ldots,K\}$. To further bound $I([2^{r_{12,1}}\msf{V}_{11,1}+2^{m_1}\msf{V}_{2\rightarrow K} -\mathbf{d}_2 ]_{\Lambda_s}; \msf{Y}_1)$, we note that the effective noise can be written as $\msf{Z}'_1 =  h_1\sqrt{\overline{\SNR}_1}\beta[\msf{V}_{12,1}-\mathbf{d}_1]_{2^{r_{12,1}}\Lambda}+\msf{Z}_1$, where $\mathbf{d}_1$ is a fixed dither to minimize energy of constellation $\mc{C}_{12,1}$. Note that we can always find such a $(\mathbf{d}_1,\mathbf{d}_2)$ pair by fixing $\mathbf{d}_1$ that minimizes the energy of $\mc{C}_{12,1}$ and let $\mathbf{d}_2=\mathbf{d}-\mathbf{d}_1$. We then scale the effective noise by
\begin{equation}
    \gamma_1=\sqrt{\frac{n}{|h_1|^2\overline{\SNR}_1\beta^2\mbb{E}[\|[\msf{V}_{12,1}-\mathbf{d}_1]_{2^{r_{12,1}}\Lambda}\|^2]+n}}.
\end{equation}
In this way, the scaled effective noise has power $\mbb{E}[\|\msf{Z}'_1\|^2]=n$.
The equivalent communication channel then becomes $\msf{Y}'_1 = \msf{X}'_1 + \msf{Z}'_1$ where
\begin{equation}
    \msf{X}'_1 = \gamma_1 h_1\sqrt{\overline{\SNR}_1} \beta \left[2^{r_{12,1}}\msf{V}_{11,1}+2^{m_1}\msf{V}_{2\rightarrow K} -\mathbf{d}_2 \right]_{\Lambda_s},
\end{equation}
and $\msf{Y}'_1 = \gamma_1\msf{Y}_1$. We then apply the established lower bound of the mutual information between a discrete random input and its noisy version shown in \cite[Lemma 6]{8291591} to obtain the following
\begin{align}\label{eq:sc155}
    &I(\left[2^{r_{12,1}}\msf{V}_{11,1}+2^{m_1}\msf{V}_{2\rightarrow K} -\mathbf{d}_2 \right]_{\Lambda_s};\msf{Y}_1)  - H(\msf{V}_{2\rightarrow K})\nonumber \\
    &\geq n r_{11,1}- \frac{n}{2}\log_2 2\pi e \left(\text{Vol}(\Lambda_1)^{-\frac{2}{n}} + \psi(\Lambda_1)\right),
\end{align}
where $\Lambda_1\defeq \gamma_1 h_1\sqrt{\overline{\SNR}_1}\beta 2^{r_{12,1}}\Lambda$.

We then establish the lower bound for the scaling factor of $\Lambda_1$ in \eqref{eq:sc2} and for $\text{Vol}(\Lambda_1)^{-\frac{2}{n}} + \psi(\Lambda_1)$ in \eqref{eq:rate_1_GBC_final}-\eqref{eq:u1_inst_rate} in Appendix~\ref{sec:u1proof}. By plugging these bounds into \eqref{eq:sc155}, we obtain the lower bound for user 1's achievable rate in bits per real dimension for a given channel realization as
\begin{align}
\frac{1}{n}I(\msf{V}_1;\msf{Y}_1) &> \min\left\{m_1,m_1 +\frac{1}{2}\log_2\left(|\hat{h}_1|^2\right)-1\right\} 
- \Psi,
\end{align}
where $\min\{\cdot\}$ here follows the constraint in \eqref{eq:r11}.

As our scheme does not invoke SIC at each receiver, the outage probability of user $k$ is calculated as $\mathbb{P}\{\frac{1}{n}I(\msf{V}_k;\msf{Y}_k)<R_k \} = \epsilon_k$, where $R_k$ is user $k$'s target transmission rate.
Given user 1's rate $R_1$ and the required outage probability $\epsilon_1$, we have
\begin{align}
 \epsilon_1 &=\mathbb{P}\left\{\frac{1}{n}I(\msf{V}_1;\msf{Y}_1) < R_1 \right\} \nonumber \\
&< \mathbb{P}\left\{m_1 +\frac{1}{2}\log_2\left(|\hat{h}_1|^2\right)-1  - \Psi < R_1 \right\} \nonumber \\
& = \mathbb{P}\left\{|\hat{h}_1|^2 < 2^{2(R_1+1-m_1+\Psi)} \right\} \nonumber \\
 &= 1 - \text{exp}\left(-2^{2(R_1+1-m_1+\Psi)} \right). 
\end{align}
The lower bound for user 1's the outage rate is
\begin{align}\label{eq:R1}
R_1 >  \min\left\{m_1,m_1+\frac{1}{2}\log_2(-\ln(1-\epsilon_1))\right\}-1-\Psi,
\end{align}
where $\min\{\cdot \}$ here follows \eqref{eq:r11}. This completes the proof for \eqref{eq:main_rate_1}.

\emph{2): }
For user $1<k < K$, we define $\msf{V}_{1\rightarrow k-1} \triangleq \msf{V}_1+ \sum_{j=2}^{k-1}2^{\sum_{i=1}^{j-1}m_i}\msf{V}_j$ by treating users $1,2,\ldots,k-1$ as a super-user and $\msf{V}_{k+1\rightarrow K} \triangleq \sum_{j=k+1}^{K}2^{\sum_{i=1}^{j-1}m_i}\msf{V}_j$ by treating users $k+1,k+2,\ldots,K$ as another super-user. Then, we decompose $\msf{V}_{1\rightarrow k-1} = [\msf{V}_{12,k}^*+2^{r_{12,k}^*}\msf{V}_{11,k}^*]_{2^{m_{1\rightarrow k-1}}\Lambda}$ and $\msf{V}_k = [\msf{V}_{k2,k} + 2^{r_{k2,k}}\msf{V}_{k1,k}]_{2^{m_k}\Lambda}$ according to \eqref{eq:C1de1} and \eqref{eq:C2de}, respectively.

We bound the mutual information for user $k$ for a channel realization as follows,
\begin{align}\label{eqn:rate_k_GBC}
        &I(\msf{V}_k;\msf{Y}_k) = h(\msf{Y}_k) - h(\msf{Y}_k|\msf{V}_k) \nonumber \\
    &= \left[h(\msf{Y}_k) - h\left(\msf{Y}_k \left|  \left[  2^{r_{12,k}^*}\msf{V}^*_{11,k}  +  2^{m_{1 \rightarrow k-1}}2^{r_{k2,k}}\msf{V}_{k1,k}+\msf{V}_{k+1\rightarrow K} - \mathbf{d}_3 \right]_{\Lambda_{s}} \right.\right)\right] \nonumber \\
    &- \left[h(\msf{Y}_k|\msf{V}_k)-h\left(\msf{Y}_k \left|  \left[  2^{r_{12,k}^*}\msf{V}^*_{11,k}  +  2^{m_{1 \rightarrow k-1}}2^{r_{k2,k}}\msf{V}_{k1,k}+\msf{V}_{k+1\rightarrow K} - \mathbf{d}_3 \right]_{\Lambda_{s}} \right.\right)\right] \nonumber \\
    &\overset{(\ref{eqn:rate_k_GBC}.a)}{=} I\left(\left[  2^{r_{12,k}^*}\msf{V}^*_{11,k}  +  2^{m_{1 \rightarrow k-1}}2^{r_{k2,k}}\msf{V}_{k1,k}+\msf{V}_{k+1\rightarrow K} - \mathbf{d}_3 \right]_{\Lambda_{s}} ; \msf{Y}_k \right) \nonumber \\
    &- [h(\msf{Y}_k|\msf{V}_k)-h(\msf{Y}_k | \msf{V}_{11,k}^*,\msf{V}_{k1,k}, \msf{V}_{k+1},\ldots,\msf{V}_K )] \nonumber \\
    & \overset{(\ref{eqn:rate_k_GBC}.b)}{\geq} I\left(\left[  2^{r_{12,k}^*}\msf{V}^*_{11,k}  +  2^{m_{1 \rightarrow k-1}+r_{k2,k}}\msf{V}_{k1,k}+\msf{V}_{k+1\rightarrow K}- \mathbf{d}_3 \right]_{\Lambda_{s}} ; \msf{Y}_k \right) \nonumber \\
    &- [h(\msf{Y}_k|\msf{V}_k)-h(\msf{Y}_k | \msf{V}_{11,k}^*,\msf{V}_k, \msf{V}_{k+1},\ldots,\msf{V}_K )]  \nonumber \\
    & = I\left(\left[  2^{r_{12,k}^*}\msf{V}^*_{11,k}  +  2^{m_{1 \rightarrow k-1}+r_{k2,k}}\msf{V}_{k1,k}+\msf{V}_{k+1\rightarrow K}- \mathbf{d}_3 \right]_{\Lambda_{s}} ; \msf{Y}_k \right) \nonumber \\
    &- I(\msf{V}_{11,k}^*, \msf{V}_{k+1},\ldots,\msf{V}_K;\msf{Y}_k | \msf{V}_k) \nonumber \\
    &\geq I\left(\left[  2^{r_{12,k}^*}\msf{V}^*_{11,k}  +  2^{m_{1 \rightarrow k-1}+r_{k2,k}}\msf{V}_{k1,k}+\msf{V}_{k+1\rightarrow K} - \mathbf{d}_3\right]_{\Lambda_{s}} ; \msf{Y}_k \right) \nonumber \\
    &- H(\msf{V}_{11,k}^*, \msf{V}_{k+1},\ldots,\msf{V}_K | \msf{V}_k) \nonumber \\
    &=I\left(\left[  2^{r_{12,k}^*}\msf{V}^*_{11,k}  +  2^{m_{1 \rightarrow k-1}+r_{k2,k}}\msf{V}_{k1,k}+\msf{V}_{k+1\rightarrow K} - \mathbf{d}_3 \right]_{\Lambda_{s}} ; \msf{Y}_k \right) \nonumber \\
    &- H(\msf{V}_{11,k}^*, \msf{V}_{k+1},\ldots,\msf{V}_K),
\end{align}
where $\mathbf{d}_3$ is a fixed dither decomposed from $\mathbf{d}$, $(\ref{eqn:rate_k_GBC}.a)$ is due to a bijective mapping between the lattice $[  2^{r_{12,k}^*}\msf{V}^*_{11,k}  +  2^{m_{1 \rightarrow k-1}}2^{r_{k2,k}}\msf{V}_{k1,k}+\msf{V}_{k+1\rightarrow K} - \mathbf{d}_3 ]_{\Lambda_{s}}$ and the term $( \msf{V}_{11,k}^*,\msf{V}_{k1,k}, \msf{V}_{k+1},\ldots,\msf{V}_K )$, and $(\ref{eqn:rate_k_GBC}.b)$ follows
\begin{align}\label{eq:inequality_1}
&h(\msf{Y}_k | \msf{V}_{11,k}^*,\msf{V}_{k1,k}, \msf{V}_{k+1},\ldots,\msf{V}_K ) 
- h(\msf{Y}_k | \msf{V}_{11,k}^*,\msf{V}_k, \msf{V}_{k+1},\ldots,\msf{V}_K )  \nonumber \\
&= h(\msf{Y}_k | \msf{V}_{11,k}^*,\msf{V}_{k1,k}, \msf{V}_{k+1},\ldots,\msf{V}_K ) 
- h(\msf{Y}_k | \msf{V}_{11,k}^*,\msf{V}_{k1,k},\msf{V}_{k2,k}, \msf{V}_{k+1},\ldots,\msf{V}_K ) \nonumber \\
& = I(\msf{V}_{k2,k};\msf{Y}_k| \msf{V}_{11,k}^*, \msf{V}_{k1,k}, \msf{V}_{k+1},\ldots,\msf{V}_K)
 \geq 0.
\end{align}
To further bound \eqref{eqn:rate_k_GBC}, we note that effective noise is $\msf{Z}'_k = \hat{h}_k\sqrt{\overline{\SNR}_k}\beta[\msf{V}^*_{12,k}+2^{m_{1 \rightarrow k-1}}\msf{V}_{k2,k}-\mathbf{d}_4]_{\Lambda_s}+\msf{Z}_k$, where $\mathbf{d}_4$ is a fixed dither decomposed from $\mathbf{d}$ and to minimize the energy of constellation $\mc{C}^*_{12,k}+2^{m_{1 \rightarrow k-1}}\mc{C}_{k2,k}$. We thus scale the effective noise by
\begin{align}
    &\gamma_k= 
    \sqrt{\frac{n}{|\hat{h}_k|^2\overline{\SNR}_k\beta^2\mbb{E}[\|[\msf{V}^*_{12,k}+2^{m_{1 \rightarrow k-1}}\msf{V}_{k2,k}-\mathbf{d}_4]_{\Lambda_s}\|^2]+n}},
\end{align}
such that $\mbb{E}[\|\msf{Z}'_k\|^2]=n$. In this way, we can similarly apply the lower bound of the mutual information between a discrete random input and its noisy version shown in \cite[Lemma 6]{8291591} to obtain
\begin{align}\label{eq:krate}
&I([  2^{r_{12,k}^*}\msf{V}^*_{11,k}  +  2^{m_{1 \rightarrow k-1}+r_{k2,k}}\msf{V}_{k1,k}+\msf{V}_{k+1\rightarrow K} 
-\mathbf{d}_4 ]_{\Lambda_{s}} ; \msf{Y}_k )- H(\msf{V}_{11,k}^*, \msf{V}_{k+1},\ldots,\msf{V}_K) \nonumber \\
&\geq n r_{k1,k}- \frac{n}{2}\log_2 2\pi e \left(\text{Vol}(\Lambda_k)^{-\frac{2}{n}} + \psi(\Lambda_k)\right),
\end{align}
where $\Lambda_k\defeq \gamma_k \hat{h}_k \sqrt{\overline{\SNR}_k}\beta \Gamma(r_{k2,k})\Lambda$, and
\begin{equation}\label{eq:special_con_k}
\Gamma(r_{k2,k}) = \begin{cases}
&2^{r^*_{12,k}}, \;\text{when} \; r_{k2,k} = 0\\
&2^{m_{1 \rightarrow k-1}+r_{k2,k}}, \; \text{when} \; r_{k2,k} \neq 0 \\
\end{cases},
\end{equation}
is the scaling factor for the minimum distance of the constellation $[  2^{r_{12,k}^*}\mc{C}^*_{11,k}  +  2^{m_{1 \rightarrow k-1}+r_{k2,k}}\mc{C}_{k1,k}+\mc{C}_{k+1\rightarrow K} ]_{\Lambda_{s}}$. The effects of changing of $r_{k2,k}$ on the constellation are illustrated in facts ix) and x) given in Section \ref{sec:dmodel_k}. We then follow the similar step as in user 1's case to obtain the lower bound on user $k$'s outage rate
\begin{align}\label{eq:R3}
&R_k > 
 \min\left\{m_k,\frac{1}{2}\log_2\left(-\ln(1-\epsilon_k)\overline{\SNR}_k\right) - m_{k+1 \rightarrow K}\right\}-\Psi,
\end{align}
where $\min\{\cdot\}$ here follows from \eqref{eq:r15} and the detail derivation is in Appendix \ref{Kuser}. This completes the proof for \eqref{eq:main_rate_k}.

\emph{3):} For user $K$, we define $\msf{V}_{1\rightarrow K-1} \triangleq \msf{V}_1+ \sum_{j=2}^{K-1}2^{\sum_{i=1}^{j-1}m_i}\msf{V}_j$ by treating users $1,2,\ldots,K-1$ as a super-user. We then similarly decompose $\msf{V}_{1\rightarrow K-1}=[\msf{V}^*_{12,K}+ 2^{r^*_{12,K}}\msf{V}^*_{11,K}]_{2^{m_{1\rightarrow K-1}}\Lambda}$ and $\msf{V}_K=[\msf{V}_{K2,K}+ 2^{r_{K2,K}}\msf{V}_{K1,K}]_{2^{m_K}\Lambda}$ according to \eqref{eq:C1de1} and \eqref{eq:C2de}, respectively. The mutual information for user $K$ is bounded by:
\begin{align}\label{eqn:rate_2_GBC}
    &I(\msf{V}_K;\msf{Y}_K)= h(\msf{Y}_K) - h(\msf{Y}_K|\msf{V}_K) \nonumber \\
    &=[h(\msf{Y}_K) - h(\msf{Y}_K|[2^{r^*_{12,K}}\msf{V}^*_{11,K} +2^{m_{1\rightarrow K-1}}2^{r_{K2,K}}\msf{V}_{K1,K} - \mathbf{d}_5]_{\Lambda_s})] \nonumber \\
    &- [h(\msf{Y}_K|\msf{V}_K)-  h(\msf{Y}_K|[2^{r^*_{12,K}}\msf{V}^*_{11,K} +2^{m_{1\rightarrow K-1}}2^{r_{K2,K}}\msf{V}_{K1,K} - \mathbf{d}_5]_{\Lambda_s})] \nonumber \\
    &\overset{(\ref{eqn:rate_2_GBC}.a)}= [h(\msf{Y}_K) - h(\msf{Y}_K|[2^{r^*_{12,K}}\msf{V}^*_{11,K} +2^{m_{1\rightarrow K-1}}2^{r_{K2,K}}\msf{V}_{K1,K} - \mathbf{d}_5]_{\Lambda_s})] \nonumber \\
    &- [h(\msf{Y}_K|\msf{V}_K)-  h(\msf{Y}_K|\msf{V}^*_{11,K},\msf{V}_{K1,K})] \nonumber \\
    & \overset{(\ref{eqn:rate_2_GBC}.b)}{\geq} I([2^{r^*_{12,K}}\msf{V}^*_{11,K} +2^{m_{1\rightarrow K-1}+r_{K2,K}}\msf{V}_{K1,K} - \mathbf{d}_5]_{\Lambda_s};\msf{Y}_K) \nonumber \\
     &- [h(\msf{Y}_K|\msf{V}_K)-  h(\msf{Y}_K|\msf{V}^*_{11,K},\msf{V}_K)] \nonumber \\
    &= I([2^{r^*_{12,K}}\msf{V}^*_{11,K} +2^{m_{1\rightarrow K-1}+r_{K2,K}}\msf{V}_{K1,K} - \mathbf{d}_5]_{\Lambda_s};\msf{Y}_K) 
     - I(\msf{V}^*_{11,K};\msf{Y}_K|\msf{V}_K) \nonumber \\
    &\geq I([2^{r^*_{12,K}}\msf{V}^*_{11,K} +2^{m_{1\rightarrow K-1}+r_{K2,K}}\msf{V}_{K1,K} - \mathbf{d}_5]_{\Lambda_s};\msf{Y}_K) - H(\msf{V}^*_{11,K}|\msf{V}_K) \nonumber \\
    & = I([2^{r^*_{12,K}}\msf{V}^*_{11,K} +2^{m_{1\rightarrow K-1}+r_{K2,K}}\msf{V}_{K1,K} - \mathbf{d}_5]_{\Lambda_s};\msf{Y}_K) - H(\msf{V}^*_{11,K})
\end{align}
where $\mathbf{d}_5$ is the fixed dither decomposed from $\mathbf{d}$, $(\ref{eqn:rate_2_GBC}.a)$ follows from the existence of a bijective mapping between $[2^{r^*_{12,K}}\msf{V}^*_{11,K} +2^{m_{1\rightarrow K-1}}2^{r_{K2,K}}\msf{V}_{K1,K} - \mathbf{d}_5]_{\Lambda_s}$ and $(\msf{V}^*_{11,K},\msf{V}_{K1,K})$, and $(\ref{eqn:rate_2_GBC}.b)$ follows from
\begin{align}\label{eq:f}
&h(\msf{Y}_K|\msf{V}^*_{11,K},\msf{V}_{K1,K}) - h(\msf{Y}_K|\msf{V}^*_{11,K},\msf{V}_K)  \nonumber \\
&= h(\msf{Y}_K|\msf{V}^*_{11,K},\msf{V}_{K1,K}) - h(\msf{Y}_K|\msf{V}^*_{11,K},\msf{V}_{K1,K},\msf{V}_{K2,K}) \nonumber \\
& = I(\msf{V}_{K2,K};\msf{Y}_K|\msf{V}^*_{11,K},\msf{V}_{K1,K}) 
 \geq 0.
\end{align}
We follow the steps as in user $1$ and $k$'s cases to further bound \eqref{eqn:rate_2_GBC} and leave the detail process in Appendix \ref{sec:u2proof}. The lower bound of user $K$'s outage rate is
\begin{align}\label{eq:R2}
R_K > \min\left\{m_K,\frac{1}{2}\log_2\left(-\ln(1-\epsilon_K)\overline{\SNR}_K\right)\right\}-\Psi ,
\end{align}
where $\min\{\cdot\}$ here follows \eqref{eq:r15}. This completes the proof for \eqref{eq:main_rate_2}.

\subsection{Outage Capacity Gap Analysis}\label{sec:gap}
In this subsection, we investigate the gap between the outage rate of our scheme and the NOMA outage capacity region. We assume that $\alpha_k \neq 0$ for $k \in \{1,\ldots,K\}$. Otherwise, the problem can be reduced to that with less users. 

\emph{1): }
When $\epsilon_1 < 0.6321$, the output of $\min\{\cdot\}$ function in \eqref{eq:R1} is smaller than $m_1$. That is:
\begin{align}\label{eq:constraint1}
m_1 >m_1+\frac{1}{2}\log_2(-\ln(1-\epsilon_1))-1.
\end{align}
The gap of user 1's outage rate to multiuser outage capacity is upper bounded by
\begin{align}\label{eq:gap_u1_c2}
C_1-R_1
&\overset{(\ref{eq:gap_u1_c2}.a)}< \frac{1}{2}\log_2\left(1+\overline{\SNR}_1F(\epsilon_1) \alpha_1 \right)  \nonumber \\
&- \left(m_1+\frac{1}{2}\log_2(-\ln(1-\epsilon_1))-1-\Psi\right) \nonumber \\
&\overset{(\ref{eq:gap_u1_c2}.b)}\leq  \frac{1}{2}\log_2(\overline{\SNR}_1F(\epsilon_1) \alpha_1 )+\frac{1}{2} \nonumber \\
&-\left(m_1+\frac{1}{2}\log_2(-\ln(1-\epsilon_1))-1-\Psi\right) \nonumber \\
&= \frac{3}{2} +\left(\frac{1}{2}\log_2(\overline{\SNR}_1\alpha_1) -m_1\right)+\Psi \nonumber \\
&< \frac{3}{2} +\left(\frac{1}{2}\log_2(1+\overline{\SNR}_1\alpha_1) -m_1\right)+\Psi \nonumber \\
&\overset{(\ref{eq:gap_u1_c2}.c)}< \frac{3}{2} +1 +\Psi 
= \Delta_1,
\end{align}
where $(\ref{eq:gap_u1_c2}.a)$ follows from \eqref{eq:constraint1}; $(\ref{eq:gap_u1_c2}.b)$ follows from
\begin{equation}\label{eq:special_condition_1}
\frac{1}{2}\log_2(1+x) \leq \begin{cases}
&\frac{1}{2}, \;\text{when} \; x < 1\\
&\frac{1}{2}\log_2(x)+\frac{1}{2}, \; \text{when} \; x\geq 1 \\
\end{cases},
\end{equation}
by letting $x = \overline{\SNR}_1F(\epsilon_1) \alpha_1$ (for $x = \overline{\SNR}_1F(\epsilon_1) \alpha_1<1$, the gap is at most $\frac{1}{2}$ bits, i.e., $C_1-R_1 \leq \frac{1}{2}$ because $C_1 \leq \frac{1}{2}$) and $(\ref{eq:gap_u1_c2}.c)$ follows from the fact that given a power allocation vector $(\alpha_1,\ldots,\alpha_K)$, one can always pick a rate tuple $(m_1,\ldots,m_K)$ satisfying \eqref{eq:dK2}-\eqref{eq:dK4}, resulting in at most 1 bit gap from the corresponding multiuser capacity $(\bar{C}_1,\ldots,\bar{C}_K)$ such that $|\bar{C}_k - m_k| \leq 1$ \cite{Avestimehr11}. This completes the proof for \eqref{eq:main_1}.

\emph{2): }
For user $1<k < K$, we let $x = \overline{\SNR}_kF(\epsilon_k)$. When $x<1$, we have $\frac{\overline{\SNR}_kF(\epsilon_k) \alpha_k}{\overline{\SNR}_kF(\epsilon_k) \sum_{i=1}^{k-1}\alpha_i+1}<1$. Thus, $C_k-R_k \leq \frac{1}{2}$ since $C_k \leq \frac{1}{2}$ according to \eqref{eq:special_condition_1}. For $x =\overline{\SNR}_kF(\epsilon_k)  \geq 1$, we need to consider two cases as the output of $\min\{\cdot\}$ function in \eqref{eq:R3} cannot be determined even with the constraint $\epsilon_k < 0.6321$.

When $\min\{\cdot\} = m_k$ in \eqref{eq:R3}, the gap of user $k$'s outage rate to multiuser outage capacity is upper bounded by
\begin{align}\label{eq:outgap_3}
 C_k-R_k 
&<  \bar{C}_k - m_k + \Psi  \nonumber \\
&< 1+\Psi,
\end{align}
where the first inequality is true for outage probabilities within our chosen range (see Lemma~\ref{lma:pout}) and the second inequality follows from (\ref{eq:gap_u1_c2}.b).

When $\min\{\cdot\} \neq m_k$ in \eqref{eq:R3}, user $k$'s gap is bounded by
\begin{align}\label{eq:gap_k_c2}
C_k-R_k&<  \frac{1}{2}\log_2 \left(1+\frac{F(\epsilon_k)\alpha_k\overline{\SNR}_k}{F(\epsilon_k)\sum_{i=1}^{k-1}\alpha_i\overline{\SNR}_k+1}\right) \nonumber \\
&+m_{k+1 \rightarrow K}-\frac{1}{2}\log_2(F(\epsilon_k)\overline{\SNR}_k)+\Psi \nonumber \\
&\overset{(\ref{eq:gap_k_c2}.a)}\leq\frac{1}{2}\log_2\left(\frac{F(\epsilon_k)\sum_{j=1}^{k}\alpha_j\overline{\SNR}_k+1}{(F(\epsilon_k)\sum_{i=1}^{k-1}\alpha_i\overline{\SNR}_k+1)F(\epsilon_k)\overline{\SNR}_k}\right) \nonumber \\ &+\frac{1}{2}\log_2\left(1+\frac{\sum_{l=k+1}^K\alpha_l\overline{\SNR}_{k+1}}{\sum_{j=1}^{k}\alpha_j\overline{\SNR}_{k+1}+1}\right) +1+\Psi \nonumber \\
&\overset{(\ref{eq:gap_k_c2}.b)}\leq\frac{1}{2}\log_2\left(\frac{F(\epsilon_k)\sum_{j=1}^{k}\alpha_j\overline{\SNR}_k+1}{(F(\epsilon_k)\sum_{i=1}^{k-1}\alpha_i\overline{\SNR}_k+1)F(\epsilon_k)\overline{\SNR}_k}\right) \nonumber \\ &+\frac{1}{2}\log_2\left(1+\frac{\sum_{l=k+1}^K\alpha_l\overline{\SNR}_{k}}{\sum_{j=1}^{k}\alpha_j\overline{\SNR}_{k}+1}\right) +1+\Psi \nonumber \\
&=\frac{1}{2}\log_2\Bigg(\frac{F(\epsilon_k)\sum_{j=1}^{k}\alpha_j\overline{\SNR}_k+1}{(F(\epsilon_k)\sum_{i=1}^{k-1}\alpha_i\overline{\SNR}_k+1)F(\epsilon_k)\overline{\SNR}_k} \nonumber\\ &\cdot\frac{\overline{\SNR}_k+1}{\sum_{j=1}^k\alpha_j\overline{\SNR}_k+1} \Bigg)  +1+\Psi \nonumber \\
&\overset{(\ref{eq:gap_k_c2}.c)}\leq\frac{1}{2}\log_2\left(\frac{\overline{\SNR}_k+1}{(F(\epsilon_k)\sum_{i=1}^{k-1}\alpha_i\overline{\SNR}_k+1)F(\epsilon_k)\overline{\SNR}_k} \right) 
+1+\Psi \nonumber \\
&<\frac{1}{2}\log_2\left(\frac{\overline{\SNR}_k+1}{F(\epsilon_k)\overline{\SNR}_k} \right)  +1+\Psi \nonumber \\
 &\overset{(\ref{eq:gap_k_c2}.d)} \leq \frac{1}{2}\log_2\left(\frac{2}{F(\epsilon_k)} \right)  +1+\Psi \nonumber \\
 &= \frac{3}{2} - \frac{1}{2}\log_2(F(\epsilon_k))+\Psi
 = \Delta_k,
\end{align}
where $(\ref{eq:gap_k_c2}.a)$ is due to treating users $k+1,\ldots,K$ as a super-user and then applying the gaps in $(\ref{eq:gap_u1_c2}.b)$; $(\ref{eq:gap_k_c2}.b)$ follows the fact that $\frac{\overline{\SNR}_{k+1}}{\sum_{j=1}^{k}\alpha_j\overline{\SNR}_{k+1}+1}$ is monotonically increasing in $\overline{\SNR}_{k+1} \in [1,\overline{\SNR}_k]$ because its first derivative is
\begin{align}
&\frac{\partial}{\partial \overline{\SNR}_{k+1}}\left(\frac{\overline{\SNR}_{k+1}+1}{\sum_{j=1}^{k}\alpha_j\overline{\SNR}_{k+1}+1}\right) \nonumber \\ =&\frac{1-\sum_{j=1}^{k}\alpha_j}{(\sum_{j=1}^{k}\alpha_j\overline{\SNR}_{k+1}+1)^2}
>0,
\end{align}
$(\ref{eq:gap_k_c2}.c)$ follows from that
\begin{align}
F(\epsilon_k)\sum_{j=1}^{k}\alpha_j\overline{\SNR}_k+1 \leq \sum_{j=1}^{k}\alpha_j\overline{\SNR}_k+1,
\end{align}
and $(\ref{eq:gap_k_c2}.d)$ follows from the fact that $\frac{\overline{\SNR}_k+1}{F(\epsilon_k)\overline{\SNR}_k}$ is monotonically decreasing for $\overline{\SNR}_k \geq 1$ because its first derivative is
\begin{align}
\frac{\partial}{\partial \overline{\SNR}_k}\left(\frac{\overline{\SNR}_k+1}{F(\epsilon_k)\overline{\SNR}_k}\right) =-\frac{1}{F(\epsilon_k)(\overline{\SNR}_k)^2}<0.
\end{align}
This completes the proof for \eqref{eq:main_k}.

\emph{3): }
For user $K$, let $x =\overline{\SNR}_KF(\epsilon_K)  \geq 1$ (since $x<1$ will result in gap at most $\frac{1}{2} $ bit), when $\min\{\cdot\} = m_K$ in \eqref{eq:R2}, the gap of user $K$'s outage rate to multiuser outage capacity is upper bounded by
\begin{align}\label{eq:outgap_2}
 C_K-R_K 
&<  \bar{C}_K - m_K + \Psi \nonumber \\
&< 1+\Psi \nonumber \\
&=\Delta_K,
\end{align}
where the first inequality is due to \eqref{eq:con21} in Lemma~\ref{lma:pout}.
For the other case, the gap is upper bounded by
\begin{align}\label{eq:gap_u2_2}
&C_K-R_K
< \frac{1}{2}\log_2\left(1+\frac{\overline{\SNR}_KF(\epsilon_K) \alpha_K}{\overline{\SNR}_KF(\epsilon_K) \sum_{i=1}^{K-1}\alpha_i+1} \right)  \nonumber \\
&- \left(\frac{1}{2}\log_2(\overline{\SNR}_K)+\frac{1}{2}\log_2(-\ln(1-\epsilon_K))-\Psi\right) \nonumber \\
 &= \frac{1}{2}\log_2\left(\frac{\overline{\SNR}_KF(\epsilon_K) +1}{(\overline{\SNR}_KF(\epsilon_K) \sum_{i=1}^{K-1}\alpha_i+1)\overline{\SNR}_KF(\epsilon_K)} \right)
+\Psi  \nonumber \\
& = \frac{1}{2}\log_2\left(\frac{x +1}{(x \sum_{i=1}^{K-1}\alpha_i+1)x} \right) +\Psi \nonumber \\
 &\overset{(\alpha_K = 1)}< \frac{1}{2}\log_2\left(\frac{x+1}{ x} \right) +\Psi \nonumber \\
&\leq \frac{1}{2}+\Psi \nonumber \\
&< \Delta_K.
\end{align}
This completes the proof for \eqref{eq:main_2}.

\subsection{Complexity Comparison}
We now discuss the complexity of our NOMA scheme with SIC and that without SIC. Consider a $K$-user downlink NOMA system, to perform SIC, the strong users have to decode other users' messages before decoding their own messages. However, each user has some probability to become the strongest channel user and may have to decode other $(K-1)$ users' messages in order to perform SIC. As a result, the decoding and detection delay introduced by SIC at the receiver can be as large as $K$ times of the decoding and detection time for our scheme without SIC. Furthermore, another encoding and modulation delay is also introduced as a result of re-encoding the decoded message and then re-mapping the codeword to the modulation. In contrast, our scheme does not require any re-encoding and re-mapping process. In addition, the detection complexity heavily depends on the dimensions of the underlying lattices. In general, the higher lattice dimension, the higher detection complexity. This is also true for any conventional power-domain NOMA scheme with high dimensional constellations. When the dimension of the underlying lattice becomes higher, our scheme can use efficient lattice decoders such as the sphere decoder \cite{771234} for detection as the superimposed constellation still preserves the nice lattice structure. Last but not least, allowing users to decode others' messages may result in a significant security problem while this can be easily avoided by using dithers for our scheme (note that the same trick cannot be used for NOMA with SIC since SIC by nature requires users to know each other's codebooks). To sum up, we have significantly reduced the complexity caused by SIC while still maintaining considerable performance in terms of individual outage rates.

\section{Simulation Results}\label{SIM}
Simulation results are provided to demonstrate the effectiveness of the proposed scheme. In Chapter~\ref{sec:rate_sim}, we use the Monte-Carlo method for simulating achievable outage rates. In Chapter~\ref{sec:pout_sim}, we implement off-the-shelf LDPC codes to realize our proposed scheme for simulating outage probability performance.

\subsection{Outage Rate Simulation}\label{sec:rate_sim}
\begin{figure}[ht!]
	\centering
\includegraphics[width=3.42in,clip,keepaspectratio]{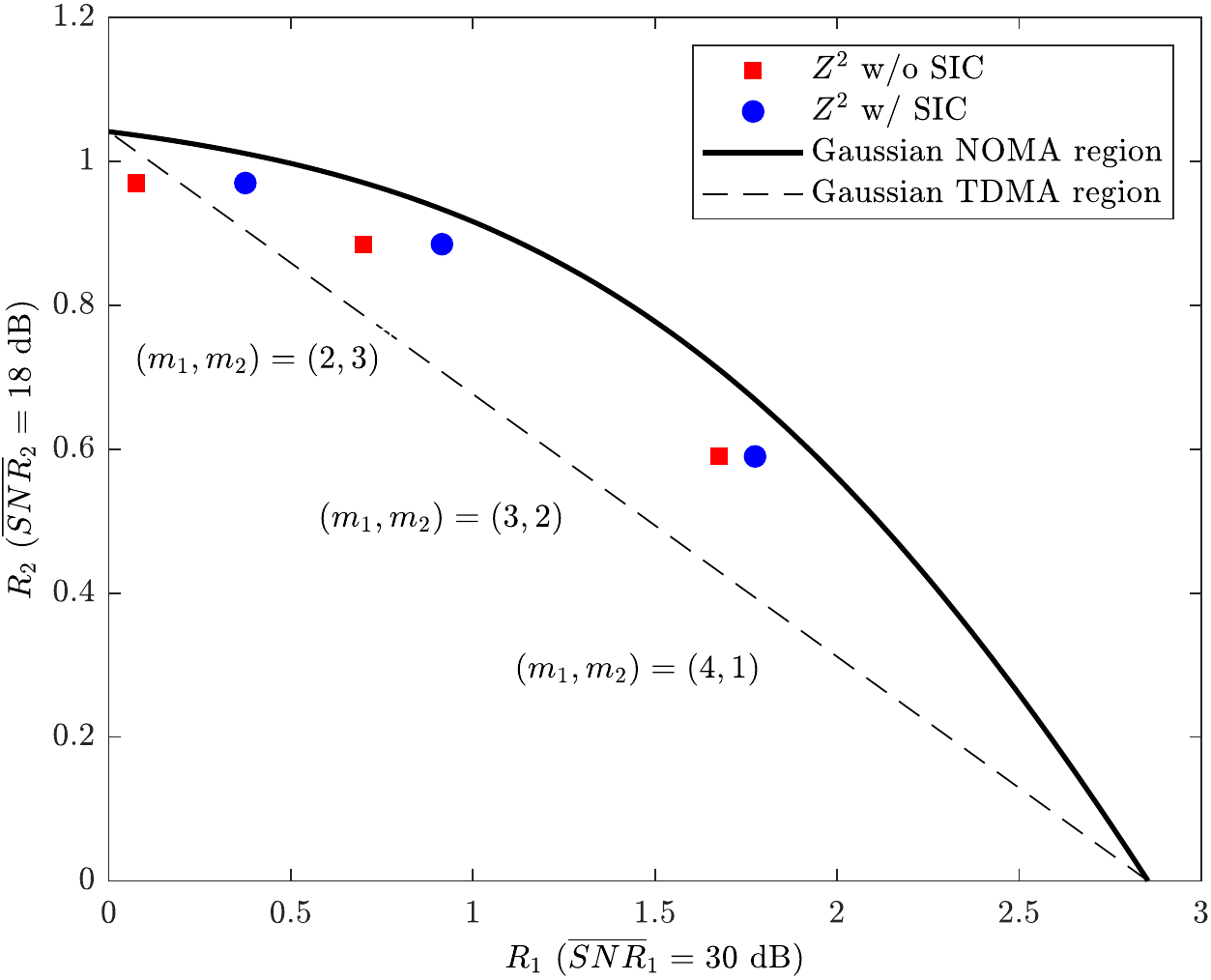}
\caption{The outage rate pairs of downlink NOMA with $(\overline{\SNR}_1, \overline{\SNR}_2)= (30,18)$ dB and $\epsilon_1=\epsilon_2=0.05$. }
\label{fig:2}
\end{figure}

\begin{figure}[ht!]
	\centering
\includegraphics[width=3.5in,clip,keepaspectratio]{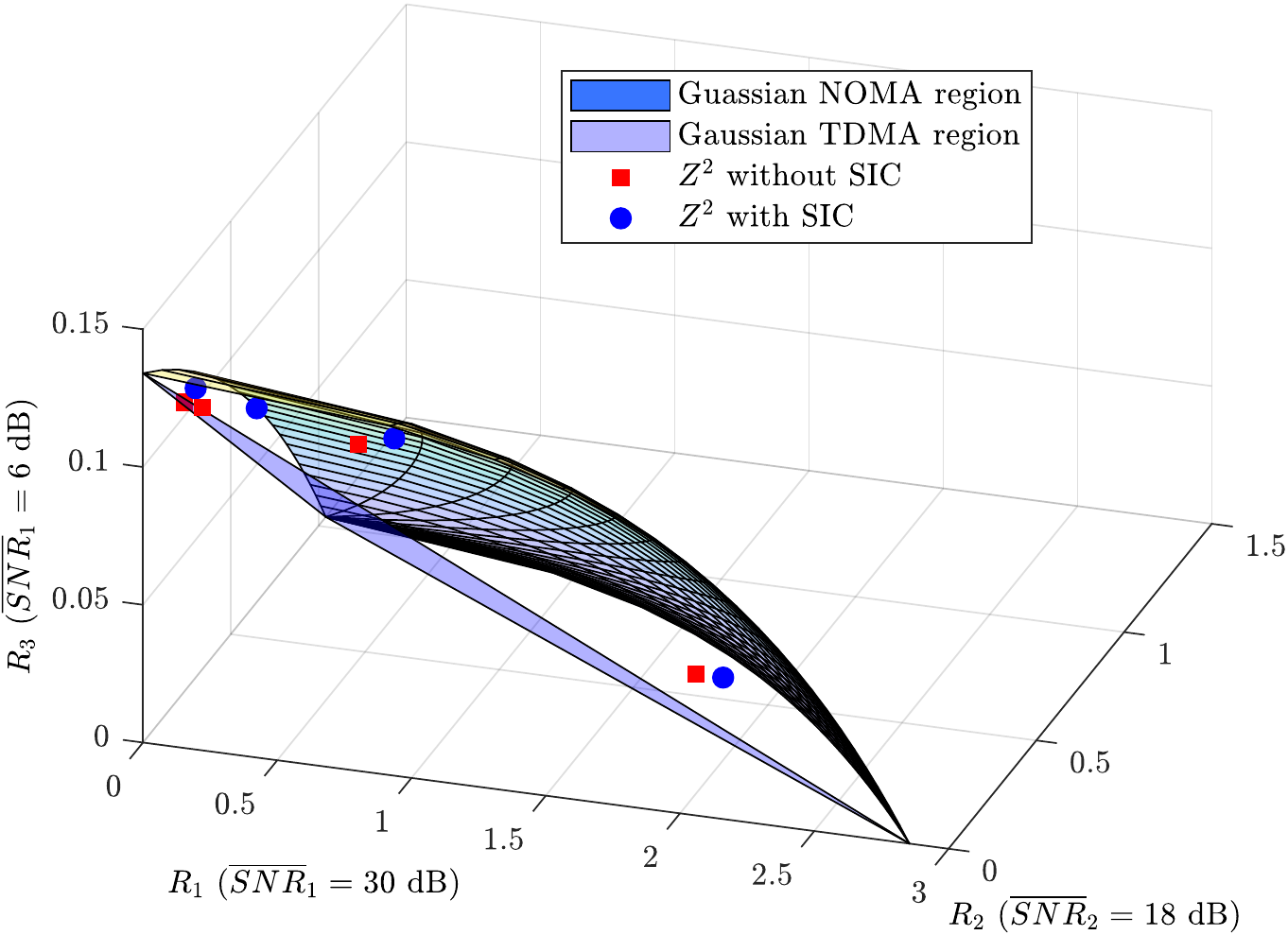}
\caption{The outage rate tuples of downlink NOMA with $(\overline{\SNR}_1,\overline{\SNR}_2,\overline{\SNR}_3) = (30,18,6)$ dB and $\epsilon_1=\epsilon_2=\epsilon_3=0.05$.}
\label{fig:3DNOMA}
\end{figure}

We construct proposed scheme over two-dimensional lattice $\mathbb{Z}^2$, which is purely for practical purposes as $\mathbb{Z}^2$ is associated with QAM modulations. In our simulations, we test $10^5$ channel realizations $\hat{h}_k$ drawn from Rayleigh distribution {\black with unit mode, i.e., $\sigma =1$}. For each realization, we evaluate the achievable rates by the Monte Carlo method with $10^6$ samples. The required outage probability is set to be 0.05\footnote{Note that we can set the outage rate to a very small value such as 0.001. However, in such cases, the achievable rate pairs for every scheme are going to be very small (at the order of 0.01 bit per real dimension or smaller), which do not seem to be interesting.}. In Fig. \ref{fig:2}, we consider two-user NOMA with $(\overline{\SNR}_1,\overline{\SNR}_2) = (30,18)$ dB, which corresponds to $(\bar{n}_1,\bar{n}_2) = (5,3)$. In Fig. \ref{fig:3DNOMA}, a three-user NOMA example is simulated with $(\overline{\SNR}_1,\overline{\SNR}_2,\overline{\SNR}_3) = (30,18,6)$ dB such that $(\bar{n}_1,\bar{n}_2,\bar{n}_3) = (5,3,1)$. In both figures, the achievable outage rates of the proposed scheme with SIC, that without SIC, the TDMA outage capacity region (with Gaussian input distributions), and the NOMA outage capacity region (with Gaussian input distributions) are plotted.

In both figures, one can see that the actual gaps between the outage capacity region and the outage rates achieved by our scheme are much smaller than the theoretical upper bounds. Specifically, the theoretical upper bounds are calculated as $(4.8396, 3.3396)$ bits and $(4.8396, 5.7275,3.3396)$ bits for two-user case and three-user case, respectively, while the simulated gaps from both figures are within 1 bit to the NOMA outage capacity region for both cases. This is because in our analysis, in order to obtain theoretic guarantees that are universal for all average SNR, many loose bounds are adopted for covering the worst case scenario. When decoded by a SIC decoder, the proposed scheme can operate at outage rate pairs that are very close to the outage capacity region. The loss is mainly due from the shaping loss incurred by having discrete input distributions. Moreover, even with single-user decoding, the proposed scheme can achieve outage rate pairs that are within 1 bit to the NOMA outage capacity region. It is worth noting that the performance of the OMA-type scheme (Gaussian TDMA region) is obtained by time-sharing between two single-user scheme with impractical Gaussian inputs. Despite that, some of the outage rate pairs of our proposed scheme with practical discrete inputs are still outside the Gaussian TDMA region, which indicates that the proposed scheme is capable of outperform OMA-type scheme even without SIC.

\subsection{Outage Probability Simulation}\label{sec:pout_sim}
Now we employ practical channel coding on top of our modulation for our proposed NOMA scheme. For illustrative purposes, we perform the simulation for a two-user case. {\black To approach the rate pair $(1.68,0.59)$ shown in Fig. \ref{fig:2} for $(m_1,m_2) = (4,1)$ and $\epsilon_1 = \epsilon_2 = 0.05$, we adopt $\mc{C}_1$ and $\mc{C}_2$ in Example~\ref{ex}. In addition to the modulation, we adopt off-the-shelf DVB-S2 LDPC codes with block length 64800 \cite{DVBS214} and group multiple coded bits for mapping to the proposed constellations. Specifically, we choose the codes with rates $\frac{2}{5}$, $\frac{3}{5}$ for users 1 and 2, respectively, such that the actual transmission rate pair is $(1.6,0.6)$, which is fairly close to the target outage rate pair.} When performing SIC, we assume that user 2's message is perfectly known at user 1. Note that this assumption is for obtaining the benchmark performance only while in practice SIC failure is deemed as an outage event. In addition, we assume that user 1 and user 2's messages are of the same length. If the lengths are different, we pad the uncoded modulated signals from the user with longer message length to the message with shorter length in order to make all the bits with interference. For a TDMA scheme to reach the rate pair of $(1.6,0.6)$, the required transmission rate pair is $(3.24,1.18)$ with time-sharing factor $(0.49,0.51)$. To approximate these target rates, we use rate $\frac{4}{5}$, $\frac{3}{5}$ DVB-S2 LDPC codes in conjunction with $\mathbb{Z}^2/16\mathbb{Z}^2$ and $\mathbb{Z}^2/4\mathbb{Z}^2$, respectively. With time-sharing factor $(0.5,0.5)$, the resultant transmission rate is $(1.6,0.6)$. The simulation is performed for $10^4$ channel realizations where the outage probability is averaged for $300$ outage samples in each realization.

\begin{figure}[ht!]
	\centering
\includegraphics[width=3.42in,clip,keepaspectratio]{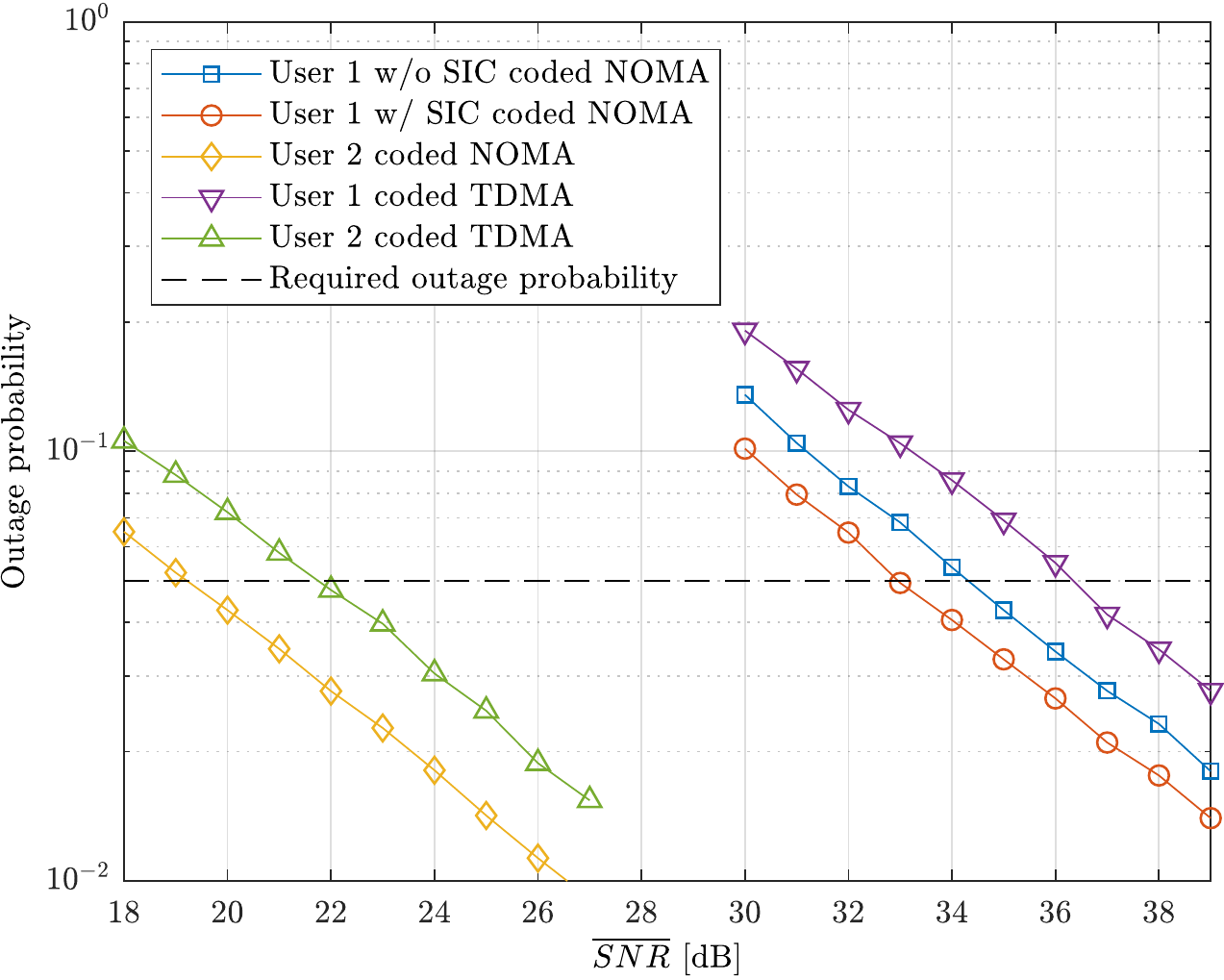}
\caption{Outage performance comparison between NOMA and TDMA.}
\label{fig:FER}
\end{figure}

From Fig. \ref{fig:FER}, we can see that user 1 without SIC saves about 2 dB power while user 2 saves about 2.5 dB power when compared with TDMA schemes. With perfect SIC, another 1 dB gain can be obtained for user 1. Note that for TDMA, user 1's transmission rate is smaller than the target rate while user 2's rate is larger than the target rate, i.e., $1.6 \cdot 2<3.24$ and $0.6\cdot2>1.18$. Thus, the gap between the rate differences for user 2 is larger than that of user 1. One may also notice that user 1 requires more power to reach $\epsilon_1 = 0.05$ than user 2. This is because user 1's constellation is large, i.e., with size 256, and using binary linear codes on top of larger-size constellations will lead to significant performance loss. Although better channel codes tailored specifically for downlink NOMA can be designed to enhance the outage performance, they are beyond the scope of this work.

\section{Summary}
In our previous work \cite{8291591}, it has been shown that lattice partition is a powerful tool for constructing coded modulation schemes for downlink NOMA with instantaneous CSIT under single-user decoding. In this work, we have further extended this result to the case where only statistical CSIT is available by 1) modifying the deterministic model; 2) analyzing the modified deterministic model; 3) proposing new lattice-partition-based schemes based on the insights obtained from the modified deterministic model. As a result, we have derived the outage rates achieved by our scheme with any base lattice. Moreover, for any outage probability below 63.21\%, we have proved that for any rate tuple lying inside the NOMA outage capacity region, there exists an instance of our scheme that is able to achieve this rate tuple within a constant gap even without SIC. Simulation results have demonstrated that the actual gaps can be much smaller than the analytic upper bounds, reaffirming that the proposed scheme is capable of achieving the near-capacity performance of downlink NOMA with only statistical CSI at the transmitter and without SIC at the receivers.

\chapter{Downlink NOMA without SIC for Block Fading Channels}\label{C7:chapter7}

\section{Introduction}
In this chapter, we continue the study of constructing practically implementable downlink NOMA schemes that perform well even under single-user decoding. In particular, we consider the scenario where the signal received at each user experiences a block fading channel \cite{MinICC19,DBLP:journals/corr/abs-1905-09514}. We again assume that the base station only has statistical CSI. For such a channel, an important performance metric is the diversity, measuring the decay rate in error probability with respect to SNR \cite{1256737,Caire06}. All the schemes \cite{Choi2016,Dong17,Fang16,Shieh16,8291591,Qiu18Globecom,8517129} previously mentioned in Chapter 6, however, achieve no diversity gains in block fading channels. To achieve full diversity order for point-to-point communication over block fading channels, it has been known for quite a while \cite{485720} that properly rotated QAM constellations will do the job. Therefore, in this chapter, we will further investigate the proposed lattice based schemes such that they can achieve diversity gains and good reliability for downlink NOMA systems over block fading channels.


\subsection{Main Contributions}

In this work, the problem of achieving full diversity order for every user in the downlink multiuser transmission over block fading channels is addressed. The main contributions of the work are summarized as follows.
\begin{itemize}
\item We propose a class of downlink NOMA schemes without SIC for block fading channels with only statistical CSI at the transmitter and full CSI at the receiver. Specifically, the proposed scheme constructs an $n$-dimensional ideal lattice from algebraic number fields and carves its coset leaders to form the constellation for each user. This class of schemes is the first attempt to use algebraic methods to provide high reliability solutions to downlink multiuser communications. Within the proposed class, we also identify a special family of schemes that are closely related to lattice partitions of the base ideal lattice.

\item To evaluate the error performance of the proposed scheme under single-user decoding, we analyze the minimum product distance of the composite constellation of the proposed scheme. We first show the equivalence between the superimposed $n$-dimensional constellation carved from any ideal lattice and the Cartesian product of $n$ identical rotated superimposed one-dimensional constellation. As a result, we then rigorously prove that the minimum product distance of the $n$-dimensional composite constellation can be upper bounded by the minimum product distance of the equivalent one-dimensional superimposed constellation and derive the analytical expression for the upper bound as a function of all users' power allocation factors and spectral efficiencies. Moreover, our bound closely captures the actual minimum product distance in the sense that all the local maximums of the actual distance coincide with our upper bound. Furthermore, it is shown numerically that the special family of schemes corresponding to lattice partition can achieve the maximal minimum product distance among all the proposed schemes.


\item We then extend our analysis to the MIMO-NOMA system with orthogonal space-time block codes (OSTBC). For such codes, the probability of error is largely determined by the minimum determinant, which can be further simplified as a function of the minimum Euclidean distance of the underlying composite constellation. Following similar steps in our analysis in the single antenna case, we obtain the exact analytical expression of the minimum determinant of the superposition coded space-time codeword with arbitrary power allocation factors and spectral efficiencies. Again, a special family of schemes corresponding to lattice partition is identified and it achieves the maximal minimum determinant.

\item Simulation results are provided to illustrate that our scheme can provide a systematic design that each user employs the same ideal lattice and same rotation is sufficient to attain full diversity with single-user decoding (i.e., without SIC). Moreover, the special family of schemes based on lattice partitions provides substantially better error performance than the benchmark NOMA schemes.
\end{itemize}


\section{System Model}\label{system_model}
In this work, we consider a downlink NOMA system where a base station wishes to broadcast $K$ messages $\mathbf{u}_1,\ldots,\mathbf{u}_K$ to $K$ users, one for each user. For $k\in\{1,\ldots,K\}$, the message $\mathbf{u}_k$ is a binary sequence of length $nm_k$, where $n$ is the dimension of the code and $m_k$ is the spectral efficiency of user $k$ in bits/s/Hz/real dimension. We emphasize here that due to the delay requirements, the channel between the transmitter and each user experiences independent block fading with a finite number of realizations within each data packet transmission duration, which is different from the slow fading model considered in \cite{8517129} where each user only gets to experience one realization within each data packet transmission duration.
For now, we assume that every device in the network is equipped with a single antenna and works in a half-duplex mode.

The base station encodes all users' messages $\mathbf{u}_1,\ldots,\mathbf{u}_K$ into a codeword $\mathbf{x} = [x[1],\ldots,x[n]]  \in \mathcal{M}$ of the codebook $\mathcal{M}\subset\mbb{R}^n$, satisfying the power constraint $\E[\| \mathbf{x} \|^2] \leq n$. We denote by $\mathbf{h}_k = [h_k[1],\ldots,h_k[n]]\in\mbb{R}^n$ the instantaneous channel coefficient vector from the base station to user $k$. Here, each fading coefficient $h_k[l]$ is drawn i.i.d. from Rayleigh distribution. 
The received signal at user $k$ is denoted by $\mathbf{y}_k=[y_k[1],\ldots,y_k[n]]$ with
\begin{align}\label{eq:sm2}
y_k[l] = \sqrt{P}h_k[l]x[l]+z_k[l], \;  l= 1,\ldots, n,
\end{align}
where $P$ is the total power constraint at the base station and $z_k[l]\sim\mc{N}(0,1)$ is the Gaussian noise experienced at user $k$. Each user $k$ is assumed to have full CSI, i.e., $\mathbf{h}_k$, while the transmitter only has the statistical CSI, i.e., the distributions of each $\mathbf{h}_k$.
We note that this channel model is quite standard and can be easily obtained by \emph{interleaving} the codeword across multiple channel coherence time periods and applying de-interleaving and \emph{coherent detection} to the received signals \cite[Chapter 3.2]{tse_book}.

We measure the \emph{reliability} by the pairwise error probability (PEP). 
Following \cite[Chapter 3.2]{tse_book}, for any two codewords $\mathbf{x}_s,\mathbf{x}_w \in \mathcal{M}$ and $\mathbf{x}_s \neq \mathbf{x}_w$, user $k$'s error probability without SIC is upper bounded by the average PEP of the composite constellation over all $(\mathbf{x}_s,\mathbf{x}_w)$ pairs, which is 
\begin{align}\label{eq:pepZ}
P_e^{(k)} &\leq \frac{1}{|\mathcal{M}|}\sum_{s \neq w}\text{Pr}(\mathbf{x}_s \rightarrow \mathbf{x}_w|\mathbf{h}_k) \nonumber \\
&\leq \frac{1}{|\mathcal{M}|}\sum_{s \neq w}\prod_{l=1}^n \frac{1}{1+\overline{\SNR}_k(x_s[l]-x_w[l])^2/4} \nonumber \\
&< \frac{1}{|\mathcal{M}|}\sum_{s \neq w}\frac{4^{L_{(s,w)}}}{d_p(\mathbf{x}_s,\mathbf{x}_w)^2\overline{\SNR}_k^{L_{(s,w)}}} \nonumber \\ 
&\leq \frac{4^{\delta_L}(|\mathcal{M}|-1)}{\overline{\SNR}_k^{\delta_L}\min\limits_{s \neq w}\{d_p(\mathbf{x}_s,\mathbf{x}_w)^2 \}},
\end{align}
where
$d_p(\mathbf{x}_s,\mathbf{x}_w) \triangleq \prod_{s \neq w}|x_s[l]-x_w[l]|$
is the \emph{product distance} of $\mathbf{x}_s$ from $\mathbf{x}_w$ that differs in $L_{(s,w)} \leq n$ components and $\overline{\SNR_k} \triangleq \E[\|\mathbf{h}_k\|^2]P$ is the average SNR. It can be seen that in the high SNR regime, the overall error probability decreases exponentially with the order of $\delta_L \triangleq \min\limits_{s \neq w}\{L_{(s,w)}\}$, which is known as the \emph{diversity order}. The code has \emph{full diversity} when $\delta_L = n$. Moreover, one would like to maximize the minimum product distance $\min\limits_{s \neq w}\{d_p(\mathbf{x}_s,\mathbf{x}_w)\}$ in a bid to minimize the overall PEP, which provides additional \emph{coding gain} on top of the diversity gain. The diversity order and the minimum product distance are important metrics for improving the reliability of communication through block fading channels. Note that although we focus solely on Rayleigh fading channels in this paper, the diversity order and product distance criterion are generalizable to other fading channels, e.g., Rician fading \cite[Chapter 3.2]{tse_book}, \cite{Vucetic:2003:SC:861866,1256737}.

In this work, we focus on the case of $K=2$ only as it is more practical for multi-carrier NOMA where each subcarrier is allocated to two users \cite{7273963,8449119}. This is also a common assumption in many works in the NOMA literature, see for example \cite{Wei17,8345745,Choi2016,Dong17}. We would like to emphasize that the schemes proposed in this work are not limited to the two-user case and can be generalized to the general $K$-user case in a straightforward manner. However, the analysis becomes quite messy for $K>2$ and is thus left for future study. Throughout Chapter 7, without loss of generality, we also assume that $\overline{\SNR_1}\geq \overline{\SNR_2}$ and thus users 1 and 2 are commonly referred to as the strong and weak users, respectively.

\section{Downlink NOMA over Block Fading Channels}\label{sec:proposed_CH7}
In this section, we first introduce the proposed class of NOMA schemes based on superpositions of codes from $n$-dimensional ideal lattices. We then identify, within the proposed class of schemes, a special family of schemes corresponding to lattice partitions of the underlying ideal lattices. The minimum product distance of the proposed schemes will be analyzed in Chapter~\ref{sec:perform_ana}.

\subsection{Proposed Downlink NOMA Schemes from Ideal Lattices}\label{A1}
Encouraged by the success of using ideal lattices for point-to-point communications over block-fading (see \cite{Oggier:2004:ANT:1166377.1166378}), we construct rotated version of multi-dimensional QAM (corresponding to $\mbb{Z}^n$ lattices) from a totally real ideal lattice. It is worth noting that the rotated versions of many other well-known lattices such as $D_4$, $E_6$, $E_8$ and $K_{12}$ that are good for block fading channels can also be constructed. Our choice of using rotated $\mbb{Z}^n$ is mainly for encoding/decoding complexity and for achieving full diversity order.

Throughout Chapter 7, we use the cyclotomic construction \cite[Chapter 7.2]{Oggier:2004:ANT:1166377.1166378} to construct ideal lattices that are equivalent to $\mathbb{Z}^n$. Consider $\zeta = e^{\frac{2\pi \sqrt{-1}}{p}}$ the $p$-th primitive root of unity for some prime number $p \geq 5$. Construct $\mbb{K}=\mbb{Q}(\zeta + \zeta^{-1})$ the maximal real sub-field of the $p$-th cyclotomic field $\mbb{Q}(\zeta)$. This $\mbb{K}$ is totally real and has degree $n = \frac{p-1}{2}$. A set of integral basis is given by $\{\zeta+\zeta^{-1},\ldots, \zeta^n+\zeta^{-n} \}$. The $n$ embeddings of $\mbb{K}$ into $\mbb{C}$ are given by
\begin{align}
    \sigma_j(\zeta^i+\zeta^{-i}) = \zeta^{ij}+\zeta^{-ij} = 2\cos\left( \frac{2\pi i j}{p}\right), \; i,j \in \{1,\ldots,n\}.
\end{align}
Then, the generator matrix is given by
\begin{align}\label{G_ideal}
\mathbf{G}_{\Lambda} = \frac{1}{\sqrt{p}}\mathbf{T}\cdot [(\sigma_j(\zeta^i+\zeta^{-i}))_{i,j=1}^n] \cdot \text{diag}(\sqrt{\sigma_1(\varsigma)},\ldots,\sqrt{\sigma_n(\varsigma)}),
\end{align}
where $\mathbf{T}$ is an upper triangular matrix with entries $t_{i,j}=1$ for $i \leq j$; $\varsigma = (1-\zeta)(1-\zeta^{-1})$ is to ensure that $\Lambda$ is equivalent to $\mathbb{Z}^n$; and $\frac{1}{\sqrt{p}}$ is to normalize the volume of $\Lambda$ such that $\text{Vol}(\Lambda)=1$.

The minimum product distance of this family of ideal lattices is
\begin{align}\label{eq:dpcal}
d_{p,\min}(\Lambda) = \sqrt{\frac{\det(\mathbf{G}_{\Lambda})}{d_\mathbb{K}}} = p^{-\frac{n-1}{2}}.
\end{align}

Having constructed the considered ideal lattice, we now introduce the encoding and decoding steps of our proposed NOMA scheme at the transmitter and the receiver, respectively.
\subsection{Transmitter Side}\label{sec:enc_CH7}
For user $k\in\{1,2\}$, a subset $\mc{C}_k$ of the ideal lattice is carved to form the constellation of the user $k$. Specifically, $\mc{C}_k$ has cardinality $2^{nm_k}$ and is the complete set of coset leaders (see Chapter~\ref{sec:2_1} for the definition) of the lattice partition $\Lambda/2^{m_k}\Lambda$. User $k$'s message $\mathbf{u}_k$ is mapped into $\mathbf{v}_k\in\mc{C}_k$. 
The transmitter then sends the superimposed signal $\mathbf{x}=\eta(\sqrt{\alpha}\mathbf{v}_1+\sqrt{1-\alpha}\mathbf{v}_2-\mathbf{d})$, where
\begin{align}\label{eq:sup1}
\mathbf{x} \in \eta(\mathcal{C}-\mathbf{d} )&= \eta(\sqrt{\alpha}\mathcal{C}_1+\sqrt{1-\alpha}\mathcal{C}_2-\mathbf{d}) \nonumber \\
&= \eta(\sqrt{\alpha}(\mathcal{C}_1-\mathbf{d}_1)+\sqrt{1-\alpha}(\mathcal{C}_2-\mathbf{d}_2)),
\end{align}
where $\mathbf{d}_1= \E[\mathcal{C}_1]$, $\mathbf{d}_2= \E[\mathcal{C}_2]$, and $\mathbf{d} = \E[\sqrt{\alpha}\mathcal{C}_1+\sqrt{1-\alpha}\mathcal{C}_2]$ are length $n$ dither vectors to ensure the constellations $\mc{C}_1$, $\mc{C}_2$, and $\mc{C}$, respectively, to have zero mean; $\eta$ is a normalize factor for ensuring power constraint $\mbb{E}[\|\mathbf{x}\|^2]\leq n$;
and $\alpha,1-\alpha \in[0,1]$ are the power allocation factors for users 1 and 2, respectively. Here, the normalization factor $\eta$ is computed by using Lemma \ref{lem:zn_P} in Appendix \ref{appendix:lemma_ch7} as
\begin{align}\label{normalize2}
\eta &= \sqrt{\frac{n}{\E[\|\mathcal{C}-\mathbf{d} \|^2]}} \nonumber \\
 &\overset{\eqref{eq:sup1}}= \sqrt{\frac{n}{\E[\|\sqrt{\alpha}(\mathcal{C}_1-\mathbf{d}_1)+\sqrt{1-\alpha}(\mathcal{C}_2-\mathbf{d}_2) \|^2]}} \nonumber \\
& = \sqrt{\frac{n}{\alpha\E[\|\mathcal{C}_1-\mathbf{d}_1\|^2]+(1-\alpha)\E\|\mathcal{C}_2-\mathbf{d}_2 \|^2]}} \nonumber \\
 &\overset{\eqref{eq:zn_P}}= \sqrt{\frac{n}{\alpha\frac{n}{12}(2^{2m_1}-1)+(1-\alpha)\frac{n}{12}(2^{2m_2}-1)}} \nonumber \\
& = \sqrt{\frac{12}{(2^{2m_1}-2^{2m_2})\alpha+2^{2m_2}-1}}.
\end{align}

\subsection{Receiver Side}
Recall that the received message at user $k\in \{1,2\}$ is denoted by $\mathbf{y}_k$ and is given in \eqref{eq:sm2}. There are two options for the decoder, depending on the implementation and application. 
If a single-user decoder is adopted (i.e., without performing SIC), the decoder of user $k$ attempts to recover $\mathbf{u}_k$ from $\mathbf{y}_k$ by treating the other user's signal as noise. If an SIC decoder is adopted, user 2 remains the same decoding procedure, while user 1 first decodes $\mathbf{u}_2$, subtracts it out, and then decodes its own message. Both single-user decoding and SIC decoding
will be included in simulations for comparison. However, our design and analysis focus solely on the case with single-user decoding as it is one of the main motivation of this work.

\begin{remark}
Similar to most works considering block Rayleigh fading channels (see \cite{Oggier:2004:ANT:1166377.1166378} and reference therein), we focus solely on diversity order and minimum product distance. It is worth mentioning that standard channel coding can be employed on top of the modulation schemes of this work to obtain additional coding gain at the cost of further lowering the spectral efficiency.
\end{remark}

\begin{remark}
Consider a $K$-user downlink NOMA system. For the conventional power-domain NOMA, each user would have to decode other $(K-1)$ users' messages to perform SIC because each user has some probability to potentially become the strongest channel user. Thus, the demodulation and decoding delay can be as large as $K$ times of that for our proposed scheme without SIC. Moreover, encoding delays are introduced by SIC as a result of re-encoding the decoded message and then re-mapping the codeword to the modulation. In contrast, re-encoding and re-mapping are not required in our scheme without SIC.
\end{remark}

\subsection{Proposed Schemes based on Lattice Partitions}\label{Sec:LP}
Now, we identify a special family of the proposed schemes within the proposed class of schemes. In this family of schemes, after the mapping process from $\mathbf{u}_k$ to $\mathbf{v}_k\in\mathcal{C}_k$ for $k\in\{1,2\}$, the transmitted signal is given by
\begin{align}\label{eqn:x_def_Ch7}
\mathbf{x}' = \eta'\left(\mathbf{v}_1+2^{m_1}\mathbf{v}_2-\mathbf{d}' \right)\in\eta'\left( \mc{C}_1 + 2^{m_1}\mc{C}_2-\mathbf{d}' \right),
\end{align}
where $\mathbf{d}'$ is a deterministic dither to ensure the composite constellation $\mc{C}'=\mc{C}_1 + 2^{m_1}\mc{C}_2$ have zero mean and
\begin{equation}\label{normalize_LP}
    \eta'=\sqrt{\frac{12}{2^{2(m_1+m_2)}-1}},
\end{equation}
is the normalization factor to ensure the power constraint $\mbb{E}[\|\mathbf{x}'\|^2]\leq n$. To see that $\eta'$ is indeed the correct normalization factor, we use Lemma \ref{lem:zn_P} in Appendix \ref{appendix:lemma_ch7} to obtain that $\mbb{E}[\|\mc{C}'-\mathbf{d}'\|^2]=\frac{n}{12}(2^{2(m_1+m_2)}-1)$. Here, the power allocation is $\alpha =  \frac{1}{1+2^{2m_1}}$. When substituting this power allocation into (6) and decomposing $\mathbf{d}' = \sqrt{\alpha}\mathbf{d}_1+\sqrt{1-\alpha}\mathbf{d}_2$, it can be easily verified that this family of schemes described in \eqref{eqn:x_def_Ch7} is a special case of the proposed class of schemes in \eqref{eq:sup1}.


The beauty of this family of schemes is that the composite constellation $\mc{C}'$ corresponds to the lattice partition $\Lambda/2^{m_1+m_2}\Lambda$ because $\Lambda/2^{m_1}\Lambda+2^{m_1}(\Lambda/2^{m_2}\Lambda) = \Lambda/2^{m_1+m_2}\Lambda$ for $\Lambda$ equivalent to $\mathbb{Z}^n$. Moreover, the relationship among $\mc{C}_1$, $\mc{C}_2$, and $\mc{C}'$ closely follows the lattice partition chain $\Lambda/2^{m_1}\Lambda/2^{m_1+m_2}\Lambda$ and hence many nice properties of the underlying ideal lattice $\Lambda$ naturally carry over to the individual and composite constellations. For example, since the superimposed constellation still preserves the nice lattice structure, efficient lattice decoders such as the sphere decoder \cite{771234} can be used at each receiver for decoding. Also, the minimum product distance of a scheme within this family can be precisely computed as shown in the following proposition.
\begin{prop}\label{prop:LP}
The lattice-partition scheme with ideal lattices as the base lattice can provide full diversity \emph{to each user} and the composite constellation $\eta'(\mc{C}'-\mathbf{d}')$ has a minimum product distance
\begin{align}\label{dminp_LP111}
d_{p,\min}(\eta'(\mc{C}'-\mathbf{d}')) = \left(\frac{12}{2^{2(m_1+m_2)}-1}\right)^{\frac{n}{2}}d_{p,\min}(\Lambda).
\end{align}
\end{prop}

\emph{\quad Proof: }
Since $\mc{C}'$ corresponds to the lattice partition $\Lambda/2^{m_1+m_2}\Lambda$, the minimum product distance of the composite constellation can be derived as
\begin{align}\label{eq:dp_lattice}
    d_{p,\min}(\eta'(\mc{C}'-\mathbf{d}'))&=d_{p,\min}(\eta'\Lambda) \nonumber \\ &= (\eta')^n\sqrt{\frac{\det(\Lambda)}{d_\mathbb{K}}} \nonumber \\ &\overset{(a)}=\left(\frac{12}{2^{2(m_1+m_2)}-1}\right)^{\frac{n}{2}}d_{p,\min}(\Lambda),
\end{align}

\par\noindent
where $(a)$ is obtained by plugging $\eta'$ from \eqref{normalize_LP}.

Now, since $ d_{p,\min}(\eta'(\mc{C}'-\mathbf{d}'))>0$, full diversity is thus guaranteed according to \eqref{eq:pepZ}. \QEDA

\section{Performance Analysis}\label{sec:perform_ana}
In this section, we analyze $d_{p,\min}(\eta(\mathcal{C}-\mathbf{d}))$, the minimum product distance of the \emph{normalized} and \emph{dithered} composite constellation $\eta(\mathcal{C}-\mathbf{d})$ defined in \eqref{eq:sup1} for any parameters $m_1$, $m_2$, $n$, $\alpha$. We emphasize that under block fading, the symbol error rate (SER) performance of the whole downlink system is closely related to $d_{p,\min}(\eta(\mathcal{C}-\mathbf{d}))$ according to \eqref{eq:pepZ}. Moreover, the analytical results of the minimum product distances will provide insights into the relationship between spectral efficiency, power allocation factor and the error performance of the proposed scheme.

\subsection{Preparations and Definitions}\label{sec:def}
we first introduce a few preparations and definitions in the following chapter.
\subsection{Layer}\label{def:layer} 
\begin{figure}[ht!]
	\centering
\includegraphics[width=3.5in,clip,keepaspectratio]{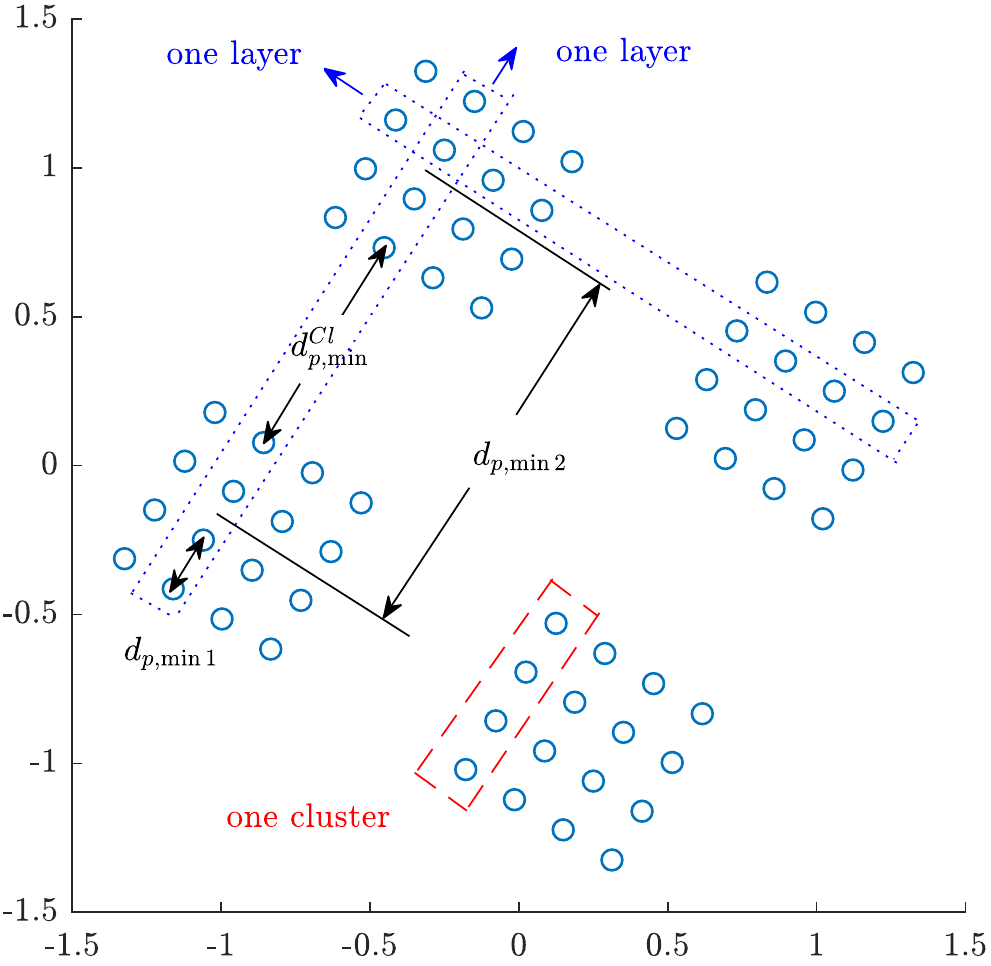}
\caption{An example of a superimposed constellation with $\Lambda$ being a two-dimensional ideal lattice and $(m_1,m_2) = (2,1)$.}
\label{fig:2_ch7}
\end{figure}

We define a \emph{layer} of $\eta(\mc{C}-\mathbf{d})$ in \eqref{eq:sup1} to be the collection of points constituting a shifted version of a rotated and dithered one-dimensional superimposed constellation
\begin{align}\label{eq:1dsup}
\eta(\mathcal{X}-d^*)\mathbf{R} = \eta(\sqrt{\alpha}(\mathcal{X}_1-d^*_1)+\sqrt{1-\alpha}(\mathcal{X}_2-d^*_2))\mathbf{R},
\end{align}
where $\mathcal{X}_k$ is a complete set of the coset leaders of the one-dimensional lattice partition $\mathbb{Z}/2^{m_k}\mathbb{Z}$, $d^*_k = \E[\mathcal{X}_k]$ is a scalar dither for $k \in\{1,2\}$, and $\mathbf{R}$ is an $n \times n$ rotation matrix such that the shifted and rotated one-dimensional constellation becomes a subset of $\eta(\mc{C}-\mathbf{d})$. In other words, a layer is given by $\{[\lambda_1,\dots,\lambda_n]\mathbf{R}| \lambda_j \in \eta(\mathcal{X}-d^*)  \}$ for some fixed $\lambda_1,\ldots,\lambda_{j-1},\lambda_{j+1},\ldots,\lambda_n\in \eta(\mathcal{X}-d^*)$. Examples of all the layers for the case of $(m_1,m_2) = (2,1)$ and $n=2$ are illustrated in Fig. \ref{fig:2_ch7} where each circle represents a constellation point of $\eta(\mc{C}-\mathbf{d})$ and there are 16 layers in total.

\subsection{Intra-Layer Minimum Product Distance}\label{def:ilmpd} We denote by $d_{p,\min}(\eta(\mathcal{X}-d^*)\mathbf{R})$ the \emph{intra-layer minimum product distance} as the minimum product distance between any pair of two distinct constellation points within a layer, i.e., within the shifted version of the (rotated) one-dimensional constellation $\eta(\mathcal{X}-d^*)\mathbf{R}$.

\subsection{Cluster}\label{def:cl} We define a \emph{cluster} to be all the points in a shifted version of user 1's constellation \emph{in one layer}, i.e., $\text{Cl}_{\nu} \triangleq \{\sqrt{\alpha}(\mathcal{X}_1-d^*_1)\mathbf{R}+\nu\}$ for a fixed $\nu \in \sqrt{1-\alpha}(\mathcal{X}_2-d^*_2)\mathbf{R}$. Each layer has $2^{m_2}$ clusters. In the example shown in Fig. \ref{fig:2_ch7}, there are 2 clusters inside a layer. With a slight abuse of notation, we define the minimum product distance between two distinct clusters $\text{Cl}_{\nu}$ and $\text{Cl}_{\mu}$ as
\begin{equation}
    d_{p,\min}(\text{Cl}_{\nu},\text{Cl}_{\mu}) \defeq \min\limits_{ \boldsymbol{\lambda}_1 \in \text{Cl}_{\nu},\boldsymbol{\lambda}_2 \in \text{Cl}_{\mu}} \{d_p(\boldsymbol{\lambda}_1,\boldsymbol{\lambda}_2)\}.
\end{equation}

\subsection{Inter-Cluster Minimum Product Distance}\label{def:icmpd} The \emph{inter-cluster minimum product distance} is then defined as
\begin{equation}\label{eq:ICPD}
  d_{p,\min}^{\text{Cl}} \triangleq \min\limits_{\nu,\mu \in \sqrt{1-\alpha}(\mathcal{X}_2-d^*_2)\mathbf{R}} \{d_{p,\min}(\text{Cl}_{\nu},\text{Cl}_{\mu}) \}.
\end{equation}
An example of $d_{p,\min}^{\text{Cl}}$ can also be found in Fig. \ref{fig:2_ch7}. 

\subsection{Minimum Product Distance Notations}\label{def:d2d2} We denote by $d_{p,\min 1}$ and $d_{p,\min 2}$ the minimum product distance of users 1 and 2's constellations \emph{in one layer}, respectively. To be specific, they are computed as
\begin{align}
&d_{p,\min1}  \triangleq d_{p,\min}(\eta\sqrt{\alpha}(\mathcal{X}_1-d^*_1)\mathbf{R})  = (\eta\sqrt{\alpha})^n d_{p,\min}((\mathcal{X}_1-d^*_1)\mathbf{R}), \label{eq:dp1_def} \\
&d_{p,\min2}  \triangleq d_{p,\min}(\eta\sqrt{1-\alpha}(\mathcal{X}_1-d^*_1)\mathbf{R}) = (\eta\sqrt{1-\alpha})^n d_{p,\min}(\mathcal{X}_1-d^*_1)\mathbf{R}). \label{eq:dp2_def}
\end{align}
Examples of $d_{p,\min1}$ and $d_{p,\min2}$ are shown in Fig. \ref{fig:2_ch7}.

\subsection{Main Result}\label{sec:Main_result}
The minimum product distance of the proposed NOMA scheme with arbitrary power allocation is upper bounded as follows.
\begin{prop}\label{prop:AP}
The minimum product distance of the proposed NOMA scheme with ideal lattices as the base lattice and with arbitrary power allocation $\alpha\in[0,1]$ is upper bounded by
\begin{align}\label{eq:case2}
&d_{p,\min}(\eta(\mathcal{C}-\mathbf{d})) \leq \nonumber \\
  &\left\{\begin{array}{ll}
d_{p,\min 1}, &\hspace {-4mm}\alpha\in [0 ,\frac{1}{1+2^{2m_1}}]\\
d_{p,\min}^{\text{Cl}(m_2=1)},
&\hspace {-4mm}\alpha \in \big(\frac{1}{1+2^{2m_1}},\frac{4}{(2^{m_1}-\frac{1}{2})+4}\big], \\
\multirow{2}{*}{$\min\limits_{\substack{\gamma \in \{ 0,\ldots,\lfloor\frac{\xi-1}{2} \rfloor\}, \\ \beta \in \{1,\ldots,\xi-1\}}} \left\{\left|\gamma\sqrt[n]{d_{p,\min 1}} -\beta\sqrt[n]{d_{p,\min}^{\text{Cl}(m_2=1)}}\right|^n  \right\},$} 
 &\hspace {-4mm}\alpha \in \Big(\frac{(\xi-1)^2}{\left(2^{m_1}-\frac{1}{2} \right)^2+(\xi-1)^2}, 
\frac{\xi^2}{\left(2^{m_1}-\frac{1}{2} \right)^2+\xi^2}  \Big],\\
&\hspace {-4mm}\xi = 3, \ldots,2^{m_2}-1 \\
\min\limits_{\substack{\gamma \in \{ 0,\ldots,\frac{2^{m_2}}{2}-1 \}, \\ \beta \in \{1,\ldots,2^{m_2}-1\}}} \left\{\left|\gamma\sqrt[n]{d_{p,\min 1}} -\beta\sqrt[n]{d_{p,\min}^{\text{Cl}(m_2=1)}}\right|^n \right\},
& \hspace {-4mm}\alpha \in \big(\frac{(2^{m_2}-1)^2}{(2^{m_1}-\frac{1}{2})^2+(2^{m_2}-1)^2},\frac{1}{2}\big]  \\
\end{array} \right.
\end{align}
where
\begin{align}
&d_{p,\min1}   = \left(\frac{12\alpha}{(2^{2m_1}-2^{2m_2})\alpha+2^{2m_2}-1}\right)^{\frac{n}{2}}d_{p,\min}(\Lambda), \label{eq:dp1} \\
&d_{p,\min2}   = \left(\frac{12(1-\alpha)}{(2^{2m_1}-2^{2m_2})\alpha+2^{2m_2}-1}\right)^{\frac{n}{2}}d_{p,\min}(\Lambda), \label{eq:dp2} \\
&d_{p,\min}^{\text{Cl}(m_2=1)} = \left\{\begin{array}{ll}
\left|\sqrt[n]{d_{p,\min 2}}- (2^{m_1}-1)\sqrt[n]{d_{p,\min 1}}\right|^n, 
&\alpha \in \big(\frac{1}{1+2^{2m_1}}, \frac{1}{(2^{m_1}-\frac{3}{2})^2+1}\big] \\
\multirow{2}{*}{$\left|\sqrt[n]{d_{p,\min 2}}- (2^{m_1}-l)\sqrt[n]{d_{p,\min 1}}\right|^n,$} &\alpha \in \big(\frac{1}{(2^{m_1}+\frac{1}{2}-l)^2+1},\frac{1}{(2^{m_1}-\frac{1}{2}-l)^2+1}\big],\\
&l = 2, \ldots,2^{m_1}-2  \\
\left(\sqrt[n]{d_{p,\min 2}} - \sqrt[n]{d_{p,\min 1}}\right)^n, 
&\alpha \in (\frac{4}{13},\frac{1}{2}] \\
\end{array} \right., \label{eq:case1}
\end{align}
and $d_{p,\min}^{\text{Cl}(m_2=1)}$ denotes $d_{p,\min}^{\text{Cl}}$ the inter-cluster minimum product distance for the case of $m_2 = 1$ for any $m_1 \in \mathbb{Z}^+$. The upper bound for $\alpha\in [\frac{1}{2},1]$ can be obtained by switching the roles of $m_1$ and $m_2$ and substituting $1-\alpha$ into $\alpha$ from \eqref{eq:case2}.
\end{prop}

The proof of this proposition is described in details in Section \ref{sec:dminp_ana}. Before that, we would like to emphasize that one can find the exact minimum product distance of $\eta(\mathcal{C}-\mathbf{d})$ for a given $\alpha \in [0,1]$ by numerically calculating all the product distances between all pairs of two constellation points in $\eta(\mathcal{C}-\mathbf{d})$ and find the minimum value among them. However, the computational complexity will dramatically increase with $m_1$, $m_2$ and $n$ increasing. We use the following example to demonstrate the effectiveness of our analytical upper bound.
\begin{figure}[ht!]
	\centering
\includegraphics[width=3.5in,clip,keepaspectratio]{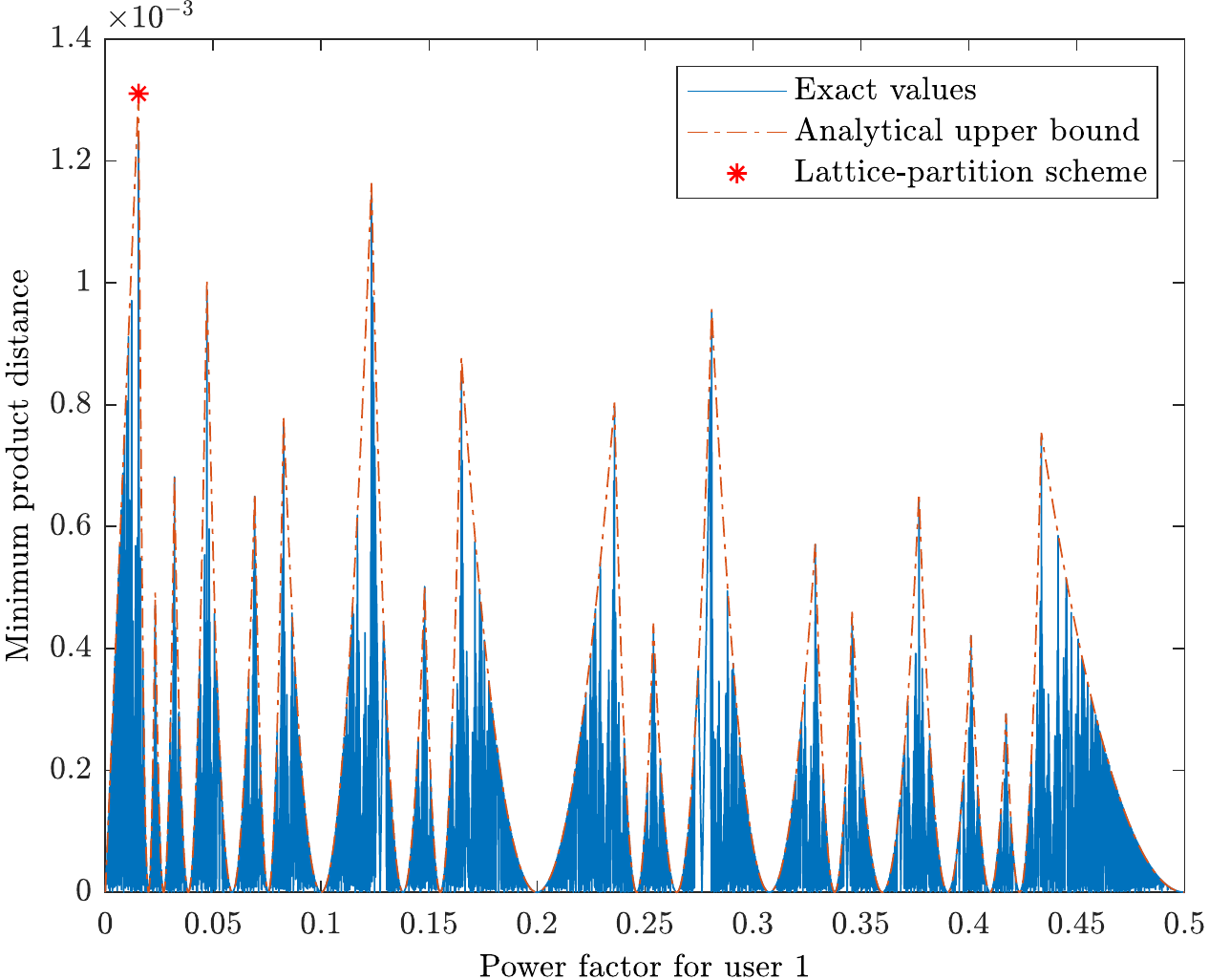}
\caption{Minimum product distances of the scheme considered in Example~\ref{exa:dpmin} with various $\alpha\in[0,0.5]$.}
\label{fig:7}
\end{figure}

\begin{exam}\label{exa:dpmin}
Consider $(m_1,m_2) = (3,3)$ and $n=2$. Both $\mc{C}_1-\mathbf{d}_1$ and $\mc{C}_2-\mathbf{d}_2$ are rotated 64-QAM constellations and $\mc{C}-\mathbf{d}$ becomes a superimposed constellation with 4096 constellation points. In Fig. \ref{fig:7}, we evaluate the upper bound of $d_{p,\min}(\eta(\mathcal{C}-\mathbf{d}))$ in \eqref{eq:case2} and the exact values of $d_{p,\min}(\eta(\mathcal{C}-\mathbf{d}))$ by computer search for $\alpha \in [0,0.5]$. The minimum product distance achieved by our scheme based on lattice partition is also plotted.


It can be observed that the derived upper bound well captures the trend of the changes in $d_{p,\min}(\eta(\mathcal{C}-\mathbf{d}))$ and fits all the local maximum points (peak values in the figure). Most notably, the proposed scheme based on lattice partition achieves the largest value (the first peak value in the figure), which shows the optimality of this scheme. Although we have not rigorously proved that this scheme is always optimal for a general pair of $(m_1,m_2)$, it is optimal for all the cases that we have tested, including every $(m_1,m_2)$ for $m_1,m_2\in\{1,\ldots,8\}$ (each user's constellation ranging from rotated 4-QAM to rotated $2^{16}$-QAM).
\end{exam}

\subsection{Proof of Proposition~\ref{prop:AP}}\label{sec:dminp_ana}
We now derive the upper bound for the minimum product distance of the superimposed constellation $d_{p,\min}(\eta(\mathcal{C}-\mathbf{d}))$. 
In what follows, we first prove in Lemma~\ref{the:1_CH7} that the $n$-dimensional superimposed constellation is an $n$-fold cartesian product of a one-dimensional \emph{superimposed} constellation. Then, we show in Lemma~\ref{the:2a} that $d_{p,\min}(\eta(\mathcal{C}-\mathbf{d}))$ can be upper bounded by the minimum product distance of this one-dimensional superimposed constellation. With these lemmas, we then bound the minimum product distance of $d_{p,\min}(\eta(\mathcal{C}-\mathbf{d}))$ by analyzing the minimum product distances of the one-dimensional superimposed constellation. 

\begin{lemm}\label{the:1_CH7}
Consider the constellation $\eta(\mathcal{C}-\mathbf{d})$ defined in \eqref{eq:sup1} for $\alpha \in [0,1]$ and the base lattice $\Lambda$ is equivalent to $\mathbb{Z}^n$. The  constellation $\eta(\mathcal{C}-\mathbf{d})$ is the rotated $n$-fold Cartesian product of the one-dimensional constellation $\eta(\sqrt{\alpha}(\mathcal{X}_1-d^*_1)+\sqrt{1-\alpha}(\mathcal{X}_2-d^*_2))$ in \eqref{eq:1dsup}.
\end{lemm}

\emph{\quad Proof: }
Following \eqref{eq:sup1}, we write the superimposed constellation as
\begin{align}
&\eta(\mathcal{C}-\mathbf{d}) = \eta(\sqrt{\alpha}(\mathcal{C}_1-\mathbf{d}_1)+\sqrt{1-\alpha}(\mathcal{C}_2-\mathbf{d}_2)) \nonumber \\
 \overset{(a)}=&  \{\eta(\sqrt{\alpha}(\mathbf{b}_1\mathbf{G}_{\mathbb{Z}^n}-\mathbf{d}^*_1)\mathbf{R}+\sqrt{1-\alpha}(\mathbf{b}_2\mathbf{G}_{\mathbb{Z}^n}-\mathbf{d}^*_2)\mathbf{R} )\} \nonumber \\
 \overset{(b)}=& \{\eta(\sqrt{\alpha}(\mathbf{b}_1-\mathbf{d}^*_1)+\sqrt{1-\alpha}(\mathbf{b}_2-\mathbf{d}^*_2))\mathbf{R} \} \nonumber \\
 \overset{(c)} =& (\eta(\sqrt{\alpha}(\mathcal{X}_1-d^*_1[1])+\sqrt{1-\alpha}(\mathcal{X}_2-d^*_2[1]))\times \nonumber \\ & \eta(\sqrt{\alpha}(\mathcal{X}_1-d^*_1[2])+\sqrt{1-\alpha}(\mathcal{X}_2-d^*_2[2]))\times \ldots \times \nonumber \\
&  \eta(\sqrt{\alpha}(\mathcal{X}_1-d^*_1[n])+\sqrt{1-\alpha}(\mathcal{X}_2-d^*_2[n])))\mathbf{R}  \nonumber \\
 \overset{(d)} =& (\eta(\sqrt{\alpha}(\mathcal{X}_1-d^*_1)+\sqrt{1-\alpha}(\mathcal{X}_2-d^*_2))\times \eta(\sqrt{\alpha}(\mathcal{X}_1-d^*_1)+\sqrt{1-\alpha}(\mathcal{X}_2-d^*_2)) \nonumber \\
&  \times \ldots \times  \eta(\sqrt{\alpha}(\mathcal{X}_1-d^*_1)+\sqrt{1-\alpha}(\mathcal{X}_2-d^*_2)))\mathbf{R},
\end{align}
where $(a)$ follows that $\mathcal{C}_k-\mathbf{d}_k$ is obtained by multiplying the dithered coset leaders of $\mathbb{Z}^n/2^{m_k}\mathbb{Z}^n$ to the rotational matrix $\mathbf{R}$ while these coset leaders are generated by $(\mathbf{b}_k\mathbf{G}_{\mathbb{Z}^n} - \mathbf{d}^*_k)$ with $\mathbf{b}_k = [b_k[1],b_k[2],\ldots,b_k[n]] \in \mathbb{Z}^n$ and $\mathbf{d}^*_k = \E[\{\mathbf{b}_k\mathbf{G}_{\mathbb{Z}^n} \}]$ for $k = 1,2$; $(b)$ follows that since $\mathbf{G}_{\mathbb{Z}^n} = \mathbf{I}_n$, thus $\mathbf{b}_k \in \{\boldsymbol{\lambda} \; \text{mod} \; 2^{m_k}\mathbb{Z}^n, \boldsymbol{\lambda} \in \mathbb{Z}^n\}$ and $\mathbf{d}^*_k = \E[\{\mathbf{b}_k \}]$ for $k = 1,2$; $(c)$ is due to that each component of $\mathbf{b}_k$ follows $b_k[i] \in \mathcal{X}_k = \{\lambda \; \text{mod} \; 2^{m_k}\mathbb{Z}, \lambda \in \mathbb{Z}\}$ for $k =1,2$ and $i = 1,\ldots,n$ because $\mathbf{b}_k$, the coset leader of $\mathbb{Z}^n/2^{m_k}\mathbb{Z}^n$, is precisely the $n$-fold Cartesian product of the coset leader of $\mathbb{Z}/2^{m_k}\mathbb{Z}$;
and $(d)$ follows that $d^*_k[1]= d^*_k[2]= \ldots = d^*_k[n] = d^*_k$ for $k = 1,2$ because
\begin{align}
[d^*_k[1],d^*_k[2],\ldots,d^*_k[n] ]&= \E[\{\mathbf{b}_k \}] \nonumber \\
&=  [\E[\{b_k[1] \}],\E[\{b_k[2] \}],\ldots,\E[\{b_k[n] \}]] \nonumber \\
&\overset{(c)} = \underbrace{[\E[\mathcal{X}_k],\E[\mathcal{X}_k],\ldots,\E[\mathcal{X}_k]]}_{\text{length~$n$}} = \underbrace{[d^*_k,d^*_k,\ldots,d^*_k]}_{\text{length~$n$}}.
\end{align}
%
Thus, $\eta(\mathcal{C}-\mathbf{d})$ is the $n$-fold Cartesian product of one-dimensional constellation $\eta(\sqrt{\alpha}(\mathcal{X}_1-d^*_1)+\sqrt{1-\alpha}(\mathcal{X}_2-d^*_2))$ with rotation. \QEDA

With Lemma \ref{the:1_CH7}, we prove an upper bound on $d_{p,\min}(\eta(\mathcal{C}-\mathbf{d}))$ in the following.

\begin{lemm}\label{the:2a}
Consider a normalized and dithered superimposed constellation $\eta(\mathcal{C}-\mathbf{d})$ defined in \eqref{eq:sup1} for $\alpha \in [0,1]$ and $\Lambda$ is equivalent to $\mathbb{Z}^n$. The minimum product distance of $\eta(\mathcal{C}-\mathbf{d})$ is upper bounded by
\begin{align}
d_{p,\min}(\eta(\mathcal{C}-\mathbf{d})) \leq d_{p,\min}(\eta(\mathcal{X}-d^*)\mathbf{R}^*),
\end{align}
where $\eta(\mathcal{X}-d^*)$ is the one-dimensional constellation defined in \eqref{eq:1dsup}; and $\mathbf{R}^*$ is an $n \times n$ rotation matrix such that $d_{p,\min}(\eta(\mathcal{X}-d^*)\mathbf{R}^*) = d_{p,\min}(\Lambda)$ when $\alpha = 0$ or 1.

%
\end{lemm}

\emph{\quad Proof: }
%
%
Given the definition of layer in Section \ref{def:layer} and based on Lemma \ref{the:1_CH7}, it is worth noting that all the layers have the same Euclidean distance profiles and thus their minimum Euclidean distances, denoted by $d_{E,\min}((\mathcal{X}-d^*)\mathbf{R})$, are the same regardless of any rotation $\mathbf{R}$. Thus, we have
\begin{align}\label{eq:dmin_1}
d_{E,\min}((\mathcal{X}-d^*)\mathbf{R}) = d_{E,\min}(\mathcal{X}-d^*).
\end{align}

When $\alpha = 0$ or 1, the superimposed constellation becomes a single user's constellation. In this case, the following relationship always holds
\begin{align}
d_{p,\min}((\mathcal{X}-d^*)\mathbf{R}) &\geq d_{p,\min}(\Lambda) = d_{p,\min}(\mathcal{C}-\mathbf{d}), \\
d_{E,\min}((\mathcal{X}-d^*)\mathbf{R}) &\overset{\eqref{eq:dmin_1}}= d_{E,\min}(\mathcal{X}-d^*) = d_{E,\min}(\Lambda), \alpha \in \{ 0,1\}.
\end{align}
Based on the above relationships and Lemma \ref{lem:dmindp} in Appendix~\ref{appendix:lemma_ch7}, there exists at least one layer such that the minimum product distance of this layer satisfies
\begin{align}\label{eq:pd_lambda_1}
d_{p,\min}((\mathcal{X}-d^*)\mathbf{R}^*) = d_{p,\min}(\Lambda) = d_{p,\min}(\mathcal{C}-\mathbf{d}), \alpha \in \{ 0,1\},
\end{align}
for some rotation matrix $\mathbf{R}^*$. By using \eqref{eq:dmin_1}-\eqref{eq:pd_lambda_1} and the relationship between minimum product distances in two different layers established in Lemma \ref{the:1a} in Appendix \ref{appendix:lemma_ch7}, we conclude that
\begin{align}
d_{p,\min}((\mathcal{X}-d^*)\mathbf{R}^*) \leq d_{p,\min}((\mathcal{X}-d^*)\mathbf{R}).
\end{align}

Now, we denoted by $d_{p,\min}(\mathcal{L})$ the minimum of the set of all product distances between all pairs of two distinct constellation points in any two \emph{different} layers. It is obvious that
\begin{align}
d_{p,\min}(\eta(\mathcal{C}-\mathbf{d})) = \min\{d_{p,\min}(\eta\mathcal{L}) ,d_{p,\min}(\eta(\mathcal{X}-d^*)\mathbf{R}^*)\} 
\leq d_{p,\min}(\eta(\mathcal{X}-d^*)\mathbf{R}^*),
\end{align}
where the normalize factor $\eta$ does not affect the equality and inequality here. \QEDA

With the upper bound in Lemma \ref{the:2a}, we now restrict the problem of bounding the minimum product distance of an $n$-dimensional constellation to analyzing the intra-layer minimum product distance $d_{p,\min}(\eta(\mathcal{X}-d^*)\mathbf{R}^*)$. This approach turns out to be sufficient for our purpose as it captures the trends of the change of the $d_{p,\min}(\eta(\mathcal{C}-\mathbf{d}))$ and fits perfectly with many local maximum values, as already shown in Example \ref{exa:dpmin}. 

\begin{remark}\label{the:2}
When analyzing the minimum product distance of the superimposed constellation $\eta(\mathcal{C}-\mathbf{d})$, we only need to analyze the case for $\alpha \in [0,\frac{1}{2}]$. Specifically, the superimposed constellation  $\eta(\sqrt{\alpha}(\mathcal{C}_1-\mathbf{d}_1)+\sqrt{1-\alpha}(\mathcal{C}_2-\mathbf{d}_2))$ for $\alpha \in [\frac{1}{2},1]$ is equivalent to $\eta(\sqrt{\alpha'}(\mathcal{C}_2-\mathbf{d}_2)+\sqrt{1-\alpha'}(\mathcal{C}_1-\mathbf{d}_1))$ for $\alpha'  = 1 - \alpha \in [\frac{1}{2},0]$. Thus, the later case is analyzed when we let $m'_1 = m_2$ and $m'_2=m_1$ such that $\eta(\mathcal{C}-\mathbf{d}) = \eta(\sqrt{\alpha'}(\mathcal{C}'_1-\mathbf{d}'_1)+\sqrt{1-\alpha'}(\mathcal{C}'_2-\mathbf{d}'_2))$, where $\mathcal{C}'_k$ corresponds to the complete set of coset leaders of $\Lambda/2^{m'_k}\Lambda$ and $\mathbf{d}'_k = \E[\mathcal{C}'_k]$ for $k = 1,2$.
\end{remark}

Based on the definitions given in Section \ref{sec:def}, the intra-layer minimum product distance is
\begin{align}
d_{p,\min}(\eta(\mathcal{X}-d^*)\mathbf{R}^*) 
= \min \{d_{p,\min1},d_{p,\min2}, d_{p,\min}^{\text{Cl}}\}.
\end{align}
Since $d_{p,\min1}$ and $d_{p,\min2}$ can be easily computed as in \eqref{eq:dp1} and \eqref{eq:dp2}, respectively, what is left is to analyze $d_{p,\min}^{\text{Cl}}$. To perform the analysis, we first consider the case of $m_1 \in \mathbb{Z}^+$ and $m_2=1$ and then use the result to analyze the general case of $m_1,m_2 \in \mathbb{Z}^+$.

\emph{1) Case I:} ($m_2=1$)
For this case, there are two clusters, each of which contains $2^{m_1}$ number of constellation points. 
Before the constellation points from two clusters start to overlap, the inter-cluster minimum product distance is the product distance between two constellation points at the edge of each cluster. This scenario is illustrated in the example shown in Fig. \ref{fig:2_ch7}. The inter-cluster minimum product distance is given by
\begin{align}\label{eq:case1a}
\sqrt[n]{d_{p,\min}^{\text{Cl}}} =\sqrt[n]{d_{p,\min2}} -(2^{m_1}-1)\sqrt[n]{d_{p,\min1}},
\end{align}
where we have used the relationship of product distances in two line segments in $\mathbb{R}^n$ established in Lemma \ref{lem:1} in Appendix \ref{appendix:lemma_ch7}. We emphasize that Lemma \ref{lem:1} will be frequently used in the rest of the proof. Since $d_{p,\min 1} \leq d_{p,\min 2}$ for $\alpha \in [0,\frac{1}{2}]$ according to \eqref{eq:dp1} and \eqref{eq:dp2}, the intra-layer minimum product distance is thus determined by comparing $d_{p,\min}^{\text{Cl}}$ and $d_{p,\min 1}$. To have $d_{p,\min}^{\text{Cl}} \geq d_{p,\min 1}$, the necessary condition to satisfy this inequality is
\begin{align}
&\sqrt[n]{d_{p,\min2}} -(2^{m_1}-1)\sqrt[n]{d_{p,\min1}} \geq \sqrt[n]{d_{p,\min 1}} \nonumber \\
\Rightarrow & \; \sqrt[n]{(\sqrt{1-\alpha})^n}  \geq (2^{m_1}-1)\sqrt[n]{(\sqrt{\alpha})^n}  
\Rightarrow \; \alpha \leq \frac{1}{1+2^{2m_1}}
\end{align}

Thus, when $\alpha_1 \in [0,\frac{1}{1+2^{2m_1}}]$, we have
\begin{align}\label{eq:b4c}
d_{p,\min}(\eta(\mathcal{C}-\mathbf{d})) \leq d_{p,\min}(\eta(\mathcal{X}-d^*)\mathbf{R}^*) = d_{p,\min1}.
\end{align}

Then, for $\alpha \in (\frac{1}{1+2^{2m_1}},\frac{1}{2})$, the intra-layer minimum product distance becomes the inter-cluster minimum product distance such that
\begin{align}
d_{p,\min}(\eta(\mathcal{X}-d^*)\mathbf{R}^*) \overset{(a)}= d_{p,\min}^{\text{Cl}},
\end{align}
where $(a)$ follows that $d_{p,\min}^{\text{Cl}} < d_{p,\min1} < d_{p,\min2}$ for $\alpha \in (\frac{1}{1+2^{2m_1}},\frac{1}{2})$. Thus, we can now focus on analyzing $d_{p,\min}^{\text{Cl}}$ for this range.

To simplify the description for the subsequent analysis, we label two clusters as clusters 1 and 2, respectively, from the left to the right of a layer. Moreover, we refer to the Voronoi cell of an element with respect to the underlying rotated $\mathbb{Z}$ lattice in cluster 1 as a \emph{cell} of cluster 1. For each cluster, there are $(2^{m_1}-1)$ cells which are labelled cell $1$ to $(2^{m_1}-1)$, respectively, from the left to the right of a cluster. With $\alpha$ increasing, two clusters are moving toward each other. When the left constellation point on cell $(2^{m_1}-1)$ in cluster 1 overlaps with the right constellation point on cell 1 in cluster 2, $d_{p,\min}^{\text{Cl}} = 0$. From \eqref{eq:case1a}, this happens when
$\alpha = \frac{1}{(2^{m_1}-1)^2+1}$.
After the overlapping, the inter-cluster minimum product distance is bounded by
\begin{align}\label{eq:con1}
\sqrt[n]{d_{p,\min}^{\text{Cl}}}  \leq \frac{1}{2}\sqrt[n]{d_{p,\min1}},
\end{align}
where $\frac{1}{2}$ is due to the fact that the maximum of the inter-cluster product distance happens when a constellation point from cluster 2 is located in the center of a cell in cluster 1.

\begin{figure}[ht!]
	\centering
\includegraphics[width=4.7in,clip,keepaspectratio]{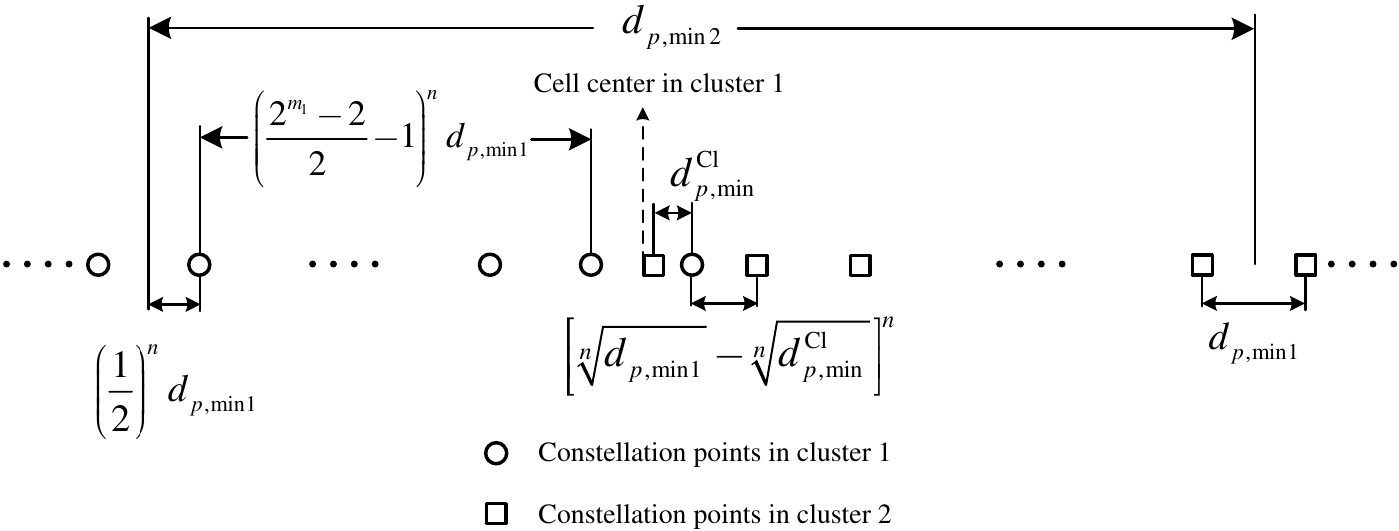}

\caption{An example of a layer in Case I.}
\label{fig:ex1}
\end{figure}

Consider the scenario where the leftmost constellation point of cell 1 of cluster 2 is in between the center and the right edge of cell $(2^{m_1}-1)$ in cluster 1. To have a clear view on this, we plot this scenario in Fig. \ref{fig:ex1}.

By counting the number of cells within clusters and inspecting the relationship between different product distances as shown in Fig. \ref{fig:ex1}, the inter-cluster minimum product distance is derived as
\begin{align}\label{eq:case1a2}
&2\left(\frac{1}{2}\sqrt[n]{d_{p,\min1}}\right)+2\left(\frac{2^{m_1}-2}{2}-1\right)\sqrt[n]{d_{p,\min1}} \nonumber \\
&+2\left(\sqrt[n]{d_{p,\min1}}-\sqrt[n]{d_{p,\min}^{\text{Cl}}}\right)+\sqrt[n]{d_{p,\min}^{\text{Cl}}} = \sqrt[n]{d_{p,\min2}} \nonumber \\
& \Rightarrow  \; d_{p,\min}^{\text{Cl}} =\left( (2^{m_1}-1)\sqrt[n]{d_{p,\min1}} - \sqrt[n]{d_{p,\min2}}\right)^n.
\end{align}

Similarly, for the case where the left constellation point in cell 1 of cluster 2 is located in between the center of cell $(2^{m_1}-l)$ and cell $(2^{m_1}-l+1)$ in cluster 1, the inter-cluster minimum product distance is
\begin{align}\label{eq:con2a}
d_{p,\min}^{\text{Cl}} =\left| \sqrt[n]{d_{p,\min2}} -(2^{m_1}-l)\sqrt[n]{d_{p,\min1}}\right|^n, \; l = 2,\ldots,2^{m_1}-2.
\end{align}
By combining the right hand side of \eqref{eq:con1}, \eqref{eq:con2a}, the boundary of $\alpha$ corresponding to the $d_{p,\min}^{\text{Cl}}$ in \eqref{eq:con2a} can be computed as
\begin{align}
\alpha = \frac{1}{(2^{m_1}+\frac{1}{2}-l)^2+1}, \; l = 2,\ldots,2^{m_1}-2.
\end{align}
Summarizing the above results, we obtain the inter-cluster minimum product distance for case I in \eqref{eq:case1} for $\alpha \in (\frac{1}{1+2^{2m_1}},\frac{1}{2}]$.

\emph{2) Case II:} ($m_2\geq 1$)
First, it is obvious that the intra-layer minimum product distance is the same as in \eqref{eq:b4c} of Case I when $\alpha_1 \in [0,\frac{1}{1+2^{2m_1}}]$. However, \eqref{eq:case1} does not hold anymore when multiple clusters start to intercept. Since there are $2^{m_2}$ clusters, different constellation points from multiple clusters can be located in a cell of any cluster. Similar to Case I, we label all the clusters as $1,\ldots,2^{m_2}$ from the left cluster to the right cluster in a layer to simplify the description in the following analysis.

We denote by $\xi$ the number of clusters intercept with \emph{each other}, i.e., there are $\xi-1$ different constellation points from $ \xi-1$ different clusters, respectively, intercept with the cells of cluster 1. When $\xi = 2$, the scenario becomes identical to Case I and the same analysis on the minimum product distance applies. For $\xi \geq 3$, the inter-cluster minimum product distance takes into account that the cells of clusters $1,\ldots,\xi$ intercepting with each other. Thus, it can be bounded by
\begin{align}\label{eq:con3}
\sqrt[n]{d_{p,\min}^{\text{Cl}}(\xi)}   \leq \frac{1}{\xi} \sqrt[n]{d_{p,\min1}},
\end{align}
where $\frac{1}{\xi}$ comes from the same reason that we have $\frac{1}{2}$ in \eqref{eq:con1}. 
Now consider any two clusters $s,w \in \{2,\ldots,\xi \}$ and we assume $s>w$ without loss of generality. In the following, for the ease of presentation, we refer to $d_{p,\min}^{\text{Cl}}$ in case I as $d_{p,\min}^{\text{Cl}(m_2=1)}$. It can be easily seen that when $s-w = 1$, $d_{p,\min}(\text{Cl}_w,\text{Cl}_s)$ and $d_{p,\min}^{\text{Cl}(m_2=1)}$ coincide. For $s-w > 1$, to determine $d_{p,\min}(\text{Cl}_w,\text{Cl}_s)$, we first need to find a set of product distances between cluster 1 and $j$ for $j\in \{s,w\}$ as follows. 
Suppose that there are $F_j$ constellation points from cluster $j$ that have intercepted with the cells of cluster 1. By applying Lemma~\ref{lem:1} multiple times, we obtain the product distance between a point in cluster 1 and the $(f_j+1)$-th constellation point from cluster $j$ intercepting with the cells of cluster 1 (called $\text{Cl}_j(f_j)$) as
\begin{align}\label{eq:con4}
&\sqrt[n]{d_{p}(\text{Cl}_1,\text{Cl}_j(f_j))} = (j-1)\sqrt[n]{d_{p,\min}^{\text{Cl}(m_2=1)}} +f_j\sqrt[n]{d_{p,\min 1}}, \nonumber \\
&f_j \in \{ 0,\ldots,F_j-1 \},j \in \{s,w\},
\end{align}
The minimum product distance $d_{p,\min}(\text{Cl}_w,\text{Cl}_s)$ is then computed as
\begin{align}\label{eq:dpij}
&d_{p,\min}(\text{Cl}_w,\text{Cl}_s) \overset{(a)}{=} \min\limits_{\substack{f_w \in \{ 0,\ldots,F_w-1 \}, \\ f_s \in \{ 0,\ldots,F_s-1 \}}}\left\{\left|\sqrt[n]{d_{p}(\text{Cl}_1,\text{Cl}_w(f_w))} -\sqrt[n]{d_{p}(\text{Cl}_1,\text{Cl}_s(f_s))} \right|^n \right\}\nonumber \\
&\overset{\eqref{eq:con4}} = \min\limits_{\substack{f_w \in \{ 0,\ldots,F_w-1 \}, \\ f_s \in \{ 0,\ldots,F_s-1 \}}} \left\{\left|(f_w-f_s)\sqrt[n]{d_{p,\min 1}} +(w-1)\sqrt[n]{d_{p,\min}^{\text{Cl}(m_2=1)}} 
- (s-1)\sqrt[n]{d_{p,\min}^{\text{Cl}(m_2=1)}}\right|^n \right\} \nonumber \\
& \overset{(b)}= \min\limits_{\gamma_{ws}\in \{0,\ldots,\lfloor\frac{s-w}{2} \rfloor\} }\left\{ \left|\gamma_{ws}\sqrt[n]{d_{p,\min1}}  
-(s-w)\sqrt[n]{d_{p,\min}^{\text{Cl}(m_2=1)}}\right|^n  \right\},
\end{align}
where $(a)$ follows from Lemma~\ref{lem:1} and $(b)$ follows from the fact that $\gamma_{ws} $ is the spacing between cluster $s$ and $w$ in terms of $\sqrt[n]{d_{p,\min1}}$  and $\lfloor\frac{s-w}{2} \rfloor$ is the maximum spacing because
\begin{align}
(s-w)\sqrt[n]{d_{p,\min}^{\text{Cl}(m_2=1)}} \overset{\eqref{eq:con1}}\leq  \frac{s-w}{2}\sqrt[n]{d_{p,\min 1}}.
\end{align}

For $\xi \in \{2, \ldots,2^{m_2}\}$, the inter-cluster minimum product distance for the scenario where the cells of clusters $1,\ldots,\xi$ intercept with each other, is obtained by finding the minimum of the product distances based on all combinations of clusters $s$ and $w$
\begin{align}
d_{p,\min}^{\text{Cl}}(\xi)&= \min_{\substack{s,w \in \{1,\ldots,\xi \},\\ s > w}} \left\{ d_{p,\min}(\text{Cl}_w,\text{Cl}_s) \right\}\nonumber \\
 &= \min_{\substack{w,s \in \{1,\ldots,\xi \},s > w \\ \gamma_{ws}\in \{0,\ldots,\lfloor\frac{s-w}{2} \rfloor\}}} \left\{\left|\gamma_{ws}\sqrt[n]{d_{p,\min 1}}
-(s-w)\sqrt[n]{d_{p,\min}^{\text{Cl}(m_2=1)}}\right|^n  \right\} \nonumber \\
& \overset{(a)}= \min\limits_{\substack{\gamma \in \{ 0,\ldots,\lfloor\frac{\xi-1}{2} \rfloor \}, \\ \beta \in \{1,\ldots,\xi-1\}}} \left\{\left|\gamma\sqrt[n]{d_{p,\min 1}}
-\beta\sqrt[n]{d_{p,\min}^{\text{Cl}(m_2=1)}}\right|^n  \right\},
\end{align}
where $(a)$ follows from that $1 \leq s-w \leq \xi -1$.

The only thing left is to find the  boundary of $\alpha$, called $\alpha(\xi)$, such that when $\alpha \geq \alpha(\xi)$, the bound in \eqref{eq:con3} is valid. This happens when the minimum product distance between cluster 1 and cluster $\xi$ satisfies the condition of $\sqrt[n]{d_{p,\min}(\text{Cl}_1,\text{Cl}_{\xi})}\leq \frac{1}{2}\sqrt[n]{d_{p,\min1}}$. Otherwise, the above scenario is reduced to the scenario of the cells of clusters $1,\ldots,(\xi-1)$ intercepting with each other because $\sqrt[n]{d_{p,\min}(\text{Cl}_1,\text{Cl}_{\xi})}> \frac{1}{2}\sqrt[n]{d_{p,\min1}} \geq \sqrt[n]{d_{p,\min}^{\text{Cl}(m_2=1)}}$ leads to
\begin{align}
d_{p,\min}^{\text{Cl}}(\xi) &= \min \{d_{p,\min}(\text{Cl}_1,\text{Cl}_{\xi}),\ldots, d_{p,\min}(\text{Cl}_{\xi-1},\text{Cl}_{\xi})\} \nonumber \\
&\overset{(a)}= \min \{d_{p,\min}(\text{Cl}_{2},\text{Cl}_{\xi}),\ldots, d_{p,\min}(\text{Cl}_{\xi-1},\text{Cl}_{\xi})\} \nonumber \\
& \overset{(b)}= \min \{d_{p,\min}(\text{Cl}_{1},\text{Cl}_{\xi-1}),\ldots, d_{p,\min}(\text{Cl}_{\xi-2},\text{Cl}_{\xi-1})\} \nonumber \\
&= d_{p,\min}^{\text{Cl}}(\xi-1),
\end{align}
where $(a)$ follows from that $d_{p,\min}(\text{Cl}_{\xi-1},\text{Cl}_{\xi}) = d_{p,\min}^{\text{Cl}(m_2=1)}<d_{p,\min}(\text{Cl}_{1},\text{Cl}_{\xi})$ and $(b)$ follows from \eqref{eq:dpij} that $d_{p,\min}(\text{Cl}_{w_2},\text{Cl}_{s_2}) = d_{p,\min}(\text{Cl}_{w_2},\text{Cl}_{s_2})$ if $s_1 - w_1 = s_2-w_2$ for any $s_1,w_1,s_2,w_2 \in \{1,\ldots,\xi \}$. Thus, the corresponding $\alpha(\xi)$ is derived by using the above condition as
\begin{align}
&(\xi-1)\sqrt[n]{d_{p,\min 2}} = \left(2^{m_1}-\frac{1}{2}\right)\sqrt[n]{d_{p,\min1}} \nonumber \\
&\Rightarrow  \alpha(\xi) = \frac{(\xi-1)^2}{\left(2^{m_1}-\frac{1}{2} \right)^2+(\xi-1)^2}.
\end{align}

\par\noindent
Note that we only needs to look at $\alpha(\xi) \leq \frac{1}{2}$ according to Remark \ref{the:2}. This completes the proof.

\section{Extension to MIMO-NOMA}\label{MIMO}
In this part, we extend the main idea and analysis to MIMO-NOMA over block fading channels for constructing good MIMO-NOMA schemes without SIC. We restrict our attention to a very popular class of codes for MIMO channel named OSTBC. Some advantages of using OSTBC include achieving full transmit diversity and efficiently detection by turning the MIMO channel into a set of non-interfering parallel subchannels. We note that a scheme of NOMA with two transmit antennas and one receive antenna for each user combined with Alamouti code \cite{730453} has been reported in \cite{8392409} where the closed form expressions for outage probabilities under Nakagami-m fading channels are derived. However, the analysis is based on Gaussian inputs. In this section, we adapt the techniques used in the previous section to analyze the error performance of MIMO-NOMA scheme with general OSTBC \cite{Vucetic:2003:SC:861866}.

\subsection{MIMO-NOMA system model}
Consider a two-user MIMO-NOMA where the base station and each user have $M_t$ and $M_r$ sufficient-spacing antennas, respectively. We again assume that the transmitter has statistical CSI while the receiver has full CSI for its own channel. The base station encodes all users' messages $\mathbf{u}_1,\mathbf{u}_2$ into a superimposed codeword $\mathbf{X} = [\mathbf{x}_1,\ldots,\mathbf{x}_T] \in \mathbb{C}^{M_t \times T}$ from the codebook $\mathcal{G}$ and broadcasts it to each user, where $T$ means the codeword spreads $T$ time slots and $\sum_{i=1}^T\E[\|\mathbf{x}_i\|^2]\leq T$ for $i = 1,\ldots,T$. We denote by $\mathbf{H}_k \in \mathbb{C}^{M_r \times M_t}$ the channel matrix for user $k \in \{1,2\}$ with i.i.d. entries. Here, we assume that $\mathbf{H}_k$ is constant during one codeword block while a transmit packet contains multiple blocks. The received signal at user $k$ for $T$ time slots is denoted by $\mathbf{Y}_k \in \mathbb{C}^{M_r \times T}$ and is given by
\begin{align}\label{STBC_Y}
\mathbf{Y}_k = \sqrt{P}\mathbf{H}_k\mathbf{X}+\mathbf{Z}_k,
\end{align}
where $P$ is the total power constraint at the base station and $\mathbf{Z}_k \in \mathbb{C}^{M_r \times T}$ is a circular-symmetric AWGN experienced at user $k$ with i.i.d. entries $\sim\mc{CN}(0,1)$.

The reliability is again measured by PEP. Consider the channels $\mathbf{H}_k$ with i.i.d. entries $h_{j,i}^{(k)}\sim\mc{CN}(0,\sigma_k^2)$. Following \cite[Chapter 2.5.1]{Vucetic:2003:SC:861866}, for any two codewords $\mathbf{X}_s,\mathbf{X}_w \in \mathcal{G}$ and $\mathbf{X}_s \neq \mathbf{X}_w$, user $k$'s error probability is upper bounded by its average PEP given by
\begin{align}\label{PEP_MIMO_NOMA}
P_e^{(k)} &\leq \frac{1}{|\mathcal{G}|}\sum_{s \neq w}\text{Pr}(\mathbf{X}_s \rightarrow \mathbf{X}_w|\mathbf{H}_k) \nonumber \\
&\leq \frac{1}{|\mathcal{G}|}\sum_{s \neq w}\det\left(\mathbf{I}_{M_t}+\overline{\SNR}_k\frac{(\mathbf{X}_s-\mathbf{X}_w)(\mathbf{X}_s-\mathbf{X}_w)^{\dag}}{4}\right)^{-M_r} \nonumber \\
&\leq \frac{1}{|\mathcal{G}|}\sum_{s \neq w} \prod_{j=1}^{r}\left(1+ \overline{\SNR}_k\frac{\phi_j}{4}\right)^{-M_r}  \nonumber \\ 
&< \left(\min\limits_{s \neq w}\left\{\prod\nolimits_{j=1}^{\min\limits_{s\neq w}\{r\}} \phi_j\right\}\right)^{-Mr}\left(\frac{4}{\overline{\SNR}_k}\right)^{\min\limits_{s\neq w} \{r\}M_r}  ,
\end{align}
where $(.)^{\dag}$ denotes the conjugate transpose, $\overline{\SNR}_k \triangleq \E[\text{tr}(\mathbf{H}_k\mathbf{H}_k^{\dag})]P$ is user $k$'s average SNR; $\{ \phi_j:j = 1,\ldots,r\}$ are the non-zero eigenvalues of $\Delta\Delta^{\dag}$ with $\Delta \triangleq \mathbf{X}_s-\mathbf{X}_w$ being the codeword difference matrix with $r = \text{rank}(\Delta)$. The diversity order of $\mathbf{X}$ is $\min\limits_{s\neq w} \{r\}M_r$. For $T \geq M_t$, the code has full rank such that $\min\limits_{s\neq w} \{r\} = M_t$ and $\prod_{j=1}^{M_t}\phi_j = \det(\Delta\Delta^{\dag})$. In this case, the code achieves full diversity, i.e., the diversity order is $M_tM_r$. To further minimize the PEP, it is important to maximize the \emph{minimum determinant} $\min\limits_{s\neq w}\{\det(\Delta\Delta^{\dag})\}$. It is noteworthy that the design criterion is generalizable to other fading channels, e.g., Rician fading \cite{Vucetic:2003:SC:861866,1256737}. Without loss of generality, we assume that $\overline{\SNR_1}\geq \overline{\SNR_2}$ and user 1 is considered as the strong user. Note that this user ordering is also adopted in \cite{8392409}.

%
%

\subsection{Proposed Scheme and Main Result}\label{sec:MIMO_scheme}

We first briefly describe the scheme of space-time coded MIMO-NOMA in the following.

\subsection{Transmitter Side}\label{MIMO_NOMA_TX}
A superimposed signal sequence $[x[1],\ldots,x[M_t]]$ is encoded into a OSTBC codeword $\mathbf{X}\hspace{-1mm} \in\hspace{-1mm} \mathbb{C}^{M_t \times T}$, where $x[l] \hspace{-1mm}\in \hspace{-1mm} \eta_T(\mathcal{C}-\mathbf{d})$ can be expressed by \eqref{eq:sup1} for $n\hspace{-1mm}=\hspace{-1mm}2 ,l = 1,\ldots,M_t$ and $\eta_T = \tau \eta$ is applied to $\mathbf{X}$ to ensure $\sum_{i=1}^T\E[\|\mathbf{x}_i\|^2]\leq T$. Here $\tau$ is an additional normalization for the space-time code on top of the normalization of the superimposed constellation $\eta$ and it depends on the specific space-time code (for example, Alamouti code has $\tau = 1$). 

\subsection{Receiver Side}
Upon receiving $\mathbf{Y}_k$ given in \eqref{STBC_Y}, the maximum-ratio combining and the space-time decoding are employed for decoding. By using the orthogonality of pairwise rows of the transmission matrix \cite[Chapter 3.6]{Vucetic:2003:SC:861866}, the decoder attempts to minimize the following metric
\begin{align}
\arg\min_{\tilde{x}[l] \in \mathcal{G}} \left\{|\tilde{x}[l]-x[l]|^2+\left(\sum\nolimits_{i = 1}^{M_t}\sum\nolimits_{j = 1}^{M_r}|h_{j,i}^{(k)}|^2-1\right)|x[l]|^2 \right\}, l \in \{1,\ldots,M_t\},
\end{align}
where $\tilde{x}[l]$ is the estimated superimposed symbol of $x[l]$. If single-user decoding is adopted, each user directly decodes their own messages 
from $\tilde{x}[l]$ for $l = 1,\ldots,M_t$ in a symbol-wise manner. For user 1 with SIC user 2's message will be decoded from $\tilde{x}[l]$ first and the corresponding codeword will be re-encoded and subtracted from the received signal.

For lattice-partition based MIMO-NOMA scheme, the superimposed signal $x[l] \in  \eta_T'(\mathcal{C}-\mathbf{d}')$ with $\eta'_T = \tau \eta'$ can be expressed by \eqref{eqn:x_def_Ch7}, which corresponds to the lattice partition chain described in Section \ref{Sec:LP}. As a result, many nice properties of the underlying lattice carry over to the individual and the superposition coded space-time codewords.

The analytical expression of the minimum Euclidean distance is similar to that of the minimum product distance given in \eqref{eq:case2}-\eqref{eq:case1}. We summarize the main result here and the proof is presented in Chapter~\ref{sec:MIMO_proof}.
\begin{prop}\label{coro:dmin}
For arbitrary power allocation, the minimum determinant is $\min\{\det(\Delta\Delta^{\dag})\} = d_{E,\min}(\eta_T(\mathcal{C}-\mathbf{d}))^{2M_t}$, where $ d_{E,\min}(\eta_T(\mathcal{C}-\mathbf{d}))$ is obtained by replacing $d_{E,\min1}$ from \eqref{eq:dmin1} to $d_{p,\min1}$ and $d_{E,\min2}$ from \eqref{eq:dmin2} to $d_{p,\min2}$ and substituting them into \eqref{eq:case2}-\eqref{eq:case1} and setting $n=1$. 
For the lattice-partition based scheme, the minimum determinant is $\min\{\det(\Delta\Delta^{\dag})\} = d_{E,\min}(\eta_T'(\mathcal{C}'-\mathbf{d}'))^{2M_t}$, where $d_{E,\min}(\eta_T'(\mathcal{C}'-\mathbf{d}')) = \tau\eta'd_{E,\min}(\Lambda)$ and $\eta'$ is given in \eqref{normalize_LP}.
\end{prop}

We would like to emphasize that although we present the results for the complex OSTBC over complex MIMO setting, the above results are valid for real OSTBC.
\begin{figure}[ht!]
	\centering
\includegraphics[width=3.5in,clip,keepaspectratio]{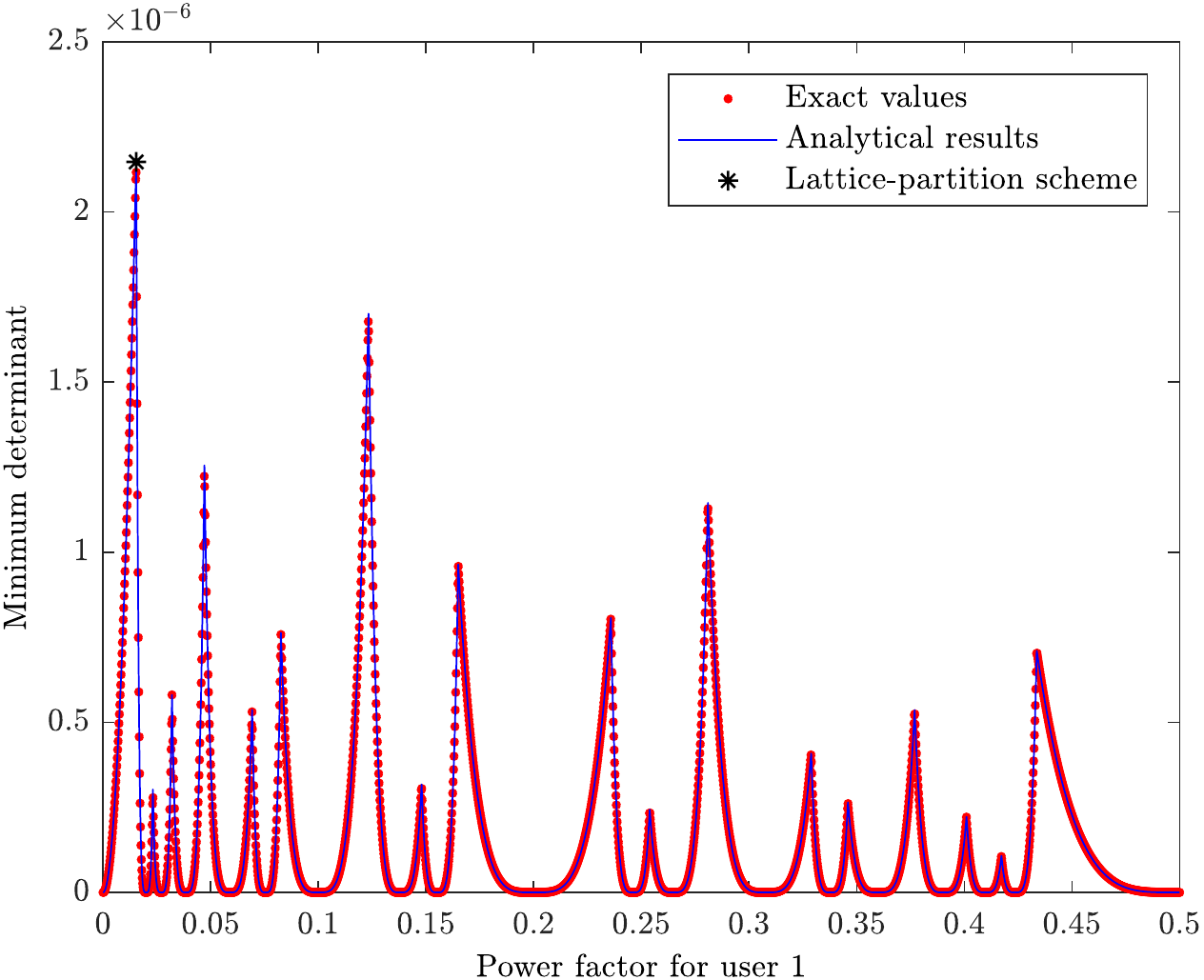}
\caption{Minimum determinant of Alamouti coded two-dimensional superimposed constellation from $(m_1,m_2) = (3,3)$}
\label{fig:14}
\end{figure}

We now give an example in the following to show the analytical results. We use the same setting as in Example \ref{exa:dpmin} and employ Alamouti code. We then obtain the exact values for $\min\{\det(\Delta\Delta^{\dag})\}$ by exhaustive search and compute the analytical results in Proposition \ref{coro:dmin}. The results are shown in Fig. \ref{fig:14}, from which one can observe that the analytical results perfectly match with the exact values of $\min\{\det(\Delta\Delta^{\dag})\}$. Moreover, the scheme based on lattice partition is again optimal in terms of minimum determinant.

\subsection{Proof of Proposition \ref{coro:dmin}}\label{sec:MIMO_proof}
Consider the transmission scheme described in Section \ref{sec:MIMO_scheme}. According to the design criterion for OSTBC \cite[Chapter 3.5]{Vucetic:2003:SC:861866}, the codeword matrix satisfies
$\mathbf{X}\mathbf{X}^{\dag} = (|x[1]|^2+\ldots+|x[M_t]|)\mathbf{I}_{M_t}$.
Hence, the minimum determinant of codeword difference matrix $\Delta \triangleq \mathbf{X}_s-\mathbf{X}_w$ is 
\begin{align}
\min\limits_{s\neq w}\{\det(\Delta\Delta^{\dag})\} &= \min\limits_{s\neq w}\{\det((|x_s[1]-x_w[1]|^2+\ldots+|x_s[M_t]-x_w[M_t]|)\mathbf{I}_{M_t})\} \nonumber \\
& = \min\limits_{s\neq w}\{(|x_s[1]-x_w[1]|^2+\ldots+|x_s[M_t]-x_w[M_t]|)^{M_t}    \} \nonumber \\
& \geq \min\limits_{l \in \{1,\ldots,M_t \}} \{ \min_{s \neq w}\{|x_s[l]-x_w[l]|^{2M_t}\} \} \nonumber \\
 &= d_{E,\min}(\eta_T(\mathcal{C}-\mathbf{d}))^{2M_t}.
\end{align}

From this point onward, the problem of analyzing minimum determinant is reduced to that of analyzing the minimum Euclidean distance of the composite constellation $\eta_T(\mathcal{C}-\mathbf{d})$. 
In what follows, we prove the following lemma about the exact minimum Euclidean distance, which in turn, gives us the exact minimum determinant.

\begin{lemm}\label{the:dmin}
Consider the constellation $\eta(\mathcal{C}-\mathbf{d})$ defined in \eqref{eq:sup1} for $\alpha \in [0,1]$ and $\Lambda$ is equivalent to $\mathbb{Z}^n$. Then the minimum Euclidean distance of $\eta(\mathcal{C}-\mathbf{d})$ is
\begin{align}\label{eq:dmin_the}
d_{E,\min}(\eta(\mathcal{C}-\mathbf{d})) = d_{E,\min}(\eta(\mathcal{X}-d^*)).
\end{align}
\end{lemm}

\emph{\quad Proof: }
From Lemmas \ref{the:1_CH7}-\ref{the:2a}, we know that each layer has the same Euclidean distance profile and thus same minimum Euclidean distance regardless of rotation. Similar to Lemma \ref{the:2a}, we have
\begin{align}
d_{E,\min}(\eta(\mathcal{C}-\mathbf{d})) = \min\{d_{E,\min}(\eta\mathcal{L}) ,d_{E,\min}(\eta(\mathcal{X}-d^*))\},
\end{align}
where $d_{E,\min}(\eta\mathcal{L})$ denotes the minimum of the set of Euclidean distances between all pairs of two distinct constellation points in any two different layers.

For any pair of non-intercepting layers from $\eta(\mathcal{C}-\mathbf{d})$, the minimum Euclidean distance between them is the length of the line segment that is orthogonal to these layers. The end points of this line segment are in fact the constellation points of a layer that is orthogonal to these layers. For any pair of intercepting layers within the composite constellation, the crossing point and two constellation points (each one is from different layer) form a right triangle. The Euclidean distance between these two constellation points is strictly larger than the Euclidean distance between the crossing point and either of those two constellation points, respectively. Thus, we conclude that
$d_{E,\min}(\eta\mathcal{L}) = d_{E,\min}(\eta(\mathcal{X}-d^*))$.
This completes the proof of \eqref{eq:dmin_the}. \QEDA

With Lemma \ref{the:dmin}, we obtain 
$\min\limits_{s\neq w}\{\det(\Delta\Delta^{\dag})\} = d_{E,\min}(\eta_T(\mathcal{X}-d^*))^{2M_t}$
by replacing the scalar with $\eta_T$.
To analyze this minimum Euclidean distance, we first denote by $d_{E,\min1}$ and $d_{E,\min2}$ the minimum Euclidean distance of constellation $\eta_T\sqrt{\alpha}(\mathcal{X}_1-d^*_1)$ and $\eta_T\sqrt{1-\alpha}(\mathcal{X}_2-d^*_2)$, respectively, where
\begin{align}
&d_{E,\min1}  \triangleq \eta_T\sqrt{\alpha} d_{E,\min}(\Lambda), \label{eq:dmin1} \\
&d_{E,\min2}  \triangleq \eta_T\sqrt{1-\alpha} d_{E,\min}(\Lambda). \label{eq:dmin2}
\end{align}
Here, $d_{E,\min}(\Lambda) = 1$ when the base lattice $\Lambda$ is equivalent to $\mathbb{Z}^n$ and $\eta_T = \tau\eta$, where $\eta$ is given in \eqref{normalize2}. Then, following the steps of our analysis in Section \ref{sec:dminp_ana} completes the proof. 

\section{Simulation Results}\label{SIM_CH7}
In this section, we provide the simulation results of our proposed scheme introduced in Chapters~\ref{sec:proposed_CH7} and \ref{MIMO} and compare them with the current state-of-the-art.

\subsection{Single Antenna Case}\label{sim_SISO}
\begin{figure}[ht!]
	\centering
\includegraphics[width=3.5in,clip,keepaspectratio]{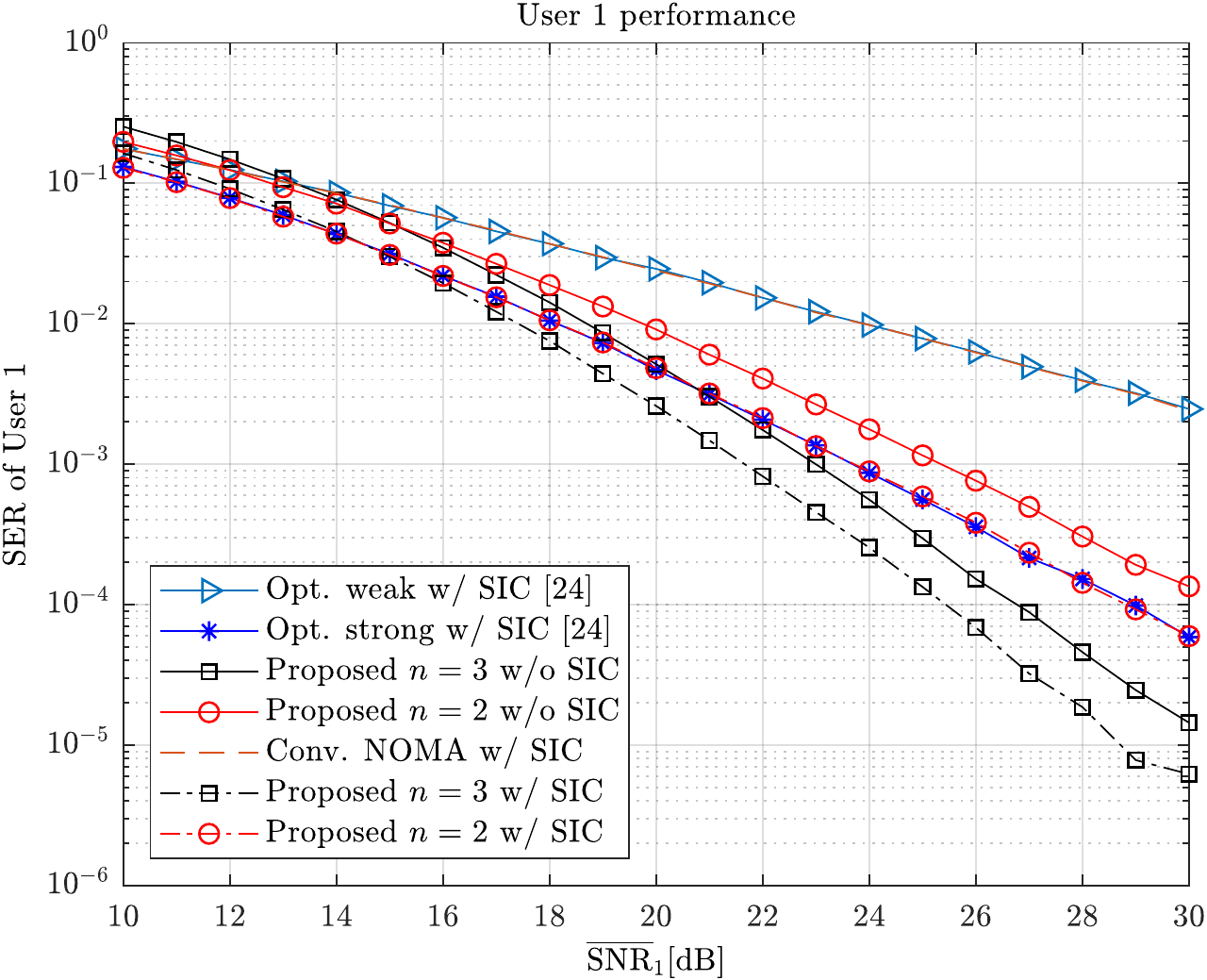}
\caption{Simulation results for user 1's SER.}
\label{fig:8a}
\end{figure}

\begin{figure}[ht!]
	\centering
\includegraphics[width=3.5in,clip,keepaspectratio]{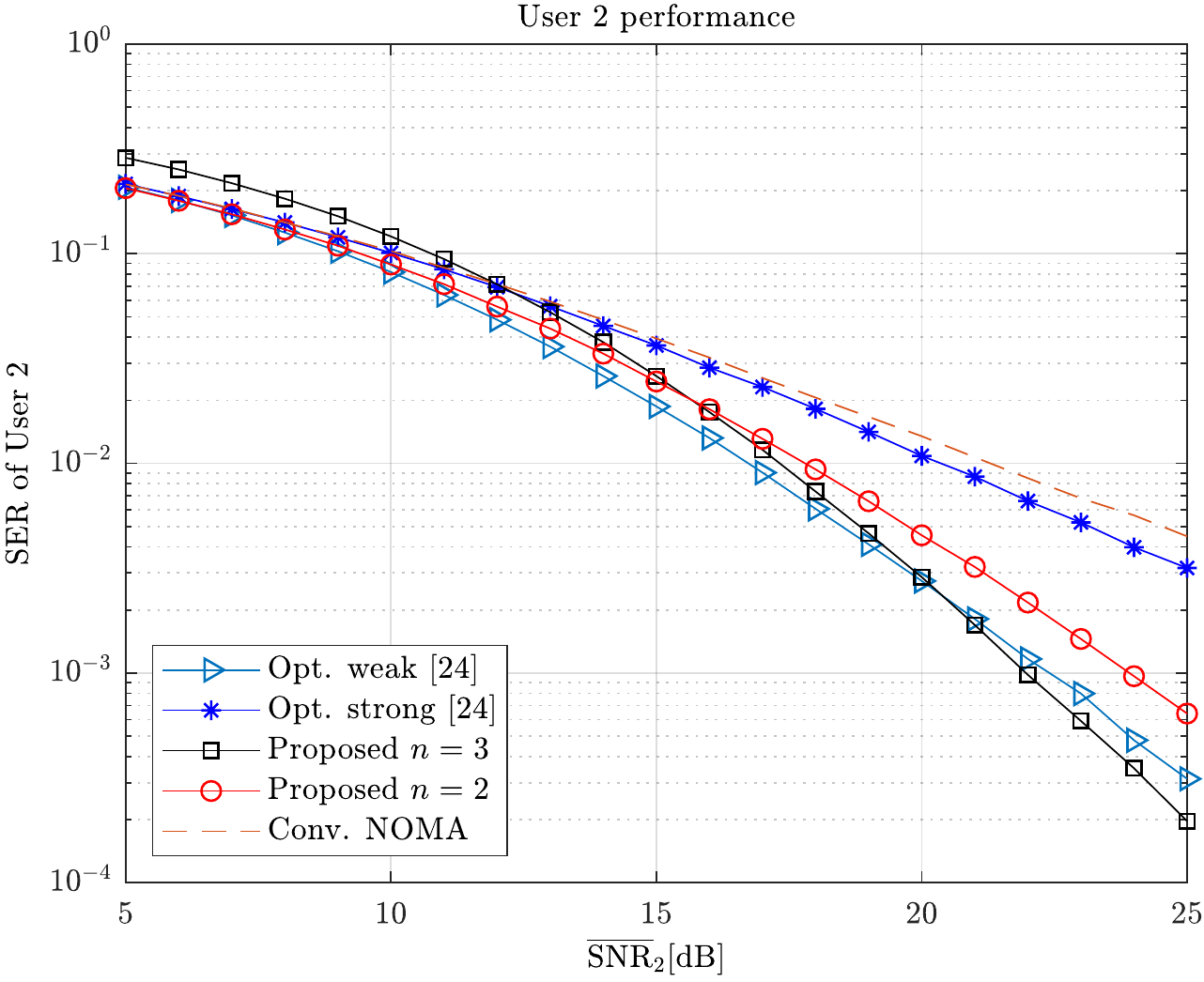}
\caption{Simulation results for user 2's SER.}
\label{fig:8b}
\end{figure}

In this subsection, we first provide simulation results of the lattice partitioned scheme for the single antenna case. The dimension of the underlying ideal lattice is set to $n= 2,3$. For illustrative purpose, we consider $(m_1,m_2) = (1,1)$ in order to make fair comparison with the scheme in \cite{7880967}. 
We use the conventional NOMA (labelled Conv. NOMA) scheme which adopts square 4-QAM (not rotated) as a benchmark. The performance of strong user (user 1) and that of weak user (user 2) are measured in terms of SER versus their average SNRs and plotted in Fig. \ref{fig:8a} and Fig. \ref{fig:8b}, respectively. In addition, the SER of the schemes in \cite{7880967} are plotted in both figures. Note that \cite{7880967} has \emph{two schemes} corresponding to optimization for strong user and optimization for weak user, respectively. We also emphasize here that the power allocations for the conventional NOMA scheme, our schemes and the schemes in \cite{7880967} are the same, i.e., $\alpha = 0.2$. In all the curves in these figures, when SIC is adopted at user 1, we assume that user 2's signals are perfectly decoded and subtracted.

It can be observed that for the proposed schemes with $n=2$ and 3, respectively, the full diversity orders of $2$ and 3, respectively, can be achieved for both users even without SIC. In particular, each user in our scheme for $n=2$ achieves comparable performance compared to the user whose constellation is optimized in \cite{7880967}.
Conversely, in \cite{7880967}, the performance at the user which is not optimized reveals no diversity gain as the conventional NOMA scheme. 
Furthermore, the maximum diversity order in scheme \cite{7880967} is only 2 while our scheme can provide higher diversity order and coding gain to both users by choosing higher-dimensional ideal lattices as the base lattices. Last but not least, our proposed scheme based on lattice partition provides a systematic way to design downlink NOMA scheme that offers full diversity gain and high coding gain, while the scheme in \cite{7880967} is based on exhaustive search.

\subsection{Multiple Antennas Case}
\begin{figure}[ht!]
	\centering
\includegraphics[width=3.5in,clip,keepaspectratio]{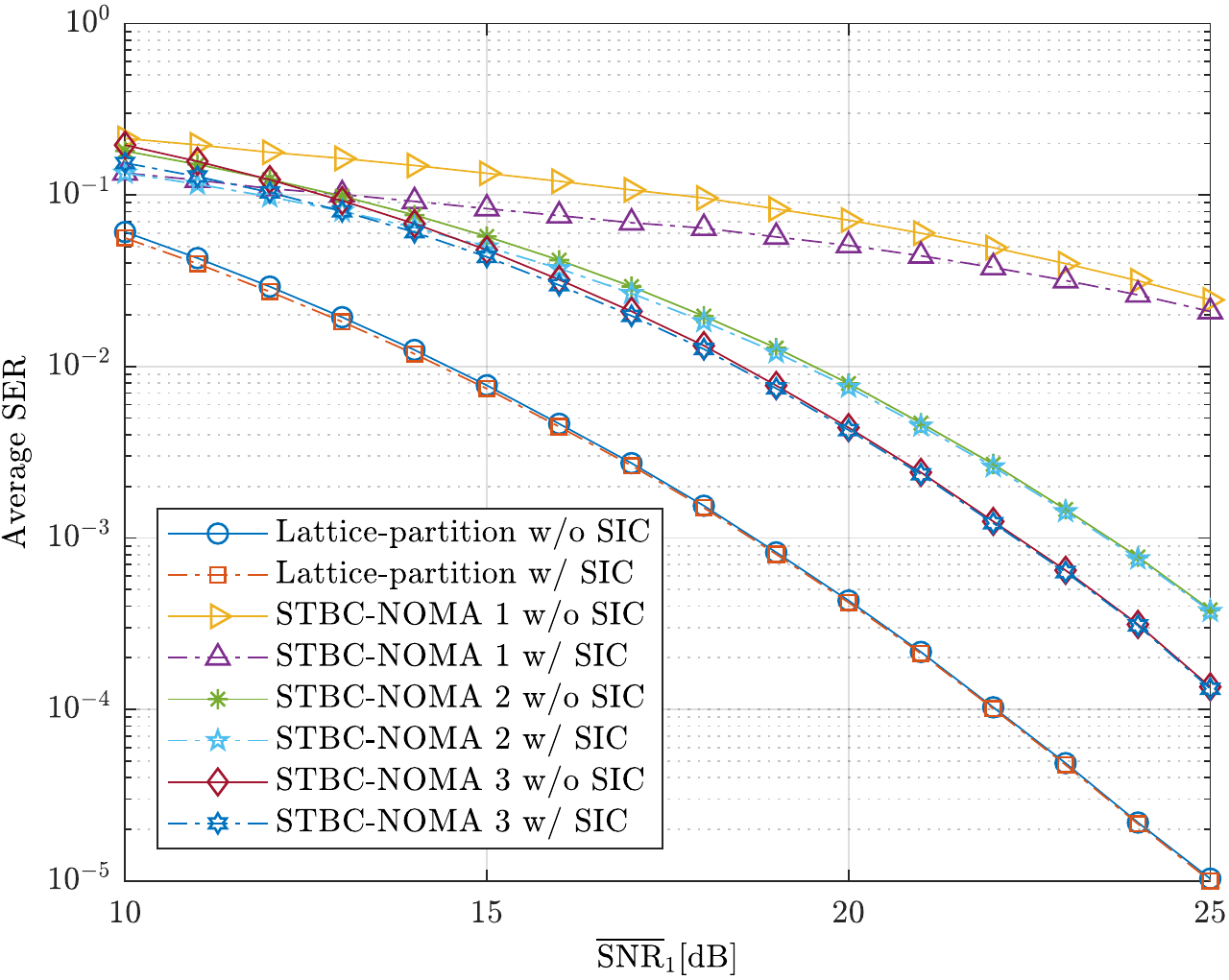}
\caption{Simulation results for average SER among two users.}
\label{fig:avg_SER}
\end{figure}

\begin{figure}[ht!]
	\centering
\includegraphics[width=3.5in,clip,keepaspectratio]{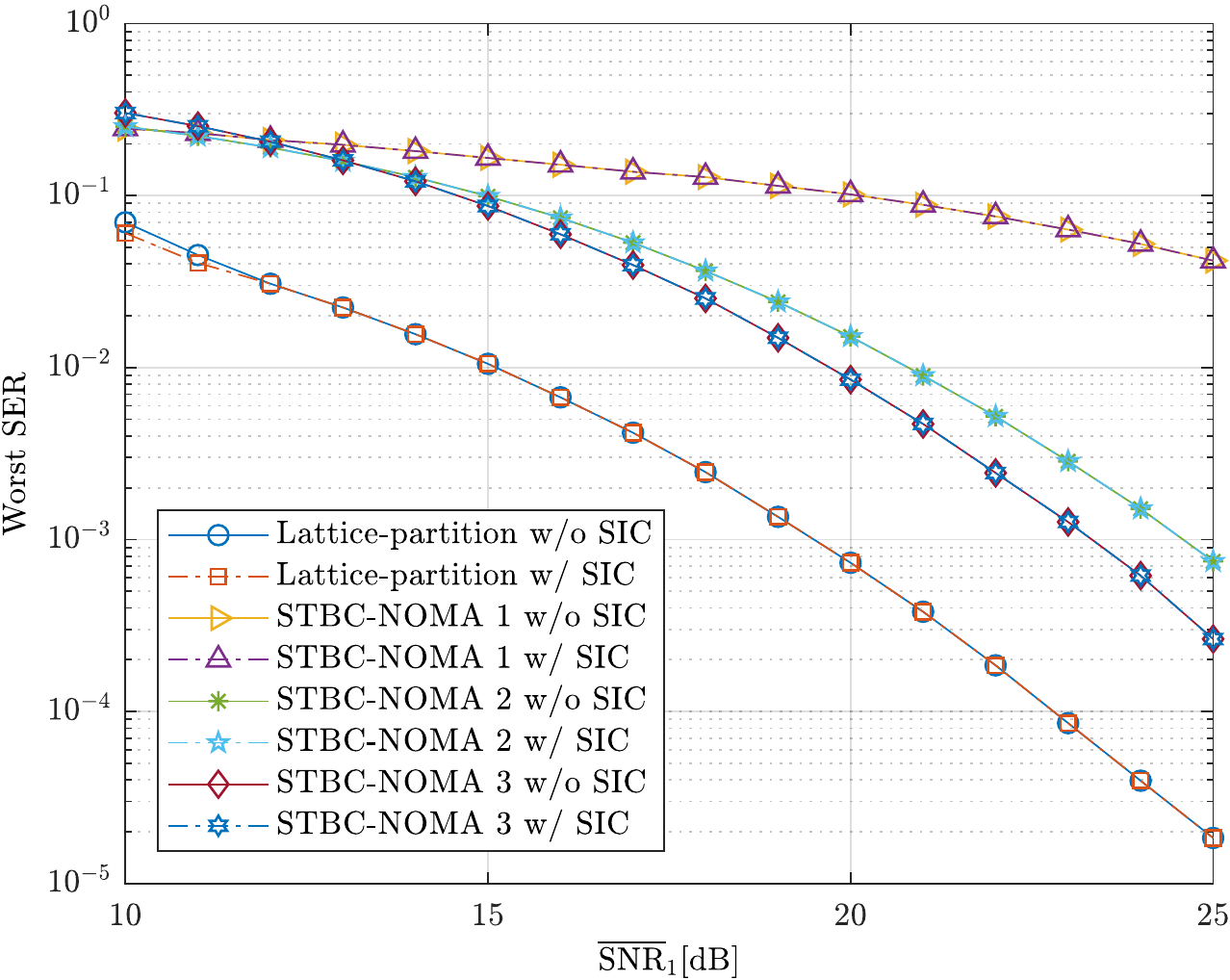}
\caption{Simulation results for worst SER among two users.}
\label{fig:max_SER}
\end{figure}

In this subsection, we provide the simulation results for the proposed MIMO-NOMA scheme where the base station and each user have two antennas and the underlying OSTBC is Alamouti code. We consider the case for $(m_1,m_2) = (2,1)$ and the channel is Rayleigh fading. The difference between $\overline{\text{SNR}}_1$ and $\overline{\text{SNR}}_2$ is 5 dB. Since we are unable to find a benchmark downlink MIMO-NOMA scheme with discrete inputs and with similar channel assumptions as ours, we thus compare the error performances of our lattice-partition scheme and a number of space-time block coded NOMA schemes with some power allocations.
Specifically, we choose $\alpha = 0.11,0.14$ and 0.31 for three schemes (labelled as STBC-NOMA 1-3, respectively) and the corresponding minimum determinants are $0.136 \cdot 10^{-4}$, $0.169 \cdot 10^{-2}$ and $0.449 \cdot 10^{-2}$, respectively. The lattice partition scheme has a minimum determinant of $0.91 \cdot 10^{-2}$. Here, the error performances are measured by average SER and worst case SER among two users versus user 1's average SNR. These results are plotted in Fig. \ref{fig:avg_SER} and Fig. \ref{fig:max_SER}, respectively. 

It can be seen that the scheme with larger minimum determinant has better error performance than that of the scheme with smaller minimum determinant. Another interesting observation is that the schemes with SIC only provide negligible gain for both average and the worst SER performance among two users. This is due to the fact that the
average/worst SER performance is largely dominated by the performance of the user with much higher SER.

\section{Summary}
In this work, we have proposed a class of downlink NOMA scheme without SIC for block fading channels. In particular, we have used algebraic lattices to design modulations such that full diversity gain and large coding gain can be attained for all users at the same time. Moreover, the minimum product distance for the superimposed constellation for arbitrary power allocation has been thoroughly investigated. Within the proposed class, a family of schemes based on lattice partitions has then been identified. It has been shown via numerical result that schemes from this special family achieve the largest minimum product distances among the proposed class. An extension of the proposed scheme to the MIMO-NOMA system with OSTBC has then been introduced. The exact minimum determinant of the proposed scheme has been derived. Simulation results have been provided, which confirms our analytical results and also demonstrates that our schemes significantly outperform the current state-of-the-art.

\chapter{Terminated Staircase Codes For NAND Flash Memories}\label{C8:chapter8}

\section{Introduction}

In addition to addressing the problem of lattice coding designs for point-to-point systems (Chapter 4) and for downlink multiuser systems (Chapters 5-7), we also address the problem of designing powerful channel codes for storage systems such as NAND flash memories in this chapter. This is also relevant and important to ultra-reliable communications for future digital systems.

NAND flash memories are non-volatile storage devices where data can be saved and retained for a long time without continuous power supply. They have become immensely popular due to their attractive features such as higher data throughput and lower power consumption when compared to traditional hard disk drives (HDDs). However, the cost-per-bit associated with NAND flash memories is higher than that of HDDs. To reduce the bit cost, a range of techniques such as multi-level cell (MLC) \cite{Micheloni17} and  triple-level cell (TLC) have been developed to increase the storage capacity density of NAND flash memories. These storage techniques enable each memory cell to store more than 1 bit information. As such, high-capacity NAND flash memories have been widely deployed in mobile phones, digital cameras, solid state drives (SSDs) and other electronic devices \cite{Huang11,Yu14}.

It is known that a flash memory is an array of cells where data is stored as electric charges. In the event of charge leakage, read and write disturbance, aging and microprogramming \cite{Li13}, it is very likely that the stored data could be in errors or erased. In addition, the stored data are even more vulnerable to noise and cell-to-cell interference with the increase in storage density. This is due to the fact that the MLC, TLC and QLC techniques and scaling technology have led to reducing noise margin and strengthening interference from adjacent memory cells \cite{Wang_Flash_11}. To overcome these issues and maintain the data integrity, i.e., reliable data storage and retrieval, it is necessary to deploy powerful error correction codes (ECCs) to protect the stored data in NAND flash memories. ECCs can be used for performing error recovery by detecting and correcting errors to ensure that the stored data can be read correctly.

In this chapter, we will introduce a class of staircase codes with improved decoding algorithm to achieve the required error floor performance for flash memory devices.

\subsection{Main Contributions}
Motivated by the success of unterminated staircase codes and in the quest of finding strong ECCs with hard-decision decoding for flash memories, we propose terminated staircase codes for flash memory devices in this work. Different from the conventional unterminated staircase codes in \cite{Smith12,6787025,Holzbaur17}, we introduce a new design in the code structure and decoding algorithm to lower the error floor. The main contributions of our work are summarized as below:
\begin{itemize}
\item We purpose terminated staircase codes and design an example of rate 0.89 terminated staircase code for flash memories with page size of 16K bytes. Specifically, the codes are terminated in a way such that all the information blocks are protected by row and column encoding. The code structure not only allows our codes inherit the properties of the unterminated staircase codes but also makes the codes satisfy the length and rate requirements. In particular, we propose a novel coding structure by performing cyclic redundancy check (CRC) encoding and decoding on the whole codeword including information bits and parity bits. The CRC bits are protected by both row and column codewords in our construction.
\item We improve the staircase code decoder. Specifically, we develop a novel CRC decoding process based on our encoding structure which allows our decoder to detect more stall patterns. A more accurate error floor estimation including the contributions of both detectable and undetectable stall patterns to the error floor, is provided based on our code structure and decoding algorithm.
\item We propose a novel iterative bit flipping algorithm which is embedded in our decoder. Theoretical analysis on the performance for our decoder is provided. Specifically, we prove and show that our decoder has the capability to solve more stall patterns, resulting in a lower error floor than that of the conventional staircase codes. Our method can be implemented on any general staircase codes and some product codes. Our error floor analysis shows that our coding scheme can satisfy the BER requirements for flash memories by lowering the error floor below $10^{-15}$. Simulation results are provided and show that our design example can outperform the conventional staircase codes and the stand-alone BCH code.

\end{itemize}

\section{Terminated Staircase Codes}\label{TSC}
\begin{figure}[ht!]
	\centering
\includegraphics[width=3.42in,clip,keepaspectratio]{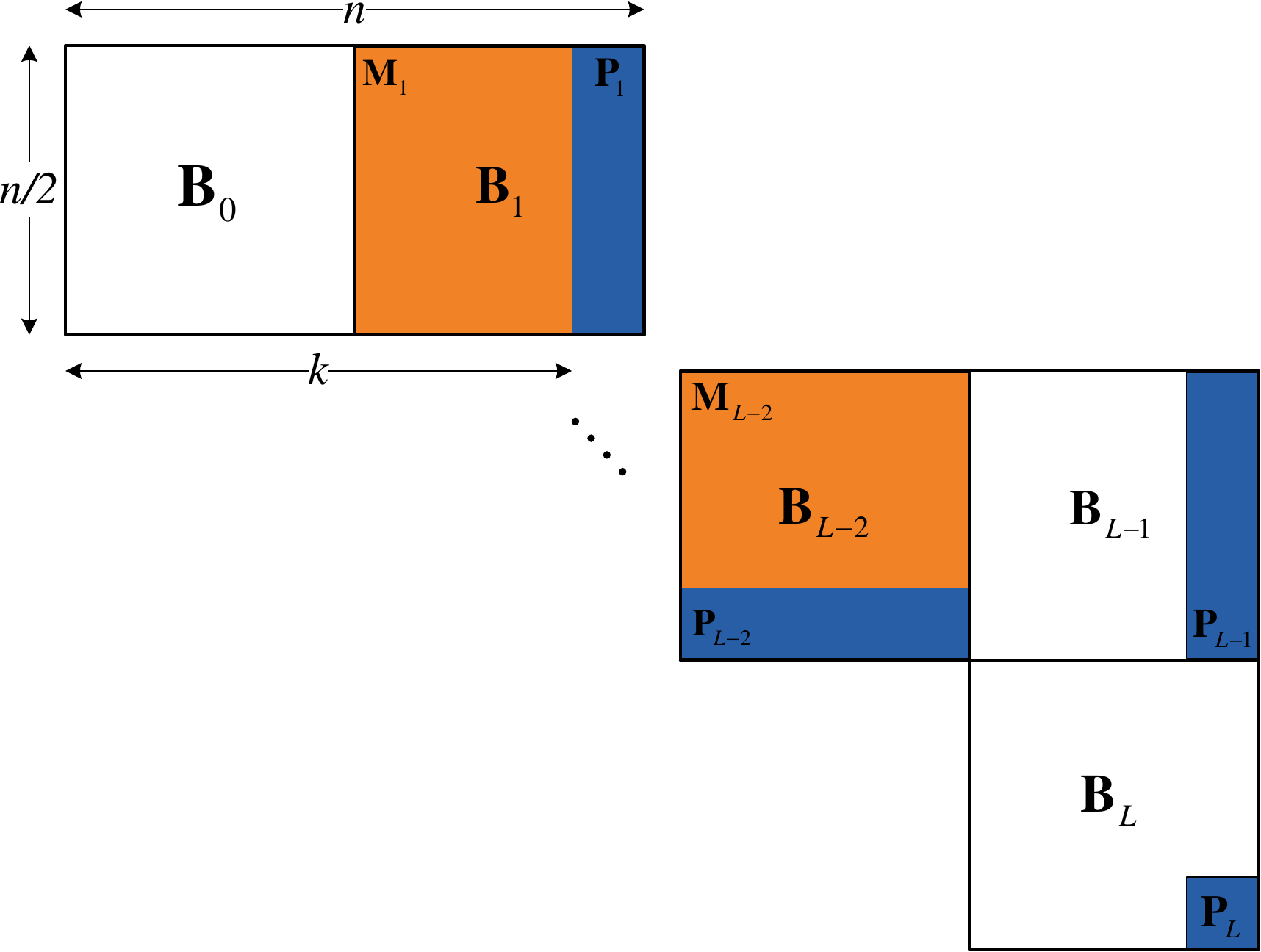}
\caption{Terminated staircase code block structure.}
\label{fig:staircase1}
\end{figure}

In this section, we present the general framework of our proposed terminated staircase codes based on unterminated staircase codes \cite{Smith12}. The code structure is depicted in Fig. \ref{fig:staircase1}.

It contains $L+1$ code blocks: $\mathbf{B}_0,\mathbf{B}_1,\cdots,\mathbf{B}_L$. The component code for constructing our terminated staircase code $\mathcal{S}$ is a linear block code $\mathcal{C}$ whose codeword length is $n$ and the information length is $k$. Note that the component code needs to be in a systematic form and its code rate $R_{\mathcal{C}} = k/n$ should satisfy $R_{\mathcal{C}} > 1/2$. As shown in Fig. \ref{fig:staircase1}, the information bits are in block $\mathbf{M}_i$ for $i = 1,2,\cdots,L-2$ while the parity bits are in block $\mathbf{P}_i$ for $i = 1,2,\cdots,L$. The blocks in white are fixed to all-zero bit-values and are assumed to be known at the encoder-decoder pair and thus will not be stored. Under this setting, the code rate for our terminated staircase codes $R_{\mathcal{S}}$ is:
\begin{equation}\label{eq:RSrate}
R_{\mathcal{S}} = \frac{\frac{n}{2}(k-\frac{n}{2})(L-2)}{\frac{n^2}{4}(L-2)+(n-k)\frac{n}{2}+(n-k)^2}.
\end{equation}
It is worth mentioning that these all-zero code blocks are used for protecting the first and the last code blocks. Although inserting zero blocks in our terminated staircase codes will lead to some code rate loss compared to direct termination of staircase codes (i.e., directly terminating the staircase codes such that code block $\mathbf{B}_{L-2}$ is the last one), this approach can ensure that all the information blocks are protected by both row and column coding. However, direct termination of staircase codes will leave the last information block only protected by either row or column codewords. In this case, the error performance of the whole staircase codes would be degraded because the error rate is largely affected by the higher error rates at the code boundaries. In this work, we focus on this simple termination approach and leave the exploration of other termination mechanisms as future work. It is also noteworthy that the choice of $\mathcal{C}$ highly depends on the requirements of the applications for our coding schemes. For the purpose of this work, we restrict all the elements in the code blocks to be binary, although the non-binary construction is analogous.

\section{Terminated Staircase Codes for NAND Flash Memories}
From now on, we will put into practice our coding scheme described in Section \ref{TSC} with the goal of designing the terminated staircase codes for flash memories. We consider the data transmission channel to be BSC \cite{Camp13} and only hard-decision channel output is available. Although more advanced noise models consider the asymmetric nature of the noise \cite{7416649}, we only focus on the BSC and leave the code design and optimization for the asymmetric channels in our future work.

In this work, we consider designing a 16K bytes staircase code for \emph{future} flash memories with 16K bytes page size. Furthermore, we consider each page of the flash memories is protected by a single ECC. The code rate of the ECC should be at least around 0.89 so that a codeword can be stored in a single page. An example that can meet the above requirements it to choose a binary primitive narrow-sense BCH code with $(n'' = 511, k''= 493 , t = 2)$ where $t$ denotes the number of correctable errors, as the component code. Furthermore, the number of information blocks is 2, which leads to $L=4$ such that the terminated staircase codes contain code block $\mathbf{B}_0,\mathbf{B}_1,\cdots,\mathbf{B}_4$. Note that this is the only possible design of the terminated staircase codes that can satisfy the length, code rate and error floor requirements simultaneously when using BCH codes as component codes. Since the length of the component code $n$ is required to be even, the original BCH code is shortened by 1 bit, resulting in $n = 510$. In the information bits, 1-bit CRC with the generator polynomial $x+1$ is included. Thus, the component code now becomes an $(n = 510, k = 491 , t = 2)$ BCH code with 19 redundant bits including 18 BCH parity bits and 1 CRC bit. Each CRC bit is used for providing even parity check for a BCH codeword and is also protected by both row and column coding. The introduction of CRC bits is crucial for any product-like codes with small $t$ component codes \cite{7352332} because the CRC can prevent additional error events that are caused by the component code decoder. In particular, it has been reported in \cite{7352332} that the staircase codes without CRC have worse performance and higher error floor compared with the staircase codes with CRC. By plugging $k,n$ and $L$ into Eq. (\ref{eq:RSrate}), our terminated staircase codes have information length $120360$ bits and code rate $R_{\mathcal{S}} = 0.8899$. One may notice that the information length is not exactly 16K bytes which is 131072 bits. For illustrative purpose, we provide a design example for terminated staircase codes to demonstrate its superior performance. To have the exact information length, one can expand the size of the information matrices $\mathbf{M}_1$, $\mathbf{M}_2$ while reducing the size of the all-zero block $\mathbf{B}_0$ and $\mathbf{B}_{L-1}$. For other code designs such as 2K bytes, 4K bytes and 8K bytes codes, it is difficult to find a suitable BCH code such that the resultant staircase code satisfies both code rate and error floor requirements. For example, one can only pick a shortened $(n = 254,k = 246,t = 1)$ BCH code and set $L = 4$ to construct a 4K bytes staircase code with rate about 0.9. However, the error floor for this code occurs at $\text{BER} \sim 10^{-5}$ due to $t=1$ according to \eqref{eq:nms3} in the later analysis. Our proposed iterative bit flipping algorithm cannot provide a huge gain to allow the code to reach $\text{BER} \leq 10^{-15}$. Thus, the $(n=510,k=491,t=2)$ shortened BCH code with 1-bit CRC is unique for the 16K bytes staircase code such that it satisfies both rate and error floor requirements. If there is no restriction on the codeword length, it is possible for one to choose a component code with larger $n$, $k$ and $t$ such that the constructed staircase code has better error floor performance and high code rate.

\subsection{Encoding of Terminated Staircase Codes}
\begin{figure}[ht!]
	\centering
\includegraphics[width=2.62in,clip,keepaspectratio]{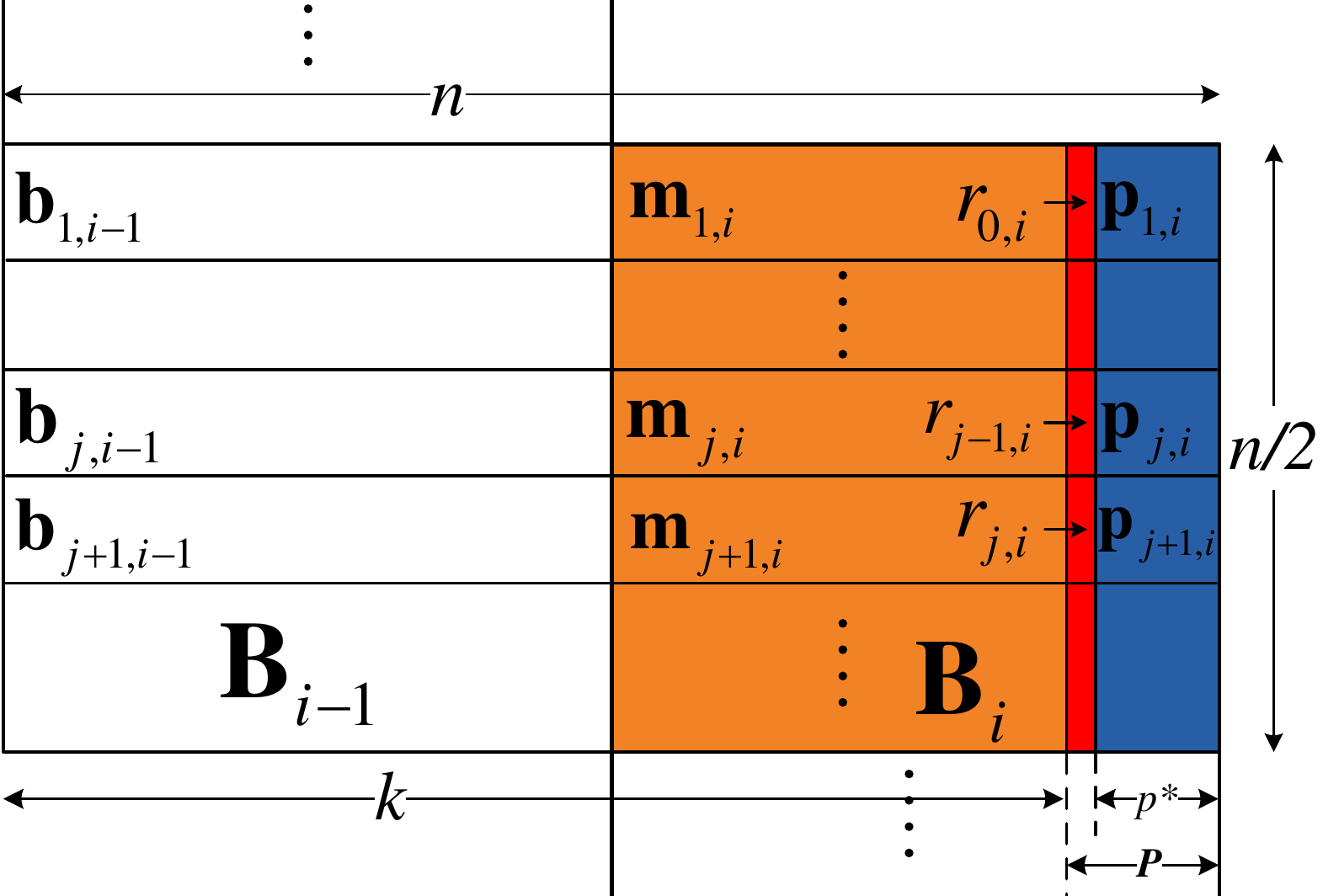}
\caption{Structure of a staircase code block.}
\label{fig:CRC1}
\end{figure}

Now consider the terminated staircase code in Fig. \ref{fig:staircase1} for $L=4$. The encoding is performed in a recursive manner by generating the code block $\mathbf{B}_i \in \mathbb{F}_2^{\frac{n}{2} \times \frac{n}{2}}$ for $i = 1,2,\cdots,L$. Our encoding algorithm is modified according to our code construction and based on the encoder of original unterminated staircase codes \cite{Smith12}. The code block $[\mathbf{B}_{i-1}^T \; \mathbf{B}_i] \in \mathbb{F}_2^{\frac{n}{2} \times n}$ of our proposed 16K bytes terminated staircase code is depicted in Fig. \ref{fig:CRC1}.

A staircase code block comprises three parts: $[\mathbf{B}_{i-1}^T \; \mathbf{B}_i]= [\mathbf{B}_{i-1}^T\;\mathbf{M}_i\;\mathbf{P}_i]$, where $\mathbf{M}_i \in \mathbb{F}_2^{\frac{n}{2} \times (k-\frac{n}{2})}$ are the orange portions and $\mathbf{P}_i \in \mathbb{F}_2^{\frac{n}{2} \times (n-k)}$ are the blue and red portions in Fig. \ref{fig:CRC1}. Let $j$ denote the row number in a code block. In this figure, for $i \in \{1,2,\cdots,L\} $ and $j \in \{1,2,\cdots,\frac{n}{2} \}$, the $j$-th row vector $[\mathbf{b}_{j,i-1} \; \mathbf{m}_{j,i} \; \mathbf{p}_{j,i}] \in \mathbb{F}_2^{1 \times n}$ is a component code codeword, where $\mathbf{b}_{j,i-1} \in \mathbb{F}_2^{1 \times \frac{n}{2}}$ is the $j$-th row of block $\mathbf{B}_{i-1}^T$; $\mathbf{m}_{j,i} \in \mathbb{F}_2^{1 \times (k-\frac{n}{2})}$ consists information bits; and $\mathbf{p}_{j,i} = [r_{j-1,i} \; \mathbf{p}^*_{j,i}]\in \mathbb{F}_2^{1 \times (n-k)} $ is the redundant bits vector which comprises the parity bits $\mathbf{p}_{j,i}^* \in \mathbb{F}_2^{1 \times (n-k-1)}$ and the CRC bit $r_{j-1,i}$ (which is highlighted in red in Fig. 4). Here, the 1-bit CRC $r_{j-1,i}$ is obtained by performing the CRC encoding on the $(j-1)$-th row vector $[\mathbf{b}_{j-1,i-1} \; \mathbf{m}_{j-1,i} \; \mathbf{p}^*_{j-1,i}] \in \mathbb{F}_2^{1 \times (n-1)}$ which excludes the CRC bit. Similarly, the CRC bit $r_{j,i}$ associated with the $j$-th row is a result of applying CRC encoding on the CRC-excluded codeword vector $[\mathbf{b}_{j,i-1} \; \mathbf{m}_{j,i} \; \mathbf{p}^*_{j,i}]$ as shown in Fig. \ref{fig:CRC1}. The detailed encoding steps for our terminated staircase codes are summarized in Algorithm \ref{TSEA} in the following.

\begin{algorithm}
\caption{Terminated Staircase Encoding}\label{TSEA}

$\;\;$

$\;\;$\textbf{\emph{Step 1} Initialization:} The CRC bit in $\mathbf{m}_{1,1}$, i.e., $r_{0,1}$, is set to zero. The code block $\mathbf{B}_0$, $\mathbf{M}_{L-1}$ and $\mathbf{M}_L$ are set to all-zero values according to Fig. \ref{fig:staircase1}.

$\;\;$

$\;\;$\textbf{\emph{Step 2} Component code encoding:} For the $j$-th row in the $i$-th code block, fill the $k-\frac{n}{2}$ bits of source information in vector $\mathbf{m}_{j,i}$. Given $\mathbf{b}_{j,i-1}$ and the CRC bit $r_{j-1,i}$, form the row vector $[\mathbf{b}_{j,i-1} \; \mathbf{m}_{j,i} \; r_{j-1,i}] \in \mathbb{F}_2^{1 \times (k+1)}$. Then apply component code encoding on $[0\; \mathbf{b}_{j,i-1} \; \mathbf{m}_{j,i} \; r_{j-1,i}] \in \mathbb{F}_2^{1 \times k+2}$ and remove the zero bit in the front to obtain the codeword vector $[\mathbf{b}_{j,i-1} \; \mathbf{m}_{j,i} \; \mathbf{p}_{j,i}] = [\mathbf{b}_{j,i-1} \; \mathbf{m}_{j,i}\;[r_{j-1,i} \; \mathbf{p}^*_{j,i}]]$.

$\;\;$

$\;\;$\textbf{\emph{Step 3} CRC encoding:} Apply CRC encoding on $[\mathbf{b}_{j,i-1}\; \mathbf{m}_{j,i} \; \mathbf{p}^*_{j,i}]$ but excluding the CRC bit $r_{j-1,i}$, to obtain the CRC bit $r_{j,i}$. In other words, the CRC bit $r_{j,i}$ is generated from the $j$-th codeword vector.

$\;\;$

$\;\;$\textbf{\emph{Step 4} Block encoding:} Repeat from Step 2 to Step 3 for $j = 1,2,\cdots,\frac{n}{2} $ to obtain the $i$-th code block $\mathbf{B}_i = [\mathbf{M}_i\;\mathbf{P}_i] \in \mathbb{F}_2^{\frac{n}{2} \times \frac{n}{2}}$. The value of the CRC bit corresponding to the $\frac{n}{2}$-th codeword in block $\mathbf{B}_i$ is assigned to the CRC bit that located in the first codeword in block $\mathbf{B}_{i+1}$ via $ r_{\frac{n}{2},i} \rightarrow r_{0,i+1}$.

$\;\;$

$\;\;$\textbf{\emph{Step 5} Recursive encoding:} Repeat from Step 2 to Step 4 for $i = 1,2,\cdots,L$ to obtain all the code blocks. Remove all-zero blocks to obtain the codeword of terminated staircase codes $\mathcal{S}$.

$\;\;$

\end{algorithm}

\begin{remark}\label{remark:diff}
When encoding proceeds to block $[\mathbf{B}_{L-1}^T\;\mathbf{B}_L]$, the all-zero information block is encoded into multiple rows of all-zero codewords. However, in practice the all-zero codeword will not be stored and thus the encoding process does not include the encoding of all-zero information block. It is also note worthy that the code block with only parity bits, e.g., $\mathbf{B}_{L-1}$, is coded with all-zero information. The introduction of the extra parity bit matrix in block $\mathbf{B}_L$ can yield to a better error performance than that of the terminated staircase codes without $\mathbf{B}_L$ \cite[Sec. III]{7938011}. This is because any $t+1$ errors occur in the parity block $\mathbf{P}_{L-1}$ will not be correctable if there are no column codewords to protect the parity block. That said, the additional parity block $\mathbf{P}_L$ only results in negligible code rate loss when $R_{\mathcal{C}}$ is high.

We also stress here that the way of using CRC in our design is different from that of conventional staircase codes designs. For example, in \cite{Holzbaur17}, the $(n = 510, k = 491, t= 2)$ BCH component code is obtained by multiplying the generator polynomial of the shortened $(n'' = 511, k'' = 493, t= 2)$ BCH codes to $(x+1)$. Although their component codes provide the same error correction capability and error detection probability as that in our design with BCH component codes, their code generation methods can only be applied to cyclic code families. In contrast, our methods of using CRC can be applied to any general component codes, providing additional error detection mechanism. As an alternative way for having an additional error detection capability, one may suggest using an extended code \cite[Sec. II-B]{hager2017approaching} as the component codeword. However, for singly-extended codes, the additional parity bits are only encoded by column codewords. As a result, when these bits are in error, the errors can only be corrected by column decoding but without row decoding.

\end{remark}

\subsection{Decoding of Terminated Staircase Codes}\label{tscdec}
The decoding of our terminated staircase codes is accomplished by using an iterative hard-decision decoder. We consider the terminated staircase code shown in Fig. \ref{fig:staircase1} except the all-zero blocks to be the received codeword where all the received bits are corrupted by a BSC. During the iterative hard-decision decoding, bounded-distance decoding (BDD) \cite[Sec. II-C]{hager2017approaching} is applied to each row of block $[\mathbf{B}_{i-1}^T \; \mathbf{B}_i]$ for $i \in \{1,2,\cdots,L\} $. An example of BDD can be the Berlekamp-Massey decoder \cite{Berlekamp:2015:ACT:2834146}. After the BDD decoding, the CRC decoding is performed on all the successfully decoded codewords. We let $l$ be the number of iterations and $l_{\text{max}}$ be the maximum number of iterations allowed. The parity-check matrix for the component code $\mathcal{C}$ is denoted by $\mathbf{H}_{\mathcal{C}}$. The decoding steps for our terminated staircase codes are summarized in Algorithm \ref{TSCDA} in the following.

\begin{algorithm}
\caption{Terminated Staircase Decoding}\label{TSCDA}

$\;\;$

$\;\;$\textbf{\emph{Step 1} Initialization:} The CRC bit $r_{0,1}$ is set to zero. Form the complete staircase code by adding the all-zero blocks $\mathbf{B}_0$, $\mathbf{M}_{L-1}$ and $\mathbf{M}_L$ to the received codeword.

$\;\;$

$\;\;$\textbf{\emph{Step 2} Syndrome Check:} For the $l$-th iteration and $i$-th code block, form the matrix $[\mathbf{B}_{i-1}^T \; \mathbf{B}_i]$ given $\mathbf{B}_{i-1}^T$. Calculate the corresponding syndrome via $\mathbf{S}_i = [\mathbf{s}_{1,i},\mathbf{s}_{2,i},\cdots,\mathbf{s}_{\frac{n}{2},i}]^T = [\mathbf{B}_{i-1}^T \; \mathbf{B}_i]\mathbf{H}_{\mathcal{C}}$. Store the syndrome $\mathbf{S}_i$. If $\mathbf{S}_i = \mathbf{0}$, proceed to the beginning of Step 2 for $i=i+1$. Otherwise, proceed to Step 3.

$\;\;$

$\;\;$\textbf{\emph{Step 3} Component code decoding:} Perform BDD on the $j$-th row in the $i$-th code block $[0 \; \mathbf{b}_{j,i-1} \; \mathbf{m}_{j,i} \; \mathbf{p}_{j,i}]$ with non-zero syndrome $\mathbf{s}_{j,i} \neq \mathbf{0}$. Calculate the syndrome for the decoded information $[\hat{\mathbf{b}}_{j,i-1} \; \hat{\mathbf{m}}_{j,i} \; \hat{\mathbf{p}}_{j,i}]$ and update the syndrome via $\mathbf{s}_{j,i} = [\hat{\mathbf{b}}_{j,i-1} \; \hat{\mathbf{m}}_{j,i} \; \hat{\mathbf{p}}_{j,i}]\mathbf{H}_{\mathcal{C}}$.

$\;\;$

$\;\;$\textbf{\emph{Step 4} CRC decoding:} For the $j$-th codeword in the $i$-th code block, if $\mathbf{s}_{j,i} = \mathbf{0}$ and $\mathbf{s}_{j+1,i} = \mathbf{0}$, apply CRC decoding on the decoded message $[\hat{\mathbf{b}}_{j,i-1} \; \hat{\mathbf{m}}_{j,i} \; \hat{r}_{j,i} \; \hat{\mathbf{p}}^*_{j,i}] \in \mathbb{F}_2^{1 \times n}$ which excludes the CRC bit $\hat{r}_{j-1,i}$ but includes the CRC bit $\hat{r}_{j,i}$. If the CRC decoding is successful, update the codeword via $[\hat{\mathbf{b}}_{j,i-1} \; \hat{\mathbf{m}}_{j,i} \; \hat{\mathbf{p}}_{j,i}] \rightarrow [\mathbf{b}_{j,i-1} \; \mathbf{m}_{j,i} \; \mathbf{p}_{j,i}]$. Otherwise, $[\mathbf{b}_{j,i-1} \; \mathbf{m}_{j,i} \; \mathbf{p}_{j,i}]$ remains unchanged.

$\;\;$

$\;\;$\textbf{\emph{Step 5} Block decoding:} Repeat from Step 3 to Step 4 for $j = 1,2,\cdots,\frac{n}{2} $ to obtain the $i$-th decoded code block $\hat{\mathbf{B}}_i$. The value of the decoded CRC bit corresponding to the $\frac{n}{2}$-th codeword in block $\hat{\mathbf{B}}_i$ is assigned to the CRC bit that is located in the first codeword of block $\mathbf{B}_{i+1}$ via $ \hat{r}_{\frac{n}{2},i} \rightarrow r_{0,i+1}$.

$\;\;$

$\;\;$\textbf{\emph{Step 6} Frame decoding:} Repeat from Step 2 to Step 5 for $i = 1,2,\cdots L$ to obtain the whole frame. Then repeat Step 1 to remove any non-zero bits that occur in the all-zero blocks.

$\;\;$

$\;\;$\textbf{\emph{Step 7} Final check:} Perform syndrome check and CRC check on all the code blocks. If there are no detected errors or the decoder reaches the $l_{\text{max}}$-th iteration, output all the information blocks $\hat{\mathbf{M}}_i$ for $i = 1,2,\cdots,L-2$. Otherwise repeat from Step 1 to Step 6 for $l = 1,2,\cdots l_{\text{max}}$ iterations.

$\;\;$

\end{algorithm}


\section{Error Floor Analysis}\label{err_fr_ana}
In this section, we first introduce the error patterns which are the main contributor of the error floor of staircase codes. We then present a method to evaluate the error floor performance for our proposed codes.

\subsection{Stall Patterns}
For staircase codes, the dominating contributions to the error floor are the error patterns that cannot be corrected by iterative decoding. These error patterns are referred to as \emph{stall patterns} \cite{Smith12}.

\begin{defi}\label{def:1}
A stall pattern is a set of error positions in a stable state such that no updates are performed by the decoder. The stall pattern that occurs in a staircase code with $t$-error-correcting component code, involves at least $t+1$ erroneous rows and columns, each of which has at least $t+1$ bits are in error.
\end{defi}

All the stall patterns are assumed to be not correctable if we use the conventional staircase code decoder \cite{Smith12}. It should be noted that even though some error events other than stall patterns can lead to failure of the iterative hard-decision decoder, these kinds of events occur less likely when the crossover probability of the BSC is low. Thus we do not count them as the cause of the error floor. When analyzing the error floor, we assume that our decoder can resolve all surrounding errors with only a stall pattern remains unsolved. Therefore, the BER of the error floor can be regarded as the probability that the stall patterns appear in the staircase code blocks. In \cite{Smith12}, a union bound technique was proposed to bound the probability of these stall patterns in order to estimate the error floor of the staircase codes. This technique has been widely accepted and adopted in \cite{hager2017approaching,7905932,Holzbaur17}. In what follows, we provide a modified error floor analysis for our 16K bytes terminated staircase codes based on this technique.

In our analysis, we consider that the stall patterns can span one block such as block $\mathbf{B}_1,\mathbf{B}_2$ or $\mathbf{P}_3$; or span two blocks such as blocks $[\mathbf{B}_1^T \; \mathbf{B}_2]$ and $[\mathbf{B}_2^T \; \mathbf{P}_3]$. Other blocks such as $\mathbf{B}_0$ and $\mathbf{M}_3$ are all-zero and are known at both encoder and decoder. For block $\mathbf{B}_4$, it is very unlikely that a stall pattern can occur in block $\mathbf{P}_4$. Even though a stall pattern may span $[\mathbf{B}_3^T \; \mathbf{B}_4]$, the errors in parity parts does not contribute to error rates of information parts. Thus, this kind of error event is not considered in the analysis.

\subsection{General Stall Pattern Analysis}\label{GSA}
First, we denote the number of erroneous rows by $E$ and the number of erroneous columns by $F$. The number of errors in the stall pattern is denoted by $\varepsilon$. According to Definition \ref{def:1}, we know that $E \geq t+1$ and $F \geq t+1$. The number of bit errors $\varepsilon$ in this $(E,F)$ stall pattern should satisfy
\begin{align}\label{eq:stall_requirement}
(t+1) \cdot \max \{E,F \} \leq \varepsilon \leq E \cdot F.
\end{align}
The number of combinations of $E$ rows and $F$ columns that occur in $[\mathbf{B}_1^T \; \mathbf{B}_2]$ is
\begin{equation}\label{eq:nms1}
A_{E,F}^{1,2}= \binom{\frac{n}{2}}{E} \binom{n}{F}  = \binom{\frac{n}{2}}{E} \cdot \sum_{\theta=0}^{F} \binom{\frac{n}{2}}{\theta}\binom{\frac{n}{2}}{F-\theta},
\end{equation}
where $\theta$ is the number of erroneous columns in block $\mathbf{B}_2$. Note that this combination includes the case where the stall pattern occurs in a single block $\mathbf{B}_1$ or $\mathbf{B}_2$ because of the second equality in \eqref{eq:nms1}. This is different from the analysis for unterminated staircase codes \cite{Smith12} where a stall pattern spanning in $\mathbf{B}_2$ is not considered because the error probability is calculated on a single code block, e.g., $\mathbf{B}_1$. When the stall pattern spans $[\mathbf{B}_2^T \; \mathbf{P}_3]$, the multiplicity becomes
\begin{equation}\label{eq:nms2}
A_{E,F}^{2,3} = \binom{\frac{n}{2}}{E} \cdot \sum_{\theta=1}^{F-1}\binom{\frac{n}{2}}{\theta}\binom{n-k}{F-\theta}.
\end{equation}
Note that here we do not need to consider the case that a stall pattern spans a single block $\mathbf{B}_2$ or $\mathbf{P}_3$, i.e., $\theta=0$, because the case of spanning in $\mathbf{B}_2$ has been covered in Eq. \eqref{eq:nms1} and there is no contribution to error rate when the whole stall pattern is in $\mathbf{P}_3$.

Among these $\varepsilon$ errors, we consider that $z$ bits are received in error. The other $\varepsilon-z$ bits of errors are caused by incorrect decoding from the component code decoder. We let $\rho$ denote the BSC crossover probability and let $\xi$ represent the probability of incorrect decoding. It has been reported in \cite{Smith12} that $\xi$ is independent of $\varepsilon$. Thus, the probability that the stall pattern has $\varepsilon$ bit errors is
\begin{align}\label{eq:nms1a}
\sum_{z=0}^{\varepsilon} \binom{\varepsilon}{z} \rho^z \xi^{\varepsilon-z} =\left(\rho + \xi \right)^{\varepsilon}.
\end{align}

We let $M_{E,F}^{\varepsilon}$ represent the number of ways to distribute $\varepsilon$ errors in an $(E,F)$ stall pattern. As reported in \cite{Holzbaur17}, $M_{E,F}^{\varepsilon}$ is overestimated in \cite{Smith12} because it includes the cases that the number of errors in one or more than one erroneous rows/columns is less than $t+1$. According to \cite{Holzbaur17}, the problem of finding $M_{E,F}^{\varepsilon}$ is equivalent to finding the number of binary matrices with size $E \times F$ and total weight $\varepsilon$. Most importantly, the minimum weight for each row and each column of these matrices should be $t+1$. To solve this combinatorial problem, we have used a reduced precise number formula for counting the number of binary matrices given in \cite{perez2002reduced}, which has a lower computational complexity than that of the method introduced in \cite{Holzbaur17}. The formula takes a vector of row weights $\mathbf{w}(\boldsymbol{\alpha}) = [w(\boldsymbol{\alpha}_1),w(\boldsymbol{\alpha}_2),\cdots,w(\boldsymbol{\alpha}_E)]$ and a vector of column weights $\mathbf{w}(\boldsymbol{\beta}) = [w(\boldsymbol{\beta}_1),w(\boldsymbol{\beta}_2),\cdots,w(\boldsymbol{\beta}_F)]$, where $w(.)$ outputs the weight and $\boldsymbol{\alpha}_e$, $\boldsymbol{\beta}_f$ represent the $e$-th erroneous row and the $f$-th erroneous column of the stall pattern, respectively. The formula then returns the number of unique binary matrices $\mathcal{A}(\mathbf{w}(\boldsymbol{\alpha}),\mathbf{w}(\boldsymbol{\beta}))$ satisfying the row and column weight requirements. For each column and row weight, the following requirements have to be met:
\begin{align}
F \overset{(a)}\geq w(\boldsymbol{\alpha}_e) \overset{(b)}\geq t+1, \;\; e \in \{1,\cdots,E \},\label{eq:cst1} \\
E \overset{(a)}\geq w(\boldsymbol{\beta}_f) \overset{(b)}\geq t+1, \;\; f \in \{1,\cdots,F \},\label{eq:cst2}
\end{align}
where $(a)$ follows that the maximum row/column weight cannot exceed the size of a stall pattern and $(b)$ follows from Definition \ref{def:1}. Since each error bit locates in one of the erroneous rows and columns, thus:
\begin{equation}\label{eq:cst3}
\sum_{e=1}^E w(\boldsymbol{\alpha}_e) = \sum_{f=1}^F w(\boldsymbol{\beta}_f) = \varepsilon.
\end{equation}
The number of stall patterns with $(E,F,\varepsilon)$ is
\begin{equation}
M_{E,F}^{\varepsilon} = \sum_{\left(\mathbf{w}(\boldsymbol{\alpha}),\mathbf{w}(\boldsymbol{\beta})\right) \in \Psi} \mathcal{A}(\mathbf{w}(\boldsymbol{\alpha}),\mathbf{w}(\boldsymbol{\beta})),
\end{equation}
where $\Psi$ is the set of all pairs of $\mathbf{w}(\boldsymbol{\alpha})$ and $\mathbf{w}(\boldsymbol{\beta})$ satisfying \eqref{eq:cst1}-\eqref{eq:cst3}. However, it has been pointed out in \cite[Section 3]{miller2013} that $\mathcal{A}(\mathbf{w}(\boldsymbol{\alpha}),\mathbf{w}(\boldsymbol{\beta}))$ is unchanged under the permutation of the entries of $\mathbf{w}(\boldsymbol{\alpha})$ and $\mathbf{w}(\boldsymbol{\beta})$. Thus, to avoid the time-consuming calculation for $\mathcal{A}(\mathbf{w}(\boldsymbol{\alpha}),\mathbf{w}(\boldsymbol{\beta}))$ over all possible pairs of $\mathbf{w}(\boldsymbol{\alpha})$ and $\mathbf{w}(\boldsymbol{\beta})$, we can have the following
\begin{equation}\label{eq:nms2a}
M_{E,F}^{\varepsilon} = \sum_{(\mathbf{w}(\boldsymbol{\alpha}),\mathbf{w}(\boldsymbol{\beta})) \in \Phi} \left( \mathcal{A}(\mathbf{w}(\boldsymbol{\alpha}),\mathbf{w}(\boldsymbol{\beta})) \cdot \frac{E !}{\prod_{e=1}^E N_{e} !}  \cdot \frac{F !}{\prod_{f=1}^F N_{f} !} \right),
\end{equation}
where $\Phi$ is a subset of $\Psi$; $N_{e} = |\{w(\boldsymbol{\alpha}_e):  w(\boldsymbol{\alpha}_e) = e\}|,e = 1,2,\cdots,E$ is the number of rows with weight $e$; and $N_{f} = |\{w(\boldsymbol{\alpha}_f):  w(\boldsymbol{\alpha}_f) = f\}|,f = 1,2,\cdots,F$ is the number of columns with weight $f$. The multipliers $\frac{E !}{\prod_{e=1}^E N_{e} !}$ and $\frac{F !}{\prod_{f=1}^F N_{f} !}$ are the number of permutations of entries inside a row vector and a column vector, respectively.

Combining \eqref{eq:nms1}-\eqref{eq:nms1a} and \eqref{eq:nms2a}, the contribution of the $(E,F,\varepsilon)$ stall patterns to the BER and PER error floors can be calculated as
\begin{align}
&\text{BER}_{E,F,\varepsilon} = (A_{E,F}^{1,2}+A_{E,F}^{2,3}) \left(\rho + \xi \right)^{\varepsilon} M_{E,F}^{\varepsilon} \frac{\varepsilon}{2(k-\frac{n}{2})\frac{n}{2}},\label{eq:nms3} \\
&\text{PER}_{E,F,\varepsilon} = (A_{E,F}^{1,2}+A_{E,F}^{2,3}) \left(\rho + \xi \right)^{\varepsilon} M_{E,F}^{\varepsilon},\label{eq:nms4}
\end{align}
where $2(k-\frac{n}{2})\frac{n}{2}$ is the number of information bits of our terminated staircase codes. Here we consider the worst case such that all the error bits of a stall pattern are inside information blocks.

Now we need to evaluate the erroneous decoding probability $\xi$ before we can calculate the BER and PER for the error floor. As pointed out in both \cite{Smith12} and \cite{Holzbaur17}, $\xi$ can only be estimated via simulations and it is related to $\rho$. Here, we use a different approach to estimate $\xi$. We run the simulation for our terminated staircase codes for a low crossover probability and record the $\text{PER}_{E,F,\varepsilon}$ for the minimal stall patterns with $(E=F=t+1,\varepsilon=(t+1)^2)$ by using the number of decoding failures caused by the minimal stall patterns divided by the total number of transmissions. Note that $M_{E,F}^{\varepsilon} = 1$ in this case. Then $\xi$ can be evaluated using Eq. \eqref{eq:nms4}.

\section{An Improved Method To Lower The Error Floor}\label{aim}
In this section, we present and analyze our proposed method for handling stall patterns. We will show that our method can lead to a considerably error floor reduction. We assume that all the errors other than the errors in a stall pattern, are solved by our iterative hard-decision decoder in Algorithm \ref{TSCDA}.

First, we have to define two types of stall patterns which will be useful for our subsequent analysis.
\begin{defi}\label{def:2}
A stall pattern is \textbf{detectable} if all the erroneous rows and columns associated with the stall patterns can be detected by using the parity-check matrix of the component code or CRC.
\end{defi}

\begin{defi}\label{def:4}
A stall pattern is \textbf{undetectable} if either the erroneous rows or columns or both associated with the stall patterns cannot be detected by using either the parity-check matrix of the component code or CRC.
\end{defi}

Stall pattern detection is crucial for correcting stall patterns. The minimal stall pattern $(E=F=3,\varepsilon = 9)$ is an example of detectable stall patterns for a staircase code with double-error-correcting component codes. Since the component code has the minimum distance of 6, any stall pattern where each erroneous row/column has less than or equal to 5 errors, can always be detected. For example, a $(E=F=6,\varepsilon = 18)$ stall pattern is always detectable. However, when the number of row/column errors are larger than 5, i.e., $(E=F=6,\varepsilon = 36)$, it may be undetectable. Furthermore, due to miscorrection, e.g., the BDD decoder outputs an incorrect codeword, some detectable stall patterns may become undetectable after some iterations. In the error floor estimation, we treat these kinds of stall patterns as undetectable stall patterns. Note that when analyzing the occurrence probability of undetectable stall patterns, the number of of $E$ rows and $F$ columns calculated in Section \ref{GSA} are replaced by the number of codewords with weight $E$ and $F$, respectively.

\subsection{Iterative Bit Flipping Algorithm (IBFA)}
We first review the low complexity bit-flip operation which was originally proposed in \cite{Holzbaur17} to solve some of the stall patterns. After a stall pattern is detected and all the location information of erroneous rows and columns is available, the bit-flip operation flips all the bits in the intersections of the erroneous rows and columns associated with the stall pattern. For a staircase code with an $(n=510,k=491,t=2)$ BCH codes as component codes, the bit-flip operation in \cite{Holzbaur17} can successfully solve the detectable stall pattern up to $(E=F=5)$. For any stall pattern with larger $E,F$ values, the algorithm flips one erroneous column. However, this approach cannot guarantee to be successful all the time.

Based on the bit-flip operation in \cite{Holzbaur17}, we propose a new post processing technique called iterative bit flipping algorithm to improve the performance \emph{by solving more stall patterns}. Later in this section we will prove that our approach can solve more stall patterns than the existing design in \cite{Holzbaur17}, leading to a significant error floor reduction.

Our iterative bit flipping operation is automatically triggered after a predefined number of decoding iterations $l_{\text{check}}$ have been completed. The proposed algorithm is embedded in Step 7 of Algorithm \ref{TSCDA} in Section \ref{tscdec}. For simplicity, we let $I$ represent the indices of erroneous rows/columns. The steps for our proposed iterative bit flipping algorithm are summarized in Algorithm \ref{IBFA_ch8} in the following.

\begin{algorithm}
\caption{Iterative Bit Flipping Algorithm (IBFA)}\label{IBFA_ch8}

$\;\;$

$\;\;$\textbf{\emph{Step 1} Locate the stall pattern:} For $i = 1,2,\cdots , L$, calculate the syndrome $\mathbf{S}_i$ for block $[\mathbf{B}_{i-1}^T \; \mathbf{B}_{i}]$. If $\mathbf{S}_i=0$, perform CRC decoding on block $[\mathbf{B}_{i-1}^T \; \mathbf{B}_{i}]$ and obtain $I$ by finding the rows that fail the CRC decoding. Otherwise, obtain $I$ by finding the rows with non-zero syndromes. The size of the stall pattern $(E,F)$ can be determined based on $I$.

$\;\;$

$\;\;$\textbf{\emph{Step 2} Operation selection:} Given $E$ and $F$, if $E,F \geq  2t+5$ or $E = 0$ or $F = 0$, the decoder stops the iteration and output all the information blocks $\hat{\mathbf{M}}_i$ for $i = 1,2,\cdots,L-2$. If $\min(E,F) \leq 2t+1$, proceed to Step 4. Otherwise, store the code block $\mathbf{B}_i$ for $i = 1,2,\cdots,L$ and proceed to Step 3.

$\;\;$

$\;\;$\textbf{\emph{Step 3} Row/column flipping:} If $E > F$, flip the first erroneous column. Otherwise, flip the first erroneous row. Then repeat from Step 1 to Step 6 of Algorithm \ref{TSCDA} for $l_{\text{bf}}$ iterations. Repeat Step 1 of Algorithm \ref{IBFA_ch8} to obtain $I'$, $E'$ and $F'$. If $E'=0$ or $F'=0$, stop the iteration and output all the information blocks $\hat{\mathbf{M}}_i$ for $i = 1,2,\cdots,L-2$. If $\max(E',F') < \max(E,F)$, directly proceed to Step 4. Otherwise, restore block $\mathbf{B}_i$ for $i = 1,2,\cdots, L$ and proceed to Step 4.

$\;\;$

$\;\;$\textbf{\emph{Step 4} All-flipping:} Flip all the bits in the intersection associated with the $(E',F')$ stall pattern. Then repeat from Step 1 to Step 6 of Algorithm \ref{TSCDA} for $l_{\text{bf}}$ iterations. Repeat Step 1 of Algorithm \ref{IBFA_ch8} to obtain $I''$, $E''$ and $F''$. If $E''=0 $ or $F''=0$, stop the iteration and output all the information blocks $\hat{\mathbf{M}}_i$ for $i = 1,2,\cdots,L-2$. Otherwise proceed to Step 5.

$\;\;$

$\;\;$\textbf{\emph{Step 5} Final check:} If $\min(E'',F'') \leq 2t+1$, repeat Step 4. If $\min(E'',F'') = 2t+2$, repeat Step 3. Otherwise, stop the iteration and output all the information blocks $\hat{\mathbf{M}}_i$ for $i = 1,2,\cdots,L-2$.

$\;\;$

\end{algorithm}

\begin{remark}
In \cite{Holzbaur17}, the bit-flip algorithm always flips an erroneous row when $E \geq 6$ and $F \geq 6$ for $t=2$. Here, we consider flipping the erroneous column or row by comparing $E$ to $F$. For example, if $E> F$, there is a higher probability that the number of errors in one column is larger than the number of errors in one row. In this case, flipping one erroneous column has a higher probability to solve more errors than flipping one erroneous row. This will become clearer in Section \ref{stall_solve}. In Step 2 of Algorithm \ref{IBFA_ch8}, the decoder does not correct any stall patterns with $E,F \geq 9$ for $t=2$. This is because the stall patterns with this size are less likely to appear and may not be solved successfully. In addition, when the crossover probability of the BSC is high, there exist non-stall-pattern error events with very large $E,F$ and applying the algorithm on these errors could introduce more errors. In Step 3, if the dimension of the stall pattern is not reduced after row/column flipping, it means the row/column flipping is not effective. We thus restore all code blocks in Step 3 and then perform the all-flipping operation. This approach has been proved to solve more stall patterns in Section \ref{stall_solve}. In addition, the restoration process can guarantee no extra errors are introduced when all decoding attempts fail. However, the bit-flip algorithm in \cite{Holzbaur17} only repeat row flipping or all flipping twice, which may not be effective for stall patterns with larger size and larger value of $\varepsilon$.
\end{remark}

\begin{exam}
Here we provide a simple example to illustrate how our decoder works. Assume that the underlying component code is with $t=2$. Now consider a $(E=F=6,\varepsilon = 21 )$ stall pattern shown in Fig. \ref{fig:stall3}.
\begin{figure}[ht!]
	\centering
\includegraphics[width=1.8in,clip,keepaspectratio]{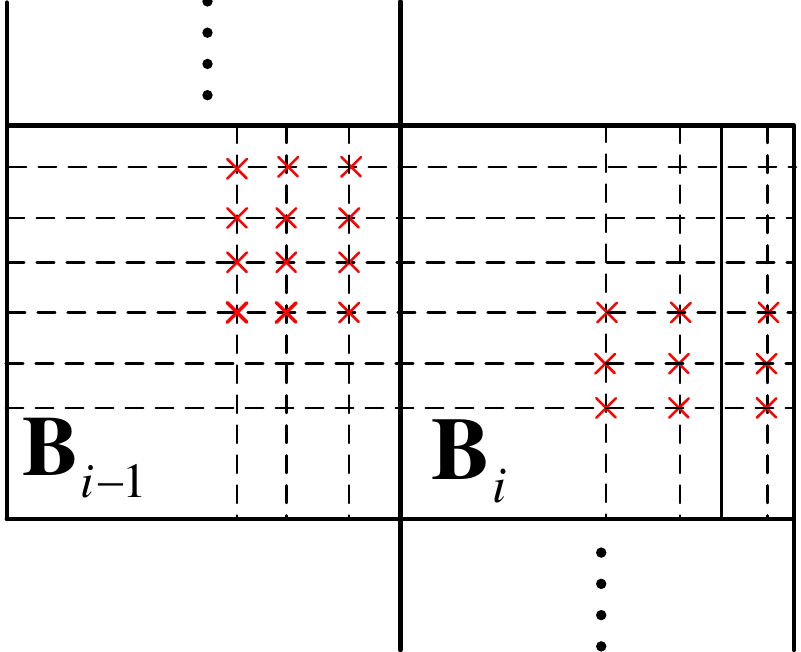}
\caption{A $(E=6,F=6,\varepsilon = 21)$ stall pattern.}
\label{fig:stall3}
\end{figure}
Our decoder flips the first erroneous row according to Step 3 of Algorithm \ref{IBFA_ch8}. Then, three erroneous columns have weights reduced to 3 and the other three columns have weights increased to 4 as shown in Fig. \ref{fig:stall4}.
\begin{figure}[ht!]
	\centering
\includegraphics[width=1.8in,clip,keepaspectratio]{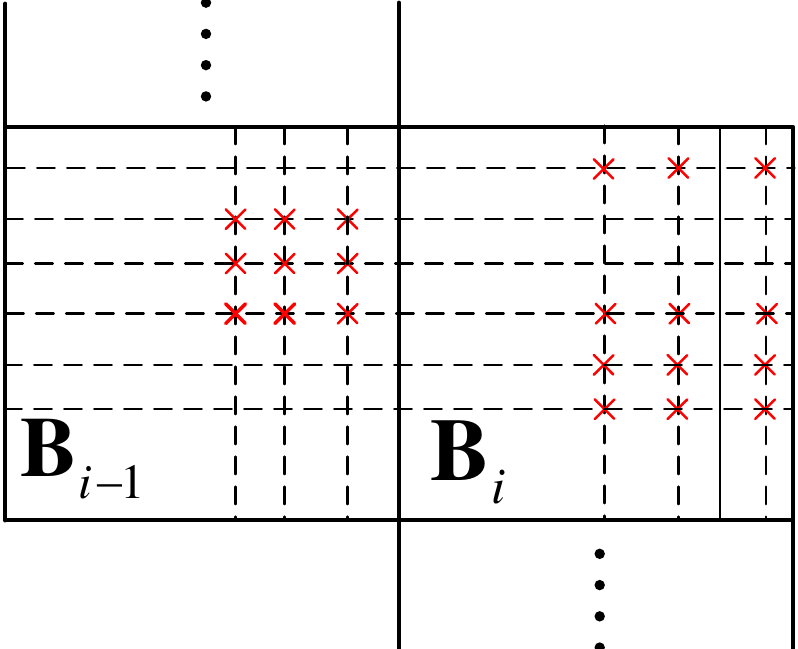}
\caption{The stall pattern after row flipping.}
\label{fig:stall4}
\end{figure}
None of the errors can be decoded by the BDD decoder since each erroneous row and column has weight larger than $t$. As the size of this stall pattern is not reduced after row flipping, i.e., $E'=E = F'=F = 6$, the decoder restores the code block such that the stall pattern becomes the one in Fig. \ref{fig:stall3} and then applies all-flipping operation in Step 4 of Algorithm \ref{IBFA_ch8}. The resultant stall pattern is shown in Fig. \ref{fig:stall5}.
\begin{figure}[ht!]
	\centering
\includegraphics[width=1.8in,clip,keepaspectratio]{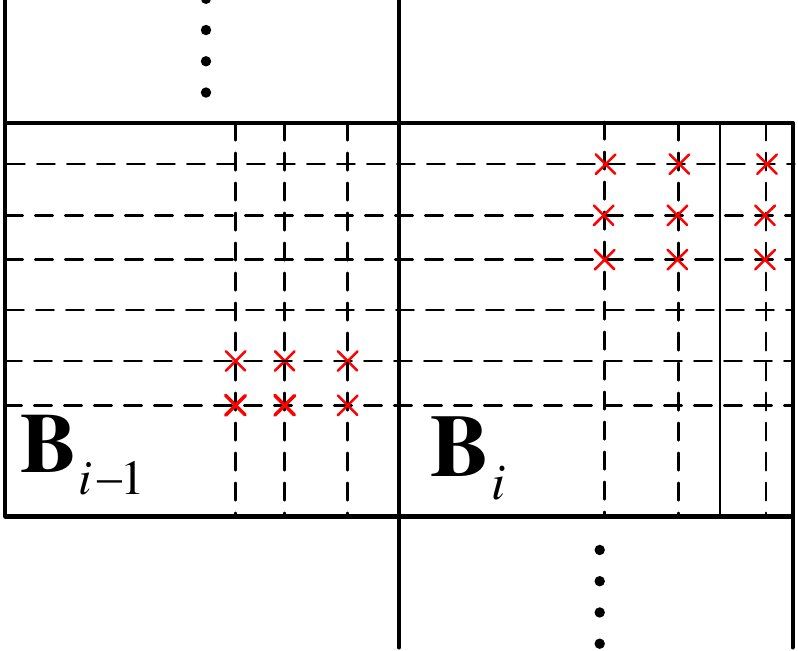}
\caption{The stall pattern after restoration and all-flipping.}
\label{fig:stall5}
\end{figure}
It can be seen that easily seen that all the errors in the first three erroneous columns (from left to right) can be corrected by the BDD decoder now. Then the stall pattern in Fig. \ref{fig:stall5} becomes a $(E'' = 3, F'' = 3,\varepsilon =9)$ stall pattern and can be solved by all-flipping operation in Step 4 of Algorithm \ref{IBFA_ch8}.

If using the bit-flip algorithm in \cite{Holzbaur17} to decode this stall pattern, only one erroneous row will be flipped twice and the stall pattern still cannot be solved.

\end{exam}

\subsection{Analysis of the Proposed Iterative Bit Flipping Algorithm}\label{stall_solve}
In this subsection, we analyze our proposed iterative bit flipping algorithm and prove that our decoder is able to solve more stall patterns than the conventional staircase codes. There results are useful for estimating the error floor later on. First, the following lemma is useful.
\begin{lemm}(Theorem 1 in \cite{Holzbaur17}) \label{the0}
Consider a terminated staircase code whose component code can correct $t$ errors. The staircase code decoder using all-flipping operation can correct all the errors for any detectable $(E,F)$ stall pattern such that $\min(E,F) \leq 2t+1$.
\end{lemm}

\emph{\quad Proof: }
See Appendix \ref{appendix:0}.
\QEDA
%

The above stall patterns only need a single all-flipping operation to correct. For any stall patterns with $(E>2t+1,F>2t+1)$, both row/column flipping and all-flipping operations are required. Based on Algorithm \ref{IBFA_ch8}, we present the main theorem of this work as follows.

\begin{theo}\label{the1}
Consider a terminated staircase code whose component code can correct $t$ errors. Our decoder with the proposed iterative bit flipping algorithm can always correct the following detectable\footnote{Here we assume that the stall patterns remain detectable in each iteration of IBFA. The undetectable stall patterns will be treated separately in the analysis in Section \ref{err_fr_est}.} stall patterns:\\
 1. $(E=F = 2t+2)$ stall patterns for any $\varepsilon$ when $t \geq 1$; \\
 2. $(E= 2t+2,F= 2t+3)$ stall patterns for any $\varepsilon$ when $t \geq 1$; \\
 3. $(E = F=2t+3,\varepsilon \leq F(t+1)+t \; \text{or} \; \varepsilon \geq F(t+2)+1 )$ stall patterns when $t \geq 1$; \\
 4. $(E= 2t+2,F= 2t+4)$ stall patterns for any $\varepsilon$ when $t \geq 1$; \\
 5. $(2t+3 \leq E \leq 2t+4,F= 2t+4, \varepsilon \leq F(t+1)+1)$ stall patterns when $t \geq 2$. \\
 Note that the value of $E$ and $F$ can be swapped.
\end{theo}
\emph{\quad Proof: }
See Appendix \ref{appendix:1_ch8} for the proof of Theorem \ref{the1}-1. \\
See Appendix \ref{appendix:2_ch8} for the proof of Theorem \ref{the1}-2. \\
See Appendix \ref{appendix:3_ch8} for the proof of Theorem \ref{the1}-3. \\
See Appendix \ref{appendix:4_ch8} for the proof of Theorem \ref{the1}-4. \\
See Appendix \ref{appendix:5} for the proof of Theorem \ref{the1}-5.
\QEDA

\begin{remark}
We have rigorously proved that not only all the stall patterns in \cite[Table 1]{Holzbaur17} for $t=2$, but also other some larger size stall patterns can always be successfully solved by using Algorithm \ref{IBFA_ch8}. For the detectable stall patterns that are not included in the above theorems, our decoder can still solve them with some probability. The reason of decoding failure is mainly due to the fact that the errors in each erroneous row and column remain larger than $t+1$ even after row/column flipping and all-flipping operations. This can happen when the size of the stall pattern is large and the number of errors inside the stall pattern is relatively small.

We point out here that the maximum number of bit-flipping required for solving stall patterns is 6. Note that among those solvable stall patterns in Theorem \ref{the1}, the stall patterns with the largest size are with $(E=F=2t+4)$. It will take up to 5 bit-flipping iterations to reduce the stall patterns with $(E=F=2t+4)$ to stall patterns with $\min\{E,F\} =2t+1$. This is because our iterative bit-flipping algorithm can solve at least one erroneous row or one erroneous column in one iteration. Otherwise, if the size of the stall pattern is not reduced after one iteration of Algorithm \ref{IBFA_ch8}, the decoder stops the iteration and outputs all the code blocks. And only one iteration of bit-flipping is required for solving the stall pattern with $\min\{E,F\} =2t+1$ according to Lemma \ref{the0}. That being said, 6 iterations only occur in the worst case while in practice the required number of iterations is smaller than that. For the bit-flip algorithm in \cite{Holzbaur17}, two iterations are required.
\end{remark}

\subsection{An Improved Error Floor Estimation}\label{err_fr_est}
In this subsection, we estimate the error floor for our design example of the terminated staircase code with double-error-correcting BCH component code introduced in Section \ref{TSC}.

When estimating the error floor, we treat the detectable and undetectable stall patterns separately. Undetectable stall patterns have row/column errors larger than 6. For example, a $(E=F=6,\varepsilon = 36)$ stall pattern can be undetectable. If this stall pattern has fewer errors, e.g., $(E=F=6,\varepsilon = 18)$, it can be successfully detected by parity-check. However, if the column errors are undetectable, this stall pattern cannot be solved by our Algorithm \ref{IBFA_ch8}. This is because the erroneous columns whose weights are increased after the row flipping, will be incorrectly decoded to wrong codewords. In such a case, the column becomes undetectable after Step 3 of Algorithm \ref{IBFA_ch8}. Therefore, we count this kind of stall pattern as undetectable stall patterns even though it is detectable before performing row/column flipping. As such, we calculate the contribution to the error floor from detectable and undetectable stall patterns separately.

We then adopt the analytical method shown in Section \ref{GSA} to calculate the error floor for our proposed codes. We first run the simulation for our terminated staircase codes with Algorithm \ref{TSCDA} (exclude Algorithm \ref{IBFA_ch8}) for a various of crossover probability $\rho$ to collect the number of decoding failures caused by the minimal stall pattern and the total number of transmissions. Given the aforementioned information, we use Eq. \eqref{eq:nms4} to obtain the incorrect decoding probability $\xi$. According to Theorem \ref{the1}, the detectable stall patterns that may not be solvable by Algorithm \ref{IBFA_ch8} include but not limited to $(E=7,F=7, 24 \leq \varepsilon \leq 28)$, $(7 \leq E \leq 8,F=8, \varepsilon \geq 26)$. The dominant undetectable stall pattern with the smallest size is $(E=3,F=6,\varepsilon=18)$. For the undetectable $(4 \leq E \leq 5,F=6)$ stall pattern, the number of errors satisfies $\varepsilon= E\cdot F$. This is because if $\varepsilon < E\cdot F$, the stall pattern can be detected in Step 2 of Algorithm \ref{IBFA_ch8} and then all the erroneous columns can be corrected after the all-flipping operations.
Therefore, the next dominant undetectable stall pattern is with $(E=F = 6$, $\varepsilon=18)$. It should be noticed that when calculating the multiplicity term in Eq. \eqref{eq:nms1} and Eq. \eqref{eq:nms2} for undetected stall patterns, we replace the multiplier $\binom{n}{F}$ with the number of valid BCH codeword with weight $F$. This number can be well approximated as $\binom{n}{F}/(n+1)^t$ by using Peterson Estimation \cite{Micheloni:2012:ISS:2412028}. We then use Eq. \eqref{eq:nms3} and Eq. \eqref{eq:nms4} to compute the BER and PER contributions of the above dominant stall patterns. An example for the contribution to the error floor for the dominant stall patterns under the setting of $\rho = 5.667 \cdot 10^{-3}$ and $\xi = 1.395 \cdot 10^{-3}$ is shown in Table \ref{table1_ch8}. We pessimistically assume that all of these stall patterns are uncorrectable. The stall patterns which have negligible contribution to the error floor are not taken into account.
\begin{table*}[tbp]
  \centering
 \caption{Contribution to error floor estimation of stall patterns.}\label{table1_ch8}
\begin{tabular}{|c|c|c|c|}
  \hline
  $(E,F,\varepsilon)$ & $\Phi: \{(\mathbf{w}(\boldsymbol{\alpha}),\mathbf{w}(\boldsymbol{\beta}))\} $ & $\text{BER}_{E,F,\varepsilon}$ & $\text{PER}_{E,F,\varepsilon}$ \\ \hline
Detectable $(7,7,24)$ &  $(4,4,4,3,3,3,3)$ & $6.043 \cdot 10^{-17}$ & $3.031 \cdot 10^{-13}$ \\ \hline
Detectable $(7,7,25)$ & $(4,4,4,4,3,3,3)$ & $2.913 \cdot 10^{-18}$ & $1.402 \cdot 10^{-14}$ \\ \hline
    \multirow{4}{*}{Detectable $(7,8,26)$} &  $(4,4,3,3,3,3,3,3)$  & \multirow{4}{*}{$2.039 \cdot 10^{-17}$} & \multirow{4}{*}{$9.438 \cdot 10^{-14}$} \\
  & $(4,5,5,3,3,3,3)$ &  &  \\
    &  $(4,4,5,4,3,3,3)$ &  &  \\
    &  $(4,4,4,4,4,3,3)$ &  &  \\ \hline
   \multirow{2}{*} {Undetectable $(3,6,18)$} &  $(3,3,3,3,3,3)$ & $7.145 \cdot 10^{-29}$ & $4.778 \cdot 10^{-25}$ \\
   &  $(6,6,6)$ &  &  \\ \hline
   Undetectable $(6,6,18)$ & $(3,3,3,3,3,3)$ & $2.765 \cdot 10^{-18}$ & $1.849 \cdot 10^{-14}$ \\ \hline
\end{tabular}
\end{table*}

When calculating $M_{E,F}^{\varepsilon}$ for these stall patterns, we have to determine the set $\Phi$ according to Eq. \eqref{eq:nms2a}. It should be noticed that not all weight vector pairs in the set $\Phi$ can produce an unsolvable stall patterns. Consider the $(E=F=7,\varepsilon=24)$ stall pattern as an example. If the weight vector pairs are $\mathbf{w}(\boldsymbol{\alpha}) = [4,5,3,3,3,3,3]$ and $\mathbf{w}(\boldsymbol{\beta}) = [4,5,3,3,3,3,3]$, then it can always be corrected by Algorithm \ref{IBFA_ch8}. This is because the erroneous row and column with weight 5 will become weight 2 after the all-flipping operation in Step 4 of Algorithm \ref{IBFA_ch8}. This will result in a stall pattern with $E=F=6$ which can be successfully solved according to Theorem \ref{the1}-1. Any permutations on this weight vector pairs will have the same result. Similarly, for $(E=F=7,\varepsilon =25)$ stall pattern, any row weight or column with weight larger than or equal to 5 will be corrected after Step 4 of Algorithm \ref{IBFA_ch8}, resulting in a stall pattern with $E=6,F=7$ in the worst case which can be successfully solved according to Theorem \ref{the1}-2. For the $(E=7,F=8,\varepsilon= 26)$ stall pattern, any row with weight larger than or equal to 6 will be corrected after Step 4 of Algorithm \ref{IBFA_ch8}. Thus it then becomes a stall pattern with $E=6,F=8$ which can be solved based on Theorem \ref{the1}-4. Therefore, we only consider the vector pairs that are associated with uncorrectable stall patterns. In this way, the error floor contribution of stall patterns can be calculated more accurately as the number of uncorrectable stall patterns is not \emph{overestimated}. These vector pairs are listed in the second column of Table \ref{table1_ch8}. It can be seen that the smallest undetectable stall pattern has negligible contribution to the error floor. All other stall patterns has the BER contribution lower than $10^{-15}$.
We then estimate the error floor via
\begin{align}
\text{BER}_{\text{floor}} = \sum_{E,F}\sum_\varepsilon\text{BER}_{E,F,\varepsilon} ,\label{eq:efber} \\
\text{PER}_{\text{floor}} = \sum_{E,F}\sum_\varepsilon\text{PER}_{E,F,\varepsilon} .\label{eq:efper}
\end{align}
This gives a BER of $8.649 \cdot 10^{-17}$ and a PER of $4.299 \cdot 10^{-13}$.

\section{Complexity Analysis}
In this section, we briefly discuss the encoding, decoding complexity and latency of the proposed code compared to the conventional staircase codes and other codes suitable for NAND flash memories.

\subsection{Encoder Complexity}
First, we investigate the encoding complexity, which includes implementation complexity, computational complexity and encoding latency of the proposed codes. Regarding the implementation complexity, our encoder requires a memory of $\frac{n}{2} \times \frac{n}{2}$ bits to store a code block as the encoded block $\mathbf{B}_{i-1}$ will be used in the encoding of $\mathbf{B}_i$. It also requires a component code encoder unit and a CRC encoder. Since the CRC is 1 bit, the CRC encoding can be performed by an adder. Compared with the conventional staircase codes \cite{Holzbaur17} whose component codes are of the same size\footnote{We say both component codes are of the same size rather than the same code due to different ways of using CRC. The difference is explained in Remark~\ref{remark:diff}.} as ours, the implementation complexity is the same because both component code encoding and CRC encoding are required. Regarding the computational complexity, both codes require the same number of component code encoding and CRC encoding for one code block. Turning to the encoding latency, we denote the time required for one time component code encoding and CRC encoding by $T_B(n,k)$ and $T_C(n)$, respectively. To produce one code block, the required time for our staircase codes is $(T_B(n,k)+T_C(n))\frac{n}{2}$. For conventional staircase codes, the required time is between $T_B(n,k)+T_C(n)$ and $\left(T_B(n,k)+T_C(n)\right)\frac{n}{2}$. When parallel encoding is available, the conventional staircase code encoder can encode at most $\frac{n}{2}$ component codewords at the same time while our encoder still has to encode each component codeword one after one. This is because the encoded CRC bit $r_{j,i}$ associated with the $j$-th row will be used in the encoding of the $(j+1)$-th row in our design. Therefore, our terminated staircase codes have a higher encoding latency than conventional staircase codes in general. That being said, during the writing access, the encoder can output a single code block $\mathbf{B}_i$ once the encoding process for this block is finished. This is more efficient than traditional linear block code encoding, e.g., encoding of BCH codes, where the encoder has to output the whole codeword only when all the information bits are encoded.

\subsection{Decoder Complexity}

We compare the decoding complexity, latency and the implementation complexity of the proposed code to those of conventional staircase codes.

Similar to the architecture of product code decoder in \cite{Smith12}, our decoder consists of a data storage unit for the product code array, a syndrome storage unit, and a BDD decoder unit and a CRC decoder. Unlike our encoder which has to encode information recursively, our decoder can decode multiple component codes simultaneously when parallel decoding is available. Thus, the computational complexity and the decoding latency of our codes are the same as that of the conventional staircase codes. However, with the proposed iterative bit flipping algorithm, additional storage is required to store the location of erroneous rows and columns associated with a stall pattern. The index of a row/column can be specified using $\lceil\log_2(n)\rceil+1$ bits. As a result, the extra storage for all the location information is $2(E+F)(\lceil\log_2(n)\rceil+1)$ bits. The factor ``2'' here is due to the storage for indices obtained from both parity-check and CRC check. Although in Step 3 of Algorithm \ref{IBFA_ch8} the code blocks are stored, the storage resource can be taken from that in Step 4 of Algorithm \ref{TSCDA} while updating the code blocks. For the internal data flow, i.e., the rate of routing/storing messages, our terminated staircase decoder shown in Algorithm \ref{TSCDA} has a similar data flow as that of the decoder in \cite{Smith12}. This data flow is much lower than that of soft message-passing decoder \cite{Smith12}.

\section{Simulation Results}\label{num}
\subsection{Error Probability}
\begin{figure}[ht!]
	\centering
\includegraphics[width=3.43in,clip,keepaspectratio]{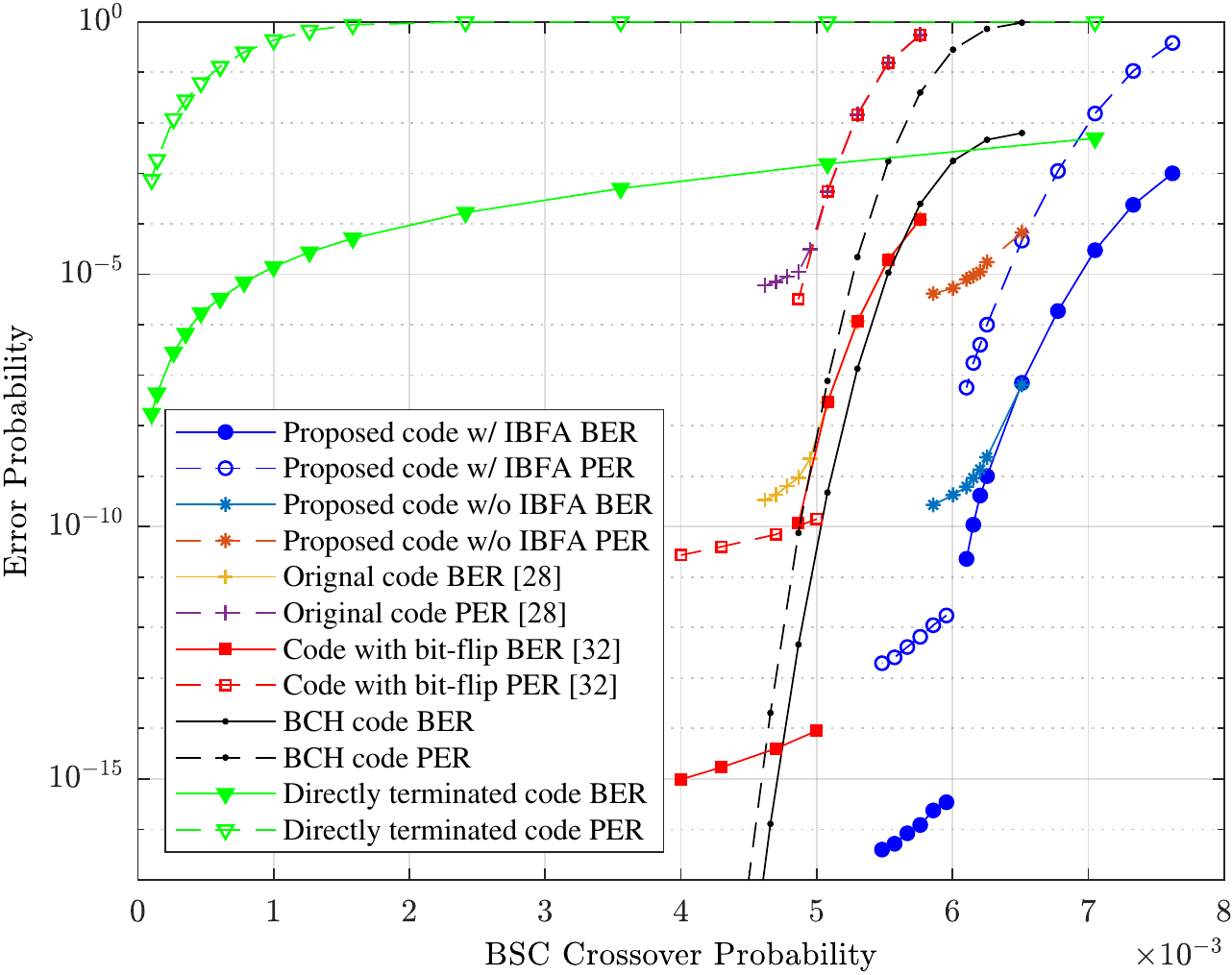}
\caption{Simulation results for BER(solid line) and PER(dash line).}
\label{fig:BER1}
\end{figure}

In this subsection, we present the simulation results for our proposed terminated staircase codes. First, to illustrate the effectiveness of introducing our iterative bit flipping algorithm, we evaluate the performance for our codes with and without Algorithm \ref{IBFA_ch8}. The performance is measured in terms of BER and PER versus the crossover probability of the BSC $\rho$ and is shown in Fig. \ref{fig:BER1}. The error floor calculated in Section \ref{err_fr_est} is plotted in Fig. \ref{fig:BER1} and its curves are labelled with ``w/ IBFA''. In the mean time, we simulate the error floor performance for our code without the iterative bit flipping algorithm whose curves are labelled with ``w/o IBFA'' in the figure. Moreover, the error probability of the staircase code with direct termination (i.e., the last code block is $\mathbf{B}_2$) is also plotted in Fig. \ref{fig:BER1}. To carry out performance comparisons, we plot the error performance of the conventional staircase codes, including the unterminated staircase codes with bit-flip operation \cite{Holzbaur17} and the original staircase codes (i.e., without solving any stall patterns) \cite{Smith12}, in Fig. \ref{fig:BER1}. Note that the error floor performance of the staircase codes with bit-flip \cite{Holzbaur17} is estimated by using the method proposed in \cite[Sec. 4]{Holzbaur17} as the simulation therein only shows the \emph{expected error floor region}. For fair comparison, we only consider that the conventional staircase codes and our proposed codes use the component codes which are of the same rate and size. We also plot the error performance for a stand-alone BCH code with the same information length and same rate. Note that the performance of this long BCH code is obtained by calculating the error probability analytically \cite[Sec. V-B]{Cho14}.

From Fig. \ref{fig:BER1}, it can be observed that our proposed 16K bytes terminated staircase code significantly outperforms the stand-alone BCH code with the same length and code rate. In particular, our code has a BER which is about six orders of magnitude lower than that of the BCH code under the same crossover probability. As more stall patterns are solved by our proposed decoder, our proposed code reaches an error floor about $10^{-16}$ for BER and $10^{-12}$ for PER which is more than one order of magnitude lower than the improved codes in \cite{Holzbaur17}. Without the iterative bit flipping algorithm, our code has almost the same error floor performance as the original staircase codes \cite{Smith12}. Thus, our staircase code with the proposed algorithm is superior to the original staircase code by reaching an error floor that is more than six orders of magnitude lower. This huge performance gain is achieved with slightly increased decoding complexity. For the staircase code with direct termination, its performance is severely degraded since it requires $\rho<10^{-4}$ to allow the BER reaching below $10^{-8}$. We also compare our code to the SC-LDPC code in \cite{7553579} with $(n, k) =  (147420, 128777)$ and with rate 0.8735 under the BSC. From \cite[Fig. 3]{7553579}, it can be seen that to reach the BER below $10^{-10}$,  the crossover probability needs to be larger than $5 \cdot 10^{-3}$ while for our code is less than $6 \cdot 10^{-3}$. Hence, our code has better decoding performance than the SC-LDPC codes in \cite{7553579} under the BSC. In addition, our code shows no error floor for BER lower than $10^{-11}$ while the SC-LDPC code only shows no error floor for BER above $10^{-10}$.

We point out here that even using the same-size component codes, the unterminated staircase codes and our terminated staircase codes have different code rates. This is because our termination method introduces a rate loss in our codes. In the above example, our terminated staircase code has a code rate of 0.8899 while the unterminated staircase code whose component code is of the same size as ours, has a code rate of 0.9255 \cite{Holzbaur17} (which is the same as that of the directly terminated staircase code). However, we emphasize that as one of the main contributions of this work is to lower the error floor of terminated staircase codes and therefore the comparisons in the error floor regime are meaningful. It might be possible to use efficient termination mechanisms to further reduce the rate loss. However, to the best of our knowledge, other termination techniques for staircase codes have not been reported in the literature yet. Hence, we only focus on this simple termination approach which is sufficient for the purpose of our work and leave the exploration of other termination mechanisms as future work.

\subsection{Computational Complexity}
\begin{figure}[ht!]
	\centering
\includegraphics[width=3.43in,clip,keepaspectratio]{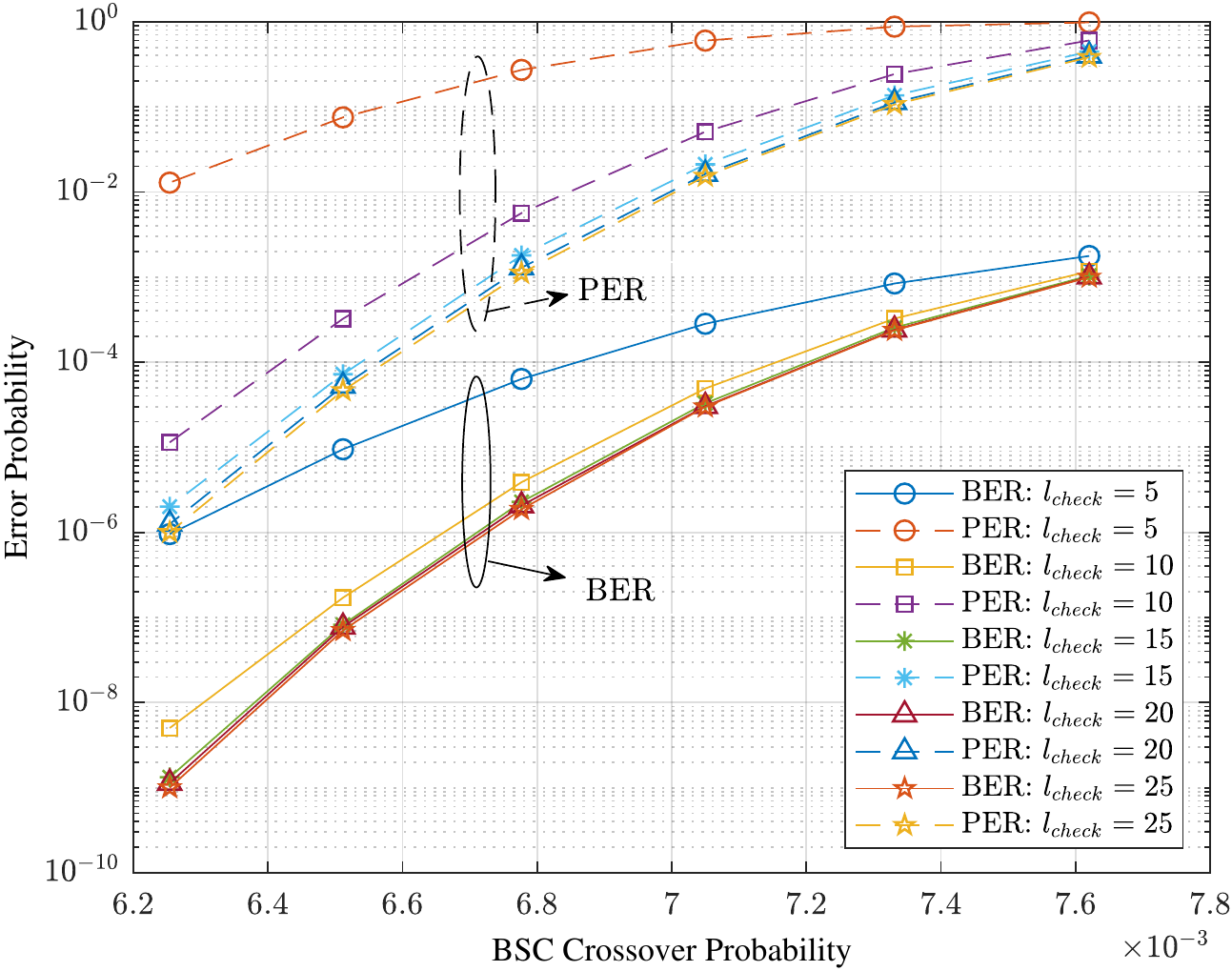}
\caption{Error performance of our code with $l_{\text{check}}= 5,10,15,20,25$.}
\label{fig:itrp}
\end{figure}

The above simulations for our terminated staircase codes are performed under $l_{\text{check}} = 25$. We now plot the error performance for our code with different number of iterations, i.e., $l_{\text{check}} =5,10,15,20,25$, in Fig. \ref{fig:itrp}.

\begin{figure}[ht!]
	\centering
\includegraphics[width=3.43in,clip,keepaspectratio]{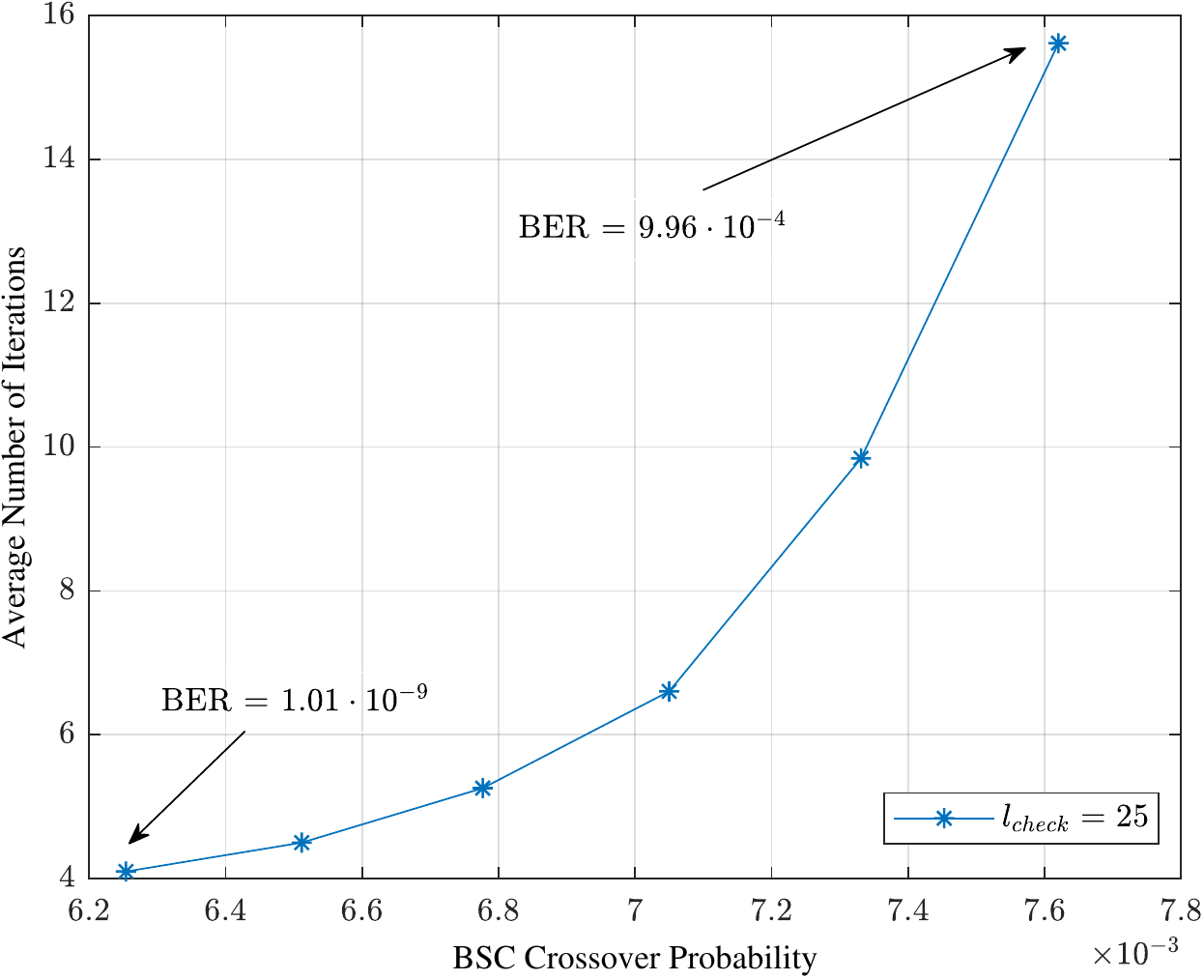}
\caption{Average number of iterations when $l_{\text{check}}=25$.}
\label{fig:itr2}
\end{figure}

\begin{figure}[ht!]
	\centering
\includegraphics[width=3.43in,clip,keepaspectratio]{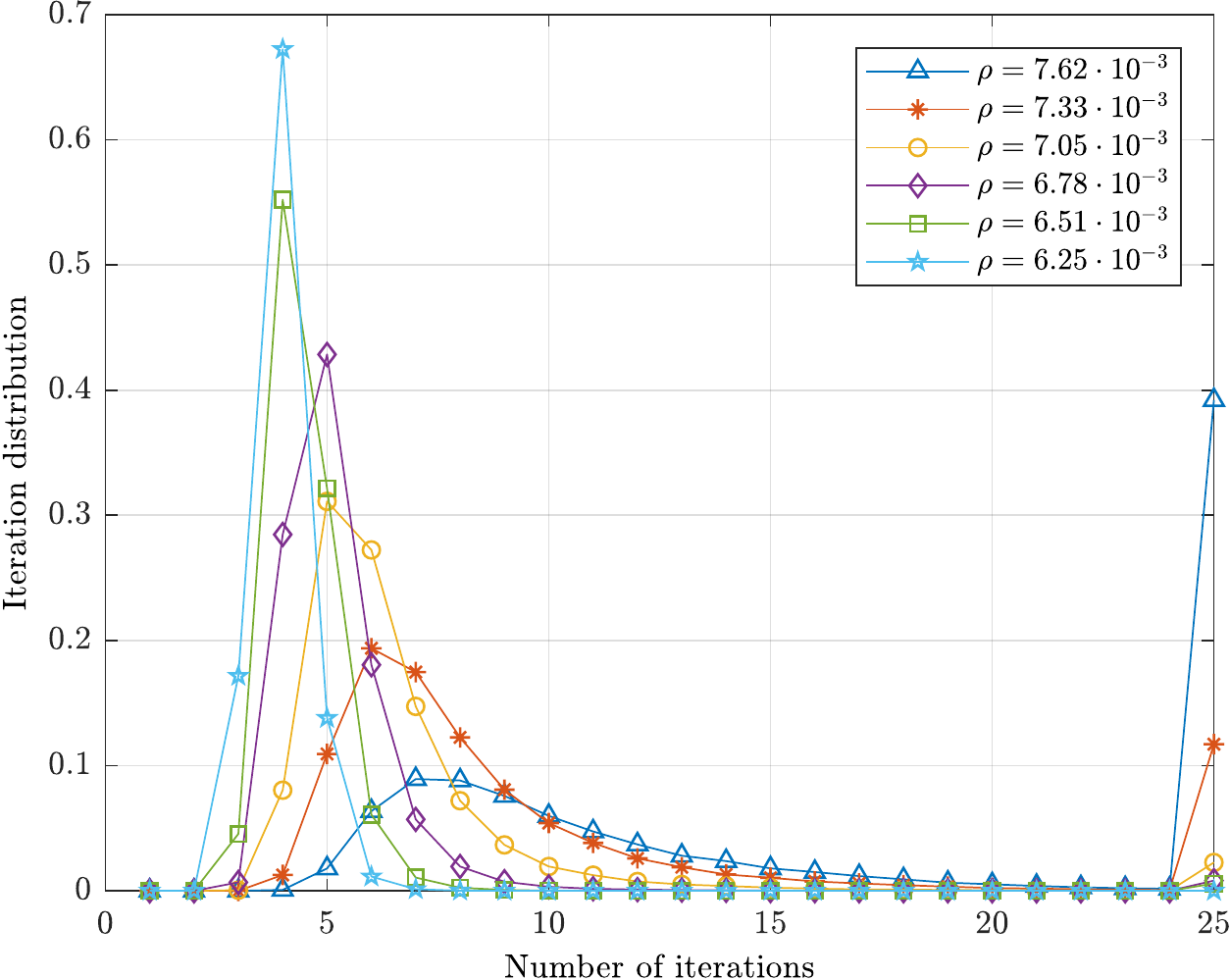}
\caption{Iteration distribution for various BSC crossover probabilities.}
\label{fig:itrdis}
\end{figure}

It can be observed that when $l_{\text{check}}\geq 15$, the performance loss is negligible. Note that in this work, we set $l_{\text{check}}=25$ in order to attain the best possible performance. In practice, we can set $10 \leq l_{\text{check}}\leq 15$ while the performance loss is still small according to Fig. \ref{fig:itrp}. To further investigate the complexity and throughput of our decoder, we plot the average number of iterations versus the BSC crossover probability and the corresponding iteration distribution in Fig. \ref{fig:itr2} and Fig. \ref{fig:itrdis}, respectively.


Here, the number of iterations in Fig. \ref{fig:itrdis} is the sum of the number of iterations of staircase code decoding and the number of iterations for bit-flipping. In addition, the range of the crossover probability corresponds to the BER from $10^{-3}$ to $10^{-9}$. It can be observed that in the low BER region, the average number of total iterations converges to 4. Since the iterative bit flipping algorithm only starts working in the low BER regime, the extra iterations introduced by our iterative bit flipping algorithm is negligible because the occurrence probability of stall patterns is very small. As a result, the throughput decrease due to using the iterative bit flipping algorithm can be deemed negligible. Compared with the decoder in \cite{Holzbaur17} when its maximum number of iterations is the same as ours, the average number of iterations is also the same as ours. This is because in the high BER regime, both decoders do not correct stall patterns while in the low BER regime the extra iteration introduced by our iterative bit flipping algorithm is negligible. Compared with other product code schemes such as the 8K bytes BCH product code in \cite{Cho14} where the maximum number of iterations could be as large as 1024 \cite[Figs. 3-4]{Cho14}, our decoder requires much less iterations to lower the error floor performance.

\section{Summary}
In this work, we proposed a class of staircase codes and designed an example of such code for NAND flash memories with 16K bytes page size. Most notably, we developed a new coding structure by performing CRC encoding and decoding to each component codeword, providing additional error detection capabilities. The CRC bits are protected by both row and column codewords. We then proposed a novel iterative bit flipping algorithm to handle stall patterns. Most compellingly, we proved and showed that the decoder with our proposed algorithm can solve more stall patterns, resulting in a lower error floor than that of the conventional staircase codes. A more accurate error floor estimation for our codes was presented. Theoretical analysis demonstrates that our proposed codes can satisfy the BER and code rate requirements for flash memory devices. Furthermore, simulation results show that our proposed codes can significantly outperform the stand-alone BCH codes and conventional staircase codes.

\chapter{Thesis Conclusions}\label{C9:chapter9}

In this thesis, the problems of point-to-point and downlink multiuser communications have been studied and addressed. Specifically, we have proposed practical lattice coding schemes to approach the capacity of the AWGN channel and the capacity of the downlink multiuser channel. In addition, we also have introduced powerful error-correction codes and a new decoding algorithm for digital systems that requires ultra-high reliability, such as optical fibre communications and data storage. We conclude this thesis in the following by summarizing our main contributions.

In Chapter 1, we have presented an overview of the future 5G communication networks followed by our motivations of this thesis. Previous works related to coding designs and downlink NOMA have been reviewed and discussed. Thesis organization and the main contributions of each work conducted in this thesis have also been provided.

In Chapter 2, we have introduced some fundamental knowledge of lattices that are useful in the later chapters. This includes many important definitions of lattices, the figures of merit and some well-known construction methods of lattices.

In Chapter 3, we have provided the necessary background knowledge on wireless communications and channel coding. In particular, we have introduced different types of wireless channel models and coding schemes that are related to our work.

In Chapter 4, we have given the detailed description of our first work: designing practical lattice codes to approach the unconstrained Shannon limit. By introducing a novel encoding structure for our multi-dimensional lattice codes, we have proved that the proposed lattice codes exhibit two properties: permutation-invariance and symmetry. These properties allow us to use one-dimensional EXIT charts to design and optimize the degree distribution of the underlying lattice codes. We have provided simulation results to show that the designed codes can approach the capacity of the AWGN channel within 0.46 dB.

In Chapter 5, we have described our proposed lattice-partition framework of $K$-user downlink NOMA without SIC. The proposed framework has many desirable properties such as explicit and systematic design, using discrete and finite inputs, and has lower complexity than conventional NOMA with SIC receivers. The individual achievable rate of the proposed framework based on any $n$-dimensional lattice has been analyzed the its gap to the multiuser capacity region has been derived. It has been proved that the upper bound of the gap is constant and universal to all SNR and $K$. The gap can be further reduced by using multi-dimensional lattices that have higher NSM. Our simulation results have verified the correctness of our analysis and the effectiveness of the proposed framework.

In Chapter 6, we have presented a new lattice-partition scheme for $K$-user downlink NOMA without SIC for slow fading channels without transmit CSI. In this work, the modulation and coding scheme have been carefully designed based on the statistical CSI. Most importantly, we have analyzed the individual outage rate of the proposed scheme based on any $n$-dimensional lattice and have derived its gap to the multiuser outage capacity. The gap can be upper bounded by a constant independent of the channel gain and the number of users. Our simulation results presented in this chapter have shown that the scheme can attain the near-capacity performance even with single-user decoding for each receiver.

In Chapter 7, a class of NOMA schemes without SIC has been presented. Different from any of our schemes in the previous chapters, we aim to achieve full diversity gain for each downlink user in the presence of block fading by carefully designing the signal constellation for each user. Within the proposed class, a special family of NOMA schemes based on lattice partitions of the underlying ideal lattices has been identified. The analysis in this chapter have shown that the lattice-partition schemes achieve the largest minimum product distances of the superimposed constellations, which guarantees full diversity and better error performance. Moreover, our designs have been extended to the multi-antenna case where similar analysis the results are also presented.

In Chapter 8, we have presented an additional work to design a class of staircase codes with improved decoding performance for storage systems. Our new designs allow the decoder to detect and correct more stall patterns which may not be correctable in the previous staircase code designs. An improved staircase code decoder that iteratively detects and corrects the stall patterns have been proposed and presented. We have proved and shown that the new decoder can solve specific types of stall patterns, leading to lower error floor. Further to this, a more accurate method used for estimating the error floor performance of the proposed staircase code by using the proposed decoder has been introduced. Numerical results have confirmed the effectiveness of our design.

\appendix
\onehalfspacing
\fancyhead[CE]{\leftmark}
\fancyhead[CO]{\leftmark}

\chapter{Proof of Theories of Chapter 4}\label{appendix_2a}

\section{Proof of Proposition \ref{linear}}
\label{appendix:1}
We divide our encoder described in Section \ref{IRAE} into two parts: the first part is from the input of the repeater to the output of the interleaver; the second part is from the input of the combiner to the output of the accumulator. To prove that our codes are linear codes, we only need to show that the second part is a linear system. This is because the first part is already linear.

A linear code has the linear property such that the linear combination of two codewords is still a valid codeword. Now suppose we have two different codewords $\mathbf{X}^{\tau}$ and $\mathbf{X}^{\upsilon}$ with length $N$. The linear combination of these two codewords is
 \begin{align}\label{eq:lc1}
 \mathbf{X}^{\tau} \oplus \mathbf{X}^{\upsilon}=& [x_1[\tau],x_2[\tau],\cdots,x_N[\tau]] \oplus \nonumber \\
  &[x_1[\upsilon],x_2[\upsilon],\cdots,x_N[\upsilon]] \nonumber \\
 =&[x_1[\tau]\oplus x_1[\upsilon] ,x_2[\tau]\oplus \nonumber \\
 &x_2[\upsilon],\cdots,x_N[\tau]\oplus x_N[\upsilon]],
 \end{align}
 where $\oplus$ is the modulo lattice addition. Now, we focus on the $n$-th component of the codeword $x_n$ for $ 1 \leq n \leq N$.
 The encoding function for the $n$-th component of the codeword is
 \begin{equation}
\left(\bigoplus_{i=0}^{j_n-1}z_{a_n+i} \oplus g_{a_n+i} \right) \oplus c_{n-1} \oplus g_n^\prime \oplus g_n^{\prime\prime}  = x_n,
\end{equation}
where $z_{a_n}$ and $z_{a_n+j_n-1}$ represent the first and last interleaved symbols to the $n$-th combiner; $c_{n-1}$ is the $(n-1)$-th output of the time-varying accumulator. Note that the random-coset is removed before iterative decoding, thus it is not considered as part of the codebook information.
We can then rewrite the above equation as
\begin{equation}
\bigoplus_{i=0}^{j_n-1}z_{a_n+i}  \oplus c_{n-1} \oplus C_{gn} = x_n,
\end{equation}
where $\bigoplus_{i=0}^{j_n-1}g_{a_n+i} \oplus g_n^\prime \oplus g_n^{\prime\prime} = C_{gn} \in \Psi$ and $C_{gn}$ is the constant associated with $x_n$. Note that the term $\bigoplus_{i=0}^{j_n-1}g_{a_n+i}$ can be extracted by using the associative law on the addition of Hurwitz integers.

Now for the $n$-th codeword component in $\mathbf{X}^{\tau}$ and $\mathbf{X}^{\upsilon}$, we have
\begin{equation}
\bigoplus_{i=0}^{j_n-1}z_{a_n+i}[\tau]  \oplus c_{n-1}[\tau] \oplus C_{gn} = x_n[\tau].
\end{equation}
\begin{equation}
\bigoplus_{i=0}^{j_n-1}z_{a_n+i}[\upsilon]  \oplus c_{n-1}[\upsilon] \oplus C_{gn} = x_n[\upsilon].
\end{equation}
Here $C_{gn}$ is \emph{deterministic} for a particular codeword position. The linear combination in Eq. (\ref{eq:lc1}) becomes Eq. (\ref{eq:lp1}) for $ 1 \leq n \leq N$.
\begin{align}\label{eq:lp1}
x_n[\tau] \oplus x_n[\upsilon] =& \left(\bigoplus_{i=0}^{j_n-1}z_{a_n+i}[\tau]  \oplus c_{n-1}[\tau]\right) \oplus \left(\bigoplus_{i=0}^{j_n-1}z_{a_n+i}[\upsilon]  \oplus c_{n-1}[\upsilon] \right) \oplus C_{gn} \oplus C_{gn} \nonumber \\
=&\left(\bigoplus_{i=0}^{j_n-1} \left(z_{a_n+i}[\tau] \oplus z_{a_n+i}[\upsilon] \right) \oplus \left(c_{n-1}[\tau]\oplus c_{n-1}[\upsilon] \right) \right) \oplus C_{gn} \oplus C_{gn}
\end{align}
The deterministic part $C_{gn} \oplus C_{gn}$ can contribute to non-linearity when $C_{gn} \oplus C_{gn} \neq C_{gn}$.
Therefore, when we let $C_{gn} = 0$, our codes are linear.

\section{Proof of Theorem \ref{OS}}\label{appendix:4}
Consider the $n$-th symbol. Let $X_n$ be the channel input. Let $Y_n$ be the $n$-th received signal with the input-output relationship given by
\setcounter{equation}{65}
\begin{equation}
Y_n = X_n+N_n  \overset{(c)}= C_n \oplus R_n+N_n,
\end{equation}
where $N_n\sim \mathcal{N}(0,\sigma_{ch}^2)$ is the noise of the AWGN channel; $(c)$ follows Eq. (\ref{eq:40}); $C_n$ is the $n$-th random variable of intended codeword before adding the random-coset and $R_n$ is the $n$-th random variable of the random-coset.

To prove that adding the random-coset can produce the output-symmetric effect, we must have
\begin{equation}\label{eq:ss1}
\text{Pr}[Y_n \not\in \mathcal{U}(X_n)| C_n = \psi_i] = \text{Pr}[Y_n \not\in \mathcal{U}(X_n)| C_n = \psi_j],
\end{equation}
where $\mathcal{U}(.)$ outputs the maximum-likelihood decision region; $\psi_i,\psi_j \in \Psi$ and $\psi_i \neq \psi_j$. In other words, the decoding error probability is the same for any transmitted codeword.

For the left term in Eq. (\ref{eq:ss1}), we have
\begin{align}
\text{Pr}&[Y_n \not\in \mathcal{U}(X_n)| C_n=  \psi_i] \nonumber \\
&= \sum_{r_i} \text{Pr}[Y_n \not\in \mathcal{U}(X_n) | R_n = r_i, C_n=\psi_i]\cdot 
\text{Pr}[R_n = r_i|C_n=\psi_i].
\end{align}
Since $R_n$ is independent of $C_n$ and $R_n$ is uniformly distributed over $\Psi$, we then have
\begin{align}\label{eq:new1}
&\text{Pr}[Y_n \not\in \mathcal{U}(X_n)| C_n = \psi_i]  \nonumber \\
&= \sum_{r_i} \text{Pr}[Y_n\not\in \mathcal{U}(X_n) | R_n = r_i, C_n=\psi_i]\cdot \text{Pr}[R_n = r_i] \nonumber \\
& = \sum_{x_i} \text{Pr}[Y_n \not\in \mathcal{U}(X_n)| X_n = x_i = r_i \oplus \psi_i ]\cdot \text{Pr}[R_n = r_i] \nonumber \\
& = \frac{1}{p^M} \sum_{x_i} \text{Pr}(Y_n \not\in \mathcal{U}(X_n)| X_n = x_i).
\end{align}
Similarly, for a different realisation of $C_n$ and $R_n$, we have
\begin{align}\label{eq:new2}
&\text{Pr}[Y_n \not\in \mathcal{U}(X_n)| C_n = \psi_j]  \nonumber \\
& =\sum_{r_j} \text{Pr}[Y_n\not\in \mathcal{U}(X_n) | R_n = r_j, C_n=\psi_j]\cdot \text{Pr}[R_n = r_j] \nonumber \\
& = \sum_{x_j} \text{Pr}[Y_n \not\in \mathcal{U}(X_n)| X_n = x_j = r_j \oplus \psi_j ]\cdot \text{Pr}[R_n = r_j] \nonumber \\
& = \frac{1}{p^M} \sum_{x_j} \text{Pr}(Y_n \not\in \mathcal{U}(X_n)| X_n = x_j).
\end{align}
Since the ranges of $x_i$ and $x_j$ are $\Psi$, therefore we can obtain that
\begin{align}\label{eq:ssl1}
\sum_{x_i} \text{Pr}(Y_n \not\in \mathcal{U}(X_n)& | X_n = x_i) 
= \sum_{x_j} \text{Pr}(Y_n \not\in \mathcal{U}(X_n)| X_n = x_j).
\end{align}
Plugging Eq. (\ref{eq:ssl1}) into Eq. (\ref{eq:new1}) and Eq. (\ref{eq:new2}), respectively, we obtain Eq. (\ref{eq:ss1}).

\section{Proof of Permutation-Invariance}
\subsection{Proof of Theorem \ref{PERI}}\label{appendix:2}
First, we define a probability-vector random variable $\mathbf{X} = [X_{\psi_0},X_{\psi_1},\ldots,X_{\psi_{p^M-1}}]$ and let $\mathbf{P} = \mathbf{X}^{+\theta}$ where $\theta$ is a random variable and uniformly chosen from $\Psi$. For the $m$-th random variable in $\mathbf{X}$, we denote a probability event by
\begin{equation}
\text{Pr}[X_{\psi_m} \in \varepsilon].
\end{equation}

Then for the $i$-th random variable in $\mathbf{P}$, we have
\begin{equation}
\text{Pr}[P_{\psi_i} \in \varepsilon] = \text{Pr}[X_{\psi_m} \in \varepsilon] \cdot \text{Pr}[\psi_m \oplus \theta = \psi_i],
\end{equation}
because $\theta$ is independent of $\mathbf{X}$.

Similarly, for the $j$-th random variable in $\mathbf{P}$, where $\psi_j \neq \psi_i$, we can obtain that:
\begin{equation}
\text{Pr}[P_{\psi_j} \in \varepsilon] = \text{Pr}[X_{\psi_m} \in \varepsilon] \cdot \text{Pr}[\psi_m \oplus \theta = \psi_j],
\end{equation}

We know $\theta$ is a random variable and uniformly chosen from $\Psi$. Thus we have:
\begin{equation}
\text{Pr}[\psi_m \oplus \theta = \psi_i] = \text{Pr}[\psi_m \oplus \theta = \psi_j] = \frac{1}{p^M}.
\end{equation}

Therefore, the distribution of any two random variables in $\mathbf{P}$ is the same. If we let $\psi_j = \psi_i \oplus \chi$ for any fixed $\chi \in \Psi$, we obtain that:
\begin{equation}
\text{Pr}[P_{\psi_i} \in \varepsilon] = \text{Pr}[P_{\psi_j} \in \varepsilon] = \text{Pr}[P_{\psi_i \oplus \chi} \in \varepsilon]= \text{Pr}[P_{\psi_i}^{ \oplus \chi} \in \varepsilon].
\end{equation}
It can be seen that every random variable in $\mathbf{P}$ is identically distributed. Therefore, we can conclude that $\mathbf{P}$ is identically distributed with $\mathbf{P}^{ \oplus \chi}$ so $\mathbf{P}$ is permutation-invariant.

\subsection{Proof of Lemma \ref{PERIW}}\label{appendix:3}
For the $m$-th LLR random variable in $\mathbf{W}$, we denote a probability event by
\begin{equation}
\text{Pr}[W_{\psi_m} \in \delta],
\end{equation}
where $\delta$ is a random event.
From (\ref{eq:LLR1}), we know that $W_{\psi_m} = \ln \left( \frac{P_{\psi_0}}{P_{\psi_m}}\right)$, thus we can obtain that
\begin{align}\label{eq:pi1}
\text{Pr}&[W_{\psi_m} \in \delta] \nonumber \\
&= \text{Pr}\Bigg[\ln \left( \frac{P_{\psi_0}}{P_{\psi_m}}\right) \in \delta \Bigg] \nonumber \\
& = \text{Pr}[ P_{\psi_0} \in e^\delta P_{\psi_m} ] \nonumber \\
& = \int_{P_{\psi_m}}\int_{e^\delta P_{\psi_m}} f_{P_{\psi_0},P_{\psi_m}}(p_{\psi_0},p_{\psi_m}) dp_{\psi_0} \, dp_{\psi_m},
\end{align}
where $f_{P_{\psi_0},P_{\psi_m}}(p_{\psi_0},p_{\psi_m})$ denotes the joint pdf of $P_{\psi_0}$ and $P_{\psi_m}$.

Similarly, for the $n$-th LLR random variable in $\mathbf{W}$ where $n \neq m$, we have
\begin{align}\label{eq:pi2}
\text{Pr}[W_{\psi_n} \in \delta]  = \int_{P_{\psi_n}}\int_{e^\delta P_{\psi_n}} f_{P_{\psi_0},P_{\psi_n}}(p_{\psi_0},p_{\psi_n}) dp_{\psi_0} \, dp_{\psi_n}.
\end{align}

We know $P_{\psi_m}$ and $P_{\psi_n}$ have the same distribution as because $\mathbf{P}$ is permutation-invariant. Thus, the joint distribution of $P_{\psi_0}$ and $P_{\psi_m}$ is the same as that of $P_{\psi_0}$ and $P_{\psi_n}$. As a result, we can obtain that:
\begin{equation}
\text{Pr}[W_{\psi_n} \in \delta] = \text{Pr}[W_{\psi_m} \in \delta].
\end{equation}
This indicates that $W_{\psi_n}$ and $W_{\psi_m}$ have the same distribution for any $n \neq m$. Therefore, $\mathbf{W}$ is permutation-invariant.

\chapter{Proof of Theories of Chapter 5}\label{appendix_2b}

\section{Two useful lemmas}\label{appendix:lemma}
\subsection{An extension of Theorem in \cite{ozarow90}}

In this appendix, we provide a lemma, which is key to our analysis of the gap to the capacity region. This lemma is an extension of the main Theorem in \cite{ozarow90} (also Proposition 1 in \cite{dytso15}) to the multi-dimensional setting.

Let $\mc{A}$ be an $n$-dimensional constellation carved from a shifted version of an $n$-dimensional lattice $\Lambda$ and $\msf{X}$ be the input random variable uniformly distributed over $\mc{A}$. We assume $\mbb{E}
[\|\msf{X}\|^2]=nP$. Let $\msf{Y}$ be the received random variable with the input-output relationship given by
\begin{equation}
    \msf{Y} = \msf{X} + \msf{Z},
\end{equation}
where $\msf{Z}$ is the noise vector which is zero-mean and has $\mbb{E}[\|\msf{Z}\|^2]=n$ independent of $\msf{X}$. We have the following lemma.

\begin{lemm}\label{lma:gap_lattice} 
    The mutual information between $\msf{X}$ and $\msf{Y}$ is lower-bounded as follows,
    \begin{align}\label{eq:th1}
        \frac{1}{n}I(\mathbf{\msf{X}};\mathbf{\msf{Y}}) &\geq \frac{1}{n}H(\mathbf{\msf{X}}) - \frac{1}{2}\log_2 2\pi e \left(\text{Vol}(\Lambda)^{-\frac{2}{n}} + \psi(\Lambda)\right).
    \end{align}
\end{lemm}

\emph{\quad Proof: }
Let $\mathbf{\msf{X'}} = \mathbf{\msf{X}} + \mathbf{\msf{U}}$ with $\mathbf{\msf{U}}$ uniformly distributed over $\mc{V}(\Lambda)$. Clearly, $\mathbf{\msf{X'}}, \mathbf{\msf{X}}, \mathbf{\msf{Y}}$ form a Markov chain in the following order
\begin{equation}
    \mathbf{\msf{X'}} \rightarrow \mathbf{\msf{X}} \rightarrow \mathbf{\msf{Y}}.
\end{equation}
Therefore, from the data processing inequality \cite{Cover:2006:EIT:1146355}, we have
\begin{align}\label{eqn:mu_I}
    I(\mathbf{\msf{X}};\mathbf{\msf{Y}})&\geq I(\mathbf{\msf{X'}};\mathbf{\msf{Y}})
    = h(\mathbf{\msf{X'}}) - h(\mathbf{\msf{X'}}|\mathbf{\msf{Y}}) \nonumber \\
    &= H(\mathbf{\msf{X}}) + h(\mathbf{\msf{U}}) - h(\mathbf{\msf{X'}}|\mathbf{\msf{Y}}) \nonumber \\
     &= H(\mathbf{\msf{X}}) + \log_2(\text{Vol}(\Lambda)) - h(\mathbf{\msf{X'}}|\mathbf{\msf{Y}}).
\end{align}
Note that
\begin{align}\label{eqn:h_x_y}
    h(\mathbf{\msf{X'}}|\mathbf{\msf{Y}}=\mathbf{y}) &= -\int p(\mathbf{x'}|\mathbf{y}) \log_2 p(\mathbf{x'}|\mathbf{y}) \text{d}\mathbf{x'} \nonumber \\
    &\leq -\int p(\mathbf{x'}|\mathbf{y}) \log_2 q_{\mathbf{y}}(\mathbf{x'}) \text{d}\mathbf{x'},
\end{align}
for any valid distribution $q_{\mathbf{y}}(\mathbf{x'})$. We pick
\begin{equation}
    q_{\mathbf{y}}(\mathbf{x'}) = \prod_{l=1}^n \left(\frac{1}{\sqrt{2\pi}s}e^{-\frac{(x'_l-k y_l)^2}{2s^2}}\right),
\end{equation}
where $x'_l$ and $y_l$ are the $l$th elements of $\mathbf{x'}$ and $\mathbf{y}$, respectively. Plugging this choice into \eqref{eqn:h_x_y} gives
\begin{align}
    &h(\mathbf{\msf{X'}}|\mathbf{\msf{Y}}=\mathbf{y}) \nonumber \\
    &\leq \left(\frac{n}{2}\ln 2\pi s^2 + \frac{1}{2s^2}\mbb{E}\left[ \|\mathbf{\msf{X'}}-k\mathbf{y}\|^2|\mathbf{\msf{Y}}=\mathbf{y}\right] \right) \log_2 e.
\end{align}
Thus,
\begin{align}\label{eqn:h_x_y2}
    h(\mathbf{\msf{X'}}|\mathbf{\msf{Y}}) \leq \left(\frac{n}{2}\ln 2\pi s^2 + \frac{1}{2s^2}\mbb{E} [\|\mathbf{\msf{X'}}-k\msf{Y}\|^2] \right) \log_2 e.
\end{align}
Now, choosing $k=\frac{P}{1+ P}$, we have
\begin{align}
    \mbb{E}[\| \mathbf{\msf{X'}}-k\mathbf{\msf{Y}}\|^2] &= \mbb{E}[\| \mathbf{\msf{X}}+\mathbf{\msf{U}}-k(\mathbf{\msf{X}}+\mathbf{\msf{Z}}) \|^2] \nonumber \\
    &= (1-k)^2n P + n\sigma^2(\Lambda) + nk^2 \nonumber \\
    &= \frac{nP}{1+P} + n\sigma^2(\Lambda).
\end{align}
Hence, \eqref{eqn:h_x_y2} becomes
\begin{align}
    h(\mathbf{\msf{X'}}|\mathbf{\msf{Y}}) \leq \left(\frac{n}{2}\ln 2\pi s^2 + \frac{n}{2s^2}  \left(\frac{P}{1+P} + \sigma^2(\Lambda)\right)\right) \log_2 e.
\end{align}
We choose $s^2 = \frac{P}{1+P} + \sigma^2(\Lambda)$ to obtain
\begin{align}\label{eqn:h_x_y3}
    h(\mathbf{\msf{X'}}|\mathbf{\msf{Y}}) \leq \frac{n}{2}\log_2 2\pi e \left(\frac{P}{1+P} + \sigma^2(\Lambda)\right).
\end{align}

Plugging \eqref{eqn:h_x_y3} into \eqref{eqn:mu_I} results in
\begin{align}
    I(\mathbf{\msf{X}};\mathbf{\msf{Y}})\geq& H(\mathbf{\msf{X}}) + \log_2(\text{Vol}(\Lambda)) \nonumber \\
     & - \frac{n}{2}\log_2 2\pi e \left(\frac{P}{1+P} + \sigma^2(\Lambda)\right) \nonumber \\
\geq& H(\mathbf{\msf{X}}) - \frac{n}{2}\log_2 2\pi e \left(\frac{1+ \sigma^2(\Lambda)}{\text{Vol}(\Lambda)^{\frac{2}{n}}}\right) \nonumber \\
=& H(\mathbf{\msf{X}}) - \frac{n}{2}\log_2 2\pi e \left(\text{Vol}(\Lambda)^{-\frac{2}{n}} + \psi(\Lambda)\right).
\end{align}
This completes the proof. \QEDA

\subsection{A corollary of Proposition 2 in \cite{Forney89}}
We now provide an upper bound on the required power of the proposed lattice constellation.
\begin{lemm}\label{lma:power}
Let $\msf{V}$ be a discrete random variable uniformly distributed over the coset representatives of the lattice partition $\Lambda/2^m\Lambda$ with any positive integer $m$. There exists a dither $\mathbf{d}\in\mc{V}(2^m\Lambda)$ such that $\msf{X} = [\msf{V}- \mathbf{d}]_{2^m\Lambda}$ has
\begin{align}\label{eq:pap}
\frac{1}{n}\E[\|\msf{X} \|^2] &\leq \sigma^2(2^m\Lambda) = 2^{2m} \text{Vol}(\Lambda)^{\frac{2}{n}}\psi(\Lambda).
\end{align}
\end{lemm}

\emph{\quad Proof: }
Let $\msf{D}$ be a random dither that is uniformly distributed over $\mc{V}(2^m\Lambda)$. From \cite[Eq. (23)]{1337105}, we have
\begin{align}\label{eq:app_p_1}
\frac{1}{n}\E[\|[\msf{V} - \msf{D}]_{2^m\Lambda}\|^2] =& \frac{1}{n}\E[\|\msf{D}\|^2]  \nonumber \\
=& \sigma^2(2^m\Lambda),
\end{align}
which says that the input power is equal to the second moment of the coarse lattice when averaged over the dither $\msf{D}$. Thus, there exists a fixed dither vector $\mathbf{d}$ such that
\begin{equation}\label{eq:app_p_2}
\E[\|[\msf{V}- \mathbf{d}]_{2^m\Lambda} \|^2] \leq  \E[\|[\msf{V} - \msf{D}]_{2^m\Lambda}\|^2].
\end{equation}
Combining (\ref{eq:app_p_1}) and (\ref{eq:app_p_2}) completes the proof for (\ref{eq:pap}).
\QEDA

\section{Proof of the 2-user case}\label{2user}
\subsection{Proof of user 1's gap}\label{pf_user1_gap}
We now bound the overall scaling factors for user 1. Recall that $\beta = \sqrt{\frac{n}{\E[\| \mathcal{C}\|^2]}}$ is to ensure $\mbb{E}[\|\mc{C}\|^2]\leq n$. The scaling factor for user 1 can be lower bounded by:
\begin{align}\label{eq:sc1}
\sqrt{\SNR_1} \beta &= \sqrt{\frac{n\SNR_1} {\E[\| \mathcal{C}\|^2]}} \nonumber \\
&\overset{\eqref{eq:lemma_f}}{\geq} \sqrt{\frac{\SNR_1} {2^{2(m_1+m_2)}\text{Vol}(\Lambda)^{\frac{2}{n}}\psi(\Lambda)}} \nonumber \\
&\overset{(b)} > \sqrt{\frac{2^{2(n_1-1)}} { 2^{2(m_1+m_2)} \text{Vol}(\Lambda)^{\frac{2}{n}}\psi(\Lambda)}} \nonumber \\
& \overset{\eqref{eq:d2}} \geq \sqrt{\frac{1}{4\text{Vol}(\Lambda)^{\frac{2}{n}} \psi(\Lambda)}}\sqrt{\frac{4^{n_1}} {4^{n_1}}} \nonumber \\
& = \sqrt{\frac{1}{4\text{Vol}(\Lambda)^{\frac{2}{n}} \psi(\Lambda)}},
\end{align}
where in (b) we have used $\SNR_1 > 2^{2(n_1-1)}$.

The gap between the user 1's achievable rate and the multiuser capacity can be further bounded by plugging \eqref{eq:sc1} into \eqref{eqn:gap_1}.
\begin{align}\label{eq:gap_12}
\Delta_1 &=  1+  \frac{1}{2}\log_2 2\pi e \left(\text{Vol}(\Lambda_1)^{-\frac{2}{n}} + \psi(\Lambda)\right) \nonumber \\
   &= 1 + \frac{1}{2}\log_2 2\pi e \left(\text{Vol}(\sqrt{\SNR_1}\beta\Lambda)^{-\frac{2}{n}} + \psi(\Lambda)\right) \nonumber \\
  & = 1 + \frac{1}{2}\log_2 2\pi e \left((\sqrt{\SNR_1}\beta)^{-2}\text{Vol}(\Lambda)^{-\frac{2}{n}} + \psi(\Lambda)\right) \nonumber \\
  & < 1 + \frac{1}{2}\log_2 2\pi e \left(4\text{Vol}(\Lambda)^{\frac{2}{n}} \psi(\Lambda)\text{Vol}(\Lambda)^{-\frac{2}{n}} + \psi(\Lambda)\right) \nonumber \\
   &= 1 + \frac{1}{2}\log_2 2\pi e \left(5\psi(\Lambda)\right).
\end{align}

\subsection{Proof of user 2's gap}\label{pf_user2_gap}
Following \eqref{eqn:gap_1}, we again obtain an upper bound of the gap of user 2's achievable rate to the multiuser capacity by using the invariant property of $\psi(\Lambda)$ as
\begin{equation}\label{eqn:gap_2}
    \Delta_2 = 1+\frac{1}{2}\log_2 2\pi e \left(\text{Vol}(\Lambda_2)^{-\frac{2}{n}} + \psi(\Lambda)\right),
\end{equation}
bits per real dimension.

We then lower bound the scaling factor for user 2 as follows:
\begin{align}\label{eq:sc2_5}
&\gamma\sqrt{\SNR_2}\beta 2^{r_{12}} = \sqrt{\frac{n\SNR_2\beta^24^{r_{12}}}{\SNR_2\beta^2 \E[\| \mathcal{C}_{12} \|^2] +n}} \nonumber \\
 &= \sqrt{\frac{1}{\frac{\E[\| \mathcal{C}_{12} \|^2]}{n4^{r_{12}}} +\frac{1}{\SNR_2\beta^24^{r_{12}}}}} \nonumber \\
& \geq \sqrt{\frac{1}{\frac{ n2^{2r_{12}}\text{Vol}(\Lambda)^{\frac{2}{n}}\psi(\Lambda)}{n4^{r_{12}}} +\frac{ 2^{2(m_1+m_2)}\text{Vol}(\Lambda)^{\frac{2}{n}}\psi(\Lambda)}{\SNR_24^{r_{12}}}}} \nonumber \\
&> \sqrt{\frac{1}{\frac{\text{Vol}(\Lambda)^{\frac{2}{n}}\psi(\Lambda)}{1}\frac{4^{r_{12}}}{4^{r_{12}}} +\frac{\text{Vol}(\Lambda)^{\frac{2}{n}}\psi(\Lambda)}{4^{-1}}\frac{4^{m_1+m_2}}{4^{n_2}4^{m_1+m_2-n_2}}}} \nonumber \\
 &= \sqrt{\frac{1}{5\text{Vol}(\Lambda)^{\frac{2}{n}}\psi(\Lambda)}}.
\end{align}

Plugging \eqref{eq:sc2_5} into \eqref{eqn:gap_2}, we have
\begin{align}\label{eq:gap_22}
\Delta_2 &=  1+  \frac{1}{2}\log_2 2\pi e \left(\text{Vol}(\Lambda_2)^{-\frac{2}{n}} + \psi(\Lambda)\right) \nonumber \\
   &= 1 + \frac{1}{2}\log_2 2\pi e \left(\text{Vol}(\gamma\sqrt{\SNR_2}\beta 2^{r_{12}}\Lambda)^{-\frac{2}{n}} + \psi(\Lambda)\right) \nonumber \\
  & < 1 + \frac{1}{2}\log_2 2\pi e \left(5\text{Vol}(\Lambda)^{\frac{2}{n}} \psi(\Lambda)\text{Vol}(\Lambda)^{-\frac{2}{n}} + \psi(\Lambda)\right) \nonumber \\
   &= 1 + \frac{1}{2}\log_2 2\pi e \left(6\psi(\Lambda)\right).
\end{align}

\section{Proof of the \emph{K}-user case}\label{Kuser_5}
Let $\msf{V}_1, \msf{V}_2, \cdots, \msf{V}_K$ be random variables uniformly over $\mc{C}_1, \mc{C}_2, \cdots, \mc{C}_K$, respectively, and let $\msf{X}=\beta\left(\left[\msf{V}_1+\sum_{k=2}^{K}2^{\sum_{i = 1}^{k-1}m_i}\msf{V}_k-\mathbf{d}\right]_{\Lambda_{s}}\right)$ be the input random variable. For analyzing the first (strongest) user, we treat users $2,\cdots,K$ as a super-user. One can thus obtain the same lower bound for the achievable rate as shown in \eqref{eqn:rate_1_GBC_5} and the same capacity upper bound in \eqref{eq:main_1_5}.

For user $k>1$, we treat users $1,2,\cdots,k-1$ as a super-user whose constellation is $\mc{C}_1^\prime$ and treat users $k, k+1,\cdots,K$ as another super-user whose constellation is $\mc{C}_k^\prime$. Similar to the two-user case, $\mc{C}_1^\prime$ and $\mc{C}_k^\prime$ are corresponding to $\Lambda/2^{m_1^\prime}\Lambda$ and $\Lambda/2^{m_k^\prime}\Lambda$, respectively. Note that $\mc{C}_1^\prime$ is the constellation for users whose signals have less power than user $k$'s signal and thus may not be correctly decoded at the user $k$. Following (\ref{eq:dKk35}) and (\ref{eq:dKk45}), we further decompose $\mc{C}_1^\prime$ into $\mc{C}_{11}^\prime$ and $\mc{C}_{1k}^\prime$ as
\begin{equation}\label{eq:k_C1}
    \mc{C}_1^\prime = \left[\mc{C}_{1k}^\prime + 2^{r_{1k}^\prime}\mc{C}_{11}^\prime\right]_{2^{m_1^\prime}\Lambda}.
\end{equation}
It is worth mentioning that $\mc{C}_{1k}^\prime$ corresponds to the part that is under the noise level as predicted by the linear deterministic model and cannot be decoded at the user $k$.

The channel output for user $k$ is then given by:
\begin{align}
\msf{Y}_k=& \sqrt{\SNR_k}\msf{X}+\msf{Z}_k \nonumber \\
=& \sqrt{\SNR_k}\beta\left(\left[\msf{V}_1+\sum_{k=2}^{K}2^{\sum_{i = 1}^{k-1}m_i}\msf{V}_k - \mathbf{d} \right]_{\Lambda_{s}} \right)+\msf{Z}_k \nonumber\\
\overset{(c)}=& \sqrt{\SNR_k}\beta\Bigg(\bigg[\left[\msf{V}_{1k}^\prime+2^{r_{1k}^\prime}\msf{V}_{11}^\prime\right]_{2^{m_1^\prime}\Lambda} 
 + \sum_{i=k}^{K}2^{\sum_{j = 1}^{i-1}m_j}\msf{V}_i - \mathbf{d} \bigg]_{\Lambda_{s}}\Bigg)+\msf{Z}_k,
\end{align}
where in (c) we have used $\msf{V}_{1^\prime} = \left[\msf{V}_{1k}^\prime+2^{r_{1k}^\prime}\msf{V}_{11}^\prime\right]_{2^{m_1^\prime}\Lambda} $ corresponding to (\ref{eq:k_C1}). Here $\msf{V}_{1k^\prime}$ and $\msf{V}_{11}^\prime$ are randomly and uniformly distributed over $\Lambda/2^{r_{1k}^\prime}\Lambda$ and $\Lambda/2^{r_{11}^\prime}\Lambda$, respectively. From user $k$'s point of view, the strong super-user's constellation $\msf{V}_{1^\prime}$ can be decomposed into two parts such that $\msf{V}_{1k}^\prime$ can be successfully received while $\msf{V}_{1k}^\prime$ is considered under noise level.

We can now bound the achievable rate for user $k$ as follows,
\begin{align}\label{eqn:rate_k_GBC_5}
        &I(\msf{V}_k;\msf{Y}_k) = h(\msf{Y}_k) - h(\msf{Y}_k|\msf{V}_k) \nonumber \\
    =& \left[h(\msf{Y}_k) - h\left(\msf{Y}_k \left|  \left[  2^{r_{1k}^\prime}\msf{V}_{11}^\prime  +  \sum_{i=k}^{K}2^{\sum_{j = 1}^{i-1}m_j}\msf{V}_i \right]_{\Lambda_{s}} \right.\right)\right] \nonumber \\
    &- \left[h(\msf{Y}_k|\msf{V}_k)-h\left(\msf{Y}_k \left|  \left[  2^{r_{1k}^\prime}\msf{V}_{11}^\prime  +  \sum_{i=k}^{K}2^{\sum_{j = 1}^{i-1}m_j}\msf{V}_i \right]_{\Lambda_{s}} \right.\right)\right] \nonumber \\
    \overset{(a)}{=}& \left[h(\msf{Y}_k) - h\left(\msf{Y}_k \left|  \left[  2^{r_{1k}^\prime}\msf{V}_{11}^\prime  +  \sum_{i=k}^{K}2^{\sum_{j = 1}^{i-1}m_j}\msf{V}_i \right]_{\Lambda_{s}} \right.\right)\right] \nonumber \\
    &- [h(\msf{Y}_k|\msf{V}_k)-h(\msf{Y}_k | \msf{V}_{11}^\prime, \msf{V}_k,\cdots,\msf{V}_K )] \nonumber \\
     =& I\left(\left[  2^{r_{1k}^\prime}\msf{V}_{11}^\prime  +  \sum_{i=k}^{K}2^{\sum_{j = 1}^{i-1}m_j}\msf{V}_i \right]_{\Lambda_{s}};\msf{Y}_k\right) 
    - I(\msf{V}_{11}^\prime, \msf{V}_{k+1},\cdots,\msf{V}_K;\msf{Y}_k | \msf{V}_k) \nonumber \\
    \geq& I\left(\left[  2^{r_{1k}^\prime}\msf{V}_{11}^\prime  +  \sum_{i=k}^{K}2^{\sum_{j = 1}^{i-1}m_j}\msf{V}_i \right]_{\Lambda_{s}};\msf{Y}_k\right) 
    - H(\msf{V}_{11}^\prime, \msf{V}_{k+1},\cdots,\msf{V}_K | \msf{V}_k) \nonumber \\
    \overset{(b)}{=}& I\left(\left[  2^{r_{1k}^\prime}\msf{V}_{11}^\prime  +  \sum_{i=k}^{K}2^{\sum_{j = 1}^{i-1}m_j}\msf{V}_i \right]_{\Lambda_{s}};\msf{Y}_k\right)  
    - H(\msf{V}_{11}^\prime, \msf{V}_{k+1},\cdots,\msf{V}_K),
\end{align}
where (a) is due to a bijective mapping between $\left[  2^{r_{1k}^\prime}\msf{V}_{11}^\prime  +  \sum_{i=k}^{K}2^{\sum_{j = 1}^{i-1}m_j}\msf{V}_i \right]_{\Lambda_s}$ and $( \msf{V}_{11}^\prime, \msf{V}_k,\cdots,\msf{V}_K)$, and (b) follows the fact that $\msf{V}_k$ is independent of $\msf{V}_{11}^\prime, \msf{V}_{k+1},$
$\cdots$, and $\msf{V}_K$. To further bound the first term in \eqref{eqn:rate_k_GBC_5}, we note that effective noise is $\sqrt{\SNR_k}\beta\msf{V}_{1k}^\prime+\msf{Z}_k$. We thus scale $\msf{Y}_k$ by
\begin{equation}
    \gamma=\sqrt{\frac{n}{\SNR_k\beta^2\mbb{E}[\|\msf{V}_{1k}^\prime\|^2]+n}},
\end{equation}
to form $\msf{Y}'_k$. The equivalent communication channel then becomes $\msf{Y}'_k = \msf{X}'_k + \msf{Z}'_k$ where
\begin{equation}
    \msf{X}'_k = \gamma\sqrt{\SNR_k} \beta \left[  2^{r_{1k}^\prime}\msf{V}_{11}^\prime  +  \sum_{i=k}^{K}2^{m_{i-1}}\msf{V}_i \right]_{\Lambda_{s}},
\end{equation}
and $\msf{Z}'_k = \gamma (\sqrt{\SNR_k}\beta\msf{V}_{1k}^\prime+\msf{Z}_k)$ with $\mbb{E}[\|\msf{Z}'_k\|^2]=n$. One can then again apply the lower bound in Appendix~\ref{appendix:lemma} to get
\begin{equation}
    I(\msf{V}_k;\msf{Y}_k) \geq n m_k - \frac{n}{2}\log_2 2\pi e \left(\text{Vol}(\Lambda_k)^{-\frac{2}{n}} + \psi(\Lambda_k)\right),
\end{equation}
where $\Lambda_k=\gamma\sqrt{\SNR_k}\beta 2^{r_{1k}^\prime}\Lambda$.

Using the invariant property of $\psi(\Lambda)$, we again obtain the gap between user $K$'s achievable rate to the capacity upper bounded by
\begin{equation}\label{eqn:gap_k}
    \Delta_k = 1+\frac{1}{2}\log_2 2\pi e \left(\text{Vol}(\Lambda_k)^{-\frac{2}{n}} + \psi(\Lambda)\right),
\end{equation}
bits per real dimension. After this, following the similar steps as those in Section~\ref{GapRate}, one can obtain the lower bound for the scaling factor for user $k$ as:
\begin{align}\label{eq:sck_ch5}
&\gamma\sqrt{\SNR_k}\beta 2^{r_{1k}^\prime} = \sqrt{\frac{n\SNR_k\beta^24^{r_{1k}^\prime}}{\SNR_k\beta^2 \E[\| \mathcal{C}_{1k}^\prime \|^2] +n}} \nonumber \\
 &= \sqrt{\frac{1}{\frac{\E[\| \mathcal{C}_{1k}^\prime \|^2]}{n4^{r_{1k}^\prime}} +\frac{1}{\SNR_k\beta^24^{r_{1k}^\prime}}}} \nonumber \\
& = \sqrt{\frac{1}{\frac{ n2^{2r_{1k}^\prime}\text{Vol}(\Lambda)^{\frac{2}{n}}\psi(\Lambda)}{n4^{r_{1k}^\prime}} +\frac{ 2^{2(\sum_{i=1}^K m_i)}\text{Vol}(\Lambda)^{\frac{2}{n}}\psi(\Lambda)}{\SNR_k4^{r_{1k}^\prime}}}} \nonumber \\
&\overset{(\ref{eq:dKk45})} > \sqrt{\frac{1}{\frac{\text{Vol}(\Lambda)^{\frac{2}{n}}\psi(\Lambda)}{1}\frac{4^{r_{1k}^\prime}}{4^{r_{1k}^\prime}} +\frac{\text{Vol}(\Lambda)^{\frac{2}{n}}\psi(\Lambda)}{4^{-1}}\frac{4^{\sum_{i=1}^K m_i}}{4^{n_k}4^{\sum_{i=1}^K m_i-n_k}}}} \nonumber \\
& = \sqrt{\frac{1}{5\text{Vol}(\Lambda)^{\frac{2}{n}}\psi(\Lambda)}}.
\end{align}

The gap to the capacity is then given by:
\begin{align}\label{eq:gap_kk}
& \Delta_k=  1+  \frac{1}{2}\log_2 2\pi e \left(\text{Vol}(\Lambda_k)^{-\frac{2}{n}} + \psi(\Lambda)\right) \nonumber \\
   &= 1 + \frac{1}{2}\log_2 2\pi e \left(\text{Vol}(\gamma\sqrt{\SNR_k}\beta 2^{r_{1k}^\prime}\Lambda)^{-\frac{2}{n}} + \psi(\Lambda)\right) \nonumber \\
  & = 1 + \frac{1}{2}\log_2 2\pi e \left((\gamma\sqrt{\SNR_k}\beta 2^{r_{1k}^\prime})^{-2}\text{Vol}(\Lambda)^{-\frac{2}{n}} + \psi(\Lambda)\right) \nonumber \\
   &< 1 + \frac{1}{2}\log_2 2\pi e \left(6\psi(\Lambda)\right),
\end{align}
which completes the proof of \eqref{eq:main_K} for $K > 2$ in Proposition \ref{main_result5}. From here we can see that the gap to capacity does not scale with the number of users.

\chapter{Proof of Theories of Chapter 6}\label{appendix_3a}

\section{A Useful Lemma}\label{apx:3}
We note that the outage capacity in Theorem~\ref{the:1} becomes infinity as $\epsilon_k$ increases to 1. This is because continuous Gaussian inputs are allowed when deriving capacity results; thereby, if the receivers can accept ridiculously high outage probabilities, the transmitter can keep tuning up the rates unboundedly. In contrast, our scheme employs discrete inputs whose achievable rates are limited by the corresponding entropies. As a consequence, the gap between the outage capacity and our achievable rate can be unbounded as $\epsilon_k$ increases. However, such high outage probabilities are of no practical significance. In the following lemma, we characterize an upper bound on the outage probability so that the outage capacity in Theorem~\ref{the:1} is contained inside the capacity region of the AWGN network whose SNRs are exactly the average SNRs. We believe that this range covers almost all the cases that are of practical interest.

\begin{lemm}\label{lma:pout}
When $\epsilon_k < 0.6321$ for $k =1,\ldots,K$, for any power allocation factors $(\alpha_1,\ldots,\alpha_K)$, the rate tuple on the boundary of the outage capacity region \eqref{eq:def1} is upper bounded by
\begin{align}\label{eq:con21}
C_k < \bar{C}_k \triangleq \frac{1}{2}\log_2\left(1+\frac{\overline{\SNR}_k\alpha_k}{\overline{\SNR}_k\sum_{i=1}^{k-1}\alpha_i+1} \right).
\end{align}
\end{lemm}
\emph{\quad Proof: }
For user $k$, $C_k \leq \bar{C}_k$ leads to
\begin{align}
& 1+\frac{\overline{\SNR}_kF(\epsilon_k) \alpha_k}{\overline{\SNR}_kF(\epsilon_k) \sum_{i=1}^{k-1}\alpha_i+1}< 1+\frac{\overline{\SNR}_k\alpha_k}{\overline{\SNR}_k\sum_{i=1}^{k-1}\alpha_i+1} \nonumber \\
\Rightarrow \;& \frac{F(\epsilon_k) }{\overline{\SNR}_kF(\epsilon_k) \sum_{i=1}^{k-1}\alpha_i+1}< \frac{1}{\overline{\SNR}_k\sum_{i=1}^{k-1}\alpha_i+1}\nonumber \\
\Rightarrow\;& \overline{\SNR}_kF(\epsilon_k) \sum_{i=1}^{k-1}\alpha_i+F(\epsilon_k) < \overline{\SNR}_kF(\epsilon_k) \sum_{i=1}^{k-1}\alpha_i+1 \nonumber \\
\Rightarrow\;& F(\epsilon_k) < 1 \nonumber \\
\Rightarrow\;&  \epsilon_k < 0.6321.
\end{align}
This completes the proof. \QEDA

\section{Proof of Individual Outage Rate}\label{appendix:ind_outage}

\subsection{Proof of User 1's Achievable Rate for a Channel Realization}\label{sec:u1proof}
We first bound the overall scaling factor for user 1's lattice constellation. Recall that $\beta = \sqrt{\frac{n}{\E[\| \mathbf{x}\|^2]}}$ is to ensure $\mbb{E}[\|\mathbf{x}\|^2]\leq n$. We then lower bound the scaling factor for user 1 as follows:
\begin{align}\label{eq:sc2}
\gamma_1 \hat{h}_1\sqrt{\overline{\SNR}_1}\beta 2^{r_{12,1}} 
&= \sqrt{\frac{n|\hat{h}_1|^2\overline{\SNR}_1\beta^24^{r_{12,1}}}{|\hat{h}_1|^2\overline{\SNR}_1\beta^2 \E[\| [\msf{V}_{12,1}-\mathbf{d}_1]_{2^{r_{12,1}}\Lambda} \|^2] +n}} \nonumber \\
&= \left(\frac{\E[\| [\msf{V}_{12,1}-\mathbf{d}_1]_{2^{r_{12,1}}\Lambda} \|^2]}{n4^{r_{12,1}}} +\frac{1}{|\hat{h}_1|^2\overline{\SNR}_1\beta^24^{r_{12,1}}}\right)^{-\frac{1}{2}} \nonumber \\
& \overset{(\ref{eq:sc2}.a)}\geq \left(\frac{ n2^{2r_{12,1}}\text{Vol}(\Lambda)^{\frac{2}{n}}\psi(\Lambda)}{n4^{r_{12,1}}} +\frac{ 2^{2(m_1+m_2)}\text{Vol}(\Lambda)^{\frac{2}{n}}\psi(\Lambda)}{|\hat{h}_1|^2\overline{\SNR}_14^{r_{12,1}}}\right)^{-\frac{1}{2}} \nonumber \\
&\overset{(\ref{eq:sc2}.b)}> \left(\frac{\text{Vol}(\Lambda)^{\frac{2}{n}}\psi(\Lambda)4^{r_{12,1}}}{4^{r_{12,1}}} +\frac{\text{Vol}(\Lambda)^{\frac{2}{n}}\psi(\Lambda)}{|\hat{h}_1|^24^{r_{12,1}}}\cdot \frac{4^{m_1+m_2}}{4^{\bar{n}_1-1}}\right)^{-\frac{1}{2}} \nonumber \\
 &\overset{(\ref{eq:sc2}.c)}{\geq} \left(\left(1+\frac{4}{|\hat{h}_1|^24^{r_{12,1}}}\right)\text{Vol}(\Lambda)^{\frac{2}{n}}\psi(\Lambda)\right)^{-\frac{1}{2}},
\end{align}
where $(\ref{eq:sc2}.a)$ follows \cite[Lemma 7]{8291591}, and $(\ref{eq:sc2}.b)$ follows from that $\SNR_k > 2^{2(n_k-1)}$ for $k \in \{1,\ldots,K\}$; and $(\ref{eq:sc2}.c)$ follows from \eqref{eq:m1n1}. Plugging \eqref{eq:sc2} into \eqref{eq:sc155} results in
\begin{align}\label{eq:rate_1_GBC_final}
    I(\msf{V}_1;\msf{Y}_1) &\geq nr_{11,1} - \frac{n}{2}\log_2 2\pi e \left(\text{Vol}(\Lambda_1)^{-\frac{2}{n}} + \psi(\Lambda_1)\right) \nonumber \\
    &= nr_{11,1} 
    - \frac{n}{2}\log_2 2\pi e \left(\text{Vol}( \gamma\sqrt{\overline{\SNR}_1}\beta 2^{r_{12,1}}\Lambda)^{-\frac{2}{n}} + \psi(\Lambda_1)\right) \nonumber \\
    & > nr_{11,1}  - \frac{n}{2}\log_2 2\pi e \left( \left(2+\frac{4}{|\hat{h}_1|^24^{r_{12,1}}} \right)\psi(\Lambda)\right).
\end{align}
Now we wish to lower bound the term $|\hat{h}_1|^24^{r_{12,1}}$. When $|\hat{h}_1|^2 \geq 1$, the achievable rate for user 1 is lower bounded by
\begin{align}\label{u1_upper_rate}
I(\msf{V}_1;\msf{Y}_1) \geq nm_1- \frac{n}{2}\log_2 2\pi e \left( \left(2+\frac{4}{|\hat{h}_1|^2} \right)\psi(\Lambda)\right),
\end{align}
as $r_{11,1} = m_1$ and $r_{12,1} = 0$ in this case. The achievable rate will approach to $nm_1$ with $|\hat{h}_1|^2$ increasing.

We now consider the case for $|\hat{h}_1|^2 < 1$ for the worst case scenario. If the channel gain is so small such that $n_1 = 0$, then user 1 cannot decode its own signal. In this case, the achievable rate becomes zero. Thus, we consider the case such that $n_1\geq 1$, i.e., $|\hat{h}_1|^2\overline{\SNR}_1 \geq 1$. The term $|\hat{h}_1|^24^{r_{12,1}}$ in \eqref{eq:rate_1_GBC_final} can be bounded by
\begin{align}\label{eq:sc3}
|\hat{h}_1|^24^{r_{12,1}} &\overset{(\ref{eq:sc3}.a)}= |\hat{h}_1|^24^{\bar{n}_1-n_1} \nonumber \\
 &= |\hat{h}_1|^24^{\left\lceil \frac{1}{2}\log_2(\overline{\SNR}_1)\right\rceil^+-\left\lceil \frac{1}{2}\log_2(|\hat{h}_1|^2\overline{\SNR}_1)\right\rceil^+} \nonumber \\
 &\overset{(\ref{eq:sc3}.b)}>|\hat{h}_1|^24^{ \frac{1}{2}\log_2(\overline{\SNR}_1)-\frac{1}{2}\log_2(|\hat{h}_1|^2\overline{\SNR}_1)-1}  \nonumber \\
 & = \frac{1}{4},
\end{align}
where $(\ref{eq:sc3}.a)$ follows from \eqref{eq:r12}; $(\ref{eq:sc3}.b)$ follows by substituting $x = \overline{\SNR}_1$ and then $x = |\hat{h}_1|^2\overline{\SNR}_1$ into
\begin{equation}\label{eq:d}
 \frac{1}{2}\log_2(x)<\left\lceil \frac{1}{2}\log_2(x)\right\rceil^+<\frac{1}{2}\log_2(x)+1, \;\; x\geq 1.
\end{equation}

The lower bound for user 1's achievable rate for a channel realization $\hat{h}_1$ is then obtained as
\begin{align}
I(\msf{V}_1;\msf{Y}_1) > nr_{11,1}  - \frac{n}{2}\log_2 2\pi e \left(18 \psi(\Lambda)\right).
\end{align}
Similarly, we can use \eqref{eq:sc3} to bound the term $r_{11,1}$ when $|\hat{h}_1|^2<1$ by using \eqref{eq:d}
\begin{align}
r_{11,1} &= m_1-(\bar{n}_1-n_1) \nonumber \\
&> m_1 - \frac{1}{2}\log_2\left(\overline{\SNR}_1\right)-1+\frac{1}{2}\log_2\left(|\hat{h}_1|^2\overline{\SNR}_1\right) \nonumber \\
&= m_1 +\frac{1}{2}\log_2\left(|\hat{h}_1|^2\right)-1.
\end{align}
Then the lower bound for user 1's achievable rate in bits per real dimension for a channel instant $\hat{h}_1$ is
\begin{align}\label{eq:u1_inst_rate}
\frac{1}{n}I(\msf{V}_1;\msf{Y}_1) &> \min\left\{m_1,m_1 +\frac{1}{2}\log_2\left(|\hat{h}_1|^2\right)-1\right\} 
- \frac{1}{2}\log_2 2\pi e \left(18 \psi(\Lambda)\right),
\end{align}
where $\min\{\cdot\}$ here follows the constraint in \eqref{eq:r11}. In order to obtain the true lower bound for the outage rate in the subsequent analysis, we use \eqref{eq:u1_inst_rate} rather than \eqref{u1_upper_rate} to be lower bound for $\frac{1}{n}I(\msf{V}_1;\msf{Y}_1)$. This is because the term $r_{11,1}$ in \eqref{eq:u1_inst_rate} captures the major fading penalty on the achievable rate while \eqref{u1_upper_rate} only shows a minor effect on the rate due to channel fading.

\subsection{Proof of User $k$'s Outage Rate}\label{Kuser}

We bound the scaling factor of $\Lambda_k$ in \eqref{eq:sck},
\begin{align}\label{eq:sck}
&\gamma_k \hat{h}_k\sqrt{\overline{\SNR}_k}\beta \Gamma(r_{k2,k})  
= \sqrt{\frac{n|\hat{h}_k|^2\overline{\SNR}_k\beta^2\Gamma(r_{k2,k})^2}{|\hat{h}_k|^2\overline{\SNR}_k\beta^2 \mbb{E}[\|[\msf{V}^*_{12,k}+2^{m_{1 \rightarrow k-1}}\msf{V}_{k2,k}-\mathbf{d}_4]_{\Lambda_s} \|^2] +n}} \nonumber \\
 &= \left(\frac{\mbb{E}[\|[\msf{V}^*_{12,k}+2^{m_{1 \rightarrow k-1}}\msf{V}_{k2,k}-\mathbf{d}_4]_{\Lambda_s} \|^2]}{n\Gamma(r_{k2,k})^2} +\frac{1}{|\hat{h}_k|^2\overline{\SNR}_2\beta^2\Gamma(r_{k2,k})^2}\right)^{-\frac{1}{2}} \nonumber \\
 & \overset{(\ref{eq:sck}.a)}=\begin{cases}
&\left(\frac{\mbb{E}[\|[\msf{V}^*_{12,k}-\mathbf{d}_4]_{\Lambda_s} \|^2]}{n2^{2(r^*_{12,k})}} +\frac{1}{|\hat{h}_k|^2\overline{\SNR}_k\beta^22^{2(r^*_{12,k})}}\right)^{-\frac{1}{2}} , \;\\ &\text{when} \; r_{k2,k} = 0\\
&\left(\frac{\mbb{E}[\|[\msf{V}_{1 \rightarrow k-1}+2^{m_{1 \rightarrow k-1}}\msf{V}_{k2,k}-\mathbf{d}_4]_{\Lambda_s} \|^2]}{n2^{2(m_{1 \rightarrow k-1}+r_{k2,k})}} +\frac{1}{|\hat{h}_k|^2\overline{\SNR}_k\beta^22^{2(m_{1 \rightarrow k-1}+r_{k2,k})}}\right)^{-\frac{1}{2}}  , \;\\ &\text{when} \; r_{k2,k} \neq 0 \\
\end{cases} \nonumber \\
 & \overset{(\ref{eq:sck}.b)}\leq \begin{cases}
&\left(\frac{n2^{2(r^*_{12,k})}\text{Vol}(\Lambda)^{\frac{2}{n}}\psi(\Lambda)}{n2^{2(r^*_{12,k})}} +\frac{2^{2(\sum_{i=1}^K m_i)}\text{Vol}(\Lambda)^{\frac{2}{n}}\psi(\Lambda)}{|\hat{h}_k|^2\overline{\SNR}_k2^{2(r^*_{12,k})}}\right)^{-\frac{1}{2}} , \; \\ &\text{when} \; r_{k2,k} = 0\\
&\left(\frac{n2^{2(m_{1 \rightarrow k-1}+r_{k2,k})}\text{Vol}(\Lambda)^{\frac{2}{n}}\psi(\Lambda)}{n2^{2(m_{1 \rightarrow k-1}+r_{k2,k})}} +\frac{2^{2(\sum_{i=1}^K m_i)}\text{Vol}(\Lambda)^{\frac{2}{n}}\psi(\Lambda)}{|\hat{h}_k|^2\overline{\SNR}_k2^{2(m_{1 \rightarrow k-1}+r_{k2,k})}}\right)^{-\frac{1}{2}}  , \; \\ &\text{when} \; r_{k2,k} \neq 0 \\
\end{cases} \nonumber \\
& \overset{(\ref{eq:sck}.c)} \leq \begin{cases}
&\left(\text{Vol}(\Lambda)^{\frac{2}{n}}\psi(\Lambda) +\frac{2^{2(\sum_{i=1}^K m_i)}\text{Vol}(\Lambda)^{\frac{2}{n}}\psi(\Lambda)}{|\hat{h}_k|^24^{\bar{n}_k-1}2^{2(r^*_{12,k})}}\right)^{-\frac{1}{2}} , \;\text{when} \; r_{k2,k} = 0\\
&\left(\text{Vol}(\Lambda)^{\frac{2}{n}}\psi(\Lambda) +\frac{2^{2(\sum_{i=1}^K m_i)}\text{Vol}(\Lambda)^{\frac{2}{n}}\psi(\Lambda)}{|\hat{h}_k|^24^{\bar{n}_k-1}2^{2(m_{1 \rightarrow k-1}+r_{k2,k})}}\right)^{-\frac{1}{2}}  , \; \text{when} \; r_{k2,k} \neq 0 \\
\end{cases} \nonumber \\
 &\overset{(\ref{eq:sck}.d)}= \left(\left(1+\frac{4}{|\hat{h}_k|^24^{\bar{n}_k-n_k}}\right)\text{Vol}(\Lambda)^{\frac{2}{n}}\psi(\Lambda)\right)^{-\frac{1}{2}}.
\end{align}
where $(\ref{eq:sck}.a)$ follows from \eqref{eq:special_con_k}; $(\ref{eq:sck}.b)$ follows from \cite[Lemma 7]{8291591}; $(\ref{eq:sck}.c)$ is obtained by using $(\ref{eq:sc2}.b)$; and $(\ref{eq:sck}.d)$ follows by using \eqref{eq:r14k} and \eqref{eq:r16k} such that
\begin{align}
\begin{cases}
\frac{2^{2(\sum_{i=1}^K m_i)}}{2^{2(r^*_{12,k})}} = \frac{4^{\sum_{i=1}^K m_i}}{4^{\sum_{i=1}^{K}m_i-n_k}} 
= \frac{1}{4^{-n_k}}, \;\text{when} \; r_{k2,k} = 0 \\
\frac{2^{2(\sum_{i=1}^K m_i)}}{2^{2(m_{1 \rightarrow k-1}+r_{k2,k})}} =\frac{4^{\sum_{i=1}^K m_i}}{4^{\sum_{i=1}^{k-1} m_i+\sum_{j=k}^K m_j-n_k}} 
= \frac{1}{4^{-n_k}} , \; \text{when} \; r_{k2,k} \neq 0 \\
\end{cases}.
\end{align}

Plugging \eqref{eq:sck} into \eqref{eq:krate} results in
\begin{align}\label{eq:rate_k_GBC_final}
    &\frac{1}{n}I(\msf{V}_k;\msf{Y}_k) \geq  r_{k1,k}- \frac{1}{2}\log_2 2\pi e \left(\text{Vol}(\Lambda_k)^{-\frac{2}{n}} + \psi(\Lambda_k)\right) \nonumber \\
    &= r_{k1,k}  
    - \frac{1}{2}\log_2 2\pi e \left(\text{Vol}\left( \gamma_k\sqrt{\overline{\SNR}_k}\beta \Gamma(r_{k2,k})\Lambda\right)^{-\frac{2}{n}} + \psi(\Lambda_k)\right) \nonumber \\
    & > r_{k1,k}  - \frac{1}{2}\log_2 2\pi e \left( \left(2+\frac{4}{|\hat{h}_k|^24^{\bar{n}_k-n_k}} \right)\psi(\Lambda)\right) \nonumber \\
    & > \min\left\{m_k,m_k-(m_k+m_{k+1 \rightarrow K}-n_k)\right\} 
    - \frac{1}{2}\log_2 2\pi e \left(18 \psi(\Lambda)\right) \nonumber \\
    & \overset{(\ref{eq:rate_k_GBC_final}.a)}>\min\left\{m_k,-m_{k+1 \rightarrow K} +\frac{1}{2}\log_2\left(|\hat{h}_k|^2\overline{\SNR}_k\right) \right\}  
    - \frac{1}{2}\log_2 2\pi e \left(18 \psi(\Lambda)\right),
\end{align}
where $(\ref{eq:rate_k_GBC_final}.a)$ follows by using \eqref{eq:d} and $\min\{\cdot\}$ here is due form the constraint in \eqref{eq:r15}.

Given a target transmission rate $R_k$ and the required outage probability $\epsilon_k$ for user $k$, we have
\begin{align}
\epsilon_k &= \mathbb{P}\left\{\frac{1}{n}I(\msf{V}_k;\msf{Y}_k) < R_k\right\} \nonumber \\
 &< \mathbb{P}\left\{-m_{k+1 \rightarrow K} +\frac{1}{2}\log_2(|\hat{h}_k|^2\overline{\SNR}_k) - \Psi < R_k\right\} \nonumber \\
& = \mathbb{P}\left\{|\hat{h}_k|^2 < \frac{2^{2(R_k+\Psi+m_{k+1 \rightarrow K})}}{\overline{\SNR}_k} \right\} \nonumber \\
 &= 1 - \text{exp}\left(-\frac{2^{2(R_k+\Psi+m_{k+1 \rightarrow K})}}{\overline{\SNR}_k} \right).
\end{align}
The lower bound for the outage rate of user $k$ is
\begin{align}
&R_k > 
 \min\left\{m_k,m_{k+1 \rightarrow K} +\frac{1}{2}\log_2\left(-\ln(1-\epsilon_k)\overline{\SNR}_k\right)\right\}-\Psi,
\end{align}
where $\min\{\cdot\}$ here follows \eqref{eq:r15}.

\subsection{Proof of User $K$'s Outage Rate}\label{sec:u2proof}
To further bound $I([2^{r^*_{12,K}}\msf{V}^*_{11,K} +2^{m_{1\rightarrow K-1}+r_{K2,K}}\msf{V}_{K1,K} - \mathbf{d}_5]_{\Lambda_s};\msf{Y}_K)$, we note that the effective noise is $\msf{Z}'_K = h_K\sqrt{\overline{\SNR}_K}\beta[\msf{V}^*_{12,K}+2^{m_{1\rightarrow K-1}}\msf{V}_{K2,K}- \mathbf{d}_6]_{\Lambda_s}+\msf{Z}_K$, where $\mathbf{d}_4$ is a fixed dither decomposed from $\mathbf{d}$ and to minimize the decomposed lattice constellation. We then scale the effective noise by
\begin{align}\label{eq:u2s}
    &\gamma_K= 
    \sqrt{\frac{n}{|h_K|^2\overline{\SNR}_K\beta^2\mbb{E}[\|[\msf{V}^*_{12,K}+2^{m_{1\rightarrow K-1}}\msf{V}_{K2,K}- \mathbf{d}_6]_{\Lambda_s}\|^2]+n}},
\end{align}
such that $\mbb{E}[\|\msf{Z}'_2\|^2]=n$. In this way,

we can then again apply the lower bound of the mutual information between a discrete random input and its noisy version shown in \cite[Lemma 6]{8291591} to obtain
\begin{align}\label{eq:sc4}
    &I([2^{r^*_{12,K}}\msf{V}^*_{11,K} +2^{m_{1\rightarrow K-1}+r_{K2,K}}\msf{V}_{K1,K} - \mathbf{d}_5]_{\Lambda_s};\msf{Y}_K) 
    - H(\msf{V}^*_{11,K})\nonumber \\
     &\geq n r_{K1,K}- \frac{n}{2}\log_2 2\pi e \left(\text{Vol}(\Lambda_K)^{-\frac{2}{n}} + \psi(\Lambda_K)\right),
\end{align}
where $\Lambda_K\defeq \gamma_K \hat{h}_K \sqrt{\overline{\SNR}_K}\beta \Gamma(r_{K2,K})\Lambda$, and
\begin{equation}\label{eq:special_condition}
\Gamma(r_{K2,K}) = \begin{cases}
&2^{r^*_{12,K}}, \;\text{when} \; r_{K2,K} = 0\\
&2^{m_{1\rightarrow K-1}+r_{K2,K}}, \; \text{when} \; r_{K2,K} \neq 0 \\
\end{cases},
\end{equation}
is the scaling factor of the minimum distance of the superimposed lattice
$[2^{r^*_{12,K}}\mc{C}^*_{11,K} +2^{m_{1\rightarrow K-1}+r_{K2,K}}\mc{C}_{K1,K}]_{\Lambda_s}$. The effects of changing of $r_{K2,K}$ on the constellation are illustrated in facts iv) and v) given in Section \ref{Assumption} by recognizing user $K$ as user 2 in the two-user case.

Then, the scaling factor for user $K$ can be bounded by directly following \eqref{eq:sck}:
\begin{align}\label{eq:sc5}
\gamma_K \hat{h}_K&\sqrt{\overline{\SNR}_K}\beta\Gamma(r_{K2,K}) 
> \left(\left(1+\frac{4}{|\hat{h}_K|^24^{\bar{n}_K-n_K}}\right)\text{Vol}(\Lambda)^{\frac{2}{n}}\psi(\Lambda)\right)^{-\frac{1}{2}}.
\end{align}

Plugging \eqref{eq:sc5} into \eqref{eq:sc4} leads to
\begin{align}\label{eq:rate_2_GBC_final}
    &\frac{1}{n}I(\msf{V}_K;\msf{Y}_K) 
    \geq  r_{K1,K}- \frac{1}{2}\log_2 2\pi e \left(\text{Vol}(\Lambda_K)^{-\frac{2}{n}} + \psi(\Lambda_K)\right) \nonumber \\
    & > \min\{m_K,n_K\}  - \frac{1}{2}\log_2 2\pi e \left(18 \psi(\Lambda)\right) \nonumber \\
    & \overset{(\ref{eq:rate_2_GBC_final}.a)}>\min\left\{m_K,\frac{1}{2}\log_2\left(|\hat{h}_K|^2\overline{\SNR}_K\right) \right\} 
    - \frac{1}{2}\log_2 2\pi e \left(18 \psi(\Lambda)\right),
\end{align}
where $(\ref{eq:rate_2_GBC_final}.a)$ follows by applying \eqref{eq:d} and $\min\{\cdot\}$ is due from the constraint in \eqref{eq:r15}.

Given a target transmission rate $R_K$ and the outage probability $\epsilon_K$ for user $K$, we have
\begin{align}
\epsilon_K& =\mathbb{P}\left\{\frac{1}{n}I(\msf{V}_K;\msf{Y}_K) < R_K \right\} \nonumber \\
&< \mathbb{P}\left\{\frac{1}{2}\log_2(|\hat{h}_K|^2\overline{\SNR}_K)
- \Psi < R_K\right\} \nonumber \\
 &= 1 - \text{exp}\left(-\frac{2^{2(R_K+\Psi)}}{\overline{\SNR}_K} \right).
\end{align}
The lower bound for user $K$'s outage rate is
\begin{align}
R_K > \min\left\{m_K,\frac{1}{2}\log_2\left(-\ln(1-\epsilon_K)\overline{\SNR}_K\right)\right\}-\Psi,
\end{align}
where $\min\{\cdot\}$ here follows \eqref{eq:r15}.

\chapter{Proof of Theories of Chapter 7}\label{appendix_3b}

\section{Useful lemmas}\label{appendix:lemma_ch7}
\begin{lemm}\label{lem:zn_P}
Let $\msf{V}$ be a discrete random variable uniformly distributed over the complete set of coset leaders of $\Lambda/2^m\Lambda$ with any $m \in \mathbb{Z}^+$. Let $\mathbf{d} = \E[\msf{V}]$ be a dither vector such that $\msf{X} = \msf{V} - \mathbf{d}$ has zero mean. When $\Lambda = \mathbb{Z}^n$, the average power of $\msf{X}$ is given by
\begin{align}\label{eq:zn_P}
\E [\| \msf{X}\|^2] = \frac{n}{12}(2^{2m}-1)
\end{align}
\end{lemm}

\emph{\quad Proof: }
Let $\msf{A}$ be a random variable that is uniformly distributed over the fundamental Voronoi cell $\mc{V}_0(\Lambda)$ and independent of $\msf{X}$. Then $\msf{X}+\msf{A}$ is a continuous random variable whose distribution is uniform over a region $\mathcal{R}$ and has zero mean. The average power of $\mathcal{R}$ is
\begin{align}
\E[\|\mathcal{R} \|^2]  &= \E[\|\msf{X}\|^2]+\E[\|\msf{A}\|^2] \nonumber \\
 &= \E[\|\msf{X}\|^2]+n\sigma^2(\Lambda),
\end{align}
where the second equality is according to \cite[Eqs (22)-(23)]{1337105}. As $|\Lambda/2^{m}\Lambda| = 2^{nm}$, the region $\mathcal{R}$ consists of $2^{nm}$ numbers of Voronor cells $\mc{V}_{\lambda}(\Lambda)$, where the coset leader is in the cell center.

Let $\msf{D}$ be a random dither that is uniformly distributed over $\mc{V}(2^m\Lambda)$. Then,
\begin{align}\label{eq:app_p_1_ch7}
\E[\|[\msf{V} - \msf{D}]\; \text{mod}\;2^m\Lambda\|^2] &= \E[\|\msf{D}\|^2]    \nonumber \\
&= n\sigma^2(2^m\Lambda).
\end{align}
Here, $[\msf{V} - \msf{D}]\; \text{mod}\;2^m\Lambda$ becomes a continuous random variable that is uniformly distributed over the fundamental Voronoi cell $\mc{V}_0(2^m\Lambda)$. We note that $\text{Vol}(\mc{V}_0(2^m\Lambda)) = 2^{nm}\text{Vol}(\mc{V}_0(\Lambda))$.

Since $\Lambda = \mathbb{Z}^n$, the lattice partition $\mathbb{Z}^n/2^{m}\mathbb{Z}^n$ is the $n$-fold Cartesian product of the one-dimensional lattice partition $\mathbb{Z}/2^{m}\mathbb{Z}$. The fundamental Voronoi cell $\mc{V}_0(2^m\mathbb{Z}^n)$ is the $n$-fold Cartesian product of the fundamental Voronoi cell $\mc{V}_0(2^m\mathbb{Z})$. Given that the fundamental Voronoi cell $\mc{V}_0(\mathbb{Z})$ is $[-\frac{1}{2},\frac{1}{2}]$, thus the region of $\mc{V}_0(2^m\mathbb{Z})$ is $[-\frac{2^m}{2},\frac{2^m}{2}]$. The coset leaders of $\mathbb{Z}^n/2^m\mathbb{Z}^n$ are the $n$-fold Cartesian product of the one-dimensional coset leaders of $\mathbb{Z}/2^m\mathbb{Z}$, i.e., $\{0,\ldots, 2^{m}-1\}$. After subtracting a fixed dither to ensure zero mean, the one-dimensional coset leaders become $\{-\frac{2^m-1}{2},\ldots, \frac{2^m-1}{2}\}$. We note that the \emph{union} of the Voronoi cells of these coset leaders is exactly $[-\frac{2^m-1}{2}-\frac{1}{2},\frac{2^m-1}{2}+\frac{1}{2}]$, same with the fundamental Voronoi cell $\mc{V}_0(2^m\mathbb{Z})$. Hence, the Cartesian products of both the union cells and the fundamental Voronoi cell $\mc{V}_0(2^m\mathbb{Z})$ lead to the same support.
Thus, the random variable uniformly distributed over these regions have the same average power, meaning that $\E[\|\mathcal{R} \|^2] = \E[\|\msf{D}\|^2]$.

As a result, the average power of $\msf{X}$ is therefore
\begin{align}
\E [\| \msf{X}\|^2] &= n(\sigma^2(2^m\Lambda)-\sigma^2(\Lambda)) \nonumber \\
&= n(2^{2m}-1)\text{Vol}(\Lambda)^{\frac{2}{n}}\psi(\Lambda)) \nonumber \\
&= \frac{n}{12}(2^{2m}-1),
\end{align}
where $\text{Vol}(\Lambda) = 1$ and $\psi(\Lambda) = \frac{1}{12}$ for $\Lambda = \mathbb{Z}^n$.
\QEDA

\begin{lemm}\label{lem:dmindp}
For the $n$-dimensional ideal lattice $\Lambda$ constructed via cyclotomic construction, there exists at least a lattice point $\boldsymbol{\lambda} \in \Lambda$ and $\boldsymbol{\lambda} \neq  \mathbf{0} $ satisfying both $d_E(\boldsymbol{\lambda},\mathbf{0}) = d_{E,\min}(\Lambda)$ and $d_{p}(\boldsymbol{\lambda},\mathbf{0}) = d_{p,\min}(\Lambda)$.
\end{lemm}

\emph{\quad Proof: }
We consider the lattice point $\boldsymbol{\lambda}$ generated from a length $n$ integer vector $\mathbf{b} = [0,0,\ldots,1]$. 
Since the generator matrix of $\Lambda$, $\mathbf{G}_{\Lambda} $ is a rotated version of $\mathbf{I}_{n}$, i.e., the rotation matrix itself, it is obvious that $d_E(\boldsymbol{\lambda},\mathbf{0}) = d_E(\mathbf{b},\mathbf{0})= d_{E,\min}(\mathbb{Z}^n) = d_{E,\min}(\Lambda)$ because rotation does not affect the Euclidean distance.

Now, we write the analytical expression for $\boldsymbol{\lambda}$ as
\begin{align}
\boldsymbol{\lambda} &= \mathbf{b} \mathbf{G}_{\Lambda} \nonumber \\
 &\overset{\eqref{G_ideal}}= [0,0,\ldots,1]\cdot\frac{1}{\sqrt{p}}\mathbf{T}\cdot [(\sigma_j(\zeta^i+\zeta^{-i}))_{i,j=1}^n] \cdot \text{diag}(\sqrt{\sigma_1(\varsigma)},\ldots,\sqrt{\sigma_n(\varsigma)}) \nonumber \\
& = [0,0,\ldots,1]\cdot\frac{1}{\sqrt{p}}\cdot [(\sigma_j(\zeta^i+\zeta^{-i}))_{i,j=1}^n] \cdot \text{diag}(\sqrt{\sigma_1(\varsigma)},\ldots,\sqrt{\sigma_n(\varsigma)}) \nonumber \\
& = \frac{1}{\sqrt{p}}[\sqrt{\sigma_1(\varsigma)}\sigma_1(\zeta^n+\zeta^{-n}),\sqrt{\sigma_2(\varsigma)}\sigma_2(\zeta^n+\zeta^{-n}),\ldots,\sqrt{\sigma_n(\varsigma)}\sigma_n(\zeta^n+\zeta^{-n})].
\end{align}
As $\zeta$ is the $p$-th root of unity $e^{\frac{2\pi\sqrt{-1}}{p}}$ of the polynomial
\begin{align}
p(z) &= z^p - 1 \nonumber \\
&=\prod_{j=0}^{p-1}(z - e^{\frac{2j\pi\sqrt{-1}}{p}}),
\end{align}
thus,
\begin{align}
-p(-z) &= z^p + 1  \nonumber \\
&= \prod_{j=1}^{p-1}\left(z+e^{\frac{2j\pi\sqrt{-1}}{p}}\right) \nonumber \\
&= (z+1)\prod_{j=1}^{\frac{p-1}{2}}\left[\left(z+e^{\frac{2j\pi\sqrt{-1}}{p}}\right)\left(z+e^{-\frac{2j\pi\sqrt{-1}}{p}}\right)\right].
\end{align}
Substituting $z=1$ into $-p(-z)$ gives
\begin{align}\label{eq:norm1}
1 &= \prod_{j=1}^{\frac{p-1}{2}}\left[\left(1+e^{\frac{2j\pi\sqrt{-1}}{p}}\right)\left(1+e^{-\frac{2j\pi\sqrt{-1}}{p}}\right)\right] \nonumber \\
&= \prod_{j=1}^{\frac{p-1}{2}}\left(e^{\frac{j\pi\sqrt{-1}}{p}}+e^{-\frac{j\pi\sqrt{-1}}{p}}\right)^2 \nonumber \\
& = \left(\prod_{j=1}^{\frac{p-1}{2}} \left|e^{\frac{j\pi\sqrt{-1}}{p}}+e^{-\frac{j\pi\sqrt{-1}}{p}}\right|\right)^2  \nonumber \\
& \overset{(a)}= \left(\prod_{j=1}^{n} \left|e^{\frac{2jn\pi\sqrt{-1}}{2n+1}}+e^{-\frac{2jn\pi\sqrt{-1}}{2n+1}}\right|\right)^2 \nonumber \\
\Rightarrow\; 1 & = \prod_{j=1}^{n} \left|\zeta^{jn}+\zeta^{-jn}\right|,
\end{align}
where $(a)$ is due to shifting the periodic function $|e^{\frac{j\pi\sqrt{-1}}{p}}+e^{-\frac{j\pi\sqrt{-1}}{p}}|$ to the right by $j\pi$ and substituting $p = 2n+1$. The product distance between $\boldsymbol{\lambda}$ and $\mathbf{0}$ is then computed as
\begin{align}
d_{p}(\boldsymbol{\lambda},\mathbf{0}) & = \prod_{j=1}^n |\lambda_j| \nonumber \\
&= \left|\left(\frac{1}{\sqrt{p}}\right)^n \sqrt{\prod_{j=1}^n \sigma_j(\varsigma)} \prod_{j = 1}^n \sigma_j(\zeta^n+\zeta^{-n})\right| \nonumber \\
& = \left(\frac{1}{\sqrt{p}}\right)^n \sqrt{N(\varsigma)}  \prod_{j = 1}^n |(\zeta^{jn}+\zeta^{-jn})| \nonumber \\
 &\overset{(a)}= \left(\frac{1}{\sqrt{p}}\right)^n \sqrt{p}   \nonumber \\
 &= p^{-\frac{n-1}{2}}  \nonumber \\ &\overset{\eqref{eq:dpcal}}=d_{p,\min}(\Lambda),
\end{align}
where $(a)$ follows that $N(\varsigma) = p$ as this is a necessary condition to obtain the $\mathbb{Z}^n$ ideal lattice \cite[Eq. (7.4)]{Oggier:2004:ANT:1166377.1166378} and $\prod_{j = 1}^n |(\zeta^{jn}+\zeta^{-jn})| = 1$ follows from \eqref{eq:norm1}.
\QEDA

\begin{lemm}\label{the:1a}
Consider any two layers: layer 1 and 2 in $\Lambda$. Let $\mathbf{a}$ and $\mathbf{b}$ be two distinct points on layer 1 and $\mathbf{e}$ and $\mathbf{f}$ be two distinct points on layer 2, such that $d_E(\mathbf{a,b}) = d_E(\mathbf{e,f})$ and $d_p(\mathbf{a,b}) \leq d_p(\mathbf{e,f})$, where $d_E$ and $d_p$ denote the Euclidean distance and product distance, respectively. Then for \emph{any} two distinct points $\mathbf{a}'$ and $\mathbf{b}'$ on layer 1 and \emph{any} two distinct points $\mathbf{e}'$ and $\mathbf{f}'$ on layer 2 satisfying $d_E(\mathbf{a',b'}) = d_E(\mathbf{e',f'})$, the following relation holds:
 $d_p(\mathbf{a',b'}) \leq d_p(\mathbf{e',f'})$.
\end{lemm}

\emph{\quad Proof: }
For layer 1 and 2, we can write the corresponding $n$-dimensional line equations as
$\mathbf{l}_1 = t_1(\mathbf{b}-\mathbf{a})+\mathbf{a}, t_1 \in \mathbb{R}$ and 
$\mathbf{l}_2 = t_2(\mathbf{f}-\mathbf{e})+\mathbf{e},t_2 \in \mathbb{R}$,
respectively. Since points $\mathbf{a}'$ and $\mathbf{b}'$ is in layer 1, we have
$\mathbf{a}' = t_a(\mathbf{b}-\mathbf{a})+\mathbf{a}, t_a \in \mathbb{R}, \; 
\mathbf{b}' = t_b(\mathbf{b}-\mathbf{a})+\mathbf{a},t_b \in \mathbb{R}$.

Similarly, for layer 2 with points $\mathbf{e}'$ and $\mathbf{f}'$ on it, we have
$\mathbf{e}' = t_e(\mathbf{f}-\mathbf{e})+\mathbf{e}, t_e \in \mathbb{R}, \; 
\mathbf{f}' = t_f(\mathbf{f}-\mathbf{e})+\mathbf{e},t_f \in \mathbb{R}$.
The product distance between points $\mathbf{a}$ and $\mathbf{b}'$ is given by
\begin{align}
d_{p}(\mathbf{a',b'}) &= \prod_{i=1}^n|a'_i-b'_i|   \nonumber \\
&= \prod_{i=1}^n|(b_i-a_i)(t_a-t_b)|  \nonumber \\
&= |t_a-t_b|^n\prod_{i=1}^n|b_i-a_i| \nonumber \\
 &= |t_a-t_b|^nd_{p}(\mathbf{a,b}).
\end{align}

Similarly, the product distance point $\mathbf{e}'$ and $\mathbf{f}'$ is given by
\begin{align}
d_{p}(\mathbf{e',f'}) &= \prod_{i=1}^n|e'_i-f'_i|   \nonumber \\
&= \prod_{i=1}^n|(f_i-e_i)(t_e-t_f)|  \nonumber \\
&= |t_e-t_f|^n\prod_{i=1}^n|f_i-e_i| \nonumber \\
 &= |t_e-t_f|^nd_{p}(\mathbf{e,f}).
\end{align}

Since $d_E(\mathbf{a',b'}) = d_E(\mathbf{e',f'})$, we have
\begin{align}\label{eq:dmin_eq}
&\sqrt{\sum_{i=1}^n(a'_i-b'_i)^2} =  \sqrt{\sum_{i=1}^n(e'_i-f'_i)^2}, \nonumber \\
\Rightarrow& \; \sqrt{\sum_{i=1}^n((b_i-a_i)(t_a-t_b))^2} =  \sqrt{\sum_{i=1}^n((f_i-e_i)(t_e-t_f))^2}, \nonumber \\
\Rightarrow&  \;|t_a-t_b|\sqrt{\sum_{i=1}^n(b_i-a_i)^2} =  |t_e-t_f|\sqrt{\sum_{i=1}^n(f_i-e_i)^2} \nonumber \\
\overset{(a)}\Rightarrow& \; |t_a-t_b| =  |t_e-t_f|,
\end{align}
where $(a)$ follows that $d_E(\mathbf{a,b}) = d_E(\mathbf{e,f})$. Since $d_p(\mathbf{a,b}) \leq d_p(\mathbf{e,f})$, and based on \eqref{eq:dmin_eq}, we have
$|t_a-t_b|^nd_{p}(\mathbf{a,b}) \leq |t_e-t_f|^nd_{p}(\mathbf{e,f})$,
which implies that $d_p(\mathbf{a',b'}) \leq d_p(\mathbf{e',f'})$.
\QEDA

\begin{lemm}\label{lem:1}
Let $\mathbf{a}$, $\mathbf{b}$ and $\mathbf{c}$ be three points on a line in $\mathbb{R}^n$. Assume that point $\mathbf{b}$ is located in between points $\mathbf{a}$ and $\mathbf{c}$. Then, the product distances of line segments $\mathbf{ab}$, $\mathbf{bc}$ and $\mathbf{ac}$ satisfy 
\begin{align}
\sqrt[n]{d_{p}(\mathbf{a,c})} = \sqrt[n]{d_{p}(\mathbf{a,b})}+\sqrt[n]{d_{p}(\mathbf{b,c})}.
\end{align}
\end{lemm}

\emph{\quad Proof: }
Let $\mathbf{a} = [a_1,\ldots,a_n]$, $\mathbf{b} = [b_1,\ldots,b_n]$ and $\mathbf{c} = [c_1,\ldots,c_n]$. The equation of the $n$-dimensional line through point $\mathbf{a}$ to point $\mathbf{b}$ is
$\mathbf{l} = t(\mathbf{b}-\mathbf{a})+\mathbf{a},t \in \mathbb{R}$.
Here, the direction of the line is from $\mathbf{a}(t=0)$ to $\mathbf{b}(t=1)$. Since point $\mathbf{c}$ is also on this line, thus point $\mathbf{c}$ satisfies

\begin{align}\label{eq:n_line}
\mathbf{c} = t'(\mathbf{b}-\mathbf{a})+\mathbf{a},t'\in \mathbb{R}.
\end{align}

Since point $\mathbf{b}$ is located in between point $\mathbf{a}$ and $\mathbf{c}$, we have $t'>1$ to ensure that the directions from $\mathbf{a}$ to $\mathbf{b}$ and from $\mathbf{b}$ to $\mathbf{c}$ are the same.
The $n$-th square root of the product distance of line segment $\mathbf{ac}$ is given by
\begin{align}\label{eq:p1}
\sqrt[n]{d_{p}(\mathbf{a,c})} = \sqrt[n]{\prod_{i=1}^n|a_i-c_i|}
\end{align}

The $n$-th square roots of product distances of line segment $\mathbf{ab}$ is
\begin{align}\label{eq:p2}
\sqrt[n]{d_{p}(\mathbf{a,b})} &= \sqrt[n]{\prod_{i=1}^n|a_i-b_i|}   \nonumber \\
&\overset{\eqref{eq:n_line}}
= \sqrt[n]{\prod_{i=1}^n|\frac{a_i-c_i}{t'}|}  \nonumber \\
&= \frac{1}{t'}\sqrt[n]{\prod_{i=1}^n|a_i-c_i|}
\end{align}

We also note that the $n$-th square roots of product distances of line segment $\mathbf{bc}$ is
\begin{align}\label{eq:p3}
\sqrt[n]{d_{p}(\mathbf{b,c})} &= \sqrt[n]{\prod_{i=1}^n|b_i-c_i|}  \nonumber \\
&\overset{\eqref{eq:n_line}}= \sqrt[n]{\prod_{i=1}^n|\frac{c_i-a_i}{t'}+a_i-c_i|} \nonumber \\
&= |\frac{1}{t'}-1|\sqrt[n]{\prod_{i=1}^n|a_i-c_i|}  \nonumber \\
&= (1-\frac{1}{t'})\sqrt[n]{\prod_{i=1}^n|a_i-c_i|}
\end{align}
Noting that \eqref{eq:p2}$+$\eqref{eq:p3}$=$\eqref{eq:p1} completes the proof.
\QEDA

\chapter{Proof of Theories of Chapter 8}

\section{Proof of Lemma \ref{the0}}\label{appendix:0}
In this section, we prove that our decoder can solve any detectable stall pattern with $\min(E,F) \leq 2t+1 $ for any $\varepsilon$.

We first consider a $(E \leq 2t+1, F \geq E)$ stall pattern. According to the condition in \eqref{eq:stall_requirement}, the number of errors in the stall pattern should satisfy $\varepsilon \geq (t+1)F$. This implies that for each erroneous column, the column weight should be at least $t+1$. Thus we have the following property:
\begin{equation}
w(\boldsymbol{\beta}_f)\geq t+1, \;\; f =1,\cdots ,F.
\end{equation}
Performing Step 4 of Algorithm \ref{IBFA_ch8} by flipping all the intersection bits of the stall pattern results in
\begin{equation}
w(\boldsymbol{\beta}_{f^{''}}) = E- w(\boldsymbol{\beta}_f) \leq t, \;\; f'' =1,\cdots,F,
\end{equation}
where $\boldsymbol{\beta}_{f''}$ denotes the $f''$-th erroneous column after the all-flipping operation. As a result, all the erroneous columns can now be decoded by the component code decoder. The proof for any $(E > F, F \leq 2t+1)$ stall pattern can be done by simply swapping $E$ with $F$ in the above argument.

\section{Proof of Theorem \ref{the1}}
For notation simplicity, we define $\mathcal{E} \triangleq \{1,\cdots,E\}$ and $\mathcal{F} \triangleq \{1,\cdots,F\}$.
\subsection{Proof of Theorem \ref{the1}-1}\label{appendix:1_ch8}
In this subsection, we prove that our decoder can solve any detectable $(E=F = 2t+2)$ stall pattern for any $\varepsilon$. The proof consists of two parts in the following.

\subsection{Case 1}\label{sec:the1-1} We first consider a detectable $(E=F=2t+2,   \varepsilon  \leq F(t+1)+t )$ stall pattern. Since $\varepsilon$ could be larger than the minimum number of errors $(t+1) \cdot \max\{E,F\}$, there could have some erroneous columns with weight larger than $t+1$. Specifically, we have
\begin{align}\label{eq:661}
&\mathcal{G}_1 \triangleq \{f_1:w(\boldsymbol{\beta}_{f_1})\geq t+2, f_1 \in \mathcal{F} \}, \; \mathcal{G}_1  \subseteq \mathcal{F}, \nonumber \\
&\mathcal{H}_1 \triangleq \{f_2:w(\boldsymbol{\beta}_{f_2})= t+1, f_2 \in \mathcal{F} \}, \; \mathcal{H}_1 \subseteq \mathcal{F} ,\nonumber \\
 & \mathcal{G}_1 \cup \mathcal{H}_1 = \mathcal{F},  \;  | \mathcal{G}_1  |+| \mathcal{H}_1  | = F,\; | \mathcal{G}_1 | \leq t,
\end{align}
where $|.|$ outputs the set size; $\boldsymbol{\beta}_{f_1}$ and $\boldsymbol{\beta}_{f_2}$ represent the $f_1$-th and $f_2$-th erroneous columns, respectively, when no flipping operations have been performed on this stall pattern yet.
According to Step 3 of Algorithm \ref{IBFA_ch8}, our decoder flips the first erroneous row $\boldsymbol{\alpha}_1$ whose weight satisfies
\begin{equation}\label{eq:a1w}
t+1 \overset{\eqref{eq:stall_requirement}}\leq w(\boldsymbol{\alpha}_1) \leq 2t+1,
\end{equation}
where the maximum weight is due to adding the additional $\varepsilon- (t+1) \cdot \max\{E,F\}$ bits of errors to the minimum weight $t+1$. After the row flipping, there are $w(\boldsymbol{\alpha}_1)$ number of erroneous columns having their weights reduced by 1. Among these erroneous columns, there is at least one column having its weight reduced from $t+1$ to $t$ while there are at most $t$ columns whose weights are larger than $t$ according to \eqref{eq:661}. Thus, we have
\begin{align}\label{eq:662}
|\{f':w(\boldsymbol{\beta}_{f'}) = w(\boldsymbol{\beta}_{f_2})-1 = t,   f'\in \mathcal{H} \}| = w(\boldsymbol{\alpha}_1) - |\mathcal{G}_1| \geq 1,
\end{align}
where $\boldsymbol{\beta}_{f'}$ denotes the $f'$-th erroneous column after the row flipping operation. Since the $f'$-th column has weight $t$, it can now be successfully decoded by the component code decoder. Therefore, at least one column is decodable.

After this step, the problem is reduced to solving a $(E'= 2t+2,F'= 2t+1)$ stall pattern. Based on Lemma \ref{the0}, this stall pattern can be successfully solved by Step 4. Otherwise, when there is no stall pattern formed, the remaining errors can be directly corrected by the iterative hard-decision decoding in Algorithm \ref{TSCDA}.

\subsection{Case 2}\label{sec:the1-2} Next, we consider a detectable $(E=F=2t+2,\varepsilon > F(t+1)+t)$ stall pattern. According to Step 3 of Algorithm \ref{IBFA_ch8}, the decoder again first flips one erroneous row $\boldsymbol{\alpha}_1$ whose weight satisfying \eqref{eq:a1w}.

If $w(\boldsymbol{\alpha}_1) \geq t+2$, the flipped row can be decoded because the number of the remaining errors after row flipping is $F - w(\boldsymbol{\alpha}_1) \leq t$. For this case, the problem is reduced to solving the stall pattern of $(E'=2t+1,F'=2t+2)$ and thus it can be solved by applying all-flipping in Step 4 of Algorithm \ref{IBFA_ch8} according to Lemma \ref{the0}.

Now consider the case such that $w(\boldsymbol{\alpha}_1) = t+1$ and none of the erroneous columns can be decoded after Step 3 of Algorithm \ref{IBFA_ch8}. Under this condition, the stall pattern has the property such that
\begin{align}\label{eq:prop1}
&\mathcal{G}_2 \triangleq \{f_1: w(\boldsymbol{\beta}_{f_1}) \geq t+2, f_1 \in \mathcal{F} \} , \;\mathcal{G}_2  \subseteq \mathcal{F}, \nonumber \\
&\mathcal{H}_2 \triangleq \{f_2: w(\boldsymbol{\beta}_{f_2}) \geq t+1, f_2 \in \mathcal{F}\} ,\; \mathcal{H}_2 \subseteq \mathcal{F}, \nonumber \\
& \mathcal{G}_2 \cup \mathcal{H}_2 = \mathcal{F},  \;  | \mathcal{G}_2  |+| \mathcal{H}_2  | = F, \; |\mathcal{G}_2|= w(\boldsymbol{\alpha}_1).
\end{align}
This indicates that after flipping the first row, all the column weights are still larger than $t+1$ and thus cannot be decoded by the component code decoder.
The decoder then restores all the code blocks and perform all-flipping operation in Step 4 of Algorithm \ref{IBFA_ch8}. This results in
\begin{align}\label{eq:prop2}
&|\{f'':w(\boldsymbol{\beta}_{f''}) = E -w(\boldsymbol{\beta}_{f_1}) \leq (2t+2) - (t+2) = t,  f'' \in \mathcal{F}\}| \nonumber \\
&\geq |\mathcal{G}_2|=t+1,
\end{align}
where $\boldsymbol{\beta}_{f''}$ denotes the $f''$-th erroneous column after the all-flipping operation. Since the $f''$-th column has maximum weight $t$, it can now be successfully decoded by the component code decoder. Therefore, at least $t+1$ columns are decodable. Then, the problem is reduced to solving the stall pattern of $(E'' = 2t+2, F'' =t+1)$. The decoder can successfully solve this stall pattern by repeating Step 4 of Algorithm \ref{IBFA_ch8} according to Lemma \ref{the0}.

Therefore, our decoder is able to correct any detectable $(E=F=2t+2)$ stall pattern with any $\varepsilon$.

\subsection{Proof of Theorem \ref{the1}-2}\label{appendix:2_ch8}
First, we prove that our decoder can solve any detectable $(E = 2t+2, F = 2t+3)$ stall pattern for any $\varepsilon$.

Since $F>E$, the decoder first flips the first erroneous row. If the flipped row can be solved by the component code decoder, the stall pattern becomes $(E' = 2t+1,F' = 2t+3)$ and thus can be corrected after Step 4 of Algorithm \ref{IBFA_ch8} according Lemma \ref{the0}. If at least one erroneous column can be decoded by the iterative hard-decision decoding, the stall pattern becomes $(E' = F' = 2t+2)$ which can be solved according to Theorem \ref{the1}-1.

Now we consider the case such that none of the erroneous rows and columns are corrected after Step 3 of Algorithm \ref{IBFA_ch8}. In other words, the row flipping does not result in any decodable rows and columns. The stall pattern thus has the property shown in \eqref{eq:prop1} with $F=2t+3$ and
\begin{equation}
t+1 \overset{\eqref{eq:stall_requirement}}\leq w(\boldsymbol{\alpha}_1) \leq  t+2.
\end{equation}
The decoder then restores all the code blocks and then proceeds to Step 4 of Algorithm \ref{IBFA_ch8}. Similar to the case in Appendix \ref{sec:the1-2}, this leads to \eqref{eq:prop2} where $F=2t+3$ after performing Step 4 of Algorithm \ref{IBFA_ch8}. As a result, at least $t+1$ erroneous columns become decodable. The problem then becomes solving a $(E'' = 2t+2, F'' = t+2)$ stall pattern. Here, we note that $F''=t+2 \leq 2t+1$ for all component codes with $t \geq 1$. Thus, the resultant stall pattern can be corrected by repeating Step 4 of Algorithm \ref{IBFA_ch8} according to Lemma \ref{the0}.

The proof for $(E = 2t+3, F = 2t+2)$ stall patterns can be done by swapping $E$ with $F$ in the above argument.

\subsection{Proof of Theorem \ref{the1}-3}\label{appendix:3_ch8}
In this subsection, we prove that our decoder can solve detectable $(E=F=2t+3)$ stall patterns for some $\varepsilon$. The proof consists of three parts in the following.

\subsection{Case 1} We first consider a $(E=F=2t+3,\varepsilon  \leq F(t+1)+t )$ stall pattern. For this case, we have the properties shown in \eqref{eq:661}-\eqref{eq:662} with $E=F=2t+3$. Similar to the case in Appendix \ref{sec:the1-1}, the same conclusion can be drawn such that there is at least one erroneous column can be decoded. Thus the stall pattern becomes a $(E'=2t+3,F' = 2t+2)$ stall pattern which can be corrected according to Theorem \ref{the1}-2.

\subsection{Case 2} We now consider a detectable $(E = F=2t+3,\varepsilon \geq F(t+2)+1)$ stall pattern. This stall pattern has the following property:
\begin{align}\label{eq:max_we1}
&\max(w(\boldsymbol{\alpha}_e)) \geq t+3 >\frac{\varepsilon}{F} ,  \; \max(w(\boldsymbol{\beta}_f))  \geq t+3>\frac{\varepsilon}{E},\nonumber \\
& |\{e: w(\boldsymbol{\alpha}_e) \geq t+3, e \in \mathcal{E},\}|\geq 1, \nonumber \\
&|\{f: w(\boldsymbol{\beta}_f) \geq t+3, f \in \mathcal{F},\}| \geq 1.
\end{align}
Here, the maximum row/column weight is strictly larger than $t+2$ because the average row/column weight is $\frac{\varepsilon}{F}=\frac{\varepsilon}{E}>t+2$. If the flipped erroneous row has weight $w(\boldsymbol{\alpha}_1)\geq t+3$, then its weight becomes $F - w(\boldsymbol{\alpha}_1) \leq  t$ after row flipping. This row can now be corrected by the component code decoder. The stall pattern then becomes a $(E' = 2t+2',F'=2t+3)$ stall pattern which can be solved according to Theorem \ref{the1}-2.

If none of the erroneous columns and rows are corrected after Step 3 of Algorithm \ref{IBFA_ch8}, the decoder again restores all the code blocks and proceeds to Step 4 of Algorithm \ref{IBFA_ch8}. As a result, any erroneous row and column whose weights were greater than $t+3$ before all-flipping now have weight $2t+3 - (t+3) = t$ after all-flipping. Thus, at least one row and one column whose weights were larger than $t+3$ \eqref{eq:max_we1} can be decoded after Step 4 of Algorithm \ref{IBFA_ch8}. The problem is reduced to solving a $(E'' =F''=2t+2)$ stall pattern which can be corrected based on Theorem \ref{the1}-2.

\subsection{Case 3}\label{app:the1part3} Now consider a detectable $(E = F=2t+3)$ stall pattern for some $\varepsilon$ satisfying \eqref{eq:stall_requirement}. This stall pattern has the following property:
\begin{align}\label{eq:a}
&\mathcal{G}_3 \triangleq \{f_1: w(\boldsymbol{\beta}_{f_1}) = t+2, f_1 \in \mathcal{F} \} , \;\mathcal{G}_3  \subseteq \mathcal{F}, \nonumber \\
&\mathcal{H}_3 \triangleq \{f_2: w(\boldsymbol{\beta}_{f_2}) = t+1, f_2 \in \mathcal{F}\},\; \mathcal{H}_3 \subseteq \mathcal{F} ,  \nonumber \\
&\mathcal{G}_3 \cup \mathcal{H}_3 = \mathcal{F},  \;  | \mathcal{G}_3  |+| \mathcal{H}_3  | = F.
\end{align}
If none of the erroneous rows and columns are decoded after Step 3 of Algorithm \ref{IBFA_ch8}, then the stall pattern satisfies the following condition
\begin{align}\label{eq:p31a}
 t+1 \leq |\mathcal{G}_3| \leq F,\; |\mathcal{G}_3|\geq w(\boldsymbol{\alpha}_1).
\end{align}
This happens when the row flipping operation corrects 1 bit error in the erroneous column whose weight was $t+2$. After Step 4 of Algorithm \ref{IBFA_ch8}, the column weights of the stall pattern are
\begin{align}
w(\boldsymbol{\beta}_{f_1})= t+1, \; w(\boldsymbol{\beta}_{f_2})= t+2.
\end{align}
No matter how many times of Step 3 and Step 4 of Algorithm \ref{IBFA_ch8} are performed on this stall pattern, none of the erroneous rows and columns are decodable by the component code decoder. For this case, the number of error bits in this stall pattern can be calculated by
\begin{align}\label{eq:p3_ch8}
\varepsilon &= w(\boldsymbol{\beta}_{f_1})|\mathcal{G}_3|+w(\boldsymbol{\beta}_{f_2})| \mathcal{H}_3 | \nonumber \\
 &\overset{\eqref{eq:a}}= w(\boldsymbol{\beta}_{f_2})F+(w(\boldsymbol{\beta}_{f_1})-w(\boldsymbol{\beta}_{f_2}))| \mathcal{G}_3 | \nonumber \\
 &\overset{\eqref{eq:a}}=(t+1)F+|\mathcal{G}_3 |.
\end{align}
Substituting \eqref{eq:p31a} into \eqref{eq:p3_ch8} results in $F(t+1)+t+1 \leq \varepsilon  \leq F(t+1)+F$. Therefore, our decoder can only guarantee to solve the stall pattern with $(E=F=2t+3)$ and $\varepsilon \leq F(t+1)+t$ or $\varepsilon \geq F(t+2)+1$.

\subsection{Proof of Theorem \ref{the1}-4}\label{appendix:4_ch8}
First, we prove that our decoder can solve any detectable $(E = 2t+2, F = 2t+4)$ stall pattern for any $\varepsilon$.

Since $F>E$, the decoder first flips the first erroneous row. If at least one erroneous row can be decoded, then the stall pattern becomes $(E' = 2t+1,F ' = 2t+4)$ and thus can be corrected by using Lemma \ref{the0}. If at least one erroneous column can be decoded, the stall pattern becomes $(E' = 2t+2, F' = 2t+3)$ which can be solved according to Theorem \ref{the1}-2.

Now we consider the case such that none of the erroneous rows and columns associated with the stall patterns are corrected after Step 3 of Algorithm \ref{IBFA_ch8}. In this case, the stall pattern has the property shown in in \eqref{eq:prop1} with $F = 2t+4$ and the flipped row whose weight satisfies
\begin{equation}\label{eq:the14}
t+1 \overset{\eqref{eq:stall_requirement}}\leq w(\boldsymbol{\alpha}_1) \leq  t+3.
\end{equation}
Similar to the case in Appendix \ref{appendix:2}, performing Step 4 of Algorithm \ref{IBFA_ch8} leads to \eqref{eq:prop2} where $F=2t+4$. Same conclusion can be drawn such that there are at least $t+1$ erroneous columns can be decoded after Step 4 of Algorithm \ref{IBFA_ch8}. The problem is then reduced to solving a $(E'' = 2t+3, F'' = t+3)$ stall pattern. Here, we note that $F'' \leq 2t+2$ for all component codes with $t \geq 1$. Thus, the resultant stall pattern can be solved according to Theorem \ref{the1}-1.

The proof for $(E = 2t+4, F = 2t+2)$ stall patterns can be done by swapping $E$ with $F$ in the above argument.

\subsection{Proof of Theorem \ref{the1}-5}\label{appendix:5}
In this subsection, we first prove that our decoder can solve any detectable $(2t+3 \leq E \leq 2t+4,F = 2t+4, \varepsilon \leq F(t+1)+1)$ stall pattern. We first note that this stall has the following property
\begin{align}\label{eq:881}
&\mathcal{G}_4 \triangleq \{f_1:w(\boldsymbol{\beta}_{f_1}) \leq t+2, f_1 \in \mathcal{F} \}, \; \mathcal{G}_4  \subseteq \mathcal{F}, \nonumber \\
&\mathcal{H}_4 \triangleq \{f_2:w(\boldsymbol{\beta}_{f_2})= t+1, f_2 \in \mathcal{F} \}, \; \mathcal{H}_4 \subseteq \mathcal{F} ,\nonumber \\
& \mathcal{G}_4 \cup \mathcal{H}_4 = \mathcal{F},  \;  | \mathcal{G}_4  |+| \mathcal{H}_4  | = F,\; | \mathcal{G}_4 | \leq 1.
\end{align}
After applying the row flipping operation in Step 3 of Algorithm \ref{IBFA_ch8} to this stall pattern, we have
\begin{align}\label{eq:882}
|\{f':w(\boldsymbol{\beta}_{f'}) = t, f' \in \mathcal{H}_4 \}| = w(\boldsymbol{\alpha}_1)-|\mathcal{G}_4| \geq t.
\end{align}
It can be seen that at least $t$ columns with weight $t$ can be successfully decoded after Step 3. The problem is then reduced to a $(2t+3 \leq E' \leq 2t+4,F' =t+4)$ stall pattern. We note that $F' \leq 2t+2$ for $t\geq 2$. Thus, the resultant stall pattern can be successfully solved according to Theorem \ref{the1}-2 and Theorem \ref{the1}-4 when the underlying component code has $t\geq2$. Note that for $t=1$, the stall pattern is $(5 \leq E' \leq 6,F' =5)$ which may not be correctable according to Appendix \ref{app:the1part3}.

For the stall pattern with $(2t+3 \leq E \leq 2t+4,F = 2t+4, \varepsilon = F(t+1)+2)$, it has the property shown in \eqref{eq:661} with $ |\mathcal{G}_1| \leq 2$. According to \eqref{eq:662}, the number of decodable columns is $w(\boldsymbol{\alpha}_1) - |\mathcal{G}_1| \geq t-1$ after Step 3 of Algorithm \ref{IBFA_ch8}. The problem is then reduced to solving a $(2t+3 \leq E' \leq 2t+4,F' =t+5)$ stall pattern. Here, we note that $F' = 2t+3$ when $t = 2$. According to Appendix \ref{app:the1part3}, whether this stall pattern can be successfully decoded depends on $\varepsilon$. Since the knowledge of $\varepsilon$ is not available to the decoder, this stall pattern may not be successfully decoded. However, we also note that $F' \leq 2t+2$ when $t \geq 3$. This implies that the stall pattern can be solved according to Theorem \ref{the1}-2 and Theorem \ref{the1}-4 if one choose the component code has $t \geq 3$. Therefore, we only claim that our decoder is able to solve any detectable $(2t+3 \leq E \leq 2t+4,F = 2t+4, \varepsilon \leq F(t+1)+1)$ stall pattern when the underlying component code has $t \geq 2$.

The proof for $(E=2t+4,2t+3 \leq F \leq 2t+4, \varepsilon \leq F(t+1)+1)$ stall pattern can be done by swapping $E$ with $F$ in the above argument.

\clearpage{\pagestyle{empty}\cleardoublepage}

\renewcommand{\bibname}{Bibliography}
\addcontentsline{toc}{chapter}{\protect\numberline{}{Bibliography}}
\singlespacing
{\bibliographystyle{IEEEtran}
\bibliography{MinQiu}
}

\end{document}